\documentclass[12pt]{article}
\usepackage{graphicx} 
\usepackage[utf8]{inputenc}
\usepackage{float}
\usepackage{geometry}
\usepackage{csvsimple}
\usepackage{booktabs}
\usepackage{subcaption}
\usepackage{longtable}
\usepackage{caption}
\usepackage{afterpage}
\usepackage{authblk}
\usepackage{mwe}
\usepackage{adjustbox}
\usepackage{multirow}
\usepackage[bottom]{footmisc}
\usepackage{subfiles}
\usepackage{xurl}
\usepackage{setspace}
\usepackage{longtable}
\usepackage{xcolor}
\DeclareRobustCommand{\bbone}{\text{\usefont{U}{bbold}{m}{n}1}} 

\usepackage{dcolumn}
\usepackage{amsmath}

\usepackage[authoryear, round]{natbib}

\usepackage{geometry}
\title{Breaking New Ground, Reinforcing Old Gaps: Gender Disparities in Access to Emerging Research Frontiers}

\author[a,1]{Carolina Biliotti\thanks{Corresponding Author. E-mail: carolina.biliotti@imtlucca.it.}}
\author[b]{Luca Verginer}
\author[a,c]{Massimo Riccaboni}

\affil[a]{AXES, IMT School for Advanced Studies, Lucca, Italy}
\affil[c]{IUSS, Pavia, Italy}
\affil[b]{Chair of Systems Design at ETH Zurich}

\begin{document}
\maketitle
\begin{abstract} 
 
\noindent
This study exploits COVID-19 as an exogenous shock in biomedical research to show how the emergence of an unexpected new research topic exacerbates gender bias in key authorship positions of scientific publications relevant to new research topics (e.g. Vaccines, Epidemiology). We determine author's gender based on the names listed on their scientific publications and analyze the changes in the composition of the scientific teams after the COVID-19 outbreak. Using a Difference-in-Differences approach, we find that although the share of female authorship has increased overall, women are less likely to be first or last authors (the most prestigious positions) on COVID-19-related research papers and more likely to be found in middle author positions. Stay-at-home mandates, the journal importance and funding opportunities do not fully account for the decline of women in key author positions. The main difference in first authorship is due to the composition of the team and the experience of the lead authors in COVID-19 related research. First authorship by women declined after teams of novices emerged, where lead authors have no prior experience in COVID-related research. Discretionality in first-author appointments for newcomers, combined with high pressure to publish quickly, may have led to discriminatory biases. Conversely, there may also be differences in risk-taking attitudes in doing research in unfamiliar domains. Monitoring gender inequality in scientific production is crucial for reducing gender inequalities and for implementing timely policies that ensure equal access to emerging research topics. \\ \\
Gender Bias $|$ COVID-19 $|$ Scientific Careers 
\end{abstract}

\newpage

\section*{Introduction}

Groundbreaking scientific advancements—such as the development of CRISPR/Cas9 gene-editing technology, mRNA vaccines, and the discovery of the SARS-CoV-2 virus—offer researchers unique opportunities to advance their careers by publishing impactful findings on these emerging topics. Early engagement in such groundbreaking research is critical, as academic careers are cumulative; future success and leadership roles often hinge on the number and impact of publications in top peer-reviewed journals \citep{petersen2012,Petersen2014InequalityAC}.

Despite significant increases in women's participation in science, gender-specific disparities persist. Gender bias in science refers to systemic prejudices that unfairly restrict or enhance access to opportunities based on gender \citep{LLORENS20212047, PHILBIN2024116456}. For instance, compared to their male counterparts, female scientists often have lower research impact and face more challenges in attaining leadership positions \citep{Huang2019HistoricalCO, 10.1371/journal.pbio.2004956}, despite no discernible differences in the quality of their work \citep{10.1093/ej/ueac032}. Social identities and demographics significantly influence research content and output \citep{RePEc:nat:nathum:v:1:y:2017:i:11:d:10.1038_s41562-017-0235-x}, as well as innovation uptake \citep{doi:10.1073/pnas.1915378117}, incurring costs in career advancement \citep{doi:10.1073/pnas.2113067119} and access to funding \citep{PHILBIN2024116456}. Critically, women are less likely to hold key authorship positions, such as first or last author, which are crucial for career progression \citep{Lerchenmuellerl6573, Bendels2017GenderEI}. These barriers not only impede career advancement but also hinder the promotion of academic diversity \citep{woolston}. A subtler issue arises when temporary barriers selectively restrict women's access to new research areas during the early stages of scientific exploration. Missing these opportunities means being absent from foundational literature, losing future citations, and forfeiting the chance to contribute to fundamental discoveries in areas of significant scientific and societal interest.

This study investigates why women's scientific contributions in COVID-19 research---an emerging high-impact field \citep{10.1371/journal.pone.0263001}---have been disproportionately affected during the pandemic compared to other research areas. We assess how these new research opportunities have influenced the competitiveness of women scientists in biomedical research, focusing on their ability to secure key authorship roles, i.e., first and last authorships.
 
The COVID-19 pandemic provides an excellent opportunity to examine the marginalization of women in scientific production, especially in new research fields. Although medical research had one of the smallest gender gaps \citep{10.7554/eLife.85427}, the pandemic has affected equal access to publication opportunities. During the crisis, research related to COVID-19 experienced an unprecedented surge in public and scientific interest \citep{10.1371/journal.pone.0263001}.
This intense focus and competition for early prominence compelled researchers to publish quickly to maximize attention and potential future impact.
This environment attracted scientists from diverse backgrounds, including authors with no prior experience on COVID-related topics \citep{sikdar2024}, whom we refer to as newcomers \citep{guimera2005}. In particular, COVID-19 may have attracted an upsurge of \emph{opportunistic teams} where both lead authors are newcomers. Conversely, these teams may be more likely to include men, widening the gender gap in access to new, emerging topics. After accounting for factors such as lockdown stringency and other potential confounding factors, COVID-19 proves to be a stress test for the gender gap in authorship on emerging topics at early stages of scientific discovery.

Our analytical framework is summarized in Figure 1. We focus on the earliest literature that rises following the advent of a new research topic, i.e. publications in 2020, comparing them with the year prior. Focusing on the very first year of publications following the new research topic allows us to capture gender inequalities when the race to publishing is most competitive --- because of the high interest towards the topic --- and most rewarding in terms of future citations and impact.

We collect published papers from PubMed from 2019 and 2020. Papers on PubMed serve as a compelling case for studying the effect of COVID-19 as a new publishing opportunity on female authorship, as the \emph{first} and \emph{last} represent the key authorship positions for future career progression in biomedical research. Each paper in PubMed is assigned a Medical Subject Heading (MeSH) indicating its research topic. We use the `major MeSH term' to determine the primary research topic of a paper.

To measure the `relatedness' of a paper to COVID-19, we calculate the conditional probability that a paper listing a major MeSH term also includes a COVID-19-related MeSH term. These terms include 'COVID-19,' 'SARS-CoV-2,' 'COVID-19 Vaccines'. We categorize papers into two groups: closely related to COVID-19 (top 10\% of our sample based on relatedness) or unrelated (bottom 10\%). Papers on vaccines and epidemiology are closely related, while publications on soil microbiology and cell movement are unrelated. We omit from the sample papers with major MeSH term given by a COVID-19 MeSH. 
Crucially, we exclude all other papers, which are dealing with MeSH terms that are only \emph{partially related} to COVID-19, to prevent bias from increased public interest in COVID-19. This design minimizes the likelihood that MeSH terms with inherently higher or lower potential outcomes are selectively included in the COVID-related category. Consequently, we leverage COVID-19 as an unforeseen, exogenous shock among COVID-related publications, to estimate the average effect of new publishing opportunities on female scientists' competitiveness in terms of authorship position in published papers in bio-medical research fields. 

This study uses the binary categories of 'woman' and 'man' to identify gender disparities. We acknowledge that gender identity is more complex and varied than these two categories suggest. Our approach is driven by the limitations of the available data and methods to infer gender from names, rather than by an intention to oversimplify gender identity. Our approach does not negate the spectrum of gender identities. We use the \emph{Genderize.io} API to assign genders to authors based on listed first names.

To examine the role of past research experience of authors on the change in women's access to key authorship in COVID-related papers, we collect published papers from 2015 to 2020 for disambiguated authors in our PubMed sample using the OpenAlex API \citep{priem2022openalex}. 

As depicted in the four quadrants of Figure 1, we aim to assess whether, after the advent of COVID-19 in 2020, the decline in women as key authors of biomedical publications occurred only in research highly related to COVID-19, and whether this decline is rooted in the surge of teams composed of newcomers, primarily featuring male scientists as lead authors.\footnote{The data and codes are available at Biliotti, C., Verginer, L., \& RICCABONI, M. (2025). Breaking New Ground, Reinforcing Old Gaps: Gender Disparities in Access to Emerging Research Frontiers. Zenodo. \url{https://doi.org/10.5281/zenodo.13839340}}

Our analysis, using a Difference-in-Differences framework, reveals a significant decline in the likelihood of women holding key authorship positions in COVID-19-related research topics. Our results suggest that women scientists are sidelined for key authorship positions on these emerging topics, often relegated to middle authorship roles that offer fewer long-term career benefits. This is despite women's increasing participation in other research areas.

We find evidence of a discriminatory effect on access to key authorship positions within emerging research topics. Women continued to publish as key authors in new research topics, i.e., areas different from their prior experience—during 2020, but only in fields not related to COVID-19. Thus, women remained able to be first authors when changing research topics during 2020, but not in the emerging COVID-19-related topics.

The decline in female first authorship is linked to the formation of these "opportunistic" research teams that lack prior experience in COVID-related studies, but organize nevertheless around these topics, in order to seize the new publishing opportunity to capitalize on the high scientific and public interest. These teams have more flexibility in selecting the first author since they lack experience and consolidated roles in COVID-related research. This occurs as teams face pressure to publish quickly, and project leaders must rapidly assemble teams under challenging conditions. This flexibility in appointing key authors, coupled with the urgency to publish quickly, could introduce discriminatory biases, additional to the distorted evaluation of merits that women normally face in academic recruitment and hiring \citep{doi:10.1073/pnas.1211286109}. 

On the other hand, women could have intentionally avoided these opportunistic teams by preferring other teams associated with less risky outcomes, such as teams with incumbent leaders. Last authors endowed with prior experience in COVID-related research tend to give first authorship to women who lack prior experience. This is especially true if the last author has already published an article with the new-coming woman. Unequal access to key authorship positions in emerging and prominent research topics has the potential to significantly hamper career prospects.

Overall, our findings indicate discriminatory biases against female authors entering new, COVID-related research topics without prior experience, particularly in securing first authorship positions.
This is driven by team composition and especially the team's key authors' research portfolio.
As more scientific work builds on and cites these early studies, this initial exclusion contributes to the long-standing gender gap in scientific production and academic rankings.

Our study contributes to two streams of literature. First, it adds to research on gender bias in academia and the barriers women face in scientific careers. Factors beyond increased family responsibilities influence women's ability to secure key authorship positions when new research opportunities arise. These barriers include biases in access to prestigious journals \citep{Filardoi847,10.7554/eLife.85427, Helmer2017GenderBI}, social capital and collaboration networks \citep{10.1371/journal.pbio.3001771, 10.7554/eLife.85427}, team size and composition \citep{LIU2022101295}, access to funding \citep{PMID:35177858}, and the peer-review process \citep{Helmer2017GenderBI}, as well as common practices in science and innovation \citep{PMID:32433639, doi:10.1073/pnas.2113067119, King2020ThePP, 10.1093/ej/ueac032, AmanoPatio2020TheUE}. In a stress-test scenario prompted by the emergence of a new, high-interest topic, we observe that new teams of non-experts were less likely to include women as first authors. We identify a discriminatory effect specifically tied to the rise of new teams focusing on emerging topics. This exclusion exacerbates women's underrepresentation in foundational literature and leads to a significant loss of future citations and opportunities to contribute to fundamental discoveries.

Second, we contribute to the literature on the impact of COVID-19 on the gender gap in academia. Prior research has used Difference-in-Differences approaches to estimate the differential impact of COVID-19 on publishing rates by gender across domains, including basic medicine, biology, chemistry, and clinical medicine \citep{10.7554/eLife.76559}. It has also been used to identify the causal impact of lockdown measures on women economists \citep{Chinetti2021AcademicPA}. Several studies report descriptive evidence on the under-representation of women in key author positions of papers published during COVID-19 including pre-prints in arXiv and bioRxiv \citep{King2020ThePP}, different countries \citep{Lerchenmullere045176, MADHIVANAN202280, RYAN2023115761}, journals \citep{10.7554/eLife.58807, Muri2020COVID19AG}, and research fields \citep{Madhivanan2022WomenRI, 10.7554/eLife.45374}. This findings are consistent with our results.
Studies report fewer journal submissions \citep{10.1371/journal.pone.0257919} and publications with women in key authorship positions \citep{Muri2020COVID19AG, 10.7554/eLife.58807, Lerchenmullere045176}.
Increased childcare responsibilities during lockdown measures have been explored as potential drivers for the widening gender gap in scientific research \citep{10.3389/fpsyg.2021.663252, 10.3389/fpubh.2022.818594, https://doi.org/10.1111/gwao.12696, doi:10.1287/msom.2021.0991, https://doi.org/10.1111/gwao.12493}.
Surveys have consistently shown a decrease in the rate at which women scientists initiate new projects \citep{Gao2021PotentiallyLE}, as well as a reduction in the time they dedicate to research when children are present \citep{NBERw28360, 10.1038/s41562-020-0921-y, doi:10.1089/jwh.2020.8710, 10.1016/j.annepidem.2022.08.033}. Despite the uneven impact of the COVID-19 pandemic across genders, the increase in family responsibilities due to stay-at-home orders do not fully explain the decline in female scientific productivity. In particular, the gender gap in publications on COVID-19 exclusively was even more pronounced than the gap in overall scientific publications during the early phases of COVID-19 \citep{Gabster2020ChallengesFT, 10.7554/eLife.58807, Muri2020COVID19AG, AmanoPatio2020TheUE, Lerchenmullere045176}. Yet, there is no clear evidence on how an emerging research topic with heightened public and scientific interest, such as COVID-19, impacts gender bias in publishing beyond childcare duties. Our approach allows us to view COVID-19 as an exogenous shock, isolating the effect of a new scientific opportunity from variations in childcare and family duties due to COVID-19 restrictions. Even after controlling for journal importance, access to funding, increased family duties and matching, the findings remain the same. 

\begin{figure}[ht]
    \centering
    \includegraphics[width=.75\linewidth]{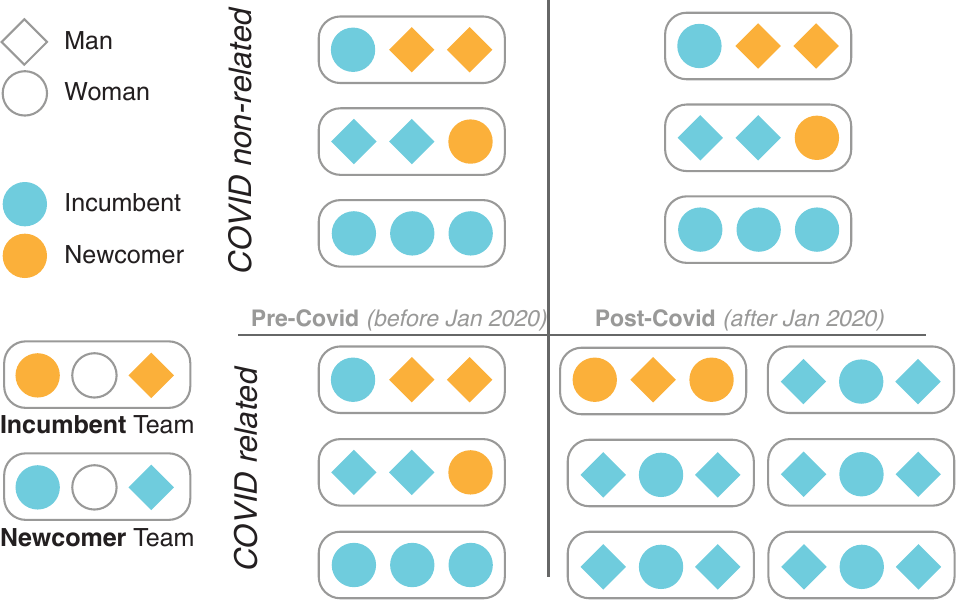}
    \caption{In a newcomer (incumbent) team, the first and last authors are newcomers (incumbent).
The gender gap is similar before 2020. After January 2020, there is a surge in newcomer teams with male lead authors among COVID-related publication.}
    \label{fig:fig_1}
\end{figure}


\section*{Data Collection}

Our dataset consists of repeated cross-sections of 772\,603 published bio-medical papers between 2018 and 2021 uploaded on PubMed. 
PubMed repository with more than 34 million biomedical citations provides biomedical publications where first and last authors are the key authorship positions, i.e. the most relevant for future career progression \citep{Bendels2017GenderEI}. 

We classify the articles into 16\,710 unique research topics based on the Medical Subject Heading (MeSH) terminology. MeSH is a controlled vocabulary used to index PubMed articles. Trained personnel assigns each paper to a list of related research topics, indexed by MeSH terms, indicating at least one major MeSH term, i.e. the research topic that identifies best the article.

MeSH terms are related in different ways to COVID-19. Research topics such as `Visitors to Patients', `Virus Shedding', `Videoconferencing', `Vaccines' or `Social isolation' were highly affected by the advent of COVID-19, while other scientific areas like `Buthanol', `Abdominal Fat' or `Brain mapping' are hardly related to COVID-19 and presumably did not experience any change in the number of publications due to the new emerging topic. Following this idea, we compute a measure of relatedness to COVID-19 of each major MeSH term as the conditional probability that a paper related to major MeSH term \emph{i} features either one of the MeSH terms pertaining to COVID-19 (`COVID-19', `SARS-CoV-2', `COVID-19 Vaccines', `COVID-19 Nucleic Acid Testing', `COVID-19 Serological Testing'):
\begin{equation*}\label{}
 \text{relatedness}_{i} = P(\text{COVID-19} | \text{MeSH}_{i})  = \frac{N(\text{MeSH}_{i} \cap \text{COVID-19})}{ N(\text{MeSH}_{i})}
\end{equation*}
where the numerator is the number of papers featuring both major MeSH term \emph{i} and a COVID-19 term, while the denominator indicates the total count of papers with major MeSH term \emph{i}.
After finding $relatedness_{i}$ for each major MeSH term, we compute the percentiles of the sample distribution of these relatedness values, such that each research topic is associated with a `COVID-relatedness' percentile. For papers identified by more than one major MeSH, we keep only the one mostly associated with COVID-19, i.e. the Mesh with higher relatedness. 

In order to exclude from the analysis articles in topics that are only \emph{partially} related to the new emerging topic, we remove from the dataset all papers with major MeSH terms falling between the 90th and the 10th percentiles of the COVID-relatedness distribution (extremes excluded).

For each paper, we observe authors' names, affiliations and countries, whether the research benefited from any grant - granted pre or post-COVID (\emph{Has New Grant} and \emph{Pre-existing Grant}), featured any clinical trial (\emph{trial}), the Journal Impact factor (\emph{JI}). We also identified the country of the majority of the team members. Section A of the Appendix reports the descriptive statistics of the sample.

We use \emph{Genderize.io} official API to identify the gender of authors in the features papers. \emph{Genderize.io} correctly assigns names to either \emph{male} or \emph{female} status with probability higher than 90\%.


Using Oxford Coronavirus Government Response Tracker (OxCGRT) we create country level monthly indicators of stringency of measures adopted against COVID-19 contagion \citep{owidcoronavirus}. See Section C.6 of the Appendix for more details.

Because we are interested in the number of scientific publications with the first or last female author within some area of research, we drop from the dataset those papers that feature less than three authors. We did not include in our sample observations coming from countries accounting for less than 10 published papers. The excluded countries are: Andorra, Burundi, Saint Vincent and the Grenadines, Maldives, Chad, Central African Republic.

We end up with a cleaner dataset of 145\,609 unique papers, accounting for 7\,325 unique major Mesh terms, and 151 countries. We remove all observations coming from the years 2018 and 2021, resulting in a sample of 91\,480 unique papers, coming from 145 countries and accounting for 6\,175 research topics. 

To identify authors' prior experience in the main topic of publication, we collect information from recent published works of authors featured in our final sample with OpenAlex official API \citep{priem2022openalex}. Authors are disambiguated utilizing the OpenAlex authors' unique indentifier. 
We keep published papers between 2015 and 2020, with information on MeSH terms from PubMed Central, and collect a total of 1,129,874 unique articles published on PubMed by 365,340 authors. Following \citep{guimera2005}, we define authors in a given authorship position (first, last, middle) as \emph{incumbents} with respect to the main research topic of the paper if the author has already published on that specific topic. If the author is not already established in the research topic, the author is deemed as \emph{newcomer}. Following this idea, if both key authors are incumbent, the paper is published by an incumbent team; if both are newcomers, then by a newcomer team. We identify an additional category, \emph{new entrant}, indicating all authors who do not have PubMed publications in the reference period of the persistence measure 2015-2020. A new entrant could either be an author publishing for the first time in 2019 or 2020 in general, or someone coming from other disciplines. 

\section*{Empirical Strategy and Identification}

We define the treatment variable $COVID-related$ as a binary variable that equals 1 if the paper's major MeSH term falls above the 90th percentile of the distribution of COVID-relatedness $relatedness_{i}$, zero if the MeSH term is below the 10th percentile. We drop all papers with major MeSH given by a COVID-18 MeSH term, for which there is no pre-covid  obtaining 45\,180  treated observations and 44\,350 control units. Covid-related research topics are the ones that have attracted much more attention since the global pandemic \citep{10.1371/journal.pone.0263001}. 
Since we are including in our analysis only paper within research fields that are \emph{naturally} related to COVID-19 and those that are absolutely not related to COVID-19, we can exclude that any manipulation into treatment has taken place among the considered MeSH terms. Only papers dealing with MeSH terms that are \emph{partially} related to COVID-19 in terms of COVID-19-relatedness could have purposely engaged in more COVID-19-related publications because of the increased public interest and popularity of the subject. As we control for any source of \emph{selection bias} coming from manipulation of research fields into treatment, we can rely on the quasi-random assignment of MeSH terms to a new research topic. 
It becomes unlikely that only those MeSH terms with a higher (or lower) potential outcome are included among the set of COVID-related publications. 

For each paper, we define a set of different binary outcomes of interest. The variable \emph{Any} identifies papers with at least one woman as author; \emph{First and Last}, whether both key authorship positions are held by women; \emph{First} (\emph{Last}), if the first (last) author of the paper is a woman; \emph{Also Middle}, if there are any women in middle positions; \emph{Only Middle}, if there are women authors in the paper, but not in first and last position. 
Let $i$ be a paper published at time $t$ for $t \in \{ 2019, 2020\}$. Define $year_{i}$ as the year of publication of paper $i$, such that $year_{i} = t$. Let $j$ be the country of listed affiliations of either the first author, the last author, or of the majority of the authors of paper $i$ -- depending on the outcome variable. 
We estimate the following linear Diff-in-Diff model with two groups and two time periods, on repeated cross-sections of papers:

\begin{align}\label{eq_1}
    y_{i} = \alpha +  \gamma_{j} + \lambda\bbone[year_{i} = 2020] +  \beta \textit{COVID-related}_{i} + \nonumber\\
    \tau(\bbone[year_{i} = 2020] \times \textit{COVID-related}_{i}) + \epsilon_{i}
\end{align}

where $y_{i}$ represents the binary dependent variable of interest observed for paper $i$ and \bbone[] the indicator function. We include a constant term $\alpha$, a treatment group indicator $\textit{COVID-related}_{i}$ and year dummy variables $\bbone[year_{i}= t]$ for $t = 2020$, with 2019 as the baseline year. We also include country fixed effects $\gamma_{j}$. The effect of the treatment is given by the coefficient on the interaction $ year_{i}= t \times treat_{i}$ for $t=2020$, $\tau$.

We estimate model (1) including paper level controls: team size (\emph{N Authors}), access to pre-existing grants, and whether a publication is related to clinical trials, which were significantly disrupted by the COVID-19 pandemic \citep{10.1371/journal.pone.0263001}. We also control for the country of the listed affiliation of either the first author, the last author, or of the majority of the team members, depending on the outcome of interest.
We decide to include these particular variables, as they most likely influence simultaneously the outcome and the treatment status, without being affected by the participation decision or the anticipation of it.


For the analysis on incumbency of authors in the topic of publication, we estimate the following model:

\begin{align}\label{pers_eq}
    y_{i} = \alpha +  \gamma_{j} +  \lambda\bbone[year_{i} = 2020] + \beta \textit{COVID-related}_{i} + \eta P(l)_{i} + \nonumber\\
    \tau(\bbone[year_{i} = 2020] \times  \textit{COVID-related}_{i}) +  \delta (year_{t} \times P(l)_{i}) +  \nonumber\\
    \mu(\bbone[year_{i} = 2020] \times \textit{COVID-related}_{i} \times P(l)_{i})   + \epsilon_{i}
\end{align}
where $P(l)_{i}$ is a three-class categorical variable indicating the incumbency in research of the author in the position of interest $l$ -- corresponding to the author position in the dependent variable $y_{i}$ -- of paper $i$. For example, when estimating the effects on first female authorship position, we include in the DiD baseline regression interactions with $P(first)_{i}=\textit{incumbent}$, that indicates an incumbent first author in paper $i$.

\section*{Results}

In our study, we leverage COVID-19 as an exogenous shock in scientific research to estimate the impact of an unexpected new publishing opportunity on women’s career prospects in science.

In Table \ref{did}, we report the Difference-in-Differences (Diff-in-Diff) estimates of equation (1) with robust standard errors. 
From column (1), the coefficient estimate on $2020$ shows that the common time trend in general female participation is significantly increasing (0.0147, SE = 0.0032). As indicated by the coefficient estimate of $COVID-related$, COVID-related topics are more likely to feature a female author before the advent of the new publishing opportunity such as COVID-19 (0.0367, SE = 0.0033), but after the event, the probability of featuring a female author ($year=2020$ $\times$ $COVID-related$) significantly decreases by 0.0350 (SE = 0.0044) among the research related to the new emerging topic. 

Looking at the effect of a new emerging topic on female participation as key authors, the shared time trend between COVID-related and non-related ($year = 2020$) indicates that the probability of having a woman in key authorship positions is increasing, as shown in column (2) (0.0131, SE = 0.0036), column (3) (0.0212, SE = 0.0049),  and column (4) (0.0189, SE = 0.0046). COVID-related papers in 2019 are more likely to feature a woman as a both key authors (0.069, SE = 0.0041), as first author (0.058, SE = 0.0053), and as last authors (0.0864, SE = 0.0051). Instead, the treatment ($year=2020$ $\times$ $COVID-related$) is significantly and negatively affecting the likelihood of observing a woman as first author (-0.0699, SE = 0.007), last author (-0.057, SE =  0.0067), and in both key authorship positions (-0.0485, SE = 0.0054). 

Considering at last the effect of COVID-19 on the \emph{middle} positions, less valuable for career-enhancing in scientific academia, from column (5) women's participation in published papers in 2020 ($year = 2020$) as middle authors is increasing (0.0177, SE = 0.0041), but among the COVID-related papers, women were less likely to be featured also in the non-relevant positions (-0.0382, SE = 0.0056). 
Instead, if we consider female middle authorship only in column (6), we see that the shared time trend ($year = 2020$) is now decreasing (-0.00936, SE = 0.0041), indicating that in 2020 women were overall less likely to be featured as middle authors, when the first and last authors are not women. Moreover, topics related to COVID-19 are less likely to feature a woman in only middle authorship position before 2020 (-0.0347, SE = 0.0044). In 2020 instead, the probability of observing a woman as a middle author in papers where key authors are men increases significantly by 0.0357 (SE = 0.0058). 


\begin{table}[ht]
\def\sym#1{\ifmmode^{#1}\else\(^{#1}\)\fi}
\caption{Linear Diff-in-Diff model estimates, with White standard errors. Country effects (omitted) of the majority of the team for Any, First and Last, Also Middle, and Only Middle; country fixed effects of the first (last) author for regression on First (Last) Female Author. \label{did}}
\tabcolsep=2pt
\resizebox{\textwidth}{!}{ 
\begin{tabular}{l c c c c c c}
\toprule 
  &\multicolumn{6}{c}{Female Author} \\
 &(1)&(2)&(3)&(4)&(5)&(6)\\
& Any &First and Last & First & Last&Also Middle & Only Middle \\
\toprule 
year=2020           &      0.0147\sym{***}&      0.0131\sym{***}&      0.0212\sym{***}&      0.0189\sym{***}&      0.0177\sym{***}&    -0.00936\sym{*}  \\
                    &      (4.60)         &      (3.62)         &      (4.32)         &      (4.13)         &      (4.36)         &     (-2.26)         \\
\addlinespace
 COVID-related            &      0.0367\sym{***}&      0.0690\sym{***}&      0.0580\sym{***}&      0.0864\sym{***}&      0.0483\sym{***}&     -0.0347\sym{***}\\
                    &     (11.10)         &     (16.73)         &     (10.95)         &     (17.09)         &     (11.32)         &     (-7.94)         \\
\addlinespace
year=2020 $\times$ COVID-related &     -0.0350\sym{***}&     -0.0485\sym{***}&     -0.0699\sym{***}&     -0.0570\sym{***}&     -0.0382\sym{***}&      0.0357\sym{***}\\
                    &     (-7.99)         &     (-8.98)         &    (-10.00)         &     (-8.54)         &     (-6.78)         &      (6.18)         \\
\addlinespace
N Authors          &      0.0124\sym{***}&    -0.00217\sym{***}&   -0.000642         &    -0.00355\sym{***}&      0.0278\sym{***}&      0.0144\sym{***}\\
                    &     (45.04)         &     (-7.97)         &     (-1.64)         &     (-9.58)         &     (50.29)         &     (34.38)         \\
\addlinespace
Trial            &      0.0257\sym{***}&     0.00556         &     0.00590         &      0.0183         &      0.0571\sym{***}&     0.00646         \\
                    &      (5.03)         &      (0.73)         &      (0.60)         &      (1.92)         &      (8.38)         &      (0.77)         \\
\addlinespace
Pre-Existing Grant      &      0.0420\sym{***}&      0.0347\sym{***}&      0.0581\sym{***}&      0.0417\sym{***}&      0.0542\sym{***}&     -0.0212\sym{***}\\
                    &     (14.63)         &      (8.47)         &     (11.29)         &      (8.44)         &     (13.77)         &     (-4.92)         \\
\addlinespace
Constant            &       0.911\sym{***}&       0.210         &       0.325\sym{***}&       0.307\sym{***}&       0.573\sym{*}  &     -0.0504\sym{**} \\
                    &     (68.43)         &      (0.94)         &      (4.93)         &      (4.49)         &      (2.50)         &     (-2.73)         \\
\midrule
Observations        &       89530         &       89530         &       83263         &       82552         &       89530         &       89530         \\
\midrule
Country FEs & Majority  & Majority  & First & Last & Majority  & Majority \\
\bottomrule
\multicolumn{7}{l}{\footnotesize \textit{t} statistics in parentheses. \sym{*} \(p<0.05\), \sym{**} \(p<0.01\), \sym{***} \(p<0.001\)}\\
\end{tabular}
}
\end{table}

Section B of the Appendix reports the estimated country fixed effects, along with 95\% confidence intervals. To check for parallel trends, we estimate the difference in outcomes prior to 2020 among the COVID-related and COVID non-related papers and find no systematic difference in monthly female authorship before 2020 --- see Section C.1 of the Appendix. In Figure \ref{fig:fig_2}, first female authorship decreases sharply around April 2020 only in COVID-related publications. Around the same time, the same publications have experience a surge in the share of women authors as middle authors, when key authors are not women.
This indicates a production lag of approximately three months in the life sciences, as the negative effect becomes visible after 3-4 months. 

The coefficient estimates remain statistically significant as we cluster the standard errors at various levels of aggregation (country, country-year, Mesh, Mesh-year, country-Mesh). See Section C.2 of the Appendix. 

The evidence strongly indicates a significant reduction in the proportion of women occupying key authorship positions, following the emergence of the new research opportunity, but only within the affected research topics. Notably, the shared trend between the treatment group (COVID-related) and the control group (COVID non-related) demonstrates that women are increasingly featured as key authors in 2020 across other research areas.
Conversely, the treatment shows a positive effect on female authorship in non-key positions, suggesting that after the emergence of the new research opportunity, women authors are more likely to be relegated to less prominent roles. This shift implies that while women's overall participation may not have decreased, their opportunities for career-advancing positions have diminished.
Further analyses reinforce these findings. In Section C.3 of the Appendix, we present the coefficient estimates of Equation (1) after modifying the definitions of the treated and control groups, confirming the robustness of our results. Additionally, in Section C.4 of the Appendix, we demonstrate that controlling for the time since an author's first publication---a proxy for academic age---does not alter the outcomes reported in Table~\ref{did}. This suggests that the observed effects are not driven by differences in career stages among authors but are indicative of systemic issues affecting women's access to key authorship roles in emerging research topics.


\begin{figure*}[ht]
    \includegraphics[width=\textwidth]{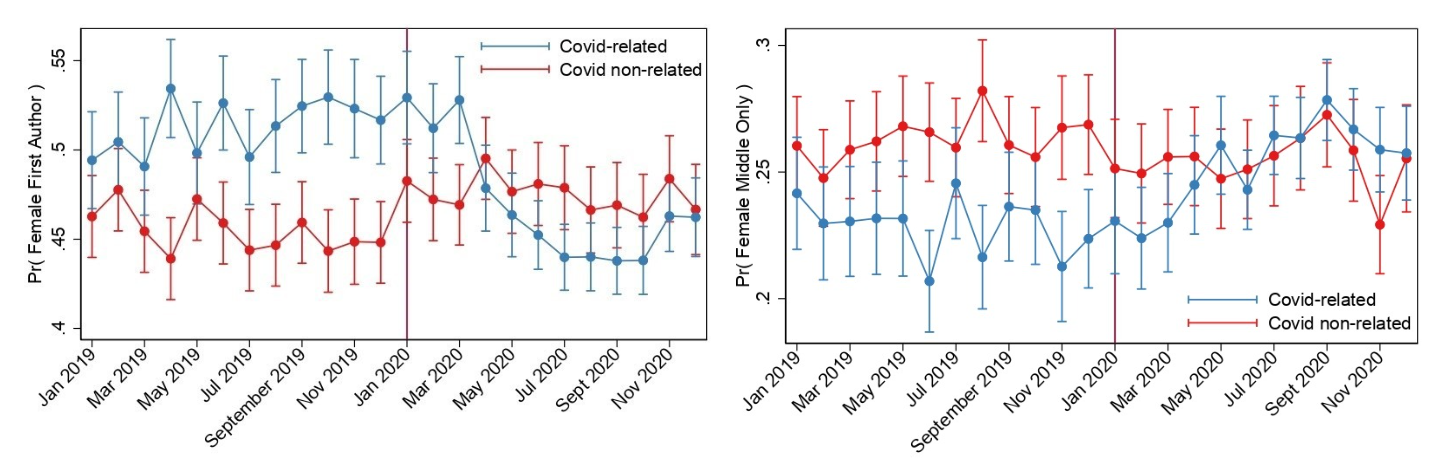}
  \caption{Monthly linear predictions for Female First Author (a) and Middle Only (b) by group. The pre-2020 monthly trends do not show any systematic difference between COVID-related and non-related. In 2020, starting from around April, there is a sharp decline in first authorship by women in COVID-related papers (left), while women participation increases as middle authors only (right). 
}
  \label{fig:fig_2}
\end{figure*}

To ensure the robustness of our findings, we assess whether the estimated treatment effect from the Difference-in-Differences (DiD) model is influenced by potential pre-existing biases related to journal impact factor, team size variability, and research funding. We re-estimate model (1) incorporating the same paper-level controls—\emph{trial}, \emph{Pre-existing Grant}, and \emph{N authors}—as previously used. Additionally, we include (i) journal impact factor, (ii) an interaction between journal impact factor and team size, and (iii) new funding obtained in 2020 as linear interactions with the treatment effect, the treatment status, and the time variable.
To ease the interpretation of the models' estimates, we proxy the impact of the Journal Impact Factor with a binary indicator, \emph{JI med}, which equals 1 if the Journal impact factor of a given paper is higher or equal to the median of the distribution of observed \emph{JI} values in our sample, and zero otherwise. \emph{Has New Grant} is a binary variable identifying  funding granted during 2020. 

We also investigate whether lockdown restrictions related to COVID-19 can fully account for the observed differences in female authorship between treated and control research topics in 2020. We consider how stay-at-home mandates might differentially affect women's outcomes, particularly through increased family duties or alterations in work schedules and routines due to these restrictions. To quantify the impact of lockdown measures, we utilize indices derived from the Oxford COVID-19 Government Response Tracker (OxCGRT) \citep{owidcoronavirus}. For authors in the US, we consider the stringency values at national level, controlling for national-level country fixed effects.

In Table \ref{tab_2}, the effect of interest (year=2020 $\times$ COVID-related) remains significant across all women's authorship outcomes, demonstrating that neither journal prestige (Model 1), team size (Model 2), nor access to new funds (Model 3) can fully explain the decrease in women's key authorship positions in publications related to the new emerging topic. Additionally, the impact of school closures (Model 4) and workplace restrictions (Model 5) does not entirely account for the observed changes over time in women's authorship in key positions between treated and control publications. 
Conversely, the effect is not significant for overall middle authorship. However, it remains significant for the appointment of women as middle authors when key author positions are held by men. See Section C.5 and Section C.6 of the Appendix for details and regression tables.

\begin{table*}[ht]
\def\sym#1{\ifmmode^{#1}\else\(^{#1}\)\fi}
\caption{Linear DiD model estimates for treatment $\tau$ (year=2020 $\times$ COVID-related) including interactions with potential confounders (rows 1-5), and PSM-DiD matching (row 6).  Results are robust to COVID-19 stringency measures, quality of the publication ranking and research funding, except overall female middle authorship, which is confounded by joint variation in journal impact factor and team size (Model 2), school closures (Model 4) and workplace closures (Model 5). The effect on overall female authorship and overall middle female authorship is not robust to PSM-DiD (Model 6). The cases in which the DiD estimates of  $\tau$ is not confirmed are highlighted in red.}\label{tab_2}
\resizebox{\textwidth}{!}{ 
\begin{tabular}{l c c c c c c}
\toprule 
  &\multicolumn{6}{c}{Female Author}\\
  &\multicolumn{1}{c}{Any}&\multicolumn{1}{c}{First and Last}&\multicolumn{1}{c}{First}&\multicolumn{1}{c}{Last}&\multicolumn{1}{c}{Middle}&\multicolumn{1}{c}{Only Middle}\\

\midrule
\textit{Model 1 with New Grants}&   -0.0335\sym{***}&     -0.0508\sym{***}&     -0.0700\sym{***}&     -0.0583\sym{***}&     -0.0361\sym{***}&      0.0363\sym{***}\\
                    &     (-7.36)         &     (-9.18)         &     (-9.74)         &     (-8.50)         &     (-6.19)         &      (6.10)         \\
\addlinespace
\hline
\addlinespace
\textit{Model 2 with Journal Impact Factor (IF)} &     -0.0344\sym{***}  &     -0.0521\sym{***}&     -0.0715\sym{***}&     -0.0593\sym{***}&     -0.0362\sym{***} &      0.0394\sym{***}\\
                    &     (-5.30)      &     (-6.86)         &     (-7.24)           &     (-6.28)            &     (-4.42)        &      (4.87)          \\
\addlinespace
\hline
\addlinespace
\textit{Model 3 with Journal IF $\times$ N authors}&       -0.0333\sym{*}  &     -0.0804\sym{***}&     -0.0850\sym{***}&     -0.0810\sym{***}&   \color{red}{0.0000906}     &      0.0517\sym{**} \\
                     &     (-2.05)               &     (-5.29)               &     (-4.15)           &     (-4.13)             &      (0.00)             &      (2.78)         \\
\addlinespace
\hline
\addlinespace
\textit{Model 4 with Monthly Max School Closures Index} 
&     -0.0206\sym{**}   &     -0.0278\sym{**}  &     -0.0433\sym{***}&      -0.0399\sym{***}  &    \color{red}{-0.0167}          &      0.0215\sym{*}  \\
                    &     (-2.67)            &     (-3.01)         &     (-3.62)               &     (-3.52)              &     (-1.72)             &      (2.14)              \\
\addlinespace
\hline
\addlinespace
\textit{Model 5 with Monthly Max Workplace Closures Index}  &     -0.0150\sym{*}  &     -0.0234\sym{**}&     -0.0429\sym{***}&     -0.0224\sym{*}      &    \color{red}{-0.00751}   &      0.0227\sym{*}  \\
                  &     (-2.00)       &     (-2.60)        &     (-3.68)           &     (-2.04)          &     (-0.79)              &      (2.32)         \\
                
\addlinespace
\hline
\addlinespace
\textit{Model 6, PSM-DiD}&    \color{red}{-0.00968}         &     -0.0350\sym{***} &     -0.0437\sym{***}&     -0.0518\sym{***}&     \color{red}{-0.0121}         &      0.0356\sym{***}\\
                    &     (-1.75)         &     (-5.06)      &     (-4.81)  &     (-6.04)        &     (-1.70)         &      (4.71)         \\
\midrule
Country FEs & Majority & Majority & First & Last & Majority & Majority \\
White SEs  & Yes & Yes & Yes & Yes & Yes & Yes\\
\bottomrule
\multicolumn{7}{l}{\footnotesize \textit{t} statistics in parentheses}\\
\multicolumn{7}{l}{\footnotesize \sym{*} \(p<0.05\), \sym{**} \(p<0.01\), \sym{***} \(p<0.001\)}\\
\end{tabular}
}
\end{table*}

Although Genderize.io is more accurate than competing tools \citep{VanHelene2024.01.30.24302027}, it still has a non-negligible error rate of about 4\% \citep{Lockhart2023NamebasedDI}. This probabilistic approach is particularly prone to mis-gendering names from certain countries, such as China. This could bias our estimated effects if countries with differing gendering error rates are not evenly distributed between treated and control groups. To address this, we employ a three-stage propensity score matching to form a \emph{pseudo panel} of treatment variables, ensuring that control papers are matched to treated papers based on paper-level controls and the country of the majority of authors, or the country of the first or last author, depending on the outcome of interest. We then re-estimate the Diff-in-Diff model on this matched sample, and our findings remain consistent on the authorship positions of most interest, first, last, on and middle only authorship (Table \ref{tab_2}, Model 6). For further details, refer to Section C.7 of the Appendix. 

The observed decline in the proportion of women as key authors in COVID-19 related research could be due to an influx of male colleagues from other areas of research. These male scientists, attracted by the new research opportunities, may disproportionately move towards publications on this emerging topic. 
We wish to assess whether the decline in first authorship by women among COVID-related research published can be explained by mechanisms underlying team assembly, such as combination of key authors with different prior research experience on the topic of publication. When pressures to publish are high, risk assessment on team composition could be a key factor in determining first authorship, potentially disadvantaging female researchers in highly competitive new research areas \citep{10.1371/journal.pone.0184601, doi:10.1177/0141076819851666}. 

We present only the impact of diverse team composition on first female authorship, which is were we observed the strongest drop in female authorship in COVID-related publications (see Table \ref{did}). Moreover, we focus on key authorship of incumbent and newcomer teams, as COVID-19 publications by teams of scientists without relevant expertise in the topic of publication were on the rise after 2020 \citep{sikdar2024}. Results related to new entrant teams and estimates of the effects of team composition on last female authorship are presented and discussed in the Appendix, Section D.1. 

\begin{figure}[H]
 \includegraphics[width=0.8\textwidth]{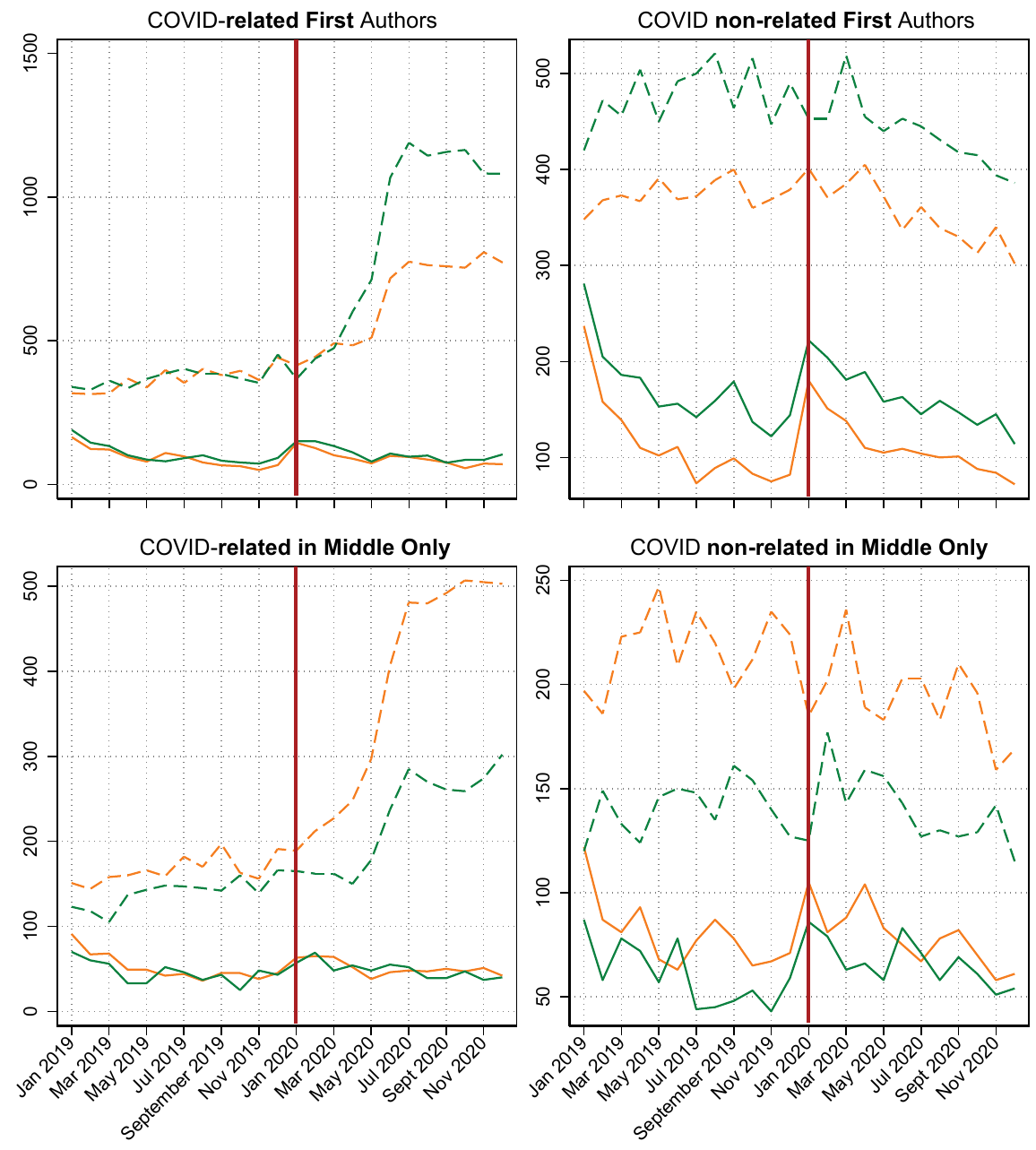}
  \caption{Monthly sample numerosity of COVID-related (left) and COVID non-related (right) papers featuring  women (orange) and men (green), with incumbent (solid) or newcomer status (dash), in first authorship (top), and as middle authors only (bottom). 
}
  \label{fig:fig_3}
\end{figure}

We can already have a good intuition of the rise of male authorship simply by considering the observed monthly trend in the number of papers by women and men newcomers among the COVID-related publications. Figure \ref{fig:fig_3} reveals how the number of papers with women newcomers as first authors are falling behind the rise of male newcomers in COVID-related research in 2020, while this is not the case among the COVID non-related. Conversely, we observe much more papers with women newcomers as middle authors in teams with male key authors in COVID-related publications, while male presence declines in parallel. As shown in Section D.1 of the Appendix, publications by new entrant first authors did not experience the same great surge as the newcomers did, which is the category that conversely reveals the largest gender gap in first authorship.  

\begin{figure}[H]
   \centering
        \includegraphics[width=0.8\textwidth]{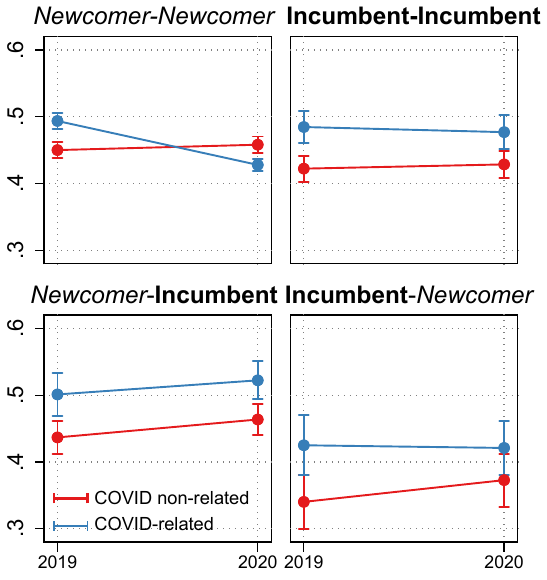}
    \caption{Predicted probability to observe a woman first author by past research experience of first and last authors; we control for country of last authors fixed effect. The decline in female first authorship is linked to the rise of teams without prior research experience on the topic with respect to established teams.}
    \label{fig:fig_4}
  \end{figure}

Figure \ref{fig:fig_4} illustrates a significant decline in female first authorship among the COVID-related research topics in 2020 when both the first and last authors are newcomers.
This suggests that the decrease in female first authorship in COVID-related publications is associated with the emergence of "opportunistic", publishing teams, that assemble around new research topics, without experts as either first or last authors. Specifically, female first authorship is less prevalent when neither key author of a COVID-related publication has prior expertise in the topic, compared to situations where both are incumbents. Instead, among COVID non-related women continue to successfully publish as first authors when both key authors are Newcomer (top-left panel), as indicated by the slightly increasing trend in 2020. This implies that the drop in women's representation in first authorship positions in COVID-related publications can be attributed to the rise of newcomers, male authors occupying these key roles. As expected, there is greater flexibility in selecting the first author when they lack prior publications on the topic, leading to less influence and familiarity with the subject, and consequently, a higher likelihood of being assigned non-key roles, compared to more experienced researchers. 

In Section D.1 of the Appendix, we report the Tables with the coefficient estimates, as well as linear predictions on all considered authorship outcomes and incumbency status, showing that the effect of new entrant teams in COVID-related research on first female authorship is in line with the impact we find among newcomers. Moreover, we show that also last female authorship in publications related to the new topic is negatively affected by new entrant teams.

We also check whether the differential in COVID-19 related papers for first female authorship in not-yet established teams is not fully explained by stringency lockdown within the country of last authors. 
Closures do play a role, but they do not explain the drop of women first authorship among COVID-related publications of teams of newcomers. 
 We find a significant positive effect on female authorship in key positions of teaming up a newcomer first author with a incumbent last author in COVID-19 publications during 2020, with closures within the country of the first author reducing this effect. Details are provided in the Appendix, Section D.2. While closures do not affect the differential created by teams of newcomers on first female authorship, they instead explain the drop in first female authorship among new entrant teams, not for women last authors, as the effect of new entrant teams is still negatively and significantly impacting last authorship (see Section D.2 of the Appendix).

\begin{figure}[H]
\includegraphics[width=\textwidth]{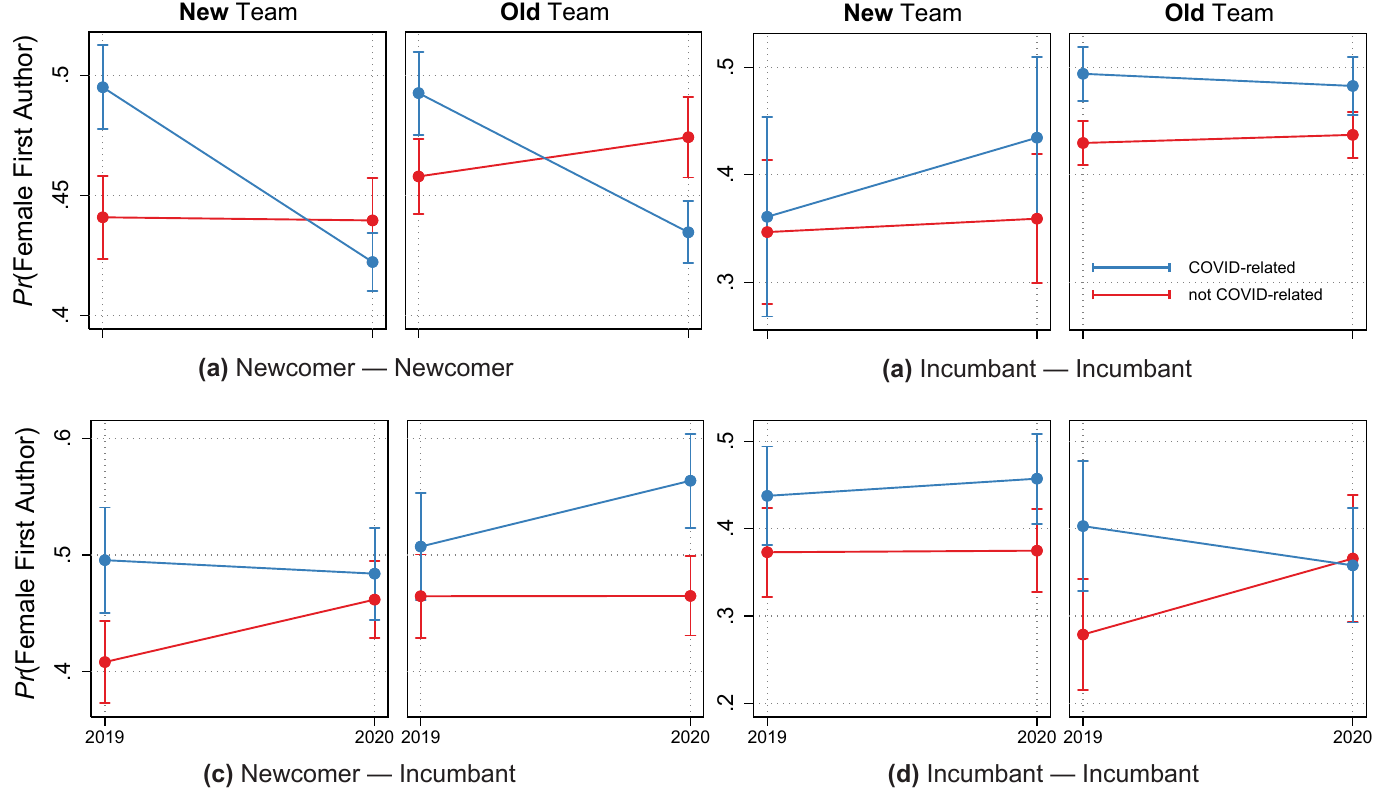}
  \caption{Linear predictions for Female First Author by incumbency in topic of publication of first and last author, within new and pre-existing teams. First female authorship is increasing over time in teams with a newcomer first author and incumbent last author collaborating in pre-existing teams, with respect to newly formed teams. 
%
}
  \label{fig:fig_5}
\end{figure}

At last, we wish to assess whether pre-existing teams favour women's appointment in key positions, with respect to first and last authors' relation to the topic (Incumbent vs Newcomer) and to their prior collaboration (New Team, if the key authors have not published any prior research together in the past, Old team if they have). Teams in which the first and the last authors do not have a pre-established collaboration, realized through previously published papers, could have higher constraints in team composition, limiting the freedom of choice upon the appointment of authors to the key authorship positions, especially is the candidate for first position also has prior publications on the research topic. 
We find that past research experience of the key authors remains the prevailing mechanism behind the drop of women's first authorship. Past publications of the key authors matter on the probability on a woman first author, as teams tend to exclude women newcomers from the first authorship position, no matter the origin of the team. From Figure \ref{fig:fig_5}, new teams include women first authors in COVID-related papers when they have a prior publication on the COVID-related topic, but not when they are new to the topic. First authorship decreases in 2020 for COVID-related topics for women newcomers in both new and old team. Old teams are less likely to have a newcomer woman first author in 2020 in COVID-related publications, except when the last author is incumbent. See Sections D.3 and D.4 of the Appendix for details and regression tables. 

\section*{Discussion}
This study exploits the quasi-random assignment of biomedical research topics to emerging scientific opportunities—specifically, the advent of COVID-19—to estimate the effect of new publishing opportunities on women's appointment to key authorship positions, namely the first and last authorship, which are fundamental for gaining recognition and competitiveness in academia.

We collect biomedical papers from 2019 and 2020 from PubMed Central and identify papers in research topics closely related to COVID-19 by constructing a \emph{relatedness} measure. 
We find a significant negative effect of new publishing opportunities on the probability of observing a woman as first (-7 percentage points), and last (-5.7 percentage points) author in research topics related to scientific novelty. However, the overall trend shows that the likelihood of women as first authors increases by 0.021, while for women last authors by 0.019 in biomedical papers.

Notably, while women’s presence in key authorship positions diminished in emerging topics, their representation in middle authorship positions—less valuable for career advancement—increased by 3.57 points in these areas, despite a general downward trend of 0.936 points in other fields. This suggests that women were more likely to be relegated to non-prominent roles in this new, high-impact research field, potentially hindering their career progression.

We conducted robustness checks to account for potential confounders such as journal impact factor, access to new funding, and increased family responsibilities due to COVID-19 restrictions. Linear interactions of these variables with our treatment effect did not fully explain the decrease in women's representation in key authorship positions; the treatment effects remained statistically significant.

In the case of COVID-19, a possible mechanism driving the drop of women featured in key authorship positions could be connected to the rise of male colleagues new to the research topic, collaborating with tenured authors who were new as well. As publishing in COVID-19-related research becomes more attractive, male scientists that do not usually publish in these topics choose to publish in the COVID-19 related research because of the increased public interest. We provide evidence of these mechanisms by using the OpenAlex API \citep{priem2022openalex} to collect the list of publications from 2015 to 2020 of each author featured in our sample of papers collected from PubMed to identify authors' past affiliation with the topic of publication. We find evidence of discriminatory effect of new research topics in first authorship position. Women are still successfully authoring papers as first authors when the main research topic of publication was not among her prior expertise during 2020, but not when moving towards COVID-related research topics.

Our findings suggest that gender dynamics within teams of newcomers, that is, where key authors are both new to the COVID-related topic, contribute to discriminatory practices in appointing female first authors. Tenured authors with prior expertise on the COVID-related research positively impacted first authorship in early COVID-related publications for women new to the topic, whereas newcomer last authors -- without prior expertise on the COVID-related research -- tend to collaborate with male authors who also are new to the research topic. 
The composition of publishing teams, particularly when the first author is new to the topic, shows discretionality in selection, but discriminatory gender biases may influence first author appointment due to high pressures for timely publication. 

Our findings reveal that women scientists in related research topics face challenges in gaining key authorship roles in early literature on a new scientific discovery, contributing to the gender gap in scientific production and academic rankings. Unlike previous literature, we isolate the impact of COVID-19 as an exogenous shock, finding evidence of discriminatory biases related to the new research topic against women authors without expertise on the topic of publication in first authorship, influenced team dynamics and social expectations.

Related to the external validity of our results, we believe that the particular case study of the advent of COVID-19 as a new research opportunity allows to gain valuable insights on the combinations of two effects that may rise with new research topics: the need to produce in a timely matter and the biases expectations on how women are expected to perform under external circumstances that greatly influence one's work schedule, a.g. a pregnancy. As under COVID-19 both of these effects were highly pronounced, our case study allows to stress-test gender discrimination on women's authorship in scientific publications highly relevant to a new, greatly publicized, research opportunity. 

Some policy implications arise from our analysis. Scientific institutions and governmental agencies should acknowledge and promote measures to mitigate discriminatory biases faced by women scientists in publications on new emerging topics, ensuring that women are given equal opportunities to contribute to and lead in emerging fields of research. 
The adoption of transparent authorship guidelines for publications in newly emerging fields could help detect and address biased patterns of collaboration and authorship assignment. 
At last, the rise of opportunistic teams under pressuring circumstances may foster an environment unfavourable to high quality of research or prone to ill scientific practices \citep{sikdar2024}. Policies promoting interdisciplinary collaborations between expert and non-experts could implement programs to assist women with the necessary resources and visibility to engage in high-impact research, and creating networks that facilitate their inclusion in emerging research teams. Moreover, journals could help promoting scientific practices that prioritizes gender equity and knowledge transfer, particularly in high-pressure contexts. 

This study has some limitations. First, our treatment status is indirectly observed; we assigned papers to the treatment group based on a constructed relatedness measure to COVID-19, focusing on research topics extremely related to the pandemic. Second, the absence of data on preprints limits our ability to assess potential selection biases in journal reviewing and acceptance processes. Future research could benefit from including preprints to examine whether similar patterns exist prior to peer review and publication.
Moreover, while our study focuses on COVID-19 as a case study, future investigations could extend this analysis to other emerging topics in scientific research, such as advancements in CRISPR/Cas9 technology. Exploring whether similar gender dynamics occur in different contexts would provide a broader understanding of the systemic issues at play.


Our findings underscore the need for proactive measures to ensure gender equity in scientific authorship, particularly in the context of emerging research opportunities. By understanding the underlying biases and structural barriers that disadvantage women, the scientific community can promote a more inclusive and diverse environment.

\section*{Data availability}
The data and codes are available at Biliotti, C., Verginer, L., \& Riccaboni, M. (2025). Breaking New Ground, Reinforcing Old Gaps: Gender Disparities in Access to Emerging Research Frontiers. Zenodo. \url{https://doi.org/10.5281/zenodo.13839340}.

\section*{Acknowledgments}
We benefited from helpful comments and suggestions from participants and discussants at ASSA 2025, International Conference on the Science of Science and Innovation (ICSSI) 2024, Royal Economic Society (RES) 2024 Annual Conference, DRUID23 Conference, WICK23 Workshop in Economics of Innovation, Complexity and Knowledge at Collegio Carlo Alberto. 

\section*{Author Contribution Statement}
C.B., L.V., and M.R. conceived the research idea and designed the identification strategy; C.B., and L.V. gathered data; C.B. conducted the formal analysis; C.B., L.V., and M.R. wrote and reviewed the manuscript. 




\newpage

\bibliographystyle{plainnat}
\bibliography{biblio}

\clearpage\appendix
\begin{center}
    \textbf{\huge Appendix} 
\end{center}

\section{Descriptive Statistics}

\begin{figure}[H]
  \begin{subfigure}[t]{0.5\textwidth}
    \includegraphics[width=\textwidth]{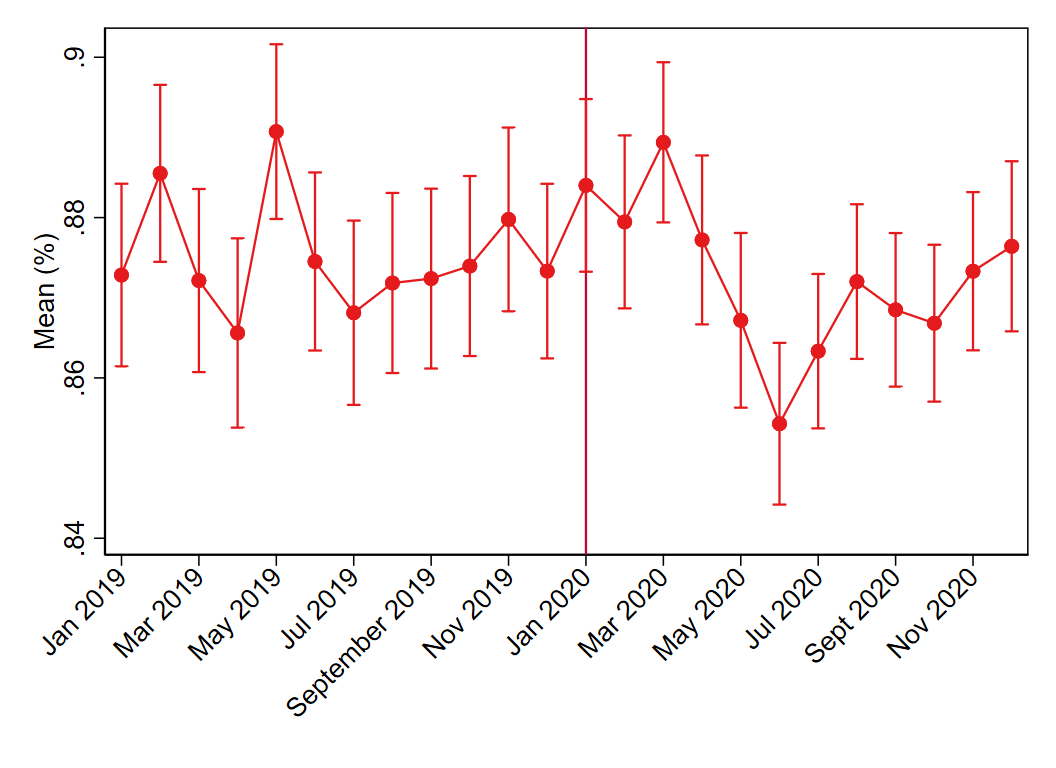}
    \caption{Female Author}
    \label{fig:f3_a}
  \end{subfigure}
  \hfill
 \begin{subfigure}[t]{0.5\textwidth}
    \includegraphics[width=\textwidth]{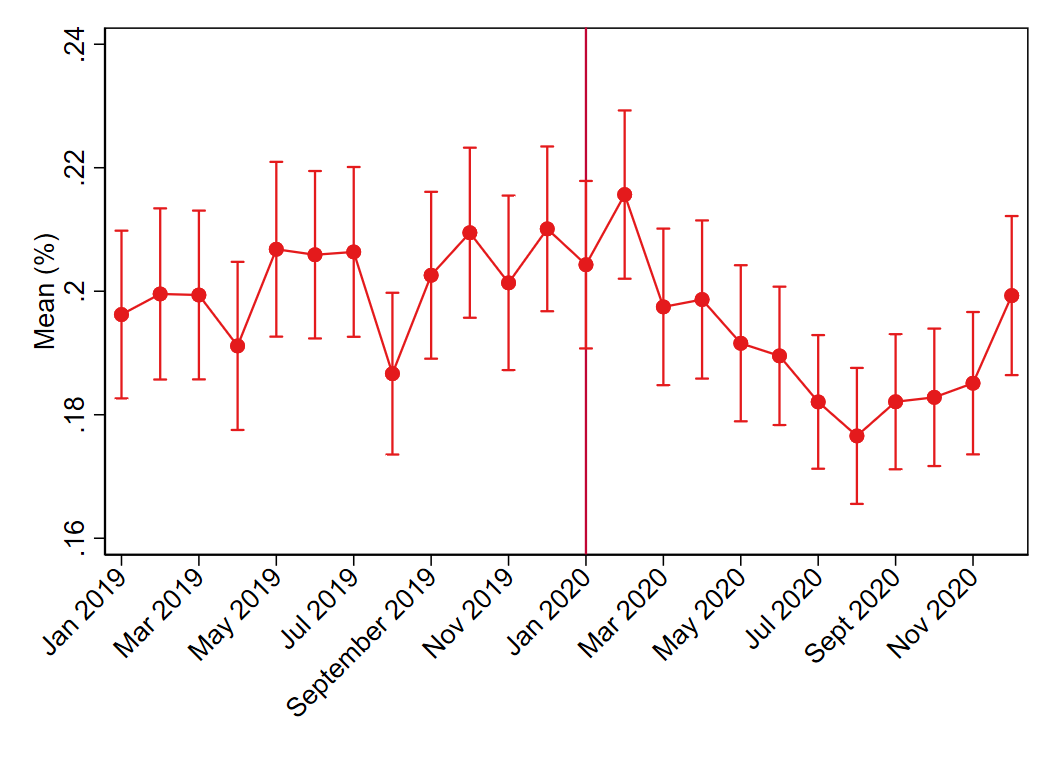}
    \caption{First and Last Female Author}
    \label{fig:f3_b}
  \end{subfigure}
    \hfill
 \begin{subfigure}[t]{0.5\textwidth}
    \includegraphics[width=\textwidth]{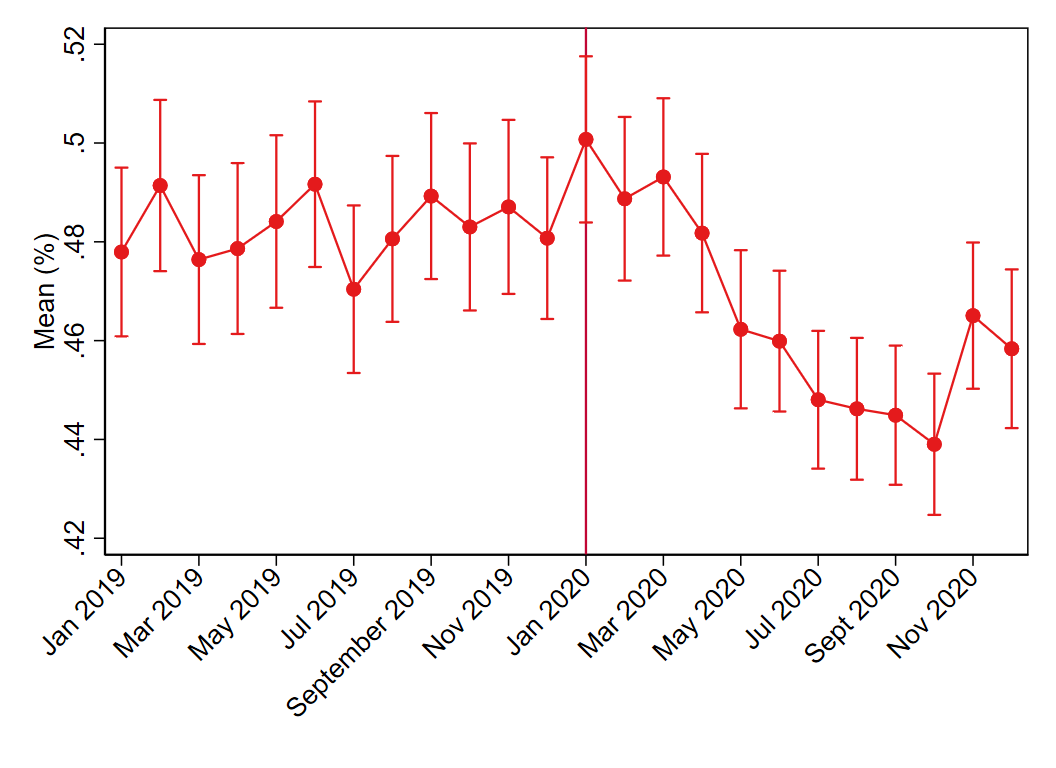}
    \caption{Female First Author}
    \label{fig:f3_b}
  \end{subfigure}
    \hfill
 \begin{subfigure}[t]{0.5\textwidth}
    \includegraphics[width=\textwidth]{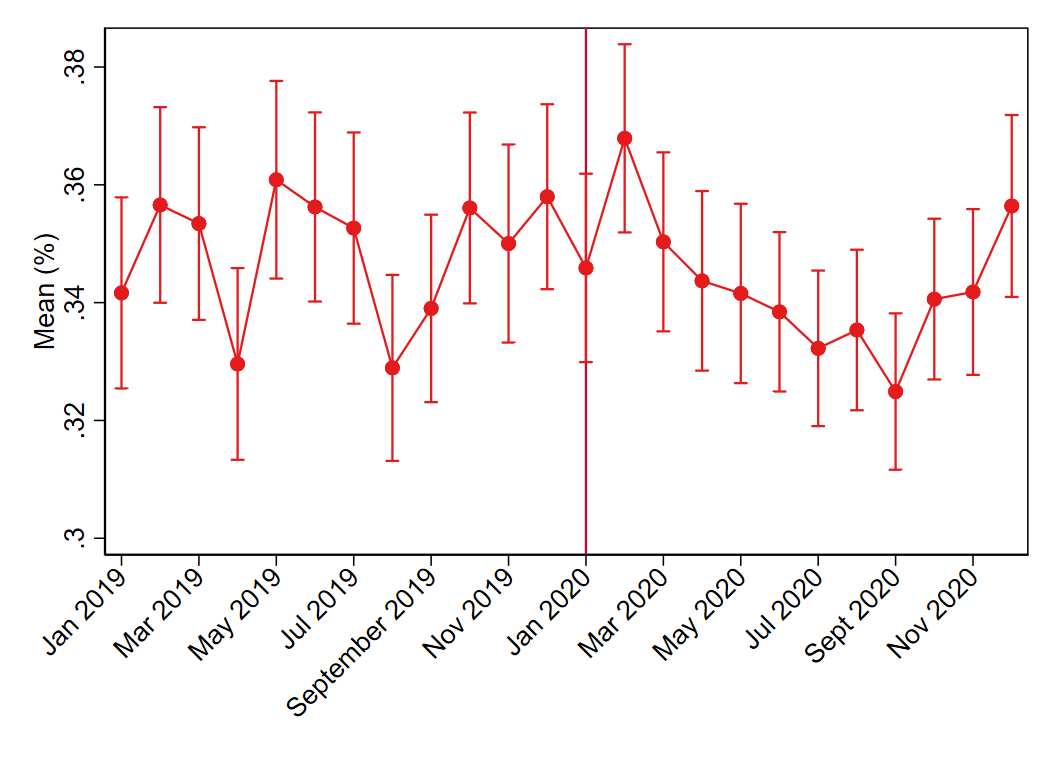}
    \caption{Female Last Author}
    \label{fig:pt_all}
  \end{subfigure}
     \begin{subfigure}[t]{0.5\textwidth}
    \includegraphics[width=\textwidth]{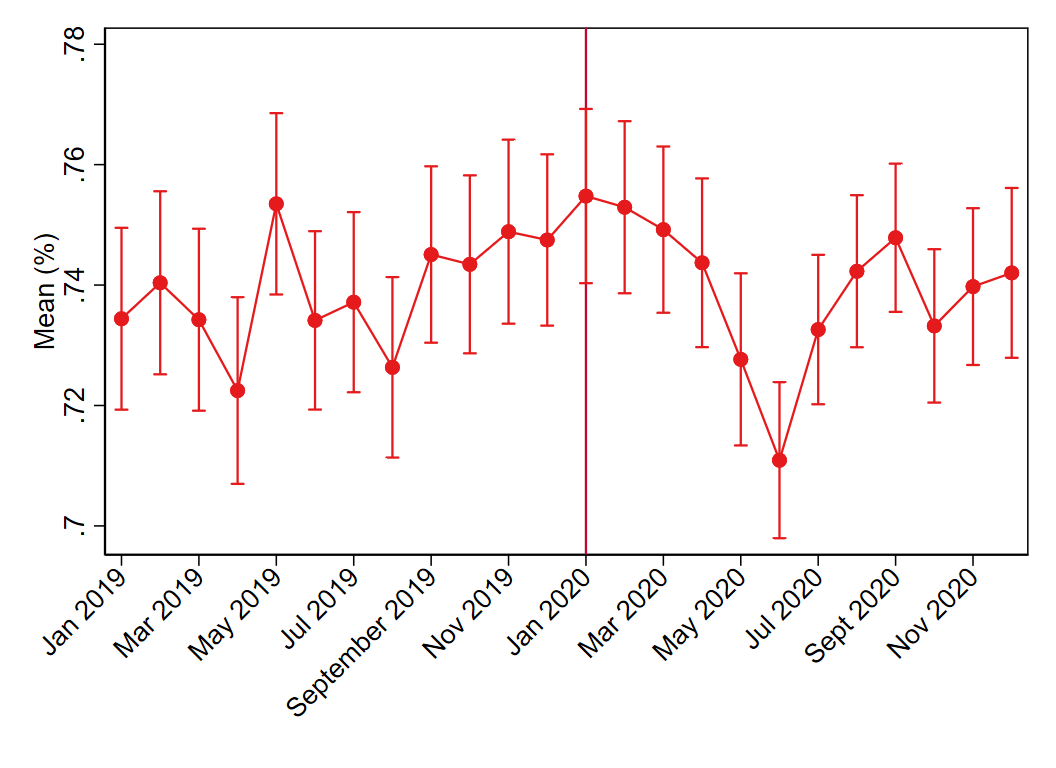}
    \caption{Middle Female Authorship}
    \label{fig:pt_all}
  \end{subfigure}
   \begin{subfigure}[t]{0.5\textwidth}
    \includegraphics[width=\textwidth]{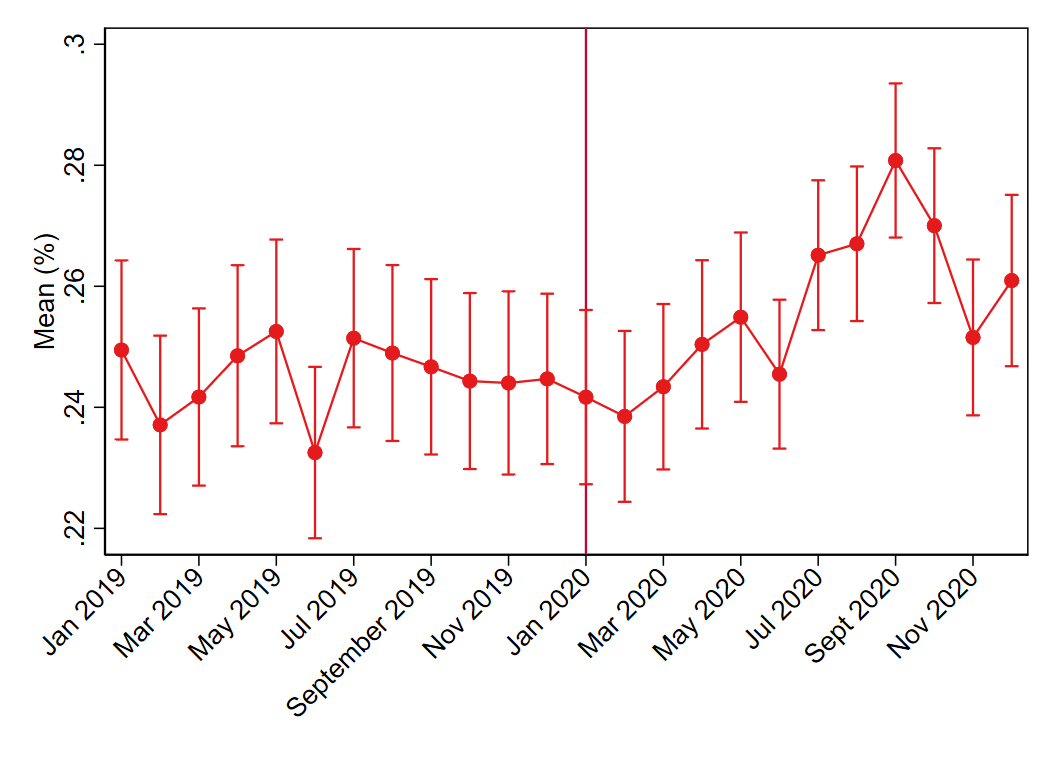}
    \caption{Middle Female Only}
    \label{fig:pt_all}
  \end{subfigure}
  \caption{Sample monthly Mean (\%), along with 95\% confidence intervals, of share of women at (a) any authorship position, (b) both first and last authors, (c) first authors only, (d) last authors only, (e) middle female authors, (f) middle female authors \emph{only}.}
  \label{fig:descr_plots}
\end{figure}

Table \ref{descr_paper} collects the main descriptive statistics of paper-level characteristics observed in our sample.  In Figure \ref{fig:descr_plots}, we plot the overall monthly trends of women's authorship outcomes from January 2019 to November 2020. We distinguish the case in which we observe a woman as middle author -- Middle Female Authorship -- from when a paper features a woman in non-relevant positions \emph{only}, i.e. without having other women as key authors -- Middle Authors Only. Figures (c) and (d) show that while there is a marked decrease in the share of women in key authorship positions, women increased their presence in the \emph{middle}, non-relevant authorship positions, when no key authors are women -- Figure (f). 

\begin{figure}[H]
   \begin{subfigure}[t]{0.5\textwidth}
        \includegraphics[width=\textwidth]{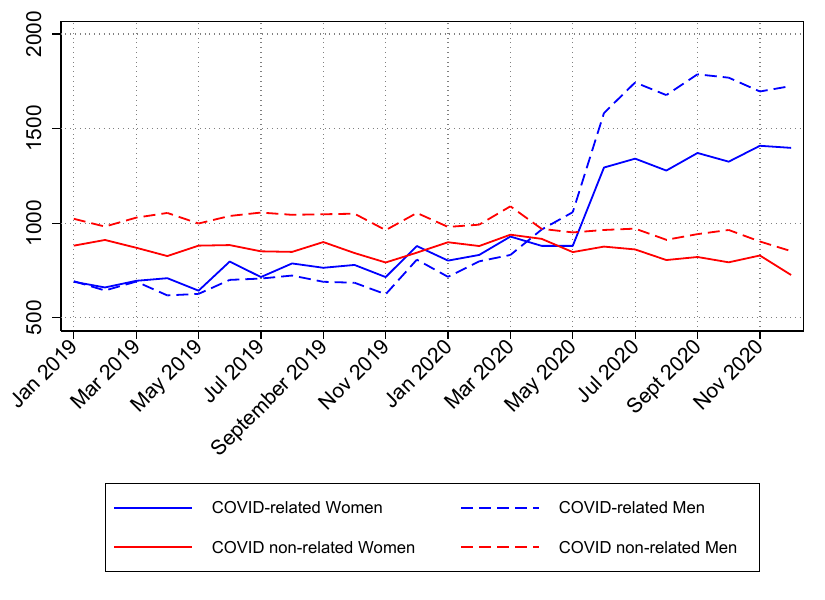}
        \caption{First Authors}
   \end{subfigure}
     \begin{subfigure}[t]{0.5\textwidth}
       \includegraphics[width=\textwidth]{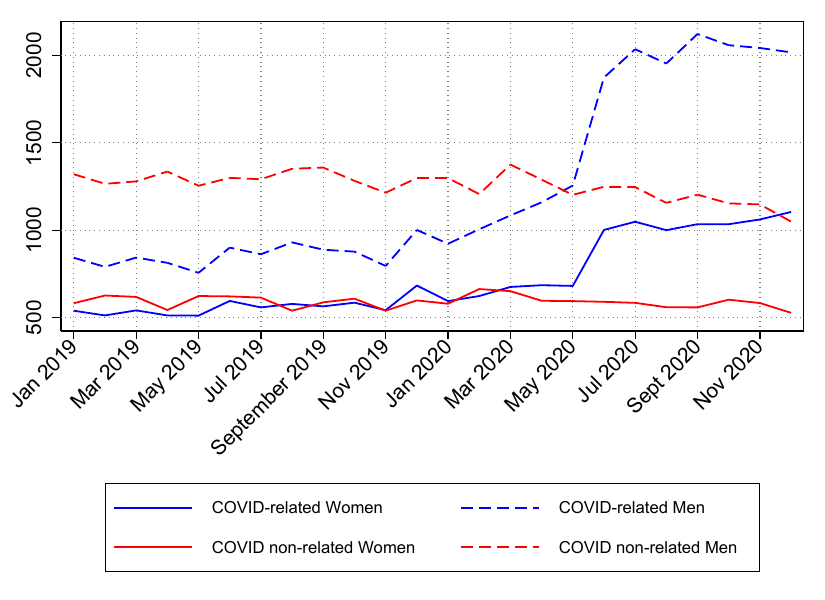}
         \caption{Last Authors}
   \end{subfigure}
         \begin{subfigure}[t]{0.5\textwidth}
        \includegraphics[width=\textwidth]{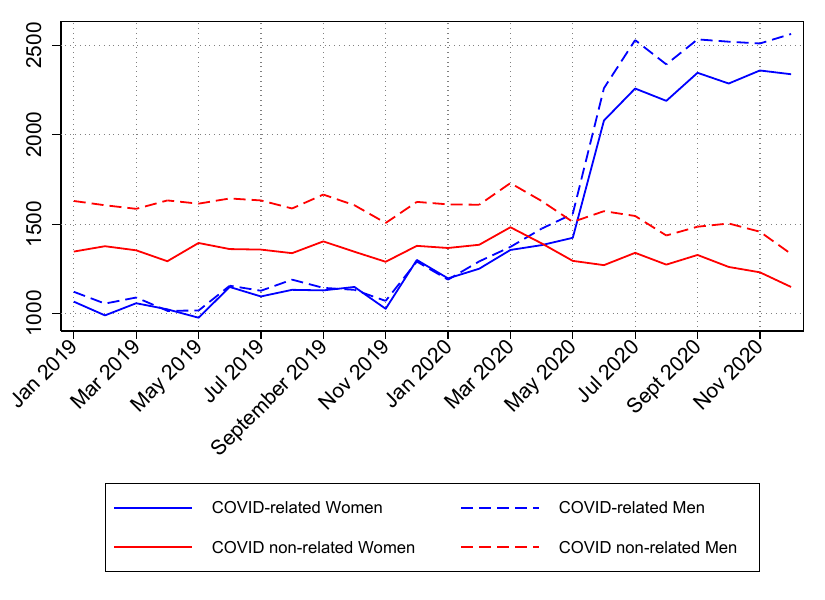}
        \caption{Middle Authors}
   \end{subfigure}
            \begin{subfigure}[t]{0.5\textwidth}
        \includegraphics[width=\textwidth]{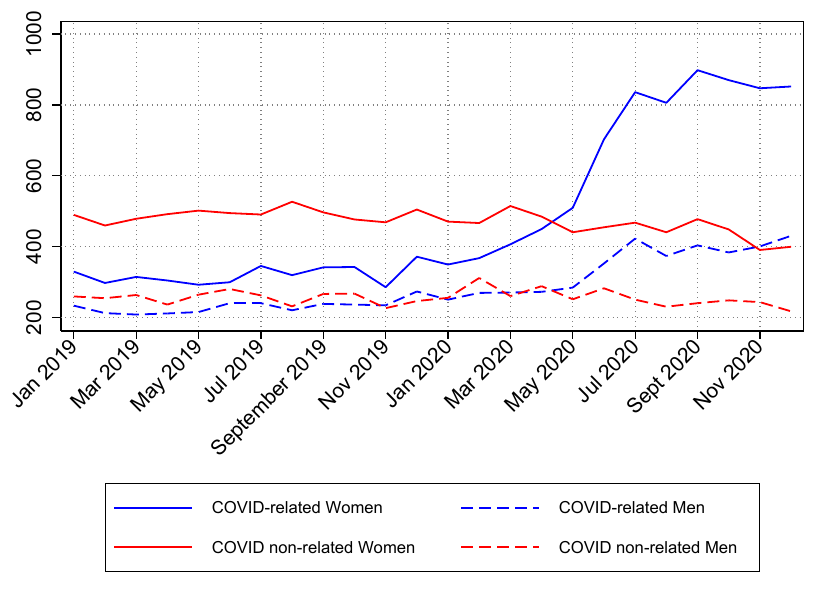}
        \caption{Middle Only}
   \end{subfigure}
    \caption{Monthly sample numerosity of publications by women and men as (a) first authors, (b) last authors, (c) middle authors, among COVID-related and non-related.
    }
    \label{fig:twoway_1}
  \end{figure}

Figure \ref{fig:twoway_1} shows the observed monthly number of COVID-related and COVID non-related publications by women and men authors in key authorship position or as middle authors.
Considering the key authorship positions of first and last, it is immediaty clear that men presence has risen dramatically in 2020, taking over the bulk of the publications that matter to the new topic, while this does not happen among the COVID non-related publications. On the other hand, the number of papers with women as middle authors \emph{only} -- when the key authors are men -- increases far more than their male counterparts. 

\begin{figure}
\centering
 \begin{subfigure}[t]{\textwidth}
    \includegraphics[width=\textwidth]{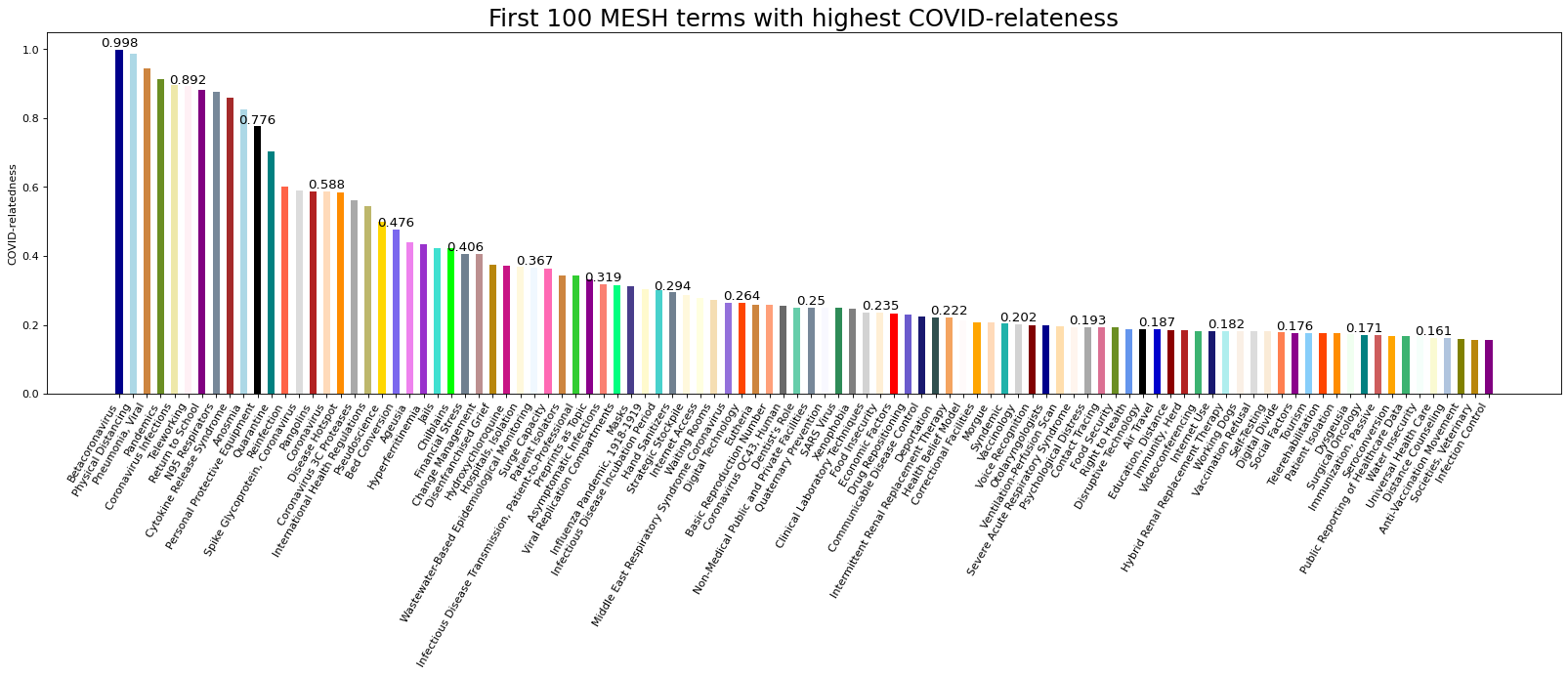}
    \caption{}
    \label{fig:treat_control_plots_1}
\end{subfigure}
 \begin{subfigure}[t]{\textwidth}
    \includegraphics[width=\textwidth]{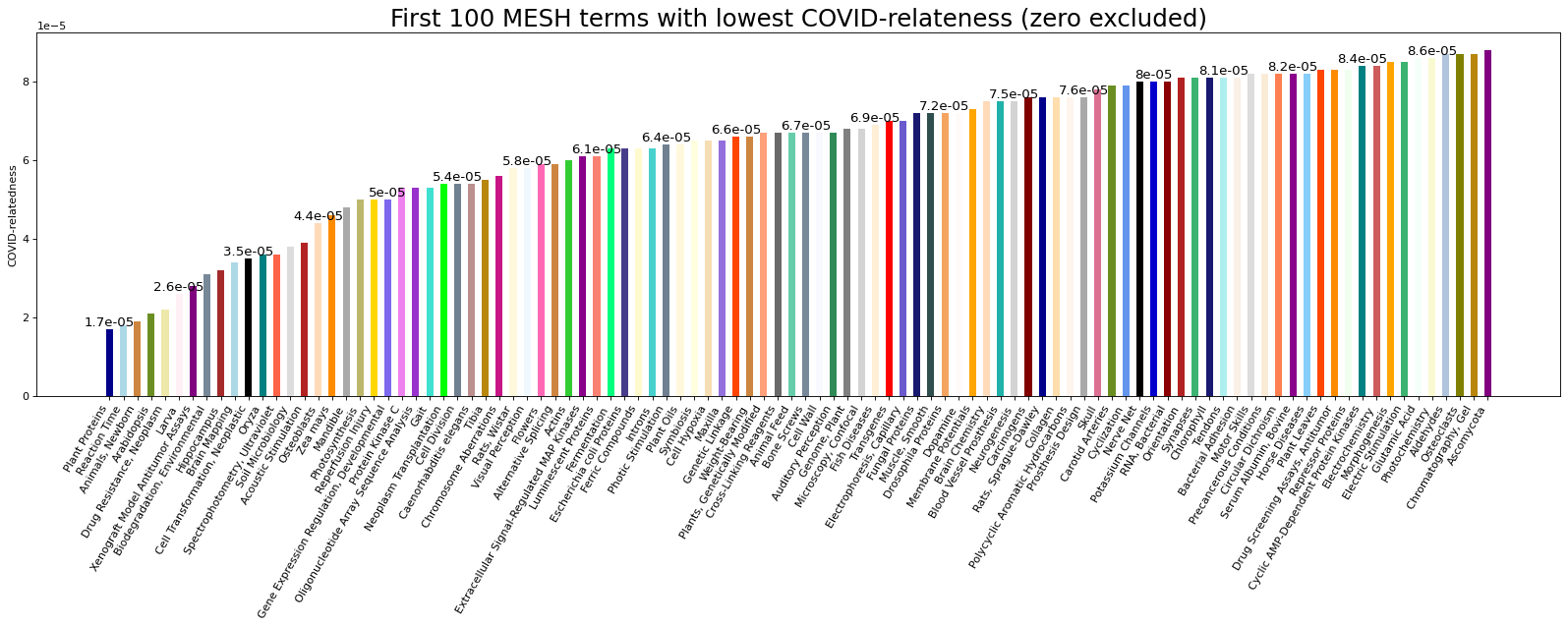}
    \caption{}
    \label{fig:treat_control_plots_1}
\end{subfigure}
       \caption{Barplots reporting (a) First 100 MESH terms with highest COVID-relatedness, and (b) First 100 MESH terms with lowest COVID-relateness (MeSH terms with zero COVID-relatedness excluded).}
    \label{fig:fig_2app}
\end{figure}

Figure \ref{fig:fig_2app} reports the first 100 MeSH terms in terms of highest COVID-relatedness values, as well as the lowest (zeros excluded). We identify 620 unique COVID-related MeSH terms and 5549 COVID non-related MeSH terms.

\begin{table}[H] \centering \renewcommand*{\arraystretch}{1.1}\caption{Summary Statistics of paper level characteristics.\label{descr_paper}}
\begin{tabular}{lrrrrrrr}
\hline
\hline &\multicolumn{4}{r}{\textbf{Year 2019}} \\
\hline
Variable & N & Mean & Std. Dev. & Min & Pctl. 25 & Pctl. 75 & Max \\ 
\hline
COVID-relatedness & 39709 & 0.028 & 0.053 & 0 & 0 & 0.044 & 1 \\ 
trial & 39709 & 0.039 & 0.19 & 0 & 0 & 0 & 1 \\ 
Has Grant & 39709 & 0.23 & 0.42 & 0 & 0 & 0 & 1 \\ 
Number of Male in team & 39709 & 3.6 & 2.9 & 0 & 2 & 5 & 63 \\ 
Number of Female in team & 39709 & 2.7 & 2.3 & 0 & 1 & 4 & 48 \\ 
Number of Unknowns in team & 39709 & 0.4 & 0.94 & 0 & 0 & 0 & 22 \\ 
JI & 39709 & 3.9 & 4.2 & 0 & 2 & 4.5 & 256 \\ 
JI - Med & 39709 & 0.48 & 0.5 & 0 & 0 & 1 & 1 \\ 
Has New Grant  & 39709 & 0 & 0 & 0 & 0 & 0 & 0 \\ 
Pre-Existing Grant & 39709 & 0.23 & 0.42 & 0 & 0 & 0 & 1 \\ 
First Female Author& 39709 & 0.48 & 0.5 & 0 & 0 & 1 & 1 \\ 
Last Female Author & 39709 & 0.35 & 0.48 & 0 & 0 & 1 & 1 \\ 
Number of authors within a team & 39709 & 6.7 & 4.3 & 3 & 4 & 8 & 97 \\ 
First and Last Female & 39709 & 0.2 & 0.4 & 0 & 0 & 0 & 1 \\ 
Middle Only  & 39709 & 0.25 & 0.43 & 0 & 0 & 0 & 1 \\ 
Also Middle & 39709 & 0.74 & 0.44 & 0 & 0 & 1 & 1 \\ 
Any Female & 39709 & 0.87 & 0.33 & 0 & 1 & 1 & 1\\ 
\hline
\hline &\multicolumn{4}{r}{\textbf{Year 2020}} \\
\hline
Variable & N & Mean & Std. Dev. & Min & Pctl. 25 & Pctl. 75 & Max \\ 
\hline
COVID-relatedness  & 51771 & 0.24 & 0.39 & 0 & 0 & 0.14 & 1 \\ 
trial & 51771 & 0.027 & 0.16 & 0 & 0 & 0 & 1 \\ 
Has Grant & 51771 & 0.2 & 0.4 & 0 & 0 & 0 & 1 \\ 
Number of Male in team & 51771 & 3.7 & 3.2 & 0 & 2 & 5 & 84 \\ 
Number of Female in team & 51771 & 2.7 & 2.5 & 0 & 1 & 4 & 54 \\ 
Number of Unknowns in team & 51771 & 0.42 & 1 & 0 & 0 & 0 & 29 \\ 
JI & 51771 & 4.3 & 5.2 & 0 & 2 & 4.8 & 256 \\ 
JI - Med & 51771 & 0.51 & 0.5 & 0 & 0 & 1 & 1 \\ 
Has New Grant & 51771 & 0.1 & 0.31 & 0 & 0 & 0 & 1 \\ 
Pre-Existing Grant& 51771 & 0.082 & 0.27 & 0 & 0 & 0 & 1 \\ 
First author female & 51771 & 0.46 & 0.5 & 0 & 0 & 1 & 1 \\ 
Last author female & 51771 & 0.34 & 0.47 & 0 & 0 & 1 & 1 \\ 
Number of authors within a team & 51771 & 6.8 & 4.9 & 3 & 4 & 8 & 96 \\ 
First and Last Female & 51771 & 0.19 & 0.39 & 0 & 0 & 0 & 1 \\ 
First or Last Female & 51771 & 0.61 & 0.49 & 0 & 0 & 1 & 1 \\ 
Middle Only & 51771 & 0.26 & 0.44 & 0 & 0 & 1 & 1 \\ 
Middle Also & 51771 & 0.74 & 0.44 & 0 & 0 & 1 & 1 \\ 
Any Female & 51771 & 0.87 & 0.33 & 0 & 1 & 1 & 1\\ 
\hline
\hline
\end{tabular}
\end{table}

\section{Country level Fixed Effect}
\begin{figure}[H]
\centering
\hspace{-2cm}
  \begin{subfigure}[t]{0.3\textwidth}
    \includegraphics[width=\textwidth]{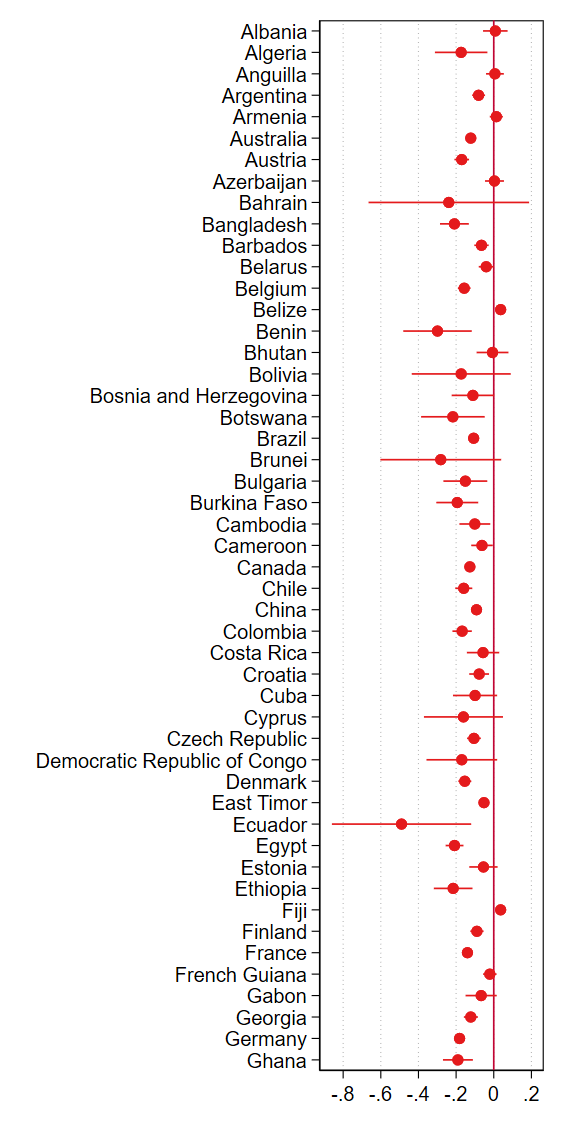}
    \caption{}
    \label{fig:cfe_1}
  \end{subfigure}
 \begin{subfigure}[t]{0.3\textwidth}
    \includegraphics[width=\textwidth]{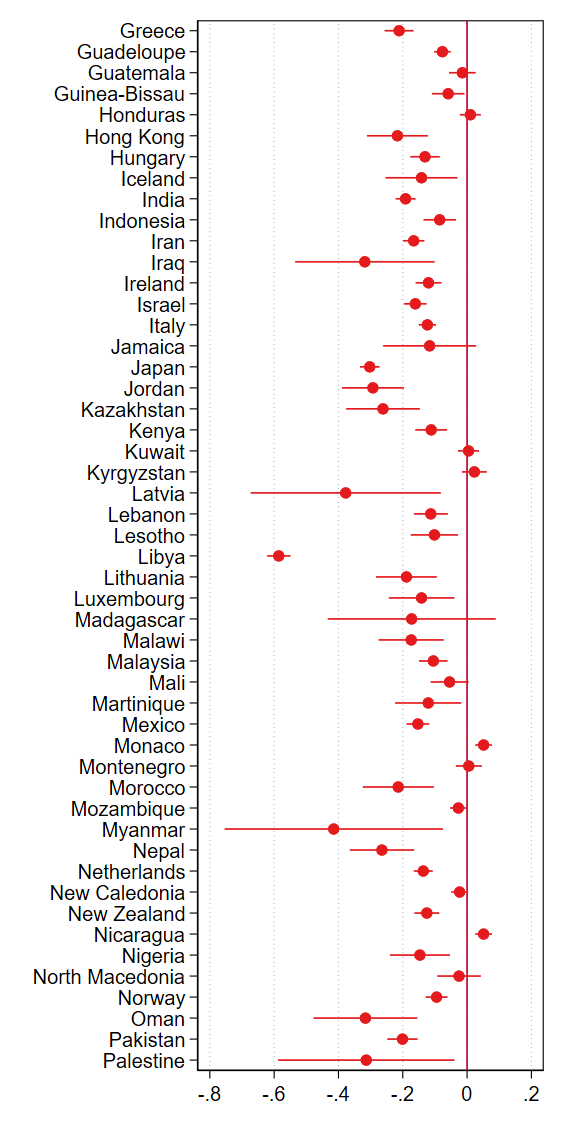}
    \caption{}
    \label{fig:cfe_2}
  \end{subfigure}
 \begin{subfigure}[t]{0.3\textwidth}
    \includegraphics[width=\textwidth]{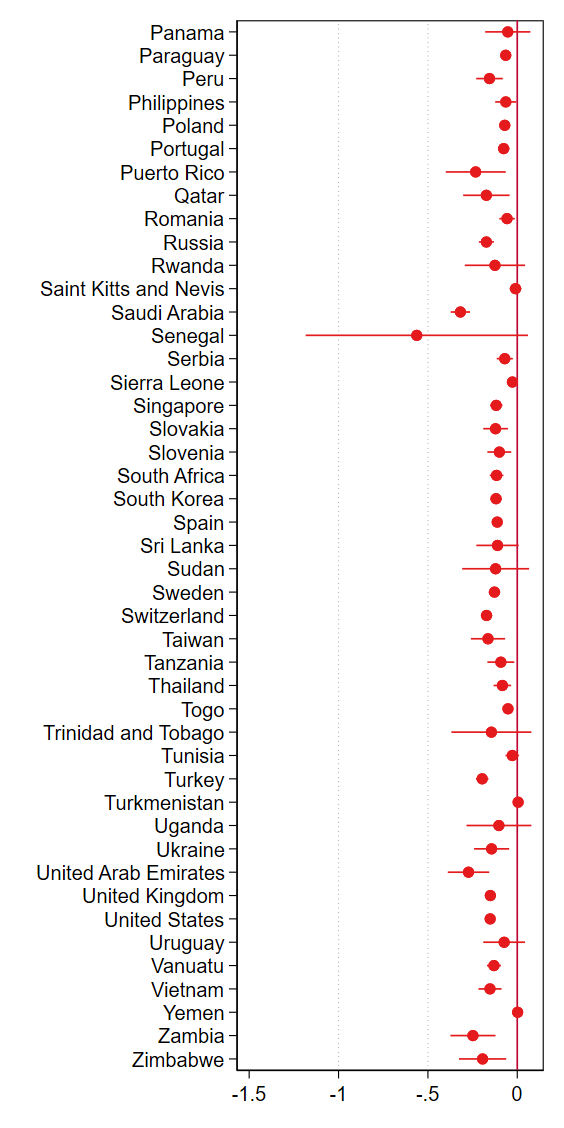}
    \caption{}
    \label{fig:cfe_4}
  \end{subfigure}
  \caption{Coefficient estimates of Country Fixed Effects, along with 95\% confidence intervals, of share of women at any authorship position; we refer to the country of the majority of the team.}
  \label{fig:fig3a}
\end{figure}

\begin{figure}[H]
\centering
\hspace{-2cm}
  \begin{subfigure}[t]{0.3\textwidth}
    \includegraphics[width=\textwidth]{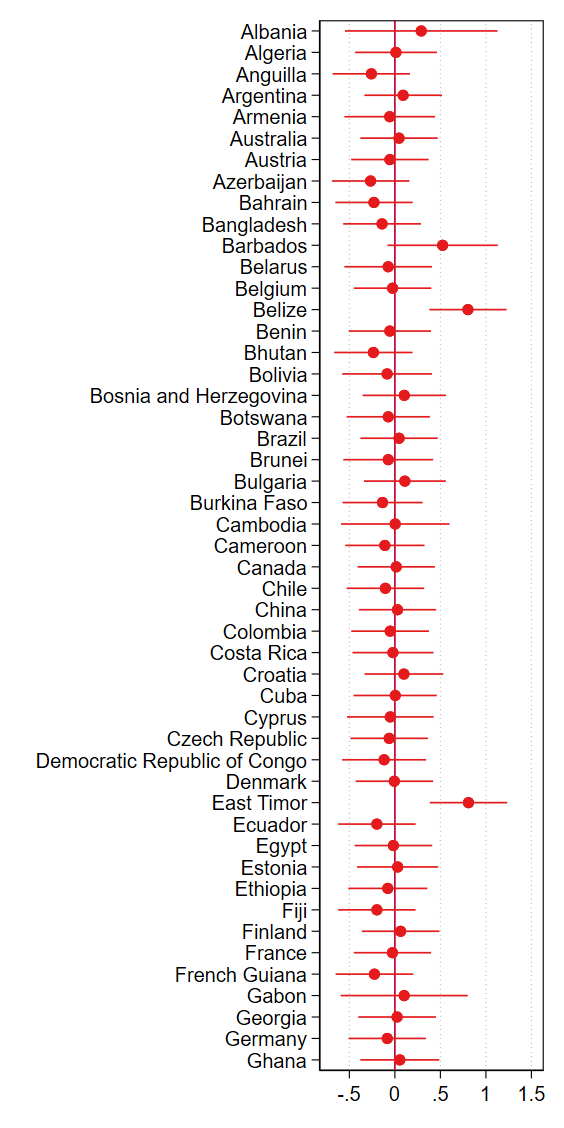}
    \caption{}
    \label{fig:cfe_1}
  \end{subfigure}
 \begin{subfigure}[t]{0.3\textwidth}
    \includegraphics[width=\textwidth]{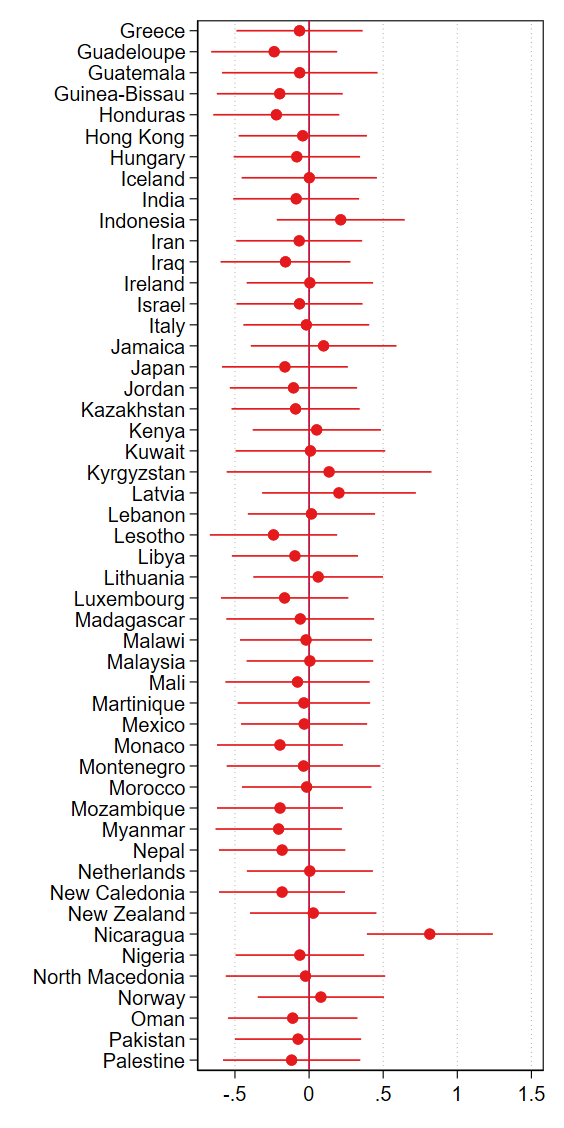}
    \caption{}
    \label{fig:cfe_2}
  \end{subfigure}
 \begin{subfigure}[t]{0.3\textwidth}
    \includegraphics[width=\textwidth]{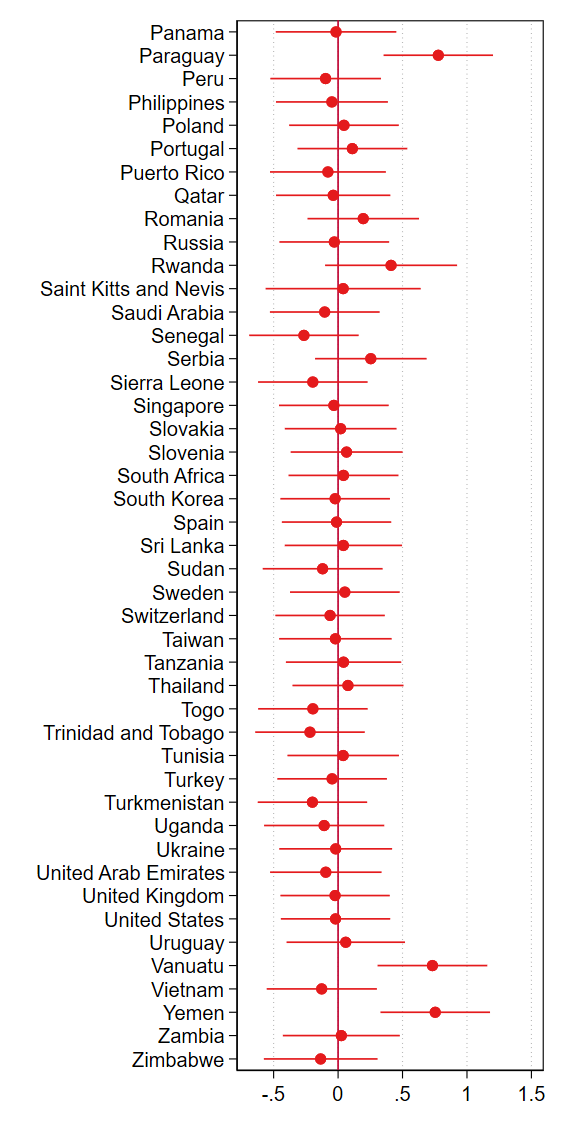}
    \caption{}
    \label{fig:cfe_4}
  \end{subfigure}
  \caption{Coefficient estimates of Country Fixed Effects, along with 95\% confidence intervals, of share of women at first and last authorship position; we refer to the country of the majority of the team.}
  \label{fig:fig3b}
\end{figure}

\begin{figure}[H]
\centering
\hspace{-2cm}
  \begin{subfigure}[t]{0.3\textwidth}
    \includegraphics[width=\textwidth]{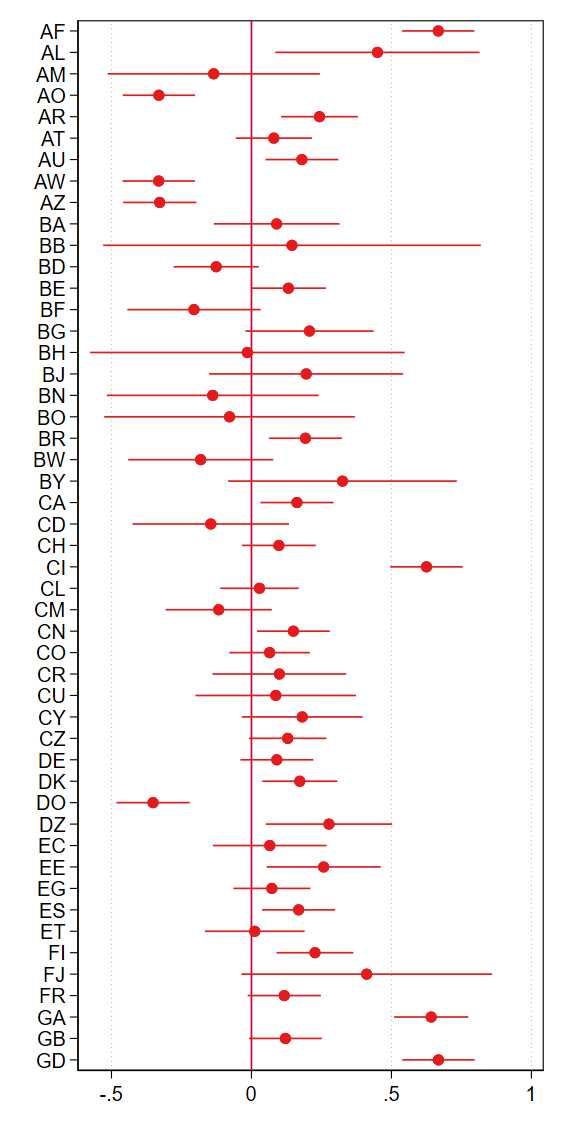}
    \caption{}
    \label{fig:cfe_1}
  \end{subfigure}
 \begin{subfigure}[t]{0.3\textwidth}
    \includegraphics[width=\textwidth]{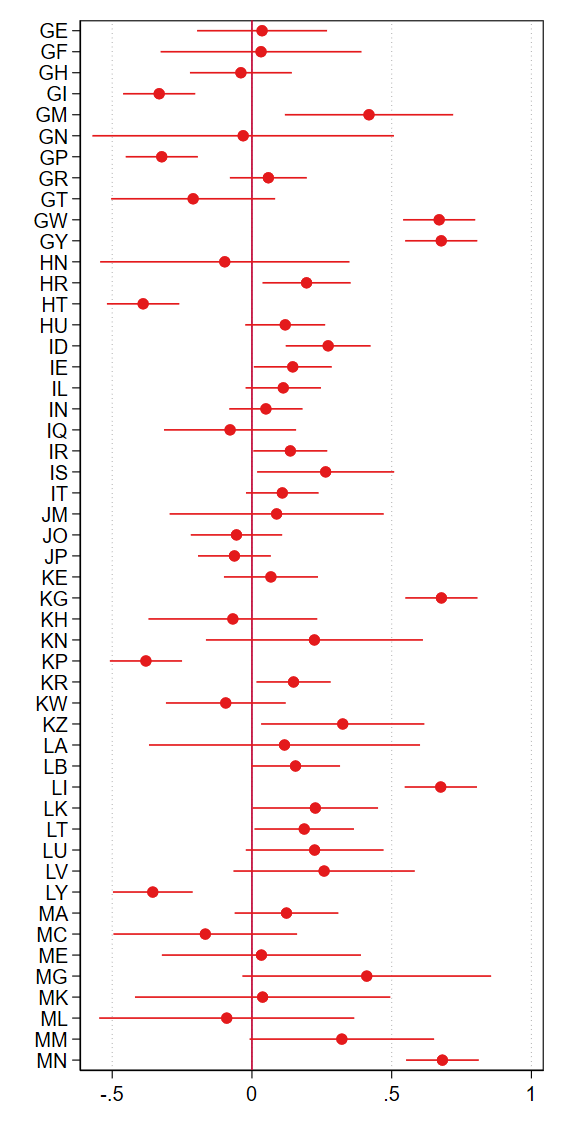}
    \caption{}
    \label{fig:cfe_2}
  \end{subfigure}
 \begin{subfigure}[t]{0.3\textwidth}
    \includegraphics[width=\textwidth]{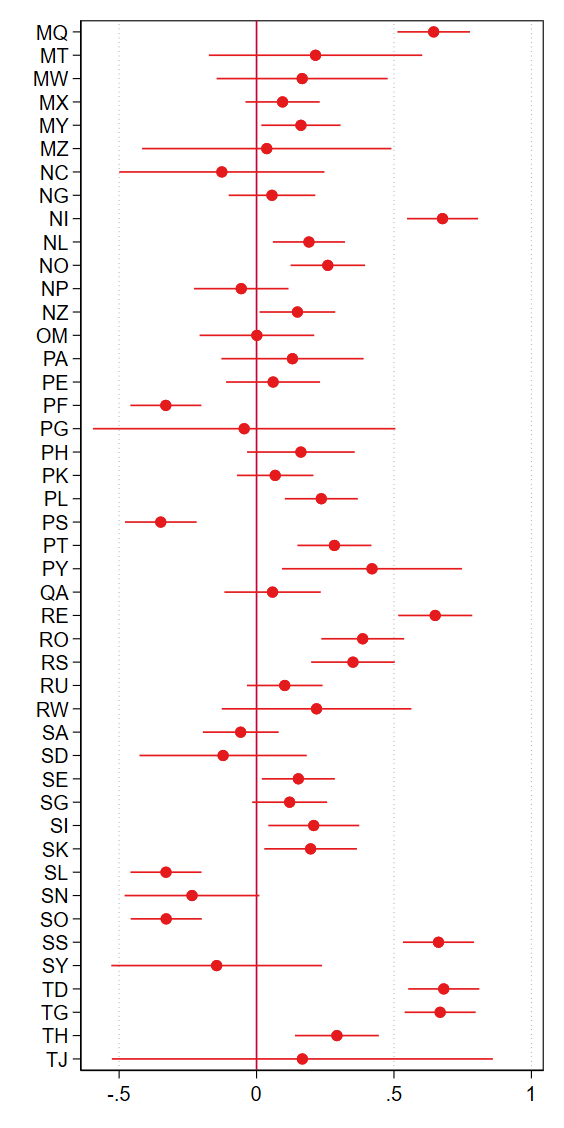}
    \caption{}
    \label{fig:cfe_4}
  \end{subfigure}
  \caption{Coefficient estimates of Country Fixed Effects, along with 95\% confidence intervals, of share of women as first authors; we refer to the country of the first (last) author. }
  \label{fig:fig3c}
\end{figure}

\begin{figure}[H]
\centering
\hspace{-2cm}
  \begin{subfigure}[t]{0.3\textwidth}
    \includegraphics[width=\textwidth]{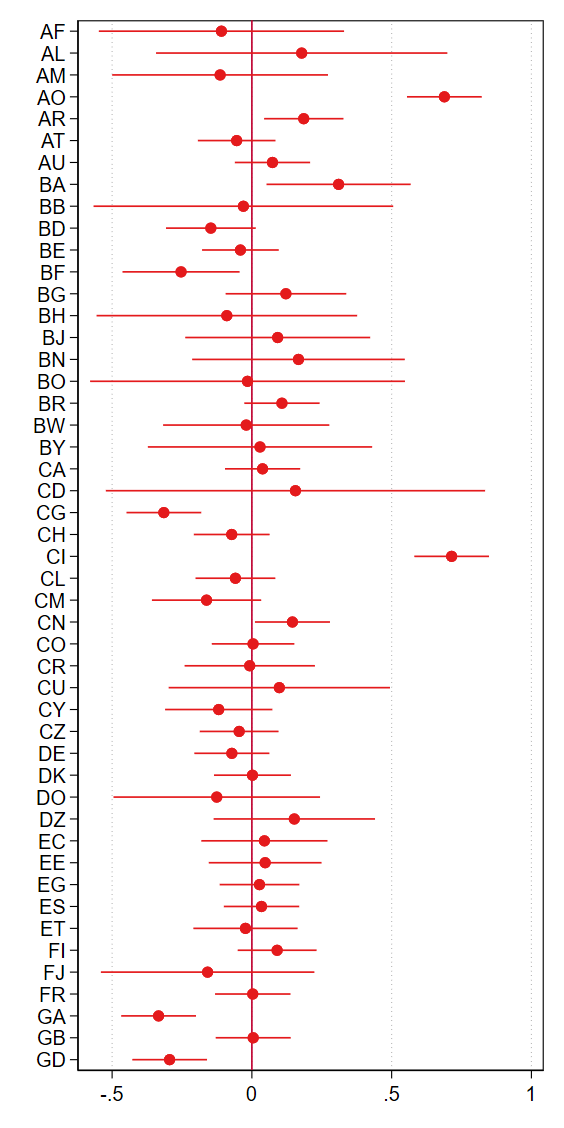}
    \caption{}
    \label{fig:cfe_1}
  \end{subfigure}
 \begin{subfigure}[t]{0.3\textwidth}
    \includegraphics[width=\textwidth]{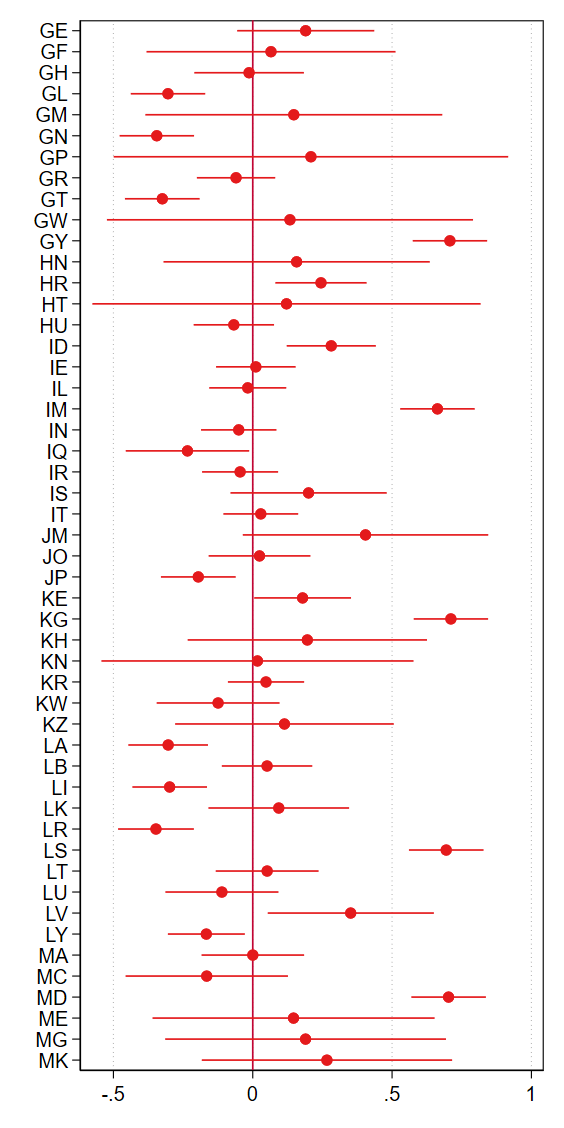}
    \caption{}
    \label{fig:cfe_2}
  \end{subfigure}
 \begin{subfigure}[t]{0.3\textwidth}
    \includegraphics[width=\textwidth]{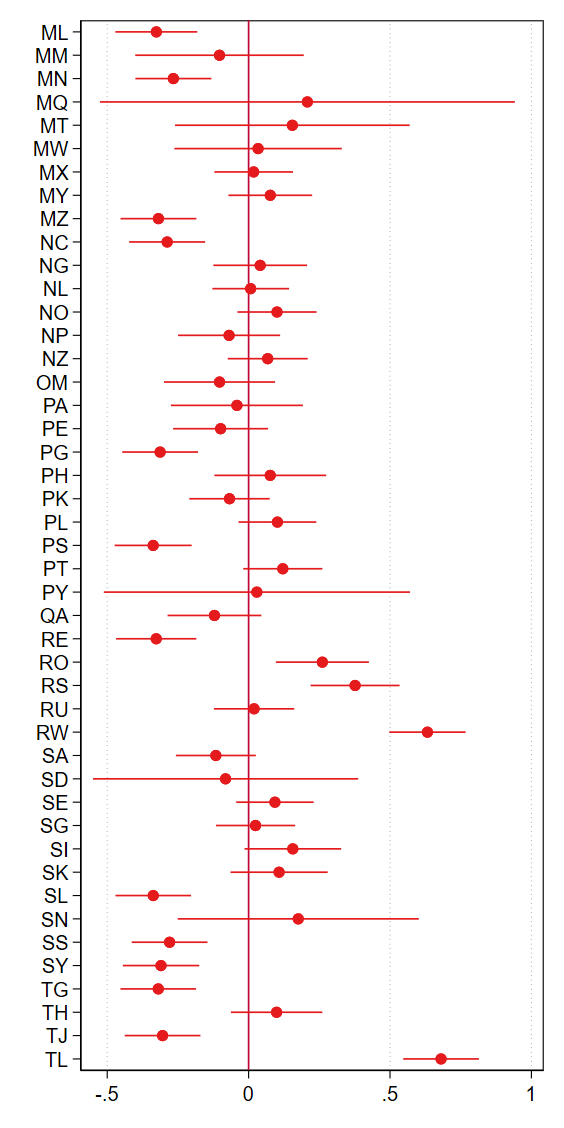}
    \caption{}
    \label{fig:cfe_4}
  \end{subfigure}
  \caption{Coefficient estimates of Country Fixed Effects, along with 95\% confidence intervals, of share of women as last authors; we refer to the country of the last author. For all other outcomes, we refer to the country of the majority of the team.}
  \label{fig:fig3d}
\end{figure}

\begin{figure}[H]
\centering
\hspace{-2cm}
  \begin{subfigure}[t]{0.3\textwidth}
    \includegraphics[width=\textwidth]{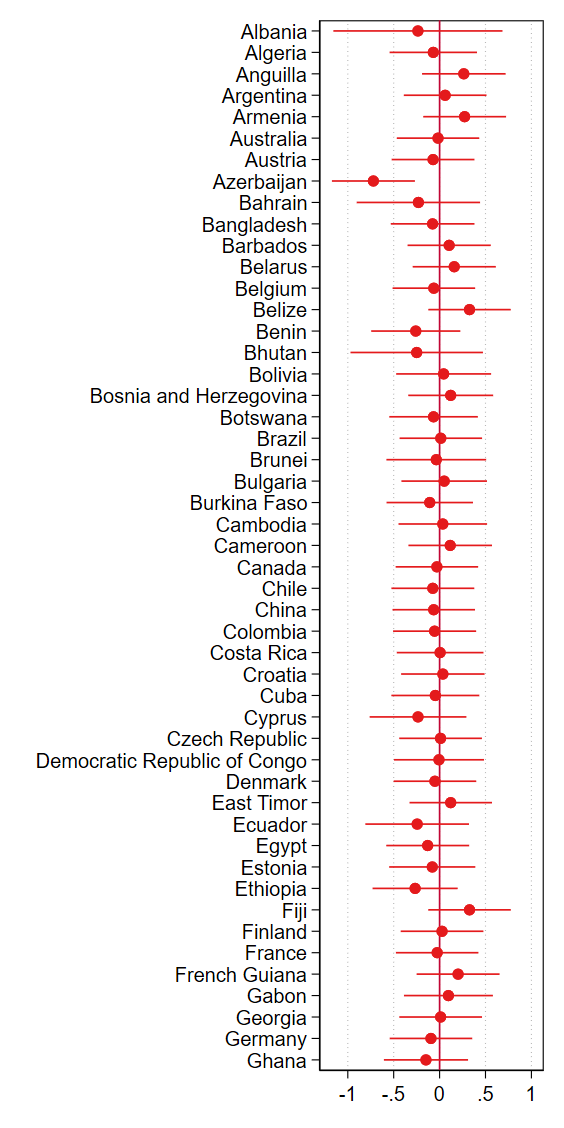}
    \caption{}
    \label{fig:cfe_1}
  \end{subfigure}
 \begin{subfigure}[t]{0.3\textwidth}
    \includegraphics[width=\textwidth]{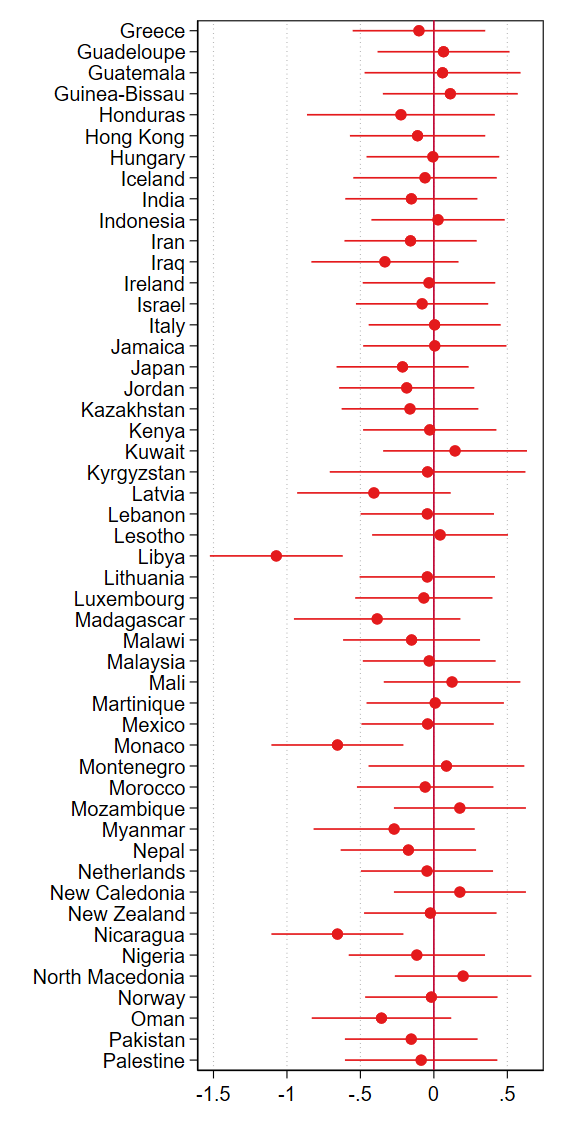}
    \caption{}
    \label{fig:cfe_2}
  \end{subfigure}
 \begin{subfigure}[t]{0.3\textwidth}
    \includegraphics[width=\textwidth]{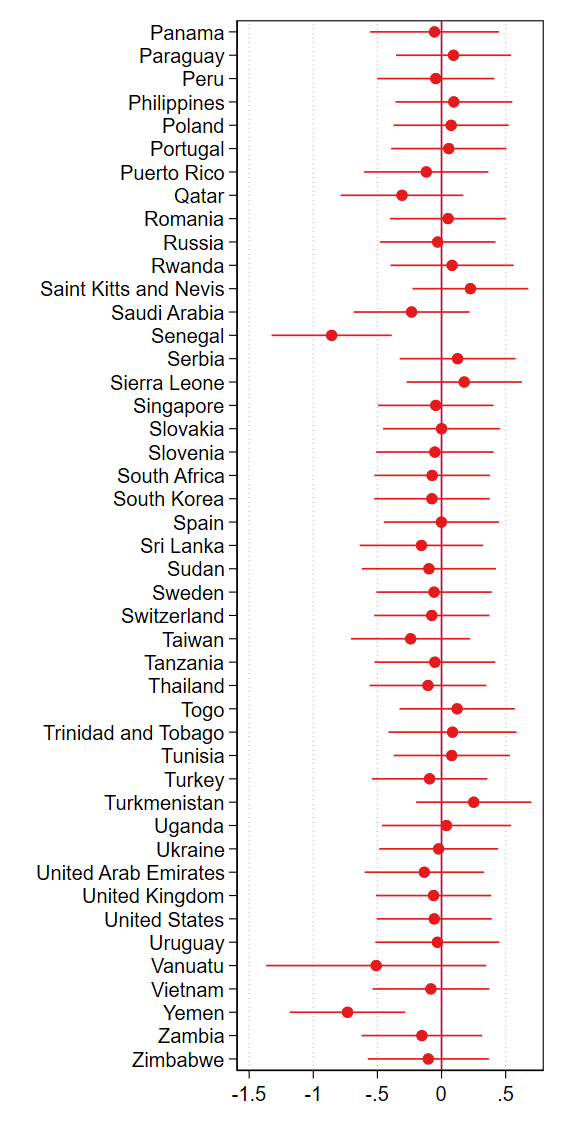}
    \caption{}
    \label{fig:cfe_4}
  \end{subfigure}
  \caption{Coefficient estimates of Country Fixed Effects, along with 95\% confidence intervals, of share of women as middle authors; we refer to the country of the majority of the team.}
  \label{fig:fig3e}
\end{figure}

\begin{figure}[H]
\centering
\hspace{-2cm}
  \begin{subfigure}[t]{0.3\textwidth}
    \includegraphics[width=\textwidth]{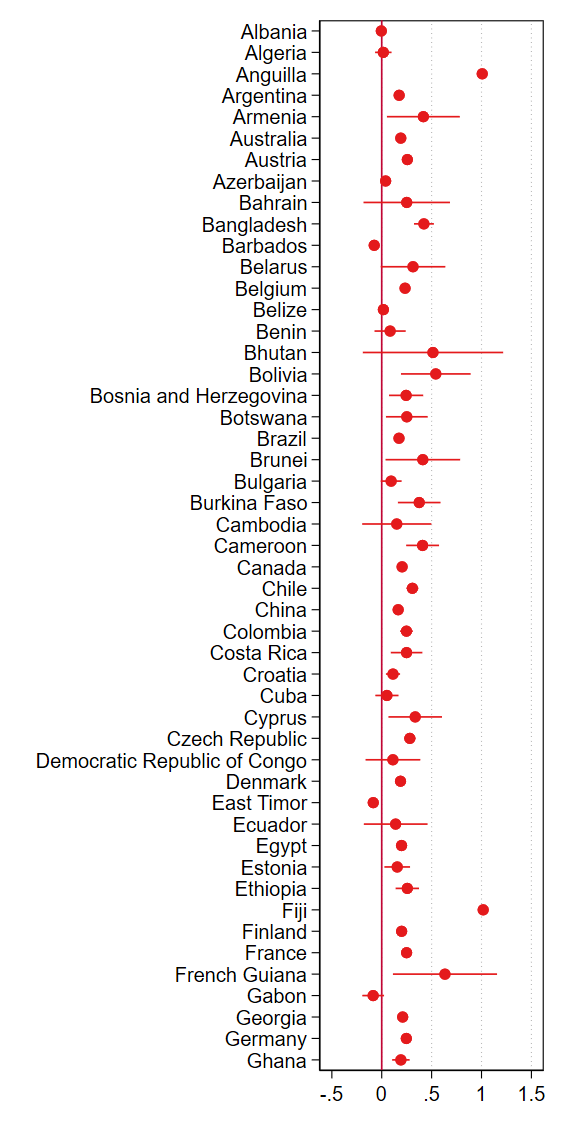}
    \caption{}
    \label{fig:cfe_1}
  \end{subfigure}
 \begin{subfigure}[t]{0.3\textwidth}
    \includegraphics[width=\textwidth]{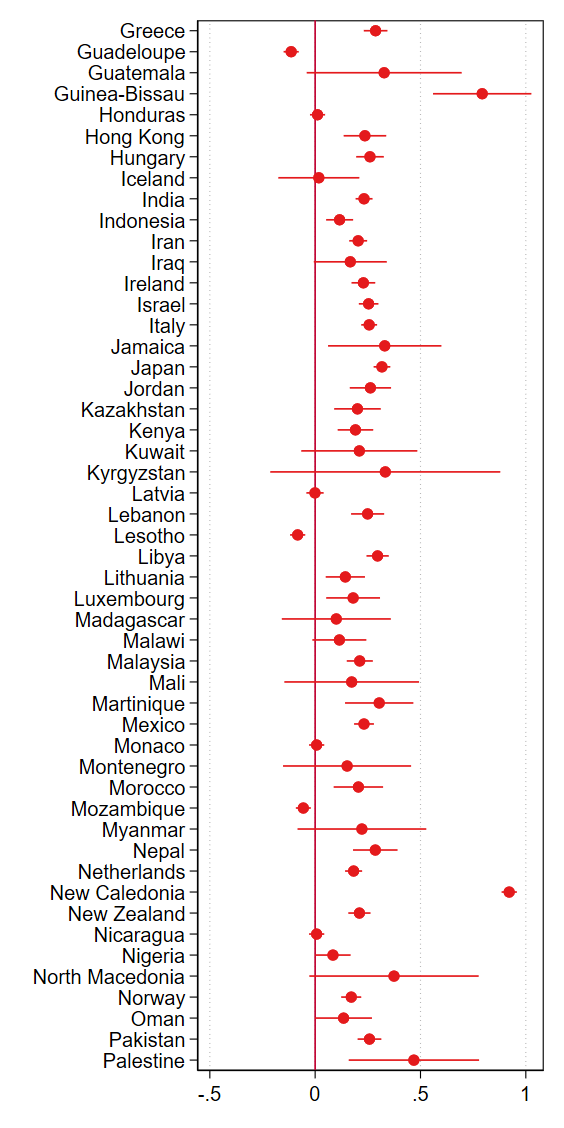}
    \caption{}
    \label{fig:cfe_2}
  \end{subfigure}
 \begin{subfigure}[t]{0.3\textwidth}
    \includegraphics[width=\textwidth]{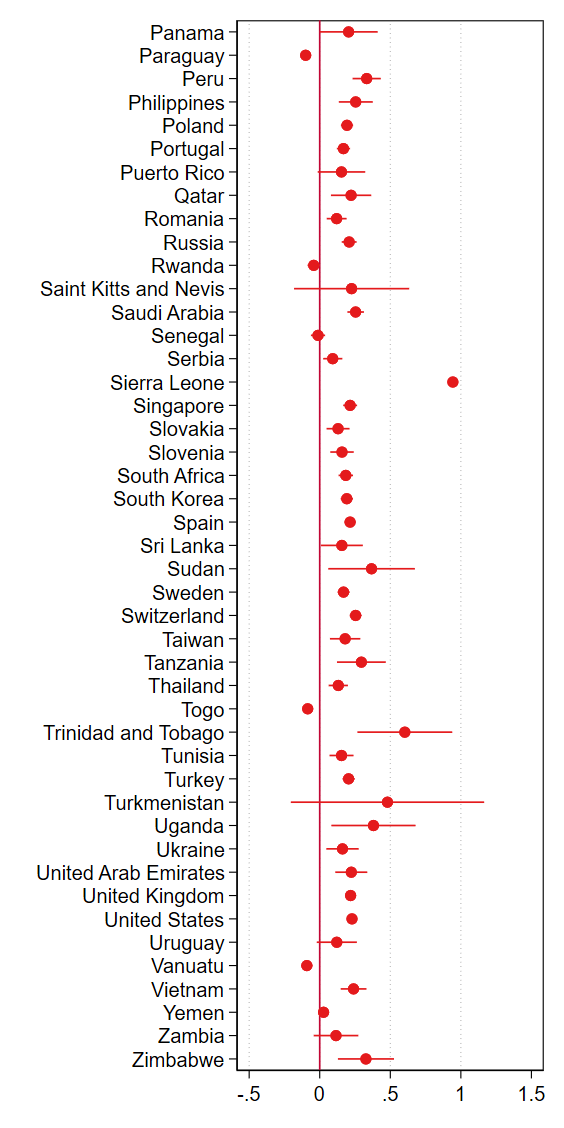}
    \caption{}
    \label{fig:cfe_4}
  \end{subfigure}
  \caption{Coefficient estimates of Country Fixed Effects, along with 95\% confidence intervals, of share of women as middle authors \emph{only} -- i.e when in team with lame key authors; we refer to the country of the majority of the team.}
  \label{fig:fig3f}
\end{figure}

In Figures \ref{fig:fig3a}-\ref{fig:fig3f}, we report the coefficient estimates of equation (1) of the main text for the country level fixed effects. When considering first (last) female authorship as outcome, we take the country of affiliation of the first (last) author. For all other authorship outcomes, we refer to the effects of the country of affiliation of the majority of the members of the publishing team.

\section{Robustness Checks} 
\subsection{Parallel Trends}
The key identification assumption of a Diff-in-Diff model is given by the \emph{parallel trends} assumptions. Under parallel trends, 
there are no systematic differences in outcome between COVID non-related and COVID-related scientific publications before the treatment (Kahn-Lang and Lang, 2020) 
The outcome of the COVID non-related group, therefore, becomes a suitable \emph{counterfactual} of the outcome for those COVID-related. Unfortunately, the parallel trends cannot be verified empirically as it relies on unobserved quantities such as counterfactuals. It becomes fundamental to give a sound and logical explanation of the validity of the choice of a specific COVID non-related group, which should be shielded against potential \emph{selection bias}. 

\begin{figure}[H]
   \begin{subfigure}[t]{0.45\textwidth}
    \includegraphics[width=\textwidth]{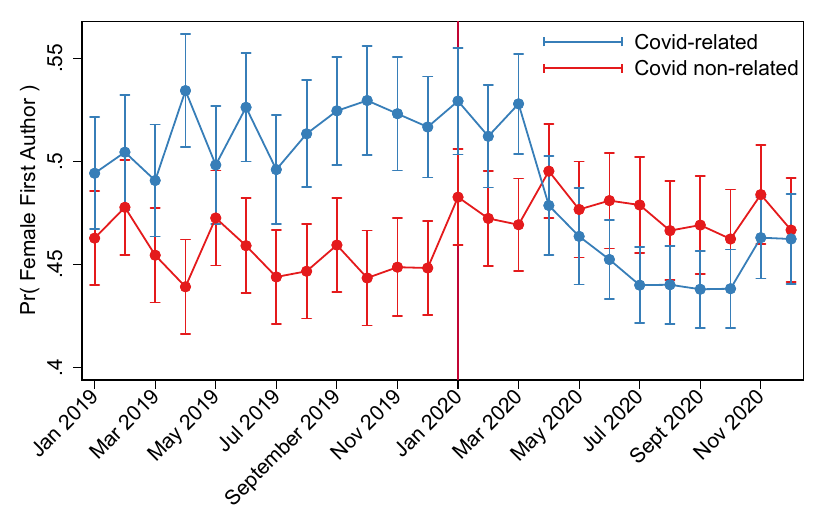}
    \caption{Female First Author}
    \label{fig:f3_b}
  \end{subfigure}
  \hfill
     \begin{subfigure}[t]{0.45\textwidth}
    \includegraphics[width=\textwidth]{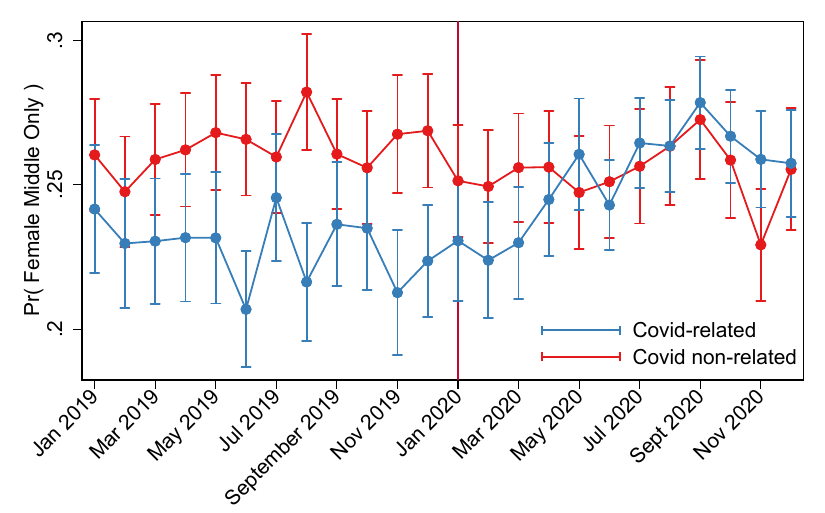}
    \caption{Middle Female Only}
    \label{fig:pt_all}
  \end{subfigure}
  \begin{subfigure}[t]{0.45\textwidth}
    \includegraphics[width=\textwidth]{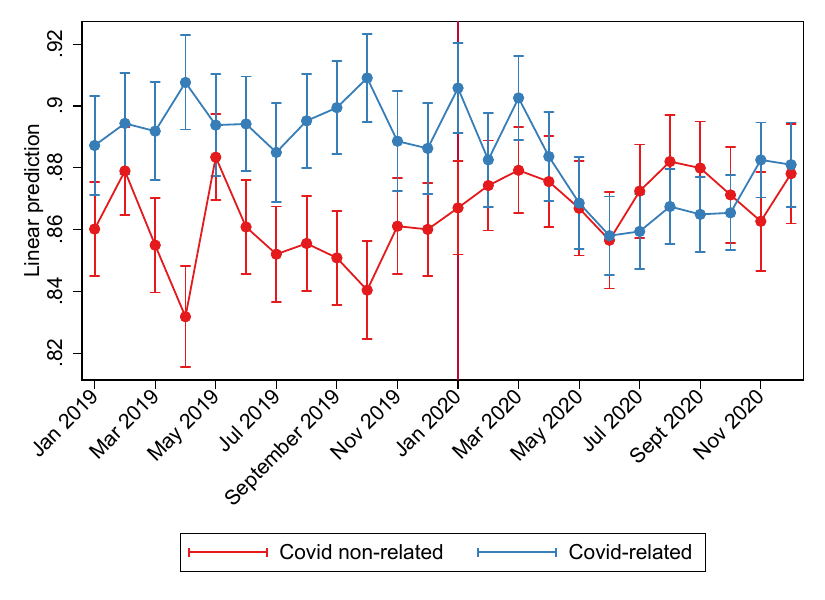}
    \caption{Female Author}
    \label{fig:f3_a}
  \end{subfigure}
 \begin{subfigure}[t]{0.45\textwidth}
    \includegraphics[width=\textwidth]{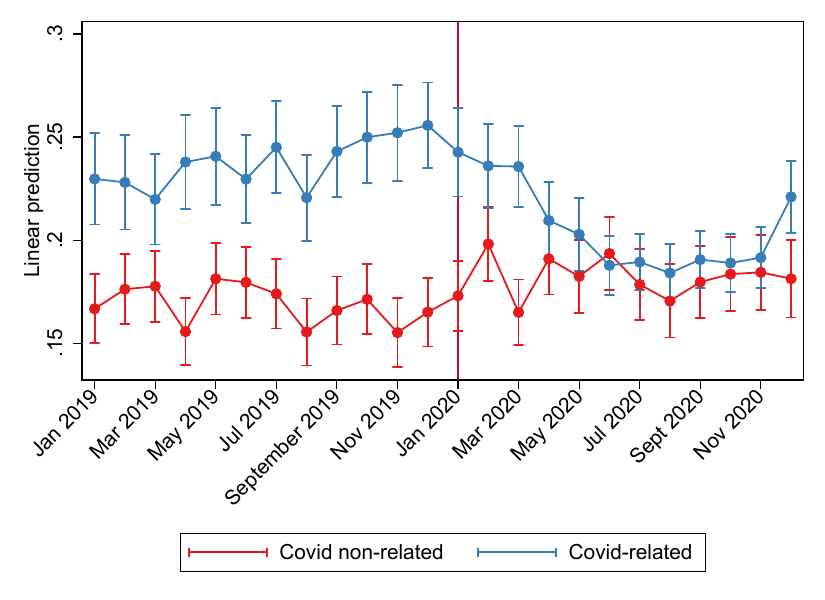}
    \caption{First and Last Female Author}
    \label{fig:f3_b}
  \end{subfigure}
 \begin{subfigure}[t]{0.45\textwidth}
    \includegraphics[width=\textwidth]{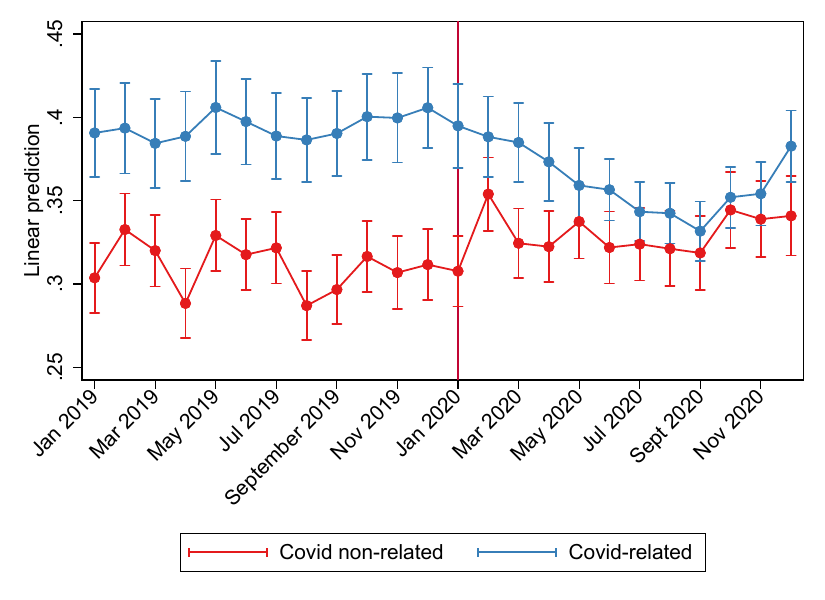}
    \caption{Female Last Author}
    \label{fig:pt_all}
  \end{subfigure}
    \hfill
   \begin{subfigure}[t]{0.45\textwidth}
    \includegraphics[width=\textwidth]{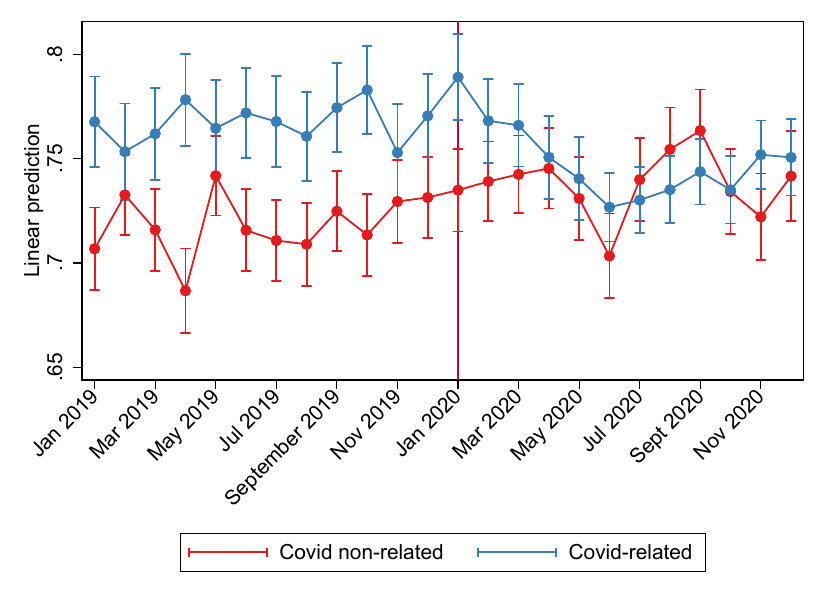}
    \caption{Middle Female Authorship}
    \label{fig:pt_all}
  \end{subfigure}
  \caption{Monthly linear predictions for female authorship at (a) any position, (b)  middle female authors \emph{only}, (c) both first and last authors, (d) first authors only, (e) last authors only, (f) middle female authors, among COVID-related and non-related publications.}
  \label{fig:pt}
\end{figure}

Common practice is to consider the estimated difference in COVID-related and COVID non-related outcomes prior to the treatment.
In Figure \ref{fig:pt}, we show the predicted probabilities of all outcomes for COVID-related and COVID non-related over time by estimating the model in equation (1) of the main text with monthly time indicators instead of yearly time dummies. The plots show no systematic difference in outcomes between COVID-related and COVID non-related research fields before 2020. After January 2020, we see a marked down-warding trend for aggregated female authorship (c) and for first and last authors (a-e). Instead, the share of female authors in \emph{middle} positions increases over time during 2020 among the COVID-related, but only when participating in publications with male first and last authors (b). Otherwise, we see that overall female participation as middle authors is decreasing (f).

\subsection{Clustered Standard Errors}
In Tables \ref{paper_2_odd}-\ref{paper_3_even}, we report the estimates of equation (1), with and without paper level COVID non-related, with clustered standard at the (i) country level, (ii) MeSH term level, (iii) country-year level and (iv) MeSH term-year level. The results confirm those obtained of the baseline model with White robust standard errors. 

\begin{table}[H]\centering
\def\sym#1{\ifmmode^{#1}\else\(^{#1}\)\fi}
\caption{DID Regression at paper level, with clustered standard errors at the country level. We include country fixed effects (omitted) of the majority of the team for Female Author, First and Last Female Authors, Middle Female Authorship and Middle Female Only; country fixed effects (omitted) of the first (last) author for regression on First (Last) Female Author.  \label{paper_2_odd}\\ }
\resizebox{\textwidth}{!}{ 
\begin{tabular}{c c c c c c c c}
\toprule  &\multicolumn{1}{c}{Female Author}&\multicolumn{1}{c}{First and Last Female}&\multicolumn{1}{c}{Female First Author}&\multicolumn{1}{c}{Female Last Author}&\multicolumn{1}{c}{Middle Female Authorship}&\multicolumn{1}{c}{Middle Female Only}\\
\toprule 
\multicolumn{1}{c}{Variables}&\multicolumn{1}{c}{(1)}&\multicolumn{1}{c}{(2)}&\multicolumn{1}{c}{(3)}&\multicolumn{1}{c}{(4)}&\multicolumn{1}{c}{(5)}&\multicolumn{1}{c}{(6)}\\
\midrule
year=2020           &      0.0147\sym{***}&      0.0131\sym{***}&      0.0212\sym{***}&      0.0189\sym{***}&      0.0177\sym{***}&    -0.00936\sym{*}  \\
                    &      (4.36)         &      (3.94)         &      (4.31)         &      (4.02)         &      (3.76)         &     (-2.52)         \\
\addlinespace
COVID-related             &      0.0367\sym{***}&      0.0690\sym{***}&      0.0580\sym{***}&      0.0864\sym{***}&      0.0483\sym{***}&     -0.0347\sym{***}\\
                    &      (3.91)         &      (4.45)         &      (3.69)         &      (4.25)         &      (3.55)         &     (-3.92)         \\
\addlinespace
year=2020 $\times$ COVID-related&     -0.0350\sym{***}&     -0.0485\sym{***}&     -0.0699\sym{***}&     -0.0570\sym{***}&     -0.0382\sym{***}&      0.0357\sym{***}\\
                    &     (-9.15)         &     (-8.78)         &     (-9.08)         &     (-7.14)         &     (-7.53)         &      (5.80)         \\
\addlinespace
\textit{N Authors}            &      0.0124\sym{***}&    -0.00217\sym{***}&   -0.000642         &    -0.00355\sym{***}&      0.0278\sym{***}&      0.0144\sym{***}\\
                    &     (20.99)         &     (-4.13)         &     (-1.09)         &     (-7.08)         &     (18.89)         &     (16.03)         \\
\addlinespace
trial               &      0.0257\sym{***}&     0.00556         &     0.00590         &      0.0183         &      0.0571\sym{***}&     0.00646         \\
                    &      (5.21)         &      (0.65)         &      (0.62)         &      (1.93)         &      (7.12)         &     (0.99)         \\
\addlinespace
Pre-existing Grant      &      0.0420\sym{***}&      0.0347\sym{***}&      0.0581\sym{***}&      0.0417\sym{***}&      0.0542\sym{***}&     -0.0212\sym{***}\\
                    &      (6.45)         &      (9.15)         &      (9.61)         &      (4.91)         &      (5.19)         &     (-5.13)         \\
\addlinespace
Constant            &       0.911\sym{***}&       0.210\sym{***}&       0.325\sym{***}&       0.307\sym{***}&       0.573\sym{***}&     -0.0504\sym{***}\\
                    &     (97.44)         &     (15.63)         &     (43.09)         &     (37.97)         &     (43.61)         &     (-6.25)         \\
\midrule
Observations        &       89530         &       89530         &       83263         &       82552         &       89530         &       89530         \\
\midrule
Country FEs & Majority  & Majority  & First & Last  & Majority  & Majority \\
\midrule
\emph{Country} Clustered SES & YES & YES & YES & YES & YES & YES \\
\bottomrule
\multicolumn{7}{l}{\footnotesize \textit{t} statistics in parentheses}\\
\multicolumn{7}{l}{\footnotesize \sym{*} \(p<0.05\), \sym{**} \(p<0.01\), \sym{***} \(p<0.001\)}\\
\end{tabular}
}
\end{table}

\begin{table}[H]\centering
\def\sym#1{\ifmmode^{#1}\else\(^{#1}\)\fi}
\caption{DID Regression at paper level, with clustered standard errors at the Mesh level. We include country fixed effects (omitted) of the majority of the team for Female Author, First and Last Female Authors, Middle Female Authorship and Middle Female Only; country fixed effects (omitted) of the first (last) author for regression on First (Last) Female Author.  \label{paper_2_even}\\ }
\resizebox{\textwidth}{!}{ 
\begin{tabular}{c c c c c c c}
\toprule  &\multicolumn{1}{c}{Female Author}&\multicolumn{1}{c}{First and Last Female}&\multicolumn{1}{c}{Female First Author}&\multicolumn{1}{c}{Female Last Author}&\multicolumn{1}{c}{Middle Female Authorship}&\multicolumn{1}{c}{Middle Female Only}\\
\toprule 
\multicolumn{1}{c}{Variables}&\multicolumn{1}{c}{(1)}&\multicolumn{1}{c}{(2)}&\multicolumn{1}{c}{(3)}&\multicolumn{1}{c}{(4)}&\multicolumn{1}{c}{(5)}&\multicolumn{1}{c}{(6)}\\
\midrule
year=2020           &      0.0147\sym{***}&      0.0131\sym{***}&      0.0212\sym{***}&      0.0189\sym{***}&      0.0177\sym{***}&    -0.00936\sym{*}  \\
                    &      (4.11)         &      (3.52)         &      (4.33)         &      (4.00)         &      (4.20)         &     (-2.28)         \\
\addlinespace
COVID-related             &      0.0367\sym{***}&      0.0690\sym{***}&      0.0580\sym{***}&      0.0864\sym{***}&      0.0483\sym{***}&     -0.0347\sym{***}\\
                    &      (4.38)         &      (6.82)         &      (4.56)         &      (7.78)         &      (4.86)         &     (-5.09)         \\
\addlinespace
year=2020 $\times$ COVID-related&     -0.0350\sym{***}&     -0.0485\sym{**} &     -0.0699\sym{***}&     -0.0570\sym{***}&     -0.0382\sym{**} &      0.0357\sym{***}\\
                    &     (-3.39)         &     (-3.28)         &     (-3.60)         &     (-3.90)         &     (-3.29)         &      (3.67)         \\
\addlinespace
\textit{N Authors}            &      0.0124\sym{***}&    -0.00217\sym{***}&   -0.000642         &    -0.00355\sym{***}&      0.0278\sym{***}&      0.0144\sym{***}\\
                    &     (28.83)         &     (-5.85)         &     (-1.41)         &     (-8.06)         &     (27.35)         &     (28.48)         \\
\addlinespace
trial               &      0.0257\sym{***}&     0.00556         &     0.00590         &      0.0183         &      0.0571\sym{***}&     0.00646         \\
                    &      (3.67)         &      (0.58)         &      (0.45)         &      (1.69)         &      (6.35)         &     (0.67)         \\
\addlinespace
Pre-existing Grant      &      0.0420\sym{***}&      0.0347\sym{***}&      0.0581\sym{***}&      0.0417\sym{***}&      0.0542\sym{***}&     -0.0212\sym{***}\\
                    &     (12.04)         &      (6.58)         &      (9.46)         &      (6.52)         &     (10.31)         &     (-4.34)         \\
\addlinespace
Constant            &       0.911\sym{***}&       0.210         &       0.325\sym{***}&       0.307\sym{***}&       0.573\sym{*}  &     -0.0504\sym{*}  \\
                    &     (58.20)         &      (0.94)         &      (4.94)         &      (4.73)         &      (2.49)         &     (-2.56)         \\
\midrule
Observations        &       89530         &       89530         &       83263         &       82552         &       89530         &       89530         \\
\midrule
Country FEs & Majority & Majority  & First & Last  & Majority  & Majority \\
\midrule
\emph{MeSH term} Clustered SES & YES & YES & YES & YES & YES & YES \\
\bottomrule
\multicolumn{7}{l}{\footnotesize \textit{t} statistics in parentheses}\\
\multicolumn{7}{l}{\footnotesize \sym{*} \(p<0.05\), \sym{**} \(p<0.01\), \sym{***} \(p<0.001\)}\\
\end{tabular}
}
\end{table}

\begin{table}[H]\centering
\def\sym#1{\ifmmode^{#1}\else\(^{#1}\)\fi}
\caption{Linear Probability DID Regression at paper level, with clustered standard errors at the country-year level. We include country fixed effects (omitted) of the majority of the team for Female Author, First and Last Female Authors, Middle Female Authorship and Middle Female Only; country fixed effects (omitted) of the first (last) author for regression on First (Last) Female Author. \label{paper_3_odd}\\ }
\resizebox{\textwidth}{!}{ 
\begin{tabular}{c c c c c c c}
\toprule  &\multicolumn{1}{c}{Female Author}&\multicolumn{1}{c}{First and Last Female}&\multicolumn{1}{c}{Female First Author}&\multicolumn{1}{c}{Female Last Author}&\multicolumn{1}{c}{Middle Female Authorship}&\multicolumn{1}{c}{Middle Female Only}\\
\toprule 
\multicolumn{1}{c}{Variables}&\multicolumn{1}{c}{(1)}&\multicolumn{1}{c}{(2)}&\multicolumn{1}{c}{(3)}&\multicolumn{1}{c}{(4)}&\multicolumn{1}{c}{(5)}&\multicolumn{1}{c}{(6)}\\
\midrule
year=2020           &      0.0147\sym{*}  &      0.0131         &      0.0212         &      0.0189         &      0.0177\sym{*}  &    -0.00936         \\
                    &      (2.19)         &      (1.27)         &      (1.91)         &      (1.62)         &      (2.05)         &     (-1.35)         \\
\addlinespace
COVID-related             &      0.0367\sym{***}&      0.0690\sym{***}&      0.0580\sym{***}&      0.0864\sym{***}&      0.0483\sym{***}&     -0.0347\sym{***}\\
                    &      (4.01)         &      (4.57)         &      (4.02)         &      (5.06)         &      (3.67)         &     (-4.01)         \\
\addlinespace
year=2020 $\times$COVID-related&     -0.0350\sym{**} &     -0.0485\sym{*}  &     -0.0699\sym{**} &     -0.0570\sym{*}  &     -0.0382\sym{*}  &      0.0357\sym{**} \\
                    &     (-2.83)         &     (-2.48)         &     (-3.32)         &     (-2.40)         &     (-2.31)         &      (2.83)         \\
\addlinespace
\textit{N Authors}            &      0.0124\sym{***}&    -0.00217\sym{***}&   -0.000642         &    -0.00355\sym{***}&      0.0278\sym{***}&      0.0144\sym{***}\\
                    &     (25.78)         &     (-5.14)         &     (-1.08)         &     (-7.32)         &     (23.21)         &     (19.28)         \\
\addlinespace
trial               &      0.0257\sym{***}&     0.00556         &     0.00590         &      0.0183\sym{*}  &      0.0571\sym{***}&     0.00646         \\
                    &      (3.93)         &      (0.66)         &      (0.54)         &      (2.12)         &      (6.86)         &      (0.86)         \\
\addlinespace
Pre-existing Grant       &      0.0420\sym{***}&      0.0347\sym{***}&      0.0581\sym{***}&      0.0417\sym{***}&      0.0542\sym{***}&     -0.0212\sym{***}\\
                    &      (8.01)         &      (6.51)         &      (9.88)         &      (6.52)         &      (7.05)         &     (-4.64)         \\
\addlinespace
Constant            &       0.911\sym{***}&       0.210         &       0.325\sym{***}&       0.307\sym{***}&       0.573\sym{**} &     -0.0504\sym{**} \\
                    &    (112.62)         &      (1.10)         &      (8.27)         &      (6.37)         &      (3.27)         &     (-3.19)         \\
\midrule
Observations        &       89530         &       89530         &       83263         &       82552         &       89530         &       89530         \\
\midrule
Country FEs & Majority & Majority  & First & Last  & Majority  & Majority \\
\bottomrule
\multicolumn{7}{l}{\footnotesize \textit{t} statistics in parentheses}\\
\multicolumn{7}{l}{\footnotesize \sym{*} \(p<0.05\), \sym{**} \(p<0.01\), \sym{***} \(p<0.001\)}\\
\end{tabular}
}
\end{table}

\begin{table}[H]\centering
\def\sym#1{\ifmmode^{#1}\else\(^{#1}\)\fi}
\caption{Linear Probability DID Regression at paper level, with clustered standard errors at the MeSH-year level. We include country fixed effects (omitted) of the majority of the team for Female Author, First and Last Female Authors, Middle Female Authorship and Middle Female Only; country fixed effects (omitted) of the first (last) author for regression on First (Last) Female Author. \label{paper_3_even}\\ }
\resizebox{\textwidth}{!}{ 
\begin{tabular}{c c c c c c c}
\toprule  &\multicolumn{1}{c}{Female Author}&\multicolumn{1}{c}{First and Last Female}&\multicolumn{1}{c}{Female First Author}&\multicolumn{1}{c}{Female Last Author}&\multicolumn{1}{c}{Middle Female Authorship}&\multicolumn{1}{c}{Middle Female Only}\\
\toprule 
\multicolumn{1}{c}{Variables}&\multicolumn{1}{c}{(1)}&\multicolumn{1}{c}{(2)}&\multicolumn{1}{c}{(3)}&\multicolumn{1}{c}{(4)}&\multicolumn{1}{c}{(5)}&\multicolumn{1}{c}{(6)}\\
\midrule
year=2020           &      0.0147\sym{*}  &      0.0131\sym{*}  &      0.0212\sym{**} &      0.0189\sym{**} &      0.0177\sym{**} &    -0.00936\sym{*}  \\
                    &      (2.56)         &      (2.52)         &      (2.78)         &      (3.00)         &      (2.69)         &     (-1.98)         \\
\addlinespace
COVID-related             &      0.0367\sym{***}&      0.0690\sym{***}&      0.0580\sym{***}&      0.0864\sym{***}&      0.0483\sym{***}&     -0.0347\sym{***}\\
                    &      (4.36)         &      (6.61)         &      (4.48)         &      (7.65)         &      (4.79)         &     (-5.01)         \\
\addlinespace
year=2020 $\times$COVID-related&     -0.0350\sym{**} &     -0.0485\sym{**} &     -0.0699\sym{**} &     -0.0570\sym{**} &     -0.0382\sym{*}  &      0.0357\sym{**} \\
                    &     (-2.62)         &     (-2.63)         &     (-2.92)         &     (-2.99)         &     (-2.45)         &      (2.97)         \\
\addlinespace
\textit{N Authors}            &      0.0124\sym{***}&    -0.00217\sym{***}&   -0.000642         &    -0.00355\sym{***}&      0.0278\sym{***}&      0.0144\sym{***}\\
                    &     (34.00)         &     (-6.45)         &     (-1.52)         &     (-8.63)         &     (29.22)         &     (31.02)         \\
\addlinespace
trial               &      0.0257\sym{***}&     0.00556         &     0.00590         &      0.0183         &      0.0571\sym{***}&     0.00646         \\
                    &      (3.80)         &      (0.64)         &      (0.49)         &      (1.79)         &      (6.79)         &      (0.74)         \\
\addlinespace
Pre-existing Grant       &      0.0420\sym{***}&      0.0347\sym{***}&      0.0581\sym{***}&      0.0417\sym{***}&      0.0542\sym{***}&     -0.0212\sym{***}\\
                    &     (12.63)         &      (7.26)         &     (10.42)         &      (7.27)         &     (10.83)         &     (-4.69)         \\
\addlinespace
Constant            &       0.911\sym{***}&       0.210         &       0.325\sym{***}&       0.307\sym{***}&       0.573\sym{*}  &     -0.0504\sym{**} \\
                    &     (60.11)         &      (0.94)         &      (5.02)         &      (4.76)         &      (2.50)         &     (-2.61)         \\
\midrule
Observations        &       89530         &       89530         &       83263         &       82552         &       89530         &       89530         \\
\midrule
Country FEs & Majority & Majority  & First & Last  & Majority  & Majority \\
\bottomrule
\multicolumn{7}{l}{\footnotesize \textit{t} statistics in parentheses}\\
\multicolumn{7}{l}{\footnotesize \sym{*} \(p<0.05\), \sym{**} \(p<0.01\), \sym{***} \(p<0.001\)}\\
\end{tabular}
}
\end{table}

\subsection{Changing Bandwidth in COVID-relatedness definition}\label{new_threshold}
Next we wish to assess whether the estimates of the DiD regression model in equation (1) of the main text are robust against the usage of different bandwidths of COVID-relatedness of paper's major MeSH term in the selection of the treatment group -- above the 90th percentile of the sample distribution of COVID-relatedness values -- and COVID non-related group of papers -- below the 10th percentile.
In order to avoid including papers with MeSH terms that are only partially related to the new research opportunity and that could have purposefully manipulated their relatedness to COVID-19 because of the high public interest, 
we  modify the definition of the percentile bandwidths by taking more extreme values of the distribution. In Tables \ref{tab1_cb} and \ref{tab2_cb}, we estimate the model in equation (1) of the main text, using as bandwidths (i) the 3rd and 97th percentiles and (ii) 1st and 99th percentiles. The results confirm the baseline estimates of the effect of interest ($year=2020$ $\times$ COVID-related). 

\begin{table}[H]\centering
\def\sym#1{\ifmmode^{#1}\else\(^{#1}\)\fi}
\caption{DiD regression model estimates changing definition of \emph{COVID-relatedness} (3rd - 97th percentiles of sample distribution of COVID-relatedness to identify control and treated units), with White standard errors. Country effects (omitted) for the majority of the team for Female Author, First and Last Female Authors, Middle Female Authorship and Middle Female Only; country fixed effects of the first (last) author for regression on First (Last) Female Author.\label{tab1_cb}}
\resizebox{\textwidth}{!}{ 
\begin{tabular}{lcccccc}
\toprule
                    \multicolumn{1}{c}{Variables}&\multicolumn{1}{c}{(1)}&\multicolumn{1}{c}{(2)}&\multicolumn{1}{c}{(3)}&\multicolumn{1}{c}{(4)}&\multicolumn{1}{c}{(5)}&\multicolumn{1}{c}{(6)}\\
                    &\multicolumn{1}{c}{Female Author}&\multicolumn{1}{c}{First and Last Female}&\multicolumn{1}{c}{Female First Author}&\multicolumn{1}{c}{Female Last Author}&\multicolumn{1}{c}{Middle Female Authorship}&\multicolumn{1}{c}{Middle Female Only}\\
\midrule
year=2020           &      0.0182\sym{***}&      0.0141\sym{**} &      0.0268\sym{***}&      0.0191\sym{***}&      0.0193\sym{***}&     -0.0107\sym{*}  \\
                    &      (4.72)         &      (3.28)         &      (4.61)         &      (3.51)         &      (3.98)         &     (-2.18)         \\
\addlinespace
COVID-related             &      0.0577\sym{***}&      0.0864\sym{***}&      0.0878\sym{***}&       0.108\sym{***}&      0.0697\sym{***}&     -0.0479\sym{***}\\
                    &     (13.98)         &     (15.61)         &     (12.68)         &     (16.10)         &     (12.74)         &     (-8.53)         \\
\addlinespace
year=2020 $\times$ COVID-related&     -0.0574\sym{***}&     -0.0694\sym{***}&      -0.107\sym{***}&     -0.0768\sym{***}&     -0.0594\sym{***}&      0.0497\sym{***}\\
                    &    (-10.72)         &    (-10.04)         &    (-12.17)         &     (-9.07)         &     (-8.50)         &      (6.92)         \\
\addlinespace
\textit{N Authors}           &      0.0123\sym{***}&    -0.00214\sym{***}&   -0.000464         &    -0.00390\sym{***}&      0.0275\sym{***}&      0.0144\sym{***}\\
                    &     (36.29)         &     (-6.57)         &     (-0.97)         &     (-8.77)         &     (40.21)         &     (28.32)         \\
\addlinespace
trial               &      0.0235\sym{***}&     0.00344         &     0.00220         &      0.0102         &      0.0578\sym{***}&      0.0128         \\
                    &      (3.76)         &      (0.37)         &      (0.18)         &      (0.89)         &      (6.96)         &      (1.26)         \\
\addlinespace
Pre-existing Grant      &      0.0421\sym{***}&      0.0356\sym{***}&      0.0563\sym{***}&      0.0438\sym{***}&      0.0592\sym{***}&     -0.0203\sym{***}\\
                    &     (11.79)         &      (6.88)         &      (8.70)         &      (7.04)         &     (12.06)         &     (-3.75)         \\
\addlinespace
Constant            &       0.899\sym{***}&       0.296         &       0.289\sym{***}&       0.263\sym{***}&       0.468         &     -0.0594\sym{**} \\
                    &     (46.16)         &      (1.05)         &      (3.83)         &      (3.40)         &      (1.62)         &     (-3.02)         \\
\midrule
Observations        &       60274         &       60274         &       55913         &       55418         &       60274         &       60274         \\
\midrule
Country FEs  & Majority  & Majority & First Author & Last Author & Majority  & Majority \\
\bottomrule
\multicolumn{7}{l}{\footnotesize \textit{t} statistics in parentheses}\\
\multicolumn{7}{l}{\footnotesize \sym{*} \(p<0.05\), \sym{**} \(p<0.01\), \sym{***} \(p<0.001\)}\\
\end{tabular}
}
\end{table}

\begin{table}[H]\centering
\def\sym#1{\ifmmode^{#1}\else\(^{#1}\)\fi}
\caption{DiD regression model estimates changing definition of \emph{COVID-relatedness} (1st - 99th percentiles of sample distribution of COVID-relatedness to identify control and treated units), with White standard errors. Country effects (omitted) for the majority of the team for Female Author, First and Last Female Authors, Middle Female Authorship and Middle Female Only; country fixed effects of the first (last) author for regression on First (Last) Female Author.\label{tab2_cb}}
\resizebox{\textwidth}{!}{ 
\begin{tabular}{lcccccc}
\toprule
                    \multicolumn{1}{c}{Variables}&\multicolumn{1}{c}{(1)}&\multicolumn{1}{c}{(2)}&\multicolumn{1}{c}{(3)}&\multicolumn{1}{c}{(4)}&\multicolumn{1}{c}{(5)}&\multicolumn{1}{c}{(6)}\\
                    &\multicolumn{1}{c}{Female Author}&\multicolumn{1}{c}{First and Last Female}&\multicolumn{1}{c}{Female First Author}&\multicolumn{1}{c}{Female Last Author}&\multicolumn{1}{c}{Middle Female Authorship}&\multicolumn{1}{c}{Middle Female Only}\\
\midrule
year=2020           &      0.0196\sym{***}&      0.0137\sym{**} &      0.0269\sym{***}&      0.0182\sym{**} &      0.0228\sym{***}&    -0.00914         \\
                    &      (4.57)         &      (2.92)         &      (4.20)         &      (3.04)         &      (4.26)         &     (-1.70)         \\
\addlinespace
COVID-related             &      0.0714\sym{***}&       0.108\sym{***}&      0.0923\sym{***}&       0.129\sym{***}&       0.101\sym{***}&     -0.0508\sym{***}\\
                    &      (7.24)         &      (6.69)         &      (4.93)         &      (6.95)         &      (7.27)         &     (-3.38)         \\
\addlinespace
year=2020 $\times$ COVID-related&     -0.0995\sym{***}&      -0.133\sym{***}&      -0.172\sym{***}&      -0.137\sym{***}&      -0.125\sym{***}&      0.0779\sym{***}\\
                    &     (-9.18)         &     (-7.86)         &     (-8.66)         &     (-6.97)         &     (-8.36)         &      (4.84)         \\
\addlinespace
\textit{N Authors}           &      0.0138\sym{***}&   -0.000953\sym{*}  &    0.000907         &    -0.00302\sym{***}&      0.0289\sym{***}&      0.0144\sym{***}\\
                    &     (27.69)         &     (-2.48)         &      (1.51)         &     (-5.56)         &     (29.78)         &     (21.95)         \\
\addlinespace
trial               &      0.0199\sym{*}  &     0.00731         &     -0.0220         &      0.0194         &      0.0676\sym{***}&      0.0265         \\
                    &      (2.01)         &      (0.59)         &     (-1.32)         &      (1.22)         &      (5.51)         &      (1.83)         \\
\addlinespace
Pre-existing Grant      &      0.0456\sym{***}&      0.0245\sym{***}&      0.0461\sym{***}&      0.0333\sym{***}&      0.0680\sym{***}&    -0.00915         \\
                    &      (8.85)         &      (3.73)         &      (5.28)         &      (4.05)         &      (9.97)         &     (-1.22)         \\
\addlinespace
Constant            &       0.967\sym{***}&      0.0139\sym{**} &       0.333\sym{***}&       0.286\sym{**} &       0.915\sym{***}&     -0.0611\sym{***}\\
                    &    (197.14)         &      (2.77)         &      (3.45)         &      (3.17)         &    (145.13)         &     (-9.83)         \\
\midrule
Observations        &       37965         &       37965         &       35059         &       34773         &       37965         &       37965         \\
\midrule
Country FEs  & Majority  & Majority & First Author & Last Author & Majority  & Majority \\
\bottomrule
\multicolumn{7}{l}{\footnotesize \textit{t} statistics in parentheses}\\
\multicolumn{7}{l}{\footnotesize \sym{*} \(p<0.05\), \sym{**} \(p<0.01\), \sym{***} \(p<0.001\)}\\
\end{tabular}
}
\end{table}

\subsection{Controlling for Career Age}\label{career_age}

We re-estimate the DiD base model including the usual paper level controls -- clinical trials, team size and for old grants -- and \emph{career age} of authors, which indicates the difference between observed year of publication in PubMed and the year of the author's first publication, as a proxy for authors' age. For each paper in our sample, we compute the career age of the first author, the last author, and for middle authors we consider the non-relevant authors' average career age. In Table \ref{tab_time_first_publ}, we see that the estimates of the treatment effect ($year=2020$ $\times$ COVID-related) are in line with the DiD model reported in Table 1 of the main text of the article.

\begin{table}[H]\centering
\def\sym#1{\ifmmode^{#1}\else\(^{#1}\)\fi}
\caption{DiD regression model estimates controlling for career age, with White-robust standards errors. Country effects (omitted) for the majority of the team for Female Author, First and Last Female Authors, Middle Female Authorship and Middle Female Only; country fixed effects of the first (last) author for regression on First (Last) Female Author.\label{tab_time_first_publ}}
\resizebox{\textwidth}{!}{ 
\begin{tabular}{lcccccc}
\toprule
                    \multicolumn{1}{c}{Variables}&\multicolumn{1}{c}{(1)}&\multicolumn{1}{c}{(2)}&\multicolumn{1}{c}{(3)}&\multicolumn{1}{c}{(4)}&\multicolumn{1}{c}{(5)}&\multicolumn{1}{c}{(6)}\\
                    &\multicolumn{1}{c}{First and Last Female}&\multicolumn{1}{c}{First and Last Female}&\multicolumn{1}{c}{Female First Author}&\multicolumn{1}{c}{Female Last Author}&\multicolumn{1}{c}{Middle Female Authorship}&\multicolumn{1}{c}{Middle Female Only}\\
\midrule
year=2020           &      0.0121\sym{**} &      0.0158\sym{***}&      0.0207\sym{***}&      0.0212\sym{***}&      0.0198\sym{***}&    -0.00994\sym{*}  \\
                    &      (3.26)         &      (4.32)         &      (4.15)         &      (4.63)         &      (4.87)         &     (-2.37)         \\
\addlinespace
COVID-related            &      0.0702\sym{***}&      0.0636\sym{***}&      0.0615\sym{***}&      0.0766\sym{***}&      0.0457\sym{***}&     -0.0357\sym{***}\\
                    &     (16.68)         &     (15.30)         &     (11.46)         &     (15.13)         &     (10.68)         &     (-8.08)         \\
\addlinespace
year=2020 $\times$ COVID-related&     -0.0467\sym{***}&     -0.0495\sym{***}&     -0.0655\sym{***}&     -0.0566\sym{***}&     -0.0402\sym{***}&      0.0369\sym{***}\\
                    &     (-8.47)         &     (-9.08)         &     (-9.25)         &     (-8.50)         &     (-7.13)         &      (6.31)         \\
\addlinespace
\textit{N Authors}            &    -0.00182\sym{***}&    -0.00170\sym{***}&   0.0000889         &    -0.00267\sym{***}&      0.0282\sym{***}&      0.0145\sym{***}\\
                    &     (-6.57)         &     (-6.06)         &      (0.23)         &     (-7.22)         &     (50.25)         &     (34.45)         \\
\addlinespace
trial               &      0.0125         &     0.00650         &      0.0154         &      0.0180         &      0.0576\sym{***}&     0.00674         \\
                    &      (1.62)         &      (0.85)         &      (1.56)         &      (1.90)         &      (8.49)         &      (0.80)         \\
\addlinespace
Pre-existing Grant      &      0.0362\sym{***}&      0.0411\sym{***}&      0.0608\sym{***}&      0.0525\sym{***}&      0.0578\sym{***}&     -0.0194\sym{***}\\
                    &      (8.71)         &      (9.99)         &     (11.70)         &     (10.67)         &     (14.67)         &     (-4.46)         \\
\addlinespace
\addlinespace
Career\_age\_first    &    -0.00213\sym{***}&                     &    -0.00495\sym{***}&                     &                     &                     \\
                    &    (-22.34)         &                     &    (-34.90)         &                     &                     &                     \\
\addlinespace
\addlinespace
Career\_age\_last     &                     &    -0.00249\sym{***}&                     &    -0.00445\sym{***}&                     &                     \\
                    &                     &    (-22.87)         &                     &    (-31.57)         &                     &                     \\
Career\_age\_middle\_avg&                     &                     &                     &                     &    -0.00496\sym{***}&    -0.00164\sym{***}\\
                    &                     &                     &                     &                     &    (-24.87)         &    (-12.02)         \\
\addlinespace
Constant            &       0.233         &       0.263         &       0.371\sym{***}&       0.325\sym{***}&       0.589\sym{**} &     -0.0453\sym{*}  \\
                    &      (1.03)         &      (1.11)         &      (5.41)         &      (4.97)         &      (2.58)         &     (-2.55)         \\
\midrule
Observations        &       84958         &       87138         &       79759         &       81293         &       87635         &       87635         \\
\midrule
Country FEs  & Majority  & Majority & First Author & Last Author & Majority  & Majority \\
\bottomrule
\multicolumn{7}{l}{\footnotesize \textit{t} statistics in parentheses}\\
\multicolumn{7}{l}{\footnotesize \sym{*} \(p<0.05\), \sym{**} \(p<0.01\), \sym{***} \(p<0.001\)}\\
\end{tabular}
}
\end{table}

\subsection{Confounding Journal and New Grant Effects}\label{robustness_1}

Tables \ref{JI} and \ref{JI_size} reports the coefficient estimates of equation (1), including interactions between the effect of interest and (i) the Journal Impact Factor, \emph{JI - med}, (ii) Journal Impact Factor and team size,  \emph{JI - med} $\times$ \emph{N authors}. The treatment effect is still significantly affecting women's appoint as key authors. There is no significant difference between COVID-related and COVID non-related's first and last female authorship during 2020 in journal impact factor ($year=2020$ $\times$ COVID-related $\times$ $JI - med=1$), nor of the joint variation of journal impact factor and team size ($year=2020$ $\times$ COVID-related $\times$ $JI - med=1$ $\times$ \emph{N authors}). Instead, in Table \ref{JI_size}, the treatment effect is no longer significantly affecting the overall appointment of women as middle authors (column (5)), but it is indeed still positively and significantly affecting female middle authorship when the key authors are men (column (6)).

In Table \ref{newgrant}, we include in equation (1) a linear interaction of treatment effect, the treatment status and time variable with new financing granted during 2020 - \emph{Has New Grant}. Once more, the treatment effect remains statistically significant on all outcomes. 

In all variations of the model, neither variables substitute the effect of the new publishing opportunity ($year=2020$ $\times$ COVID-related), which still holds a significant effect on women's appointment as key authors, and as middle authors when collaborating with male key authors. This means that the estimated coefficient for the effect of a new publishing opportunity on female authorship is robust against any potential confounding effect coming from a selection process created by journals' relevance or access to new funding.

\begin{table}[H]\centering
\def\sym#1{\ifmmode^{#1}\else\(^{#1}\)\fi}
\caption{Linear Regression estimates including interactions with \emph{JI Med}, with White standard errors. Country effects (omitted) for the majority of the team for Female Author, First and Last Female Authors, First or Last Female Authors, Middle Female Authorship, and Middle Female Only; country fixed effects of the first (last) author for regression on First (Last) Female Author.\label{JI}}
\resizebox{\textwidth}{!}{ 
\begin{tabular}{l c c c c c c }
\toprule  &\multicolumn{1}{c}{Female Author}&\multicolumn{1}{c}{First and Last Female}&\multicolumn{1}{c}{Female First Author}&\multicolumn{1}{c}{Female Last Author}&\multicolumn{1}{c}{Middle Female Authorship}&\multicolumn{1}{c}{Middle Female Only}\\
\toprule 
                    \multicolumn{1}{c}{Variables}&\multicolumn{1}{c}{(1)}&\multicolumn{1}{c}{(2)}&\multicolumn{1}{c}{(3)}&\multicolumn{1}{c}{(4)}&\multicolumn{1}{c}{(5)}&\multicolumn{1}{c}{(6)}\\

\midrule
year=2020           &      0.0114\sym{*}  &     0.00643         &      0.0121         &      0.0115         &      0.0180\sym{**} &    -0.00390         \\
                    &      (2.24)         &      (1.25)         &      (1.70)         &      (1.73)         &      (2.91)         &     (-0.65)         \\
\addlinespace
COVID-related             &      0.0636\sym{***}&      0.0938\sym{***}&      0.0991\sym{***}&       0.114\sym{***}&      0.0814\sym{***}&     -0.0525\sym{***}\\
                    &     (13.55)         &     (16.50)         &     (13.64)         &     (16.27)         &     (13.54)         &     (-8.90)         \\
\addlinespace
year=2020 $\times$COVID-related&     -0.0344\sym{***}&     -0.0521\sym{***}&     -0.0715\sym{***}&     -0.0593\sym{***}&     -0.0362\sym{***}&      0.0394\sym{***}\\
                    &     (-5.30)         &     (-6.86)         &     (-7.24)         &     (-6.28)         &     (-4.42)         &      (4.87)         \\
\addlinespace
JI\_med=1           &      0.0286\sym{***}&      0.0108\sym{*}  &      0.0332\sym{***}&      0.0157\sym{*}  &      0.0349\sym{***}&     -0.0133\sym{*}  \\
                    &      (6.27)         &      (2.15)         &      (4.83)         &      (2.46)         &      (6.07)         &     (-2.27)         \\
\addlinespace
year=2020 $\times$ JI\_med=1&     0.00417         &      0.0118         &      0.0141         &      0.0124         &    -0.00277         &    -0.00908         \\
                    &      (0.65)         &      (1.65)         &      (1.45)         &      (1.37)         &     (-0.34)         &     (-1.10)         \\
\addlinespace
COVID-related $\times$ JI\_med=1&     -0.0557\sym{***}&     -0.0546\sym{***}&     -0.0860\sym{***}&     -0.0579\sym{***}&     -0.0688\sym{***}&      0.0379\sym{***}\\
                    &     (-8.60)         &     (-6.67)         &     (-8.24)         &     (-5.78)         &     (-8.20)         &      (4.40)         \\
\addlinespace
year=2020 $\times$COVID-related $\times$ JI\_med=1&     0.00418         &      0.0138         &      0.0128         &      0.0116         &     0.00154         &     -0.0119         \\
                    &      (0.48)         &      (1.28)         &      (0.91)         &      (0.87)         &      (0.14)         &     (-1.02)         \\
\addlinespace
\textit{N Authors}            &      0.0123\sym{***}&    -0.00215\sym{***}&   -0.000778\sym{*}  &    -0.00359\sym{***}&      0.0277\sym{***}&      0.0145\sym{***}\\
                    &     (44.41)         &     (-7.85)         &     (-1.97)         &     (-9.61)         &     (49.85)         &     (34.22)         \\
\addlinespace
trial               &      0.0275\sym{***}&     0.00703         &     0.00823         &      0.0197\sym{*}  &      0.0594\sym{***}&     0.00543         \\
                    &      (5.37)         &      (0.93)         &      (0.84)         &      (2.07)         &      (8.71)         &      (0.65)         \\
\addlinespace
Pre-existing Grant       &      0.0408\sym{***}&      0.0355\sym{***}&      0.0576\sym{***}&      0.0421\sym{***}&      0.0531\sym{***}&     -0.0208\sym{***}\\
                    &     (14.18)         &      (8.59)         &     (11.12)         &      (8.45)         &     (13.47)         &     (-4.80)         \\
\addlinespace
Constant            &       0.893\sym{***}&       0.201         &       0.310\sym{***}&       0.299\sym{***}&       0.548\sym{*}  &     -0.0437\sym{*}  \\
                    &     (75.84)         &      (0.90)         &      (4.65)         &      (4.38)         &      (2.47)         &     (-2.12)         \\
\midrule
Observations        &       89530         &       89530         &       83263         &       82552         &       89530         &       89530         \\
\midrule
Country FEs & Majority  & Majority & First & Last & Majority  & Majority\\
\bottomrule
\multicolumn{7}{l}{\footnotesize \textit{t} statistics in parentheses}\\
\multicolumn{7}{l}{\footnotesize \sym{*} \(p<0.05\), \sym{**} \(p<0.01\), \sym{***} \(p<0.001\)}\\
\end{tabular}
}
\end{table}

\begin{table}[H]\centering
\def\sym#1{\ifmmode^{#1}\else\(^{#1}\)\fi}
\caption{Linear Regression estimates including interactions with \emph{JI Med} $\times$ \emph{N authors}, with White standard errors. Country effects (omitted) for the majority of the team for Female Author, First and Last Female Authors, First or Last Female Authors, Middle Female Authorship, and Middle Female Only; country fixed effects of the first (last) author for regression on First (Last) Female Author.\label{JI_size}}
\resizebox{\textwidth}{!}{ 
\begin{tabular}{l c c c c c c }
\toprule  &\multicolumn{1}{c}{Female Author}&\multicolumn{1}{c}{First and Last Female}&\multicolumn{1}{c}{Female First Author}&\multicolumn{1}{c}{Female Last Author}&\multicolumn{1}{c}{Middle Female Authorship}&\multicolumn{1}{c}{Middle Female Only}\\
\toprule 
                    \multicolumn{1}{c}{Variables}&\multicolumn{1}{c}{(1)}&\multicolumn{1}{c}{(2)}&\multicolumn{1}{c}{(3)}&\multicolumn{1}{c}{(4)}&\multicolumn{1}{c}{(5)}&\multicolumn{1}{c}{(6)}\\

\midrule

year=2020           &     0.00501         &     0.00383         &      0.0114         &     0.00737         &     0.00419         &    -0.00503         \\
                    &      (0.36)         &      (0.35)         &      (0.73)         &      (0.51)         &      (0.19)         &     (-0.35)         \\
\addlinespace
COVID-related             &       0.131\sym{***}&       0.145\sym{***}&       0.145\sym{***}&       0.153\sym{***}&       0.176\sym{***}&     -0.0243         \\
                    &     (10.82)         &     (12.99)         &      (9.69)         &     (10.79)         &      (8.57)         &     (-1.83)         \\
\addlinespace
year=2020 $\times$COVID-related&     -0.0333\sym{*}  &     -0.0804\sym{***}&     -0.0850\sym{***}&     -0.0810\sym{***}&   0.0000906         &      0.0517\sym{**} \\
                    &     (-2.05)         &     (-5.29)         &     (-4.15)         &     (-4.13)         &      (0.00)         &      (2.78)         \\
\addlinespace
JI\_med=1           &      0.0959\sym{***}&     0.00813         &      0.0386\sym{**} &     0.00356         &       0.167\sym{***}&      0.0578\sym{***}\\
                    &      (7.90)         &      (0.85)         &      (2.79)         &      (0.28)         &      (8.01)         &      (4.53)         \\
\addlinespace
year=2020 $\times$ JI\_med=1&      0.0219         &      0.0269         &      0.0227         &      0.0300         &      0.0207         &    -0.00782         \\
                    &      (1.32)         &      (1.90)         &      (1.14)         &      (1.62)         &      (0.75)         &     (-0.43)         \\
\addlinespace
COVID-related $\times$ JI\_med=1&      -0.109\sym{***}&     -0.0800\sym{***}&      -0.100\sym{***}&     -0.0732\sym{***}&      -0.127\sym{***}&     -0.0142         \\
                    &     (-6.97)         &     (-5.32)         &     (-5.03)         &     (-3.79)         &     (-4.89)         &     (-0.77)         \\
\addlinespace
year=2020 $\times$COVID-related $\times$ JI\_med=1&    -0.00636         &      0.0206         &    0.000668         &      0.0137         &     -0.0344         &    -0.00264         \\
                    &     (-0.31)         &      (1.03)         &      (0.02)         &      (0.53)         &     (-1.01)         &     (-0.11)         \\
\addlinespace
\textit{N Authors}            &      0.0236\sym{***}&    0.000292         &     0.00258         &    -0.00252         &      0.0498\sym{***}&      0.0234\sym{***}\\
                    &     (15.58)         &      (0.27)         &      (1.60)         &     (-1.79)         &     (16.96)         &     (13.77)         \\
\addlinespace
year=2020 $\times$ \textit{N Authors} &    0.000965         &    0.000425         &   0.0000969         &    0.000685         &     0.00210         &   0.0000945         \\
                    &      (0.49)         &      (0.26)         &      (0.04)         &      (0.32)         &      (0.58)         &      (0.04)         \\
\addlinespace
COVID-related $\times$ \textit{N Authors} &     -0.0116\sym{***}&    -0.00852\sym{***}&    -0.00758\sym{***}&    -0.00654\sym{**} &     -0.0166\sym{***}&    -0.00511\sym{*}  \\
                    &     (-6.87)         &     (-5.48)         &     (-3.48)         &     (-3.22)         &     (-5.12)         &     (-2.36)         \\
\addlinespace
year=2020 $\times$COVID-related $\times$ \textit{N Authors} &   -0.000148         &     0.00465\sym{*}  &     0.00231         &     0.00358         &    -0.00586         &    -0.00201         \\
                    &     (-0.07)         &      (2.18)         &      (0.78)         &      (1.27)         &     (-1.43)         &     (-0.67)         \\
\addlinespace
JI\_med=1 $\times$ \textit{N Authors} &     -0.0115\sym{***}&   -0.000173         &    -0.00144         &     0.00142         &     -0.0225\sym{***}&     -0.0115\sym{***}\\
                    &     (-6.94)         &     (-0.14)         &     (-0.76)         &      (0.84)         &     (-6.99)         &     (-5.88)         \\
\addlinespace
year=2020 $\times$ JI\_med=1 $\times$ \textit{N Authors} &    -0.00253         &    -0.00209         &    -0.00114         &    -0.00249         &    -0.00348         &   -0.000168         \\
                    &     (-1.15)         &     (-1.09)         &     (-0.41)         &     (-0.99)         &     (-0.84)         &     (-0.06)         \\
\addlinespace
COVID-related $\times$ JI\_med=1 $\times$ \textit{N Authors} &     0.00957\sym{***}&     0.00497\sym{**} &     0.00328         &     0.00319         &      0.0116\sym{**} &     0.00828\sym{**} \\
                    &      (4.89)         &      (2.63)         &      (1.25)         &      (1.26)         &      (3.07)         &      (3.05)         \\
\addlinespace
year=2020 $\times$COVID-related $\times$ JI\_med=1 $\times$ \textit{N Authors} &     0.00149         &    -0.00174         &     0.00121         &   -0.000891         &     0.00598         &   -0.000859         \\
                    &      (0.58)         &     (-0.69)         &      (0.34)         &     (-0.26)         &      (1.23)         &     (-0.23)         \\
\addlinespace
trial               &      0.0281\sym{***}&     0.00797         &     0.00925         &      0.0205\sym{*}  &      0.0602\sym{***}&     0.00501         \\
                    &      (5.52)         &      (1.05)         &      (0.94)         &      (2.15)         &      (8.91)         &      (0.60)         \\
\addlinespace
Pre-existing Grant       &      0.0403\sym{***}&      0.0358\sym{***}&      0.0578\sym{***}&      0.0423\sym{***}&      0.0515\sym{***}&     -0.0215\sym{***}\\
                    &     (14.00)         &      (8.65)         &     (11.16)         &      (8.48)         &     (13.08)         &     (-4.97)         \\
\addlinespace
Constant            &       0.827\sym{***}&       0.181         &       0.292\sym{***}&       0.294\sym{***}&       0.423         &     -0.0912\sym{***}\\
                    &     (53.16)         &      (0.81)         &      (4.34)         &      (4.28)         &      (1.87)         &     (-3.65)         \\
\midrule
Observations        &       89530         &       89530         &       83263         &       82552         &       89530         &       89530         \\
\midrule
Country FEs & Majority  & Majority & First & Last & Majority  & Majority\\
\bottomrule
\multicolumn{7}{l}{\footnotesize \textit{t} statistics in parentheses}\\
\multicolumn{7}{l}{\footnotesize \sym{*} \(p<0.05\), \sym{**} \(p<0.01\), \sym{***} \(p<0.001\)}\\
\end{tabular}
}
\end{table}

\begin{table}[H]\centering
\def\sym#1{\ifmmode^{#1}\else\(^{#1}\)\fi}
\caption{Linear Regression estimates including interactions with \emph{Has New Grant}, with White standard errors.  Country effects (omitted) for the majority of the team for Female Author, First and Last Female Authors, First or Last Female Authors,  Middle Female Authorship, and Middle Female Only; country fixed effects of the first (last) author for regression on First (Last) Female Author.\label{newgrant}}
\resizebox{\textwidth}{!}{ 
\begin{tabular}{lcccccc}
\toprule
\multicolumn{1}{c}{Variables}&\multicolumn{1}{c}{(1)}&\multicolumn{1}{c}{(2)}&\multicolumn{1}{c}{(4)}&\multicolumn{1}{c}{(3)}&\multicolumn{1}{c}{(5)}&\multicolumn{1}{c}{(6)}\\
&\multicolumn{1}{c}{Female Author}&\multicolumn{1}{c}{First and Last Female}&\multicolumn{1}{c}{Female First Author}&\multicolumn{1}{c}{Female Last Author}&\multicolumn{1}{c}{Middle Female Authorship}&\multicolumn{1}{c}{Middle Female Only}\\
\midrule
year=2020           &     0.00949\sym{**} &     0.00953\sym{*}  &      0.0134\sym{**} &      0.0142\sym{**} &      0.0110\sym{**} &    -0.00636         \\
                    &      (2.82)         &      (2.54)         &      (2.63)         &      (2.99)         &      (2.59)         &     (-1.48)         \\
\addlinespace
COVID-related             &      0.0368\sym{***}&      0.0691\sym{***}&      0.0581\sym{***}&      0.0865\sym{***}&      0.0485\sym{***}&     -0.0348\sym{***}\\
                    &     (11.13)         &     (16.76)         &     (10.98)         &     (17.10)         &     (11.34)         &     (-7.96)         \\
\addlinespace
year=2020 $\times$ COVID-related&     -0.0335\sym{***}&     -0.0508\sym{***}&     -0.0700\sym{***}&     -0.0583\sym{***}&     -0.0361\sym{***}&      0.0363\sym{***}\\
                    &     (-7.36)         &     (-9.18)         &     (-9.74)         &     (-8.50)         &     (-6.19)         &      (6.10)         \\
\addlinespace
has\_new\_grant=1     &      0.0400\sym{***}&      0.0285\sym{***}&      0.0583\sym{***}&      0.0354\sym{***}&      0.0511\sym{***}&     -0.0233\sym{**} \\
                    &      (7.09)         &      (3.66)         &      (5.90)         &      (3.76)         &      (6.59)         &     (-2.80)         \\
\addlinespace
COVID-related $\times$ has\_new\_grant=1&      0.0106         &      0.0467\sym{***}&      0.0409\sym{**} &      0.0393\sym{**} &      0.0101         &     -0.0234         \\
                    &      (1.33)         &      (3.81)         &      (2.75)         &      (2.72)         &      (0.89)         &     (-1.88)         \\
\addlinespace
\textit{N Authors}           &      0.0123\sym{***}&    -0.00235\sym{***}&   -0.000916\sym{*}  &    -0.00376\sym{***}&      0.0276\sym{***}&      0.0145\sym{***}\\
                    &     (44.69)         &     (-8.60)         &     (-2.34)         &    (-10.10)         &     (50.04)         &     (34.53)         \\
\addlinespace
trial               &      0.0254\sym{***}&     0.00490         &     0.00537         &      0.0177         &      0.0568\sym{***}&     0.00682         \\
                    &      (4.98)         &      (0.65)         &      (0.55)         &      (1.87)         &      (8.35)         &      (0.82)         \\
\addlinespace
Pre-existing Grant      &      0.0447\sym{***}&      0.0377\sym{***}&      0.0627\sym{***}&      0.0450\sym{***}&      0.0574\sym{***}&     -0.0232\sym{***}\\
                    &     (15.43)         &      (9.17)         &     (12.15)         &      (9.06)         &     (14.51)         &     (-5.37)         \\
\addlinespace
Constant            &       0.914\sym{***}&       0.213         &       0.325\sym{***}&       0.310\sym{***}&       0.576\sym{*}  &     -0.0527\sym{**} \\
                    &     (68.41)         &      (0.95)         &      (4.91)         &      (4.54)         &      (2.51)         &     (-2.76)         \\
\midrule
Observations        &       89530         &       89530         &       83263         &       82552         &       89530         &       89530         \\
\midrule
Country FEs & Majority    & Majority  & First  & Last  & Majority  & Majority   \\
\bottomrule
\multicolumn{7}{l}{\footnotesize \textit{t} statistics in parentheses}\\
\multicolumn{7}{l}{\footnotesize \sym{*} \(p<0.05\), \sym{**} \(p<0.01\), \sym{***} \(p<0.001\)}\\
\end{tabular}
}
\end{table}

\subsection{Confounding effect of Lockdown Restrictions}\label{robustness_2} 

We use the Oxford Coronavirus Government Response Tracker (OxCGRT) \citep{owidcoronavirus} to verify the confounding effect of stringency norms of stay-at-home mandates implemented by the country of the affiliation of authors to tackle the spread of the novel Coronavirus. 
For each paper, we consider the maximum value of monthly school and workplace stringency measures in the month of publication of the paper and in the country of affiliation of authors. Table \ref{ds_c} reports the main descriptive statistics of school and workplace stringency measures observed at the country level, as well as lagged monthly values. Stringency metrics are observed only in 2020.

\begin{table}[H] \centering \renewcommand*{\arraystretch}{1.1}\caption{Main descriptive statistics of Stringency measures at country level for the year 2020.}\label{ds_c}
\begin{tabular}{lrrrrrrr}
\hline
\hline
Variable & N & Mean & Std. Dev. & Min & Pctl. 25 & Pctl. 75 & Max \\ 
\hline
SchoolClosuresMax & 125 & 2.1 & 1.1 & 0 & 1 & 3 & 3 \\ 
WorkplaceClosuresMax & 125 & 1.7 & 1 & 0 & 1 & 2 & 3 \\ 
SchoolClosuresMax\_lag1 & 125 & 2.1 & 1.2 & 0 & 1 & 3 & 3 \\ 
WorkplaceClosuresMax\_lag1 & 125 & 1.7 & 1.1 & 0 & 1 & 3 & 3 \\ 
SchoolClosuresMax\_lag2 & 125 & 1.9 & 1.3 & 0 & 0 & 3 & 3 \\ 
WorkplaceClosuresMax\_lag2 & 125 & 1.5 & 1.2 & 0 & 0 & 2 & 3 \\ 
SchoolClosuresMax\_lag3 & 125 & 1.7 & 1.4 & 0 & 0 & 3 & 3 \\ 
WorkplaceClosuresMax\_lag3 & 125 & 1.3 & 1.2 & 0 & 0 & 2 & 3\\ 
\hline
\hline
\end{tabular}
\end{table}

We estimate the following model including interactions of the treatment effect of interest (year=2020 $\times$COVID-related) with \emph{stringency} indicators of country $j$ of the listed affiliation of either the first author, of the last author, or of the majority of team members at time $t$ -- depending on the dependent variable:

\begin{equation}\label{eq_2}
\begin{split}
    y_{i} = \alpha+\gamma_{j}+\lambda\bbone[year_{i} = 2020]+\beta treat_{i}+\tau(\bbone[year_{i} = 2020]\times treat_{i})+ \\\delta stringency_{j,t}+\eta(stringency_{j,t}\times treat_{i})+\epsilon_{i}
\end{split}
\end{equation}

We use monthly maximum values of (i) \emph{school closures} and (ii) \emph{workplace closures} indices as proxies for increased family duties for the adult female population.
For authors in the US, we consider the stringency values at national level. Consequentially, for the stringency analysis we control for country fixed effects at the national level for authors within the US. 

In Table \ref{reg_fam_duties_1_school} and \ref{reg_fam_duties_1_work}, we report the coefficient estimates for the augmented model with (i) school closures and (ii) workplace closures. Lockdown closures and workplace restrictions do not fully account for the decline in women's authorship in key positions in research affected by the new emerging topic. Instead, stringency of workplace closures are explaining the drop in middle female authorship, when no consideration is placed on the gender of the key authors, but, from column (6), we see that the confounding role of workplace closures is limited to the case of female key authors. In fact, the treatment effect has once again a positive and significant impact on the appointment of women as non-relevant author in related publications, when there are only men as key authors.

Next, we estimate model (3) by using lagged values of monthly maximum \emph{school closures} and \emph{workplace closures} as proxy of increased family duties for women in scientific academia following the restrictions against COVID-19 contagion put in place by countries in 2020. Again, depending on the outcome of interest, we refer to the stringency norms of the country of the listed affiliation of (i) the majority of the team members, (ii) of the first author, (iii) of the last author. 
We utilize the values of the stringency measures lagged by one, two and three months and display the regression results in Tables (\ref{lag1s})-(\ref{lag3w}). In all model's specifications, the treatment effect is still significantly affecting all outcomes, except for joint first and last authorship in column (2) and Middle authorship positions in column (5) and (6) of Table \ref{lag1w}, where the first lagged value of monthly maximum workplace closure index significantly explains the difference between COVID-related and COVID non-related in 2020 (\emph{COVID-related} $\times$ \emph{WorkplaceClosures\_max\_lag1}).

\begin{table}[H]\centering
\def\sym#1{\ifmmode^{#1}\else\(^{#1}\)\fi}
\caption{Linear regression estimates including interaction monthly maximum \emph{school closures}. Closures refer to the country of majority of authors for models in columns (1), (2),(5), (6); to country of first author in columns (3); to country of last author in columns (4). We control for old grants, number of authors, \emph{trial} with White-robust standard errors.  Country effects (omitted) for the majority of the team for Female Author, First and Last Female Authors,  Middle Female Authorship, and Middle Female Only; country fixed effects of the first (last) author for regression on First (Last) Female Author.\label{reg_fam_duties_1_school}} 
\resizebox{\textwidth}{!}{ 
\begin{tabular}{l c c c c c c}
\toprule  &\multicolumn{1}{c}{Female Author}&\multicolumn{1}{c}{First and Last Female}&\multicolumn{1}{c}{Female First Author}&\multicolumn{1}{c}{Female Last Author}&\multicolumn{1}{c}{Middle Female Authorship}&\multicolumn{1}{c}{Middle Female Only}\\
\toprule 
                    \multicolumn{1}{c}{Variables}&\multicolumn{1}{c}{(1)}&\multicolumn{1}{c}{(2)}&\multicolumn{1}{c}{(3)}&\multicolumn{1}{c}{(4)}&\multicolumn{1}{c}{(5)}&\multicolumn{1}{c}{(6)}\\
\midrule
year=2020           &     0.00946         &      0.0129\sym{*}  &      0.0210\sym{*}  &      0.0193\sym{*}  &      0.0163\sym{*}  &     -0.0120         \\
                    &      (1.62)         &      (2.01)         &      (2.42)         &      (2.40)         &      (2.25)         &     (-1.61)         \\
\addlinespace
COVID-related             &      0.0373\sym{***}&      0.0700\sym{***}&      0.0590\sym{***}&      0.0877\sym{***}&      0.0492\sym{***}&     -0.0352\sym{***}\\
                    &     (11.21)         &     (16.97)         &     (11.10)         &     (17.30)         &     (11.46)         &     (-8.01)         \\
\addlinespace
year=2020 $\times$ COVID-related&     -0.0206\sym{**} &     -0.0278\sym{**} &     -0.0433\sym{***}&     -0.0399\sym{***}&     -0.0167         &      0.0215\sym{*}  \\
                    &     (-2.67)         &     (-3.01)         &     (-3.62)         &     (-3.52)         &     (-1.72)         &      (2.14)         \\
\addlinespace
SchoolClosures &     0.00144         &    -0.00151         &                     &                     &   -0.000170         &     0.00143         \\
                    &      (0.68)         &     (-0.63)         &                     &                     &     (-0.06)         &      (0.52)         \\
\addlinespace
COVID-related $\times$ SchoolClosures&    -0.00681\sym{*}  &    -0.00984\sym{**} &                     &                     &    -0.00994\sym{**} &     0.00662         \\
                    &     (-2.43)         &     (-2.94)         &                     &                     &     (-2.81)         &      (1.81)         \\
\addlinespace
\textit{N Authors}            &      0.0124\sym{***}&    -0.00214\sym{***}&   -0.000663         &    -0.00355\sym{***}&      0.0278\sym{***}&      0.0143\sym{***}\\
                    &     (44.96)         &     (-7.85)         &     (-1.69)         &     (-9.58)         &     (50.20)         &     (34.20)         \\
\addlinespace
trial               &      0.0250\sym{***}&     0.00550         &     0.00489         &      0.0175         &      0.0562\sym{***}&     0.00647         \\
                    &      (4.89)         &      (0.72)         &      (0.50)         &      (1.84)         &      (8.23)         &      (0.78)         \\
\addlinespace
Pre-existing Grant      &      0.0418\sym{***}&      0.0345\sym{***}&      0.0578\sym{***}&      0.0415\sym{***}&      0.0539\sym{***}&     -0.0207\sym{***}\\
                    &     (14.52)         &      (8.41)         &     (11.20)         &      (8.37)         &     (13.67)         &     (-4.80)         \\
\addlinespace
SchoolClosures\_first&                     &                     &    -0.00116         &                     &                     &                     \\
                    &                     &                     &     (-0.36)         &                     &                     &                     \\
\addlinespace
COVID-related $\times$ SchoolClosures\_first&                     &                     &     -0.0128\sym{**} &                     &                     &                     \\
                    &                     &                     &     (-2.93)         &                     &                     &                     \\
\addlinespace
SchoolClosures\_last&                     &                     &                     &    -0.00205         &                     &                     \\
                    &                     &                     &                     &     (-0.69)         &                     &                     \\
\addlinespace
COVID-related $\times$ SchoolClosures\_last&                     &                     &                     &    -0.00871\sym{*}  &                     &                     \\
                    &                     &                     &                     &     (-2.10)         &                     &                     \\
\addlinespace
Constant            &       0.907\sym{***}&       0.201         &       0.323\sym{***}&       0.305\sym{***}&       0.565\sym{*}  &     -0.0432\sym{**} \\
                    &     (64.30)         &      (0.91)         &      (4.88)         &      (4.48)         &      (2.48)         &     (-2.70)         \\
\midrule
Observations        &       89267         &       89267         &       83235         &       82534         &       89267         &       89267         \\
\midrule
Country FEs & Majority  & Majority  & First & Last   & Majority  & Majority \\
\bottomrule
\multicolumn{7}{l}{\footnotesize \textit{t} statistics in parentheses}\\
\multicolumn{7}{l}{\footnotesize \sym{*} \(p<0.05\), \sym{**} \(p<0.01\), \sym{***} \(p<0.001\)}\\
\end{tabular}
}
\end{table}

\begin{table}[H]\centering
\def\sym#1{\ifmmode^{#1}\else\(^{#1}\)\fi}
\caption{Linear regression estimates including interaction monthly maximum \emph{workplace closures}. Closures refer to the country of majority of authors for models in columns (1), (2),(5), (6); to country of first author in columns (3); to country of last author in columns (4). We control for old grants, number of authors, \emph{trial} with White-robust standard errors.  Country effects (omitted) for the majority of the team for Female Author, First and Last Female Authors,  Middle Female Authorship, and Middle Female Only; country fixed effects of the first (last) author for regression on First (Last) Female Author.\label{reg_fam_duties_1_work}} 
\resizebox{\textwidth}{!}{ 
\begin{tabular}{l c c c c c c}
\toprule  &\multicolumn{1}{c}{Female Author}&\multicolumn{1}{c}{First and Last Female}&\multicolumn{1}{c}{Female First Author}&\multicolumn{1}{c}{Female Last Author}&\multicolumn{1}{c}{Middle Female Authorship}&\multicolumn{1}{c}{Middle Female Only}\\
\toprule 
                    \multicolumn{1}{c}{Variables}&\multicolumn{1}{c}{(1)}&\multicolumn{1}{c}{(2)}&\multicolumn{1}{c}{(3)}&\multicolumn{1}{c}{(4)}&\multicolumn{1}{c}{(5)}&\multicolumn{1}{c}{(6)}\\
\midrule

year=2020           &     0.00590         &      0.0116         &      0.0222\sym{**} &     0.00896         &      0.0106         &     -0.0116         \\
                    &      (1.02)         &      (1.85)         &      (2.61)         &      (1.14)         &      (1.48)         &     (-1.59)         \\
\addlinespace
COVID-related             &      0.0373\sym{***}&      0.0701\sym{***}&      0.0590\sym{***}&      0.0876\sym{***}&      0.0492\sym{***}&     -0.0352\sym{***}\\
                    &     (11.20)         &     (16.98)         &     (11.10)         &     (17.29)         &     (11.46)         &     (-8.02)         \\
\addlinespace
year=2020 $\times$ COVID-related&     -0.0150\sym{*}  &     -0.0234\sym{**} &     -0.0429\sym{***}&     -0.0224\sym{*}  &    -0.00751         &      0.0227\sym{*}  \\
                    &     (-2.00)         &     (-2.60)         &     (-3.68)         &     (-2.04)         &     (-0.79)         &      (2.32)         \\
\addlinespace
WorkplaceClosures&     0.00350         &   -0.000839         &                     &                     &     0.00289         &     0.00129         \\
                    &      (1.54)         &     (-0.32)         &                     &                     &      (1.01)         &      (0.43)         \\
\addlinespace
COVID-related $\times$ WorkplaceClosures&     -0.0105\sym{***}&     -0.0134\sym{***}&                     &                     &     -0.0159\sym{***}&     0.00693         \\
                    &     (-3.51)         &     (-3.67)         &                     &                     &     (-4.17)         &      (1.75)         \\
\addlinespace
\textit{N Authors}            &      0.0124\sym{***}&    -0.00214\sym{***}&   -0.000662         &    -0.00354\sym{***}&      0.0278\sym{***}&      0.0143\sym{***}\\
                    &     (44.97)         &     (-7.84)         &     (-1.69)         &     (-9.55)         &     (50.21)         &     (34.21)         \\
\addlinespace
trial               &      0.0250\sym{***}&     0.00548         &     0.00496         &      0.0174         &      0.0562\sym{***}&     0.00641         \\
                    &      (4.88)         &      (0.72)         &      (0.51)         &      (1.83)         &      (8.23)         &      (0.77)         \\
\addlinespace
Pre-existing Grant      &      0.0419\sym{***}&      0.0346\sym{***}&      0.0579\sym{***}&      0.0415\sym{***}&      0.0540\sym{***}&     -0.0208\sym{***}\\
                    &     (14.53)         &      (8.43)         &     (11.22)         &      (8.38)         &     (13.69)         &     (-4.82)         \\
\addlinespace
WorkplaceClosures\_first&                     &                     &    -0.00157         &                     &                     &                     \\
                    &                     &                     &     (-0.45)         &                     &                     &                     \\
\addlinespace
COVID-related $\times$ WorkplaceClosures\_first&                     &                     &     -0.0148\sym{**} &                     &                     &                     \\
                    &                     &                     &     (-3.13)         &                     &                     &                     \\
\addlinespace
WorkplaceClosures\_last&                     &                     &                     &     0.00327         &                     &                     \\
                    &                     &                     &                     &      (1.01)         &                     &                     \\
\addlinespace
COVID-related $\times$ WorkplaceClosures\_last&                     &                     &                     &     -0.0188\sym{***}&                     &                     \\
                    &                     &                     &                     &     (-4.17)         &                     &                     \\
\addlinespace
Constant            &       0.905\sym{***}&       0.196         &       0.321\sym{***}&       0.304\sym{***}&       0.560\sym{*}  &     -0.0420\sym{*}  \\
                    &     (67.46)         &      (0.89)         &      (4.85)         &      (4.46)         &      (2.48)         &     (-2.49)         \\
\midrule
Observations        &       89267         &       89267         &       83235         &       82534         &       89267         &       89267         \\
\midrule
Country FEs & Majority  & Majority  & First & Last   & Majority  & Majority \\
\bottomrule
\multicolumn{7}{l}{\footnotesize \textit{t} statistics in parentheses}\\
\multicolumn{7}{l}{\footnotesize \sym{*} \(p<0.05\), \sym{**} \(p<0.01\), \sym{***} \(p<0.001\)}\\
\end{tabular}
}
\end{table}

\begin{table}[H]\centering
\def\sym#1{\ifmmode^{#1}\else\(^{#1}\)\fi}
\caption{Linear regression estimates including interaction with one month lagged monthly maximum \emph{school closures}, controlling for old grants, number of authors, \emph{trial} with White-robust standard errors.  Country effects (omitted) for the majority of the team for Female Author, First and Last Female Authors, Middle Female Authorship and Middle Female Only; country fixed effects of the first (last) author for regression on First (Last) Female Author.\label{lag1s}}
\resizebox{\textwidth}{!}{ 
\begin{tabular}{l c c c c c c}
\toprule  &\multicolumn{1}{c}{Female Author}&\multicolumn{1}{c}{First and Last Female}&\multicolumn{1}{c}{Female First Author}&\multicolumn{1}{c}{Female Last Author}&\multicolumn{1}{c}{Middle Female Authorship}&\multicolumn{1}{c}{Middle Female Only}\\
\toprule 
                    \multicolumn{1}{c}{Variables}&\multicolumn{1}{c}{(1)}&\multicolumn{1}{c}{(2)}&\multicolumn{1}{c}{(3)}&\multicolumn{1}{c}{(4)}&\multicolumn{1}{c}{(5)}&\multicolumn{1}{c}{(6)}\\
\midrule
year=2020           &      0.0152\sym{**} &     0.00598         &      0.0184\sym{*}  &      0.0116         &      0.0198\sym{**} &    -0.00737         \\
                    &      (3.09)         &      (1.08)         &      (2.44)         &      (1.67)         &      (3.18)         &     (-1.15)         \\
\addlinespace
COVID-related             &      0.0374\sym{***}&      0.0700\sym{***}&      0.0591\sym{***}&      0.0877\sym{***}&      0.0493\sym{***}&     -0.0352\sym{***}\\
                    &     (11.24)         &     (16.96)         &     (11.11)         &     (17.29)         &     (11.48)         &     (-8.01)         \\
\addlinespace
year=2020 $\times$ COVID-related&     -0.0207\sym{**} &     -0.0193\sym{*}  &     -0.0356\sym{***}&     -0.0294\sym{**} &     -0.0197\sym{*}  &      0.0184\sym{*}  \\
                    &     (-3.11)         &     (-2.35)         &     (-3.35)         &     (-2.92)         &     (-2.30)         &      (2.08)         \\
\addlinespace
SchoolClosures\_lag1&    -0.00138         &     0.00177         &                     &                     &    -0.00201         &   -0.000718         \\
                    &     (-0.75)         &      (0.84)         &                     &                     &     (-0.86)         &     (-0.30)         \\
\addlinespace
COVID-related $\times$ SchoolClosures\_lag1&    -0.00696\sym{**} &     -0.0145\sym{***}&                     &                     &    -0.00893\sym{**} &     0.00848\sym{**} \\
                    &     (-2.83)         &     (-4.88)         &                     &                     &     (-2.84)         &      (2.61)         \\
\addlinespace
\textit{N Authors}            &      0.0124\sym{***}&    -0.00214\sym{***}&   -0.000661         &    -0.00354\sym{***}&      0.0278\sym{***}&      0.0143\sym{***}\\
                    &     (44.93)         &     (-7.86)         &     (-1.68)         &     (-9.57)         &     (50.17)         &     (34.21)         \\
\addlinespace
trial               &      0.0245\sym{***}&     0.00482         &     0.00388         &      0.0168         &      0.0557\sym{***}&     0.00683         \\
                    &      (4.78)         &      (0.64)         &      (0.40)         &      (1.77)         &      (8.15)         &      (0.82)         \\
\addlinespace
Pre-existing Grant      &      0.0415\sym{***}&      0.0343\sym{***}&      0.0573\sym{***}&      0.0412\sym{***}&      0.0536\sym{***}&     -0.0206\sym{***}\\
                    &     (14.40)         &      (8.35)         &     (11.11)         &      (8.31)         &     (13.59)         &     (-4.78)         \\
\addlinespace
SchoolClosures\_lag1\_first&                     &                     &  -0.0000474         &                     &                     &                     \\
                    &                     &                     &     (-0.02)         &                     &                     &                     \\
\addlinespace
COVID-related $\times$ SchoolClosures\_lag1\_first&                     &                     &     -0.0172\sym{***}&                     &                     &                     \\
                    &                     &                     &     (-4.40)         &                     &                     &                     \\
\addlinespace
SchoolClosures\_lag1\_last&                     &                     &                     &     0.00161         &                     &                     \\
                    &                     &                     &                     &      (0.61)         &                     &                     \\
\addlinespace
COVID-related $\times$ SchoolClosures\_lag1\_last&                     &                     &                     &     -0.0143\sym{***}&                     &                     \\
                    &                     &                     &                     &     (-3.86)         &                     &                     \\
\addlinespace
Constant            &       0.905\sym{***}&       0.201         &       0.324\sym{***}&       0.306\sym{***}&       0.565\sym{*}  &     -0.0440\sym{**} \\
                    &     (62.28)         &      (0.91)         &      (4.88)         &      (4.50)         &      (2.48)         &     (-2.72)         \\
\midrule
Observations        &       89267         &       89267         &       83235         &       82534         &       89267         &       89267         \\
\midrule
Country FEs & Majority  & Majority  & First & Last   & Majority  & Majority \\
\bottomrule
\multicolumn{7}{l}{\footnotesize \textit{t} statistics in parentheses}\\
\multicolumn{7}{l}{\footnotesize \sym{*} \(p<0.05\), \sym{**} \(p<0.01\), \sym{***} \(p<0.001\)}\\
\end{tabular}
}
\end{table}

\begin{table}[H]\centering
\def\sym#1{\ifmmode^{#1}\else\(^{#1}\)\fi}
\caption{Linear regression estimates including interaction with one month lagged monthly maximum \emph{workplace closures}, controlling for old grants, number of authors, \emph{trial} with White-robust standard errors.  Country effects (omitted) for the majority of the team for Female Author, First and Last Female Authors, Middle Female Authorship and Middle Female Only; country fixed effects of the first (last) author for regression on First (Last) Female Author.\label{lag1w}}
\resizebox{\textwidth}{!}{ 
\begin{tabular}{l c c c c c c}
\toprule  &\multicolumn{1}{c}{Female Author}&\multicolumn{1}{c}{First and Last Female}&\multicolumn{1}{c}{Female First Author}&\multicolumn{1}{c}{Female Last Author}&\multicolumn{1}{c}{Middle Female Authorship}&\multicolumn{1}{c}{Middle Female Only}\\
\toprule 
                    \multicolumn{1}{c}{Variables}&\multicolumn{1}{c}{(1)}&\multicolumn{1}{c}{(2)}&\multicolumn{1}{c}{(3)}&\multicolumn{1}{c}{(4)}&\multicolumn{1}{c}{(5)}&\multicolumn{1}{c}{(6)}\\
\midrule
year=2020           &      0.0113\sym{*}  &     0.00303         &      0.0154\sym{*}  &     0.00598         &      0.0163\sym{**} &    -0.00729         \\
                    &      (2.28)         &      (0.55)         &      (2.07)         &      (0.87)         &      (2.64)         &     (-1.15)         \\
\addlinespace
COVID-related             &      0.0375\sym{***}&      0.0701\sym{***}&      0.0591\sym{***}&      0.0877\sym{***}&      0.0494\sym{***}&     -0.0352\sym{***}\\
                    &     (11.25)         &     (16.98)         &     (11.12)         &     (17.29)         &     (11.49)         &     (-8.03)         \\
\addlinespace
year=2020 $\times$ COVID-related&     -0.0153\sym{*}  &     -0.0130         &     -0.0275\sym{**} &     -0.0221\sym{*}  &     -0.0135         &      0.0153         \\
                    &     (-2.32)         &     (-1.61)         &     (-2.64)         &     (-2.24)         &     (-1.60)         &      (1.77)         \\
\addlinespace
WorkplaceClosures\_lag1&    0.000877         &     0.00389         &                     &                     &  -0.0000499         &    -0.00100         \\
                    &      (0.45)         &      (1.67)         &                     &                     &     (-0.02)         &     (-0.38)         \\
\addlinespace
COVID-related $\times$ WorkplaceClosures\_lag1&     -0.0111\sym{***}&     -0.0203\sym{***}&                     &                     &     -0.0139\sym{***}&      0.0115\sym{**} \\
                    &     (-4.18)         &     (-6.13)         &                     &                     &     (-4.02)         &      (3.21)         \\
\addlinespace
\textit{N Authors}            &      0.0124\sym{***}&    -0.00214\sym{***}&   -0.000655         &    -0.00354\sym{***}&      0.0278\sym{***}&      0.0143\sym{***}\\
                    &     (44.96)         &     (-7.84)         &     (-1.67)         &     (-9.56)         &     (50.20)         &     (34.19)         \\
\addlinespace
trial               &      0.0244\sym{***}&     0.00467         &     0.00369         &      0.0167         &      0.0555\sym{***}&     0.00696         \\
                    &      (4.77)         &      (0.62)         &      (0.38)         &      (1.76)         &      (8.13)         &      (0.83)         \\
\addlinespace
Pre-existing Grant      &      0.0416\sym{***}&      0.0344\sym{***}&      0.0574\sym{***}&      0.0414\sym{***}&      0.0538\sym{***}&     -0.0207\sym{***}\\
                    &     (14.44)         &      (8.38)         &     (11.14)         &      (8.35)         &     (13.63)         &     (-4.78)         \\
\addlinespace
WorkplaceClosures\_lag1\_first&                     &                     &     0.00207         &                     &                     &                     \\
                    &                     &                     &      (0.67)         &                     &                     &                     \\
\addlinespace
COVID-related $\times$ WorkplaceClosures\_lag1\_first&                     &                     &     -0.0244\sym{***}&                     &                     &                     \\
                    &                     &                     &     (-5.69)         &                     &                     &                     \\
\addlinespace
WorkplaceClosures\_lag1\_last&                     &                     &                     &     0.00532         &                     &                     \\
                    &                     &                     &                     &      (1.82)         &                     &                     \\
\addlinespace
COVID-related $\times$ WorkplaceClosures\_lag1\_last&                     &                     &                     &     -0.0206\sym{***}&                     &                     \\
                    &                     &                     &                     &     (-5.02)         &                     &                     \\
\addlinespace
Constant            &       0.902\sym{***}&       0.196         &       0.322\sym{***}&       0.306\sym{***}&       0.560\sym{*}  &     -0.0405\sym{*}  \\
                    &     (66.86)         &      (0.89)         &      (4.85)         &      (4.49)         &      (2.48)         &     (-2.45)         \\
\midrule
Observations        &       89267         &       89267         &       83235         &       82534         &       89267         &       89267         \\
\midrule
Country FEs & Majority  & Majority  & First & Last   & Majority  & Majority \\
\bottomrule
\multicolumn{7}{l}{\footnotesize \textit{t} statistics in parentheses}\\
\multicolumn{7}{l}{\footnotesize \sym{*} \(p<0.05\), \sym{**} \(p<0.01\), \sym{***} \(p<0.001\)}\\
\end{tabular}
}
\end{table}

\begin{table}[H]\centering
\def\sym#1{\ifmmode^{#1}\else\(^{#1}\)\fi}
\caption{Linear regression estimates including interaction two-months lagged monthly maximum \emph{school closures}, controlling for old grants, number of authors, \emph{trial} with White-robust standard errors.  Country effects (omitted) for the majority of the team for Female Author, First and Last Female Authors, Middle Female Authorship and Middle Female Only; country fixed effects of the first (last) author for regression on First (Last) Female Author.\label{lag2s}}
\resizebox{\textwidth}{!}{ 
\begin{tabular}{l c c c c c c}
\toprule  &\multicolumn{1}{c}{Female Author}&\multicolumn{1}{c}{First and Last Female}&\multicolumn{1}{c}{Female First Author}&\multicolumn{1}{c}{Female Last Author}&\multicolumn{1}{c}{Middle Female Authorship}&\multicolumn{1}{c}{Middle Female Only}\\
\toprule 
                    \multicolumn{1}{c}{Variables}&\multicolumn{1}{c}{(1)}&\multicolumn{1}{c}{(2)}&\multicolumn{1}{c}{(3)}&\multicolumn{1}{c}{(4)}&\multicolumn{1}{c}{(5)}&\multicolumn{1}{c}{(6)}\\
\midrule
year=2020           &      0.0162\sym{***}&      0.0110\sym{*}  &      0.0232\sym{***}&      0.0132\sym{*}  &      0.0211\sym{***}&    -0.00736         \\
                    &      (3.64)         &      (2.16)         &      (3.37)         &      (2.07)         &      (3.74)         &     (-1.27)         \\
\addlinespace
COVID-related             &      0.0375\sym{***}&      0.0701\sym{***}&      0.0592\sym{***}&      0.0877\sym{***}&      0.0493\sym{***}&     -0.0352\sym{***}\\
                    &     (11.25)         &     (16.98)         &     (11.13)         &     (17.29)         &     (11.49)         &     (-8.02)         \\
\addlinespace
year=2020 $\times$ COVID-related&     -0.0212\sym{***}&     -0.0246\sym{**} &     -0.0358\sym{***}&     -0.0374\sym{***}&     -0.0206\sym{**} &      0.0168\sym{*}  \\
                    &     (-3.48)         &     (-3.24)         &     (-3.66)         &     (-4.02)         &     (-2.62)         &      (2.07)         \\
\addlinespace
SchoolClosures\_lag2&    -0.00214         &   -0.000932         &                     &                     &    -0.00304         &   -0.000767         \\
                    &     (-1.26)         &     (-0.48)         &                     &                     &     (-1.41)         &     (-0.35)         \\
\addlinespace
COVID-related $\times$ SchoolClosures\_lag2&    -0.00695\sym{**} &     -0.0124\sym{***}&                     &                     &    -0.00876\sym{**} &     0.00979\sym{**} \\
                    &     (-3.07)         &     (-4.51)         &                     &                     &     (-3.02)         &      (3.27)         \\
\addlinespace
\textit{N Authors}            &      0.0124\sym{***}&    -0.00213\sym{***}&   -0.000639         &    -0.00354\sym{***}&      0.0278\sym{***}&      0.0143\sym{***}\\
                    &     (44.95)         &     (-7.82)         &     (-1.63)         &     (-9.55)         &     (50.18)         &     (34.19)         \\
\addlinespace
trial               &      0.0241\sym{***}&     0.00427         &     0.00286         &      0.0167         &      0.0551\sym{***}&     0.00729         \\
                    &      (4.70)         &      (0.56)         &      (0.29)         &      (1.75)         &      (8.07)         &      (0.87)         \\
\addlinespace
Pre-existing Grant      &      0.0412\sym{***}&      0.0339\sym{***}&      0.0566\sym{***}&      0.0412\sym{***}&      0.0533\sym{***}&     -0.0204\sym{***}\\
                    &     (14.30)         &      (8.25)         &     (10.97)         &      (8.30)         &     (13.49)         &     (-4.72)         \\
\addlinespace
SchoolClosures\_lag2\_first&                     &                     &    -0.00284         &                     &                     &                     \\
                    &                     &                     &     (-1.09)         &                     &                     &                     \\
\addlinespace
COVID-related $\times$ SchoolClosures\_lag2\_first&                     &                     &     -0.0176\sym{***}&                     &                     &                     \\
                    &                     &                     &     (-4.89)         &                     &                     &                     \\
\addlinespace
SchoolClosures\_lag2\_last&                     &                     &                     &    0.000880         &                     &                     \\
                    &                     &                     &                     &      (0.36)         &                     &                     \\
\addlinespace
COVID-related $\times$ SchoolClosures\_lag2\_last&                     &                     &                     &     -0.0110\sym{**} &                     &                     \\
                    &                     &                     &                     &     (-3.23)         &                     &                     \\
\addlinespace
Constant            &       0.907\sym{***}&       0.204         &       0.324\sym{***}&       0.308\sym{***}&       0.567\sym{*}  &     -0.0457\sym{**} \\
                    &     (56.65)         &      (0.93)         &      (4.89)         &      (4.52)         &      (2.48)         &     (-2.99)         \\
\midrule
Observations        &       89267         &       89267         &       83235         &       82534         &       89267         &       89267         \\
\midrule
Country FEs & Majority  & Majority  & First & Last   & Majority  & Majority \\
\bottomrule
\multicolumn{7}{l}{\footnotesize \textit{t} statistics in parentheses}\\
\multicolumn{7}{l}{\footnotesize \sym{*} \(p<0.05\), \sym{**} \(p<0.01\), \sym{***} \(p<0.001\)}\\
\end{tabular}
}
\end{table}

\begin{table}[H]\centering
\def\sym#1{\ifmmode^{#1}\else\(^{#1}\)\fi}
\caption{Linear regression estimates including interaction two-months lagged monthly maximum \emph{workplace closures}, controlling for old grants, number of authors, \emph{trial} with White-robust standard errors.  Country effects (omitted) for the majority of the team for Female Author, First and Last Female Authors, Middle Female Authorship and Middle Female Only; country fixed effects of the first (last) author for regression on First (Last) Female Author.\label{lag2w}}
\resizebox{\textwidth}{!}{ 
\begin{tabular}{l c c c c c c}
\toprule  &\multicolumn{1}{c}{Female Author}&\multicolumn{1}{c}{First and Last Female}&\multicolumn{1}{c}{Female First Author}&\multicolumn{1}{c}{Female Last Author}&\multicolumn{1}{c}{Middle Female Authorship}&\multicolumn{1}{c}{Middle Female Only}\\
\toprule 
                    \multicolumn{1}{c}{Variables}&\multicolumn{1}{c}{(1)}&\multicolumn{1}{c}{(2)}&\multicolumn{1}{c}{(3)}&\multicolumn{1}{c}{(4)}&\multicolumn{1}{c}{(5)}&\multicolumn{1}{c}{(6)}\\
\midrule
year=2020           &      0.0160\sym{***}&     0.00835         &      0.0229\sym{***}&     0.00909         &      0.0210\sym{***}&    -0.00585         \\
                    &      (3.60)         &      (1.66)         &      (3.36)         &      (1.44)         &      (3.74)         &     (-1.02)         \\
\addlinespace
COVID-related             &      0.0376\sym{***}&      0.0702\sym{***}&      0.0593\sym{***}&      0.0877\sym{***}&      0.0494\sym{***}&     -0.0352\sym{***}\\
                    &     (11.28)         &     (17.00)         &     (11.15)         &     (17.30)         &     (11.51)         &     (-8.02)         \\
\addlinespace
year=2020 $\times$ COVID-related&     -0.0189\sym{**} &     -0.0209\sym{**} &     -0.0354\sym{***}&     -0.0293\sym{**} &     -0.0188\sym{*}  &      0.0170\sym{*}  \\
                    &     (-3.13)         &     (-2.81)         &     (-3.68)         &     (-3.21)         &     (-2.42)         &      (2.13)         \\
\addlinespace
WorkplaceClosures\_lag2&    -0.00211         &    0.000880         &                     &                     &    -0.00314         &    -0.00198         \\
                    &     (-1.15)         &      (0.40)         &                     &                     &     (-1.32)         &     (-0.82)         \\
\addlinespace
COVID-related $\times$ WorkplaceClosures\_lag2&    -0.00955\sym{***}&     -0.0169\sym{***}&                     &                     &     -0.0114\sym{***}&      0.0114\sym{***}\\
                    &     (-3.84)         &     (-5.47)         &                     &                     &     (-3.53)         &      (3.46)         \\
\addlinespace
\textit{N Authors}            &      0.0124\sym{***}&    -0.00213\sym{***}&   -0.000645         &    -0.00354\sym{***}&      0.0278\sym{***}&      0.0143\sym{***}\\
                    &     (44.92)         &     (-7.84)         &     (-1.64)         &     (-9.55)         &     (50.16)         &     (34.20)         \\
\addlinespace
trial               &      0.0239\sym{***}&     0.00421         &     0.00303         &      0.0165         &      0.0550\sym{***}&     0.00717         \\
                    &      (4.67)         &      (0.55)         &      (0.31)         &      (1.73)         &      (8.05)         &      (0.86)         \\
\addlinespace
Pre-existing Grant      &      0.0412\sym{***}&      0.0341\sym{***}&      0.0568\sym{***}&      0.0412\sym{***}&      0.0534\sym{***}&     -0.0205\sym{***}\\
                    &     (14.31)         &      (8.29)         &     (11.00)         &      (8.30)         &     (13.52)         &     (-4.76)         \\
\addlinespace
WorkplaceClosures\_lag2\_first&                     &                     &    -0.00261         &                     &                     &                     \\
                    &                     &                     &     (-0.90)         &                     &                     &                     \\
\addlinespace

COVID-related $\times$ WorkplaceClosures\_lag2\_first&                     &                     &     -0.0210\sym{***}&                     &                     &                     \\
                    &                     &                     &     (-5.25)         &                     &                     &                     \\
\addlinespace
WorkplaceClosures\_lag2\_last&                     &                     &                     &     0.00390         &                     &                     \\
                    &                     &                     &                     &      (1.42)         &                     &                     \\
\addlinespace
COVID-related $\times$ WorkplaceClosures\_lag2\_last&                     &                     &                     &     -0.0179\sym{***}&                     &                     \\
                    &                     &                     &                     &     (-4.66)         &                     &                     \\
\addlinespace
Constant            &       0.908\sym{***}&       0.205         &       0.322\sym{***}&       0.306\sym{***}&       0.568\sym{*}  &     -0.0467\sym{**} \\
                    &     (53.94)         &      (0.94)         &      (4.84)         &      (4.50)         &      (2.48)         &     (-3.02)         \\
\midrule
Observations        &       89267         &       89267         &       83235         &       82534         &       89267         &       89267         \\
\midrule
Country FEs & Majority  & Majority  & First & Last   & Majority  & Majority \\
\bottomrule
\multicolumn{7}{l}{\footnotesize \textit{t} statistics in parentheses}\\
\multicolumn{7}{l}{\footnotesize \sym{*} \(p<0.05\), \sym{**} \(p<0.01\), \sym{***} \(p<0.001\)}\\
\end{tabular}
}
\end{table}

\begin{table}[H]\centering
\def\sym#1{\ifmmode^{#1}\else\(^{#1}\)\fi}
\caption{Linear regression estimates including interaction three-months lagged monthly maximum \emph{school closures}, controlling for old grants, number of authors, \emph{trial} with White-robust standard errors.  Country effects (omitted) for the majority of the team for Female Author, First and Last Female Authors, Middle Female Authorship and Middle Female Only; country fixed effects of the first (last) author for regression on First (Last) Female Author.\label{lag3s}}
\resizebox{\textwidth}{!}{ 
\begin{tabular}{l c c c c c c}
\toprule  &\multicolumn{1}{c}{Female Author}&\multicolumn{1}{c}{First and Last Female}&\multicolumn{1}{c}{Female First Author}&\multicolumn{1}{c}{Female Last Author}&\multicolumn{1}{c}{Middle Female Authorship}&\multicolumn{1}{c}{Middle Female Only}\\
\toprule 
                    \multicolumn{1}{c}{Variables}&\multicolumn{1}{c}{(1)}&\multicolumn{1}{c}{(2)}&\multicolumn{1}{c}{(3)}&\multicolumn{1}{c}{(4)}&\multicolumn{1}{c}{(5)}&\multicolumn{1}{c}{(6)}\\
\midrule
year=2020           &      0.0158\sym{***}&      0.0106\sym{*}  &      0.0237\sym{***}&      0.0142\sym{*}  &      0.0193\sym{***}&    -0.00958         \\
                    &      (3.78)         &      (2.22)         &      (3.69)         &      (2.36)         &      (3.64)         &     (-1.76)         \\
\addlinespace
COVID-related             &      0.0374\sym{***}&      0.0701\sym{***}&      0.0591\sym{***}&      0.0877\sym{***}&      0.0493\sym{***}&     -0.0352\sym{***}\\
                    &     (11.24)         &     (16.97)         &     (11.12)         &     (17.30)         &     (11.47)         &     (-8.02)         \\
\addlinespace
year=2020 $\times$ COVID-related&     -0.0249\sym{***}&     -0.0299\sym{***}&     -0.0462\sym{***}&     -0.0386\sym{***}&     -0.0266\sym{***}&      0.0231\sym{**} \\
                    &     (-4.35)         &     (-4.20)         &     (-5.03)         &     (-4.41)         &     (-3.61)         &      (3.03)         \\
\addlinespace
SchoolClosures\_lag3&    -0.00208         &   -0.000784         &                     &                     &    -0.00221         &    0.000549         \\
                    &     (-1.29)         &     (-0.42)         &                     &                     &     (-1.07)         &      (0.26)         \\
\addlinespace
COVID-related $\times$ SchoolClosures\_lag3&    -0.00543\sym{*}  &     -0.0105\sym{***}&                     &                     &    -0.00621\sym{*}  &     0.00696\sym{*}  \\
                    &     (-2.53)         &     (-4.03)         &                     &                     &     (-2.25)         &      (2.45)         \\
\addlinespace
\textit{N Authors}            &      0.0124\sym{***}&    -0.00213\sym{***}&   -0.000635         &    -0.00353\sym{***}&      0.0278\sym{***}&      0.0143\sym{***}\\
                    &     (44.92)         &     (-7.81)         &     (-1.62)         &     (-9.53)         &     (50.16)         &     (34.20)         \\
\addlinespace
trial               &      0.0242\sym{***}&     0.00446         &     0.00334         &      0.0166         &      0.0555\sym{***}&     0.00714         \\
                    &      (4.73)         &      (0.59)         &      (0.34)         &      (1.74)         &      (8.13)         &      (0.86)         \\
\addlinespace
Pre-existing Grant      &      0.0413\sym{***}&      0.0340\sym{***}&      0.0567\sym{***}&      0.0410\sym{***}&      0.0535\sym{***}&     -0.0204\sym{***}\\
                    &     (14.31)         &      (8.27)         &     (10.99)         &      (8.26)         &     (13.55)         &     (-4.72)         \\
\addlinespace
SchoolClosures\_lag3\_first&                     &                     &    -0.00354         &                     &                     &                     \\
                    &                     &                     &     (-1.41)         &                     &                     &                     \\
\addlinespace
COVID-related $\times$ SchoolClosures\_lag3\_first&                     &                     &     -0.0131\sym{***}&                     &                     &                     \\
                    &                     &                     &     (-3.83)         &                     &                     &                     \\
\addlinespace
SchoolClosures\_lag3\_last&                     &                     &                     &    0.000356         &                     &                     \\
                    &                     &                     &                     &      (0.15)         &                     &                     \\
\addlinespace
COVID-related $\times$ SchoolClosures\_lag3\_last&                     &                     &                     &     -0.0110\sym{***}&                     &                     \\
                    &                     &                     &                     &     (-3.41)         &                     &                     \\
\addlinespace
Constant            &       0.909\sym{***}&       0.206         &       0.325\sym{***}&       0.308\sym{***}&       0.570\sym{*}  &     -0.0469\sym{**} \\
                    &     (58.45)         &      (0.94)         &      (4.91)         &      (4.52)         &      (2.49)         &     (-2.94)         \\
\midrule
Observations        &       89267         &       89267         &       83235         &       82534         &       89267         &       89267         \\
\midrule
Country FEs & Majority  & Majority  & First & Last   & Majority  & Majority \\
\bottomrule
\multicolumn{7}{l}{\footnotesize \textit{t} statistics in parentheses}\\
\multicolumn{7}{l}{\footnotesize \sym{*} \(p<0.05\), \sym{**} \(p<0.01\), \sym{***} \(p<0.001\)}\\
\end{tabular}
}
\end{table}

\begin{table}[H]\centering
\def\sym#1{\ifmmode^{#1}\else\(^{#1}\)\fi}
\caption{Linear regression estimates including interaction three-months lagged monthly maximum \emph{workplace closures}, controlling for old grants, number of authors, \emph{trial} with White-robust standard errors.  Country effects (omitted) for the majority of the team for Female Author, First and Last Female Authors, Middle Female Authorship and Middle Female Only; country fixed effects of the first (last) author for regression on First (Last) Female Author.\label{lag3w}}
\resizebox{\textwidth}{!}{ 
\begin{tabular}{l c c c c c c}
\toprule  &\multicolumn{1}{c}{Female Author}&\multicolumn{1}{c}{First and Last Female}&\multicolumn{1}{c}{Female First Author}&\multicolumn{1}{c}{Female Last Author}&\multicolumn{1}{c}{Middle Female Authorship}&\multicolumn{1}{c}{Middle Female Only}\\
\toprule 
                    \multicolumn{1}{c}{Variables}&\multicolumn{1}{c}{(1)}&\multicolumn{1}{c}{(2)}&\multicolumn{1}{c}{(3)}&\multicolumn{1}{c}{(4)}&\multicolumn{1}{c}{(5)}&\multicolumn{1}{c}{(6)}\\
\midrule
year=2020           &      0.0140\sym{***}&     0.00905         &      0.0214\sym{***}&      0.0123\sym{*}  &      0.0178\sym{***}&    -0.00870         \\
                    &      (3.38)         &      (1.92)         &      (3.35)         &      (2.05)         &      (3.39)         &     (-1.61)         \\
\addlinespace
COVID-related             &      0.0375\sym{***}&      0.0702\sym{***}&      0.0593\sym{***}&      0.0877\sym{***}&      0.0493\sym{***}&     -0.0352\sym{***}\\
                    &     (11.26)         &     (17.00)         &     (11.15)         &     (17.30)         &     (11.49)         &     (-8.03)         \\
\addlinespace
year=2020 $\times$ COVID-related&     -0.0214\sym{***}&     -0.0248\sym{***}&     -0.0392\sym{***}&     -0.0334\sym{***}&     -0.0235\sym{**} &      0.0211\sym{**} \\
                    &     (-3.79)         &     (-3.55)         &     (-4.34)         &     (-3.88)         &     (-3.22)         &      (2.82)         \\
\addlinespace
WorkplaceClosures\_lag3&   -0.000998         &    0.000413         &                     &                     &    -0.00135         &   -0.000131         \\
                    &     (-0.57)         &      (0.19)         &                     &                     &     (-0.59)         &     (-0.06)         \\
\addlinespace
COVID-related $\times$ WorkplaceClosures\_lag3&    -0.00868\sym{***}&     -0.0156\sym{***}&                     &                     &    -0.00937\sym{**} &     0.00943\sym{**} \\
                    &     (-3.67)         &     (-5.34)         &                     &                     &     (-3.05)         &      (2.99)         \\
\addlinespace
\textit{N Authors}            &      0.0124\sym{***}&    -0.00212\sym{***}&   -0.000628         &    -0.00353\sym{***}&      0.0278\sym{***}&      0.0143\sym{***}\\
                    &     (44.94)         &     (-7.79)         &     (-1.60)         &     (-9.52)         &     (50.17)         &     (34.19)         \\
\addlinespace
trial               &      0.0242\sym{***}&     0.00434         &     0.00327         &      0.0165         &      0.0555\sym{***}&     0.00713         \\
                    &      (4.73)         &      (0.57)         &      (0.33)         &      (1.73)         &      (8.13)         &      (0.85)         \\
\addlinespace
Pre-existing Grant      &      0.0413\sym{***}&      0.0339\sym{***}&      0.0567\sym{***}&      0.0410\sym{***}&      0.0535\sym{***}&     -0.0204\sym{***}\\
                    &     (14.33)         &      (8.26)         &     (10.98)         &      (8.27)         &     (13.56)         &     (-4.73)         \\
\addlinespace
WorkplaceClosures\_lag3\_first&                     &                     &    -0.00202         &                     &                     &                     \\
                    &                     &                     &     (-0.72)         &                     &                     &                     \\
\addlinespace
COVID-related $\times$ WorkplaceClosures\_lag3\_first&                     &                     &     -0.0200\sym{***}&                     &                     &                     \\
                    &                     &                     &     (-5.24)         &                     &                     &                     \\
\addlinespace
WorkplaceClosures\_lag3\_last&                     &                     &                     &     0.00205         &                     &                     \\
                    &                     &                     &                     &      (0.77)         &                     &                     \\
\addlinespace
COVID-related $\times$ WorkplaceClosures\_lag3\_last&                     &                     &                     &     -0.0164\sym{***}&                     &                     \\
                    &                     &                     &                     &     (-4.49)         &                     &                     \\
\addlinespace
Constant            &       0.909\sym{***}&       0.206         &       0.323\sym{***}&       0.307\sym{***}&       0.570\sym{*}  &     -0.0472\sym{**} \\
                    &     (56.00)         &      (0.94)         &      (4.86)         &      (4.50)         &      (2.49)         &     (-3.01)         \\
\midrule
Observations        &       89267         &       89267         &       83235         &       82534         &       89267         &       89267         \\
\midrule
Country FEs & Majority  & Majority  & First & Last   & Majority  & Majority \\
\bottomrule
\multicolumn{7}{l}{\footnotesize \textit{t} statistics in parentheses}\\
\multicolumn{7}{l}{\footnotesize \sym{*} \(p<0.05\), \sym{**} \(p<0.01\), \sym{***} \(p<0.001\)}\\
\end{tabular}
}
\end{table}

\subsection{Propensity Score Matching with Diff-In-Diff}\label{psm_did}

We perform a propensity-score matching (PSM) and Diff-in-Diff estimation on the matched sample to check whether our estimated treatment effect is robust against \emph{selection on observable variables} that may systematically differ between treatment and COVID non-related, causing a bias in our estimated results. The fundamental assumption behind PSM is that units are assigned to treatment independent of potential outcome, once we condition on observable characteristics unaffected by treatment (\emph{conditional independence}). 
We wish to match COVID-related observations with COVID non-related on the basis of a propensity score, i.e. the probability of treatment assignment given a vector of observed covariates (Rosenbaum and Rubin, 1983). 
The choice of characteristics to include in the propensity score is crucial, as one should include only variables that influence simultaneously the treatment and the outcome, i.e. potential confounders, but that cannot be affected by an anticipation of the treatment. 
Because the dataset consists of repeated cross-sections of observations at the paper level, we need to carry out a three steps procedure (Binci et al., 2018):

\begin{enumerate}
    \item Exact matching of COVID-related units at baseline (2019) with COVID-related units at midline (2020).
    \item 1-to-1 Propensity Score (PS) matching with Nearest Neighbor of COVID non-related units at baseline with COVID-related units at baseline selected in step 1 (caliper 0.05).
    \item 1-to-1 PS matching with Nearest Neighbor of COVID non-related units at midline with COVID-related units at midline selected in step 1 (caliper 0.05). 
\end{enumerate}

We follow step 1 and decide to construct the \emph{pseudo panel} of COVID-related unit by finding the exact match on major MeSH term and country. For all outcomes, except first female authorship and last female authorship, we refer to the country of the majority of team members. For first (last) female authorship, we refer to the country of the first (last) author. We get a total of 30392 observations in the pseudo panel sample matched on MeSH terms and majority country; 27941 (27639) COVID-related matched observations for first (last) female authorship. 

In step 2, we perform 1-to-1 Nearest Neighbor matching on the propensity score to match COVID non-related units in 2019 with 2019 treatment units in the \emph{pseudo panel}. We use a Probit regression to estimate the propensity to treatment, with \emph{country}, \emph{trial}, \emph{has old grant} and \emph{N authors} as potential confounders. We obtain a matched sample for 2019 featuring 26012 observations for matching on majority country; 24058 (24062) for matching on first (last) authors' countries. 

At last, we perform 1-to-1 Nearest Neighbor matching on the propensity score and pair each 2020 treatment unit selected in step 1 with the 2020 COVID non-related units with closest propensity score (caliper = 0.05). Again, the propensity to treatment is estimated with a Probit regression on country, \emph{number of authors}, \emph{has old grant} and \emph{trial}. The matched sample for 2020 ends up with 26146 observations when matching on majority country, 24060 (24386) when using country of the first (last) author. 

Finally, we obtain a matched sample of 52158 observations for the matching procedure with majority country, 48118 for country of the first author and 48448 for the last author. 

\begin{figure}[H]
 \begin{subfigure}[t]{0.5\textwidth}
    \includegraphics[width=\textwidth]{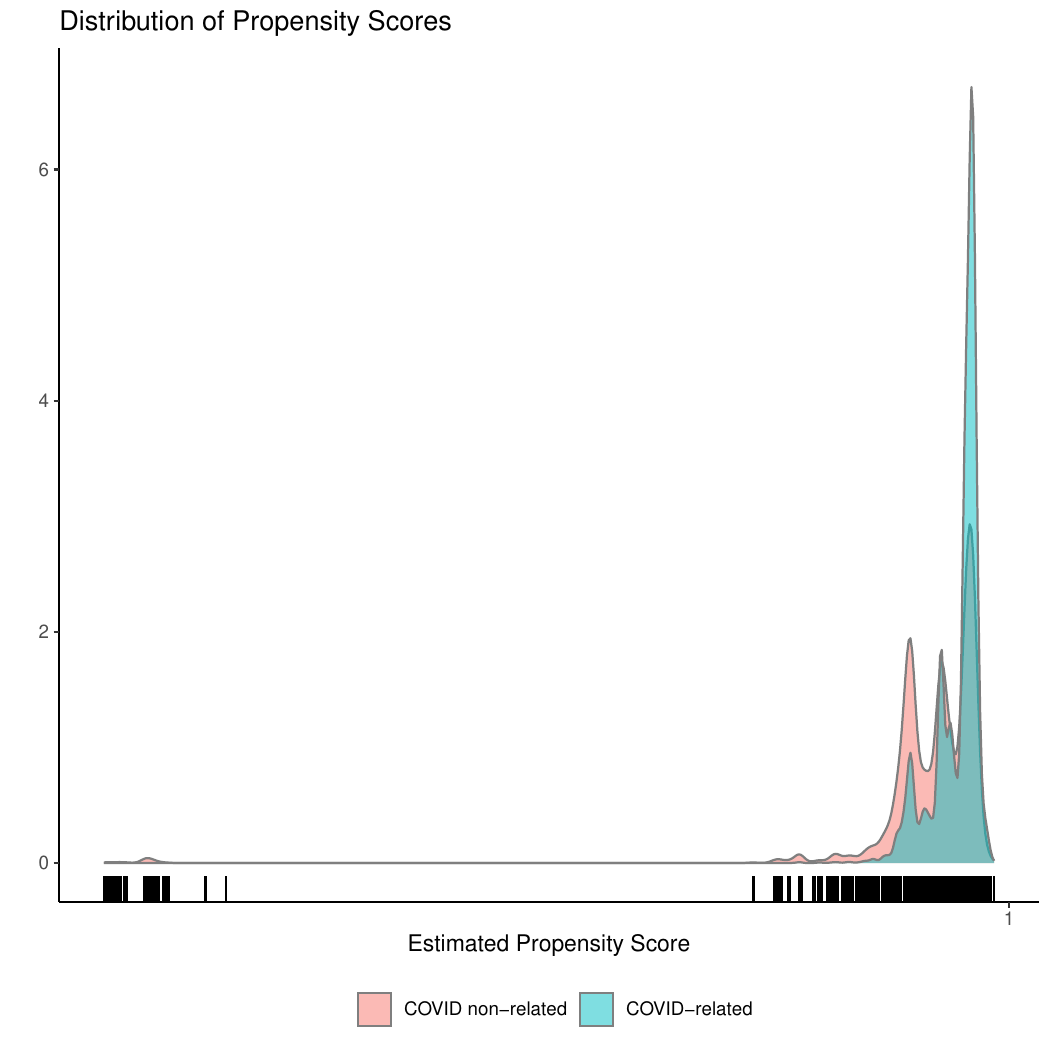}
    \caption{Unbalanced - Baseline}
    \label{fig:ps_1}
  \end{subfigure}
    \hfill
 \begin{subfigure}[t]{0.5\textwidth}
    \includegraphics[width=\textwidth]{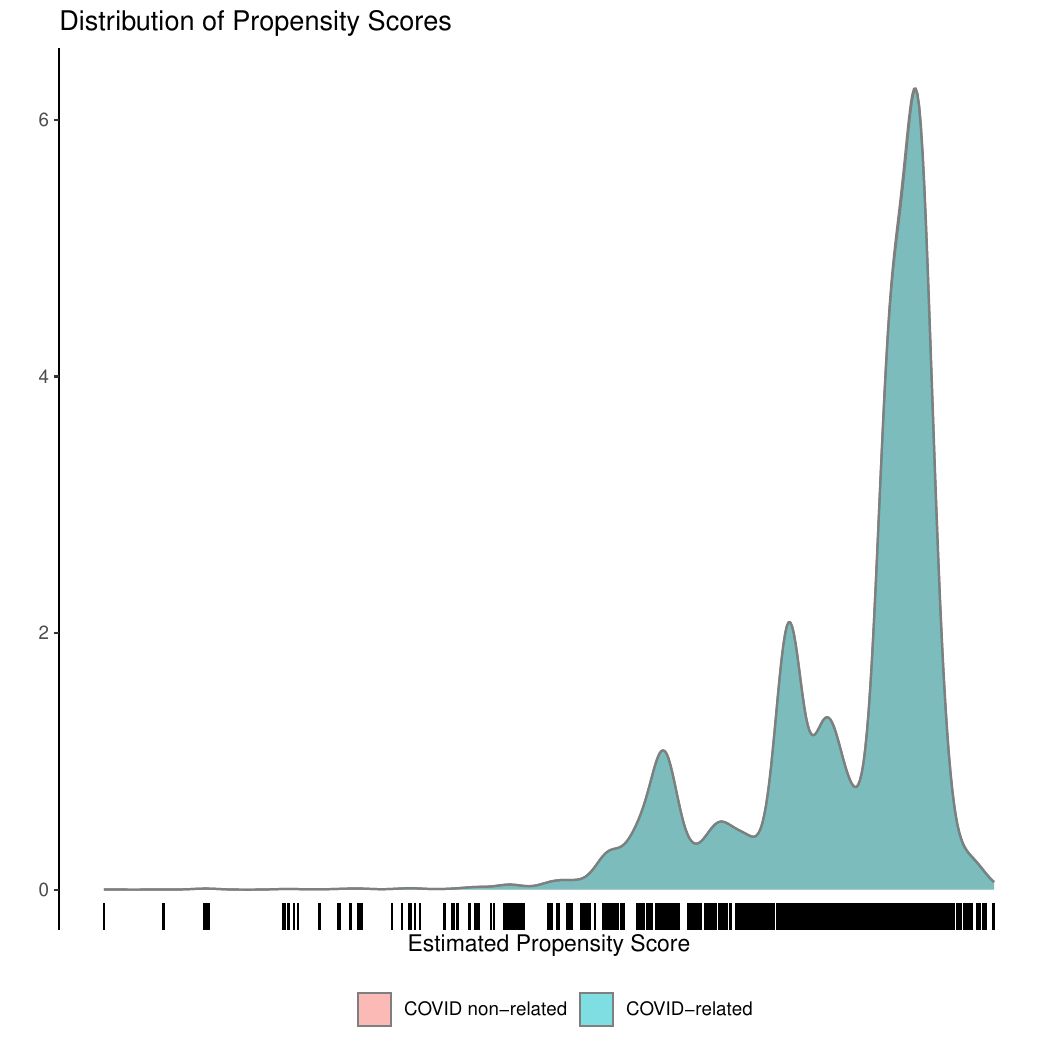}
    \caption{Balanced - Baseline}
    \label{fig:ps_2}
  \end{subfigure}
    \hfill
 \begin{subfigure}[t]{0.5\textwidth}
    \includegraphics[width=\textwidth]{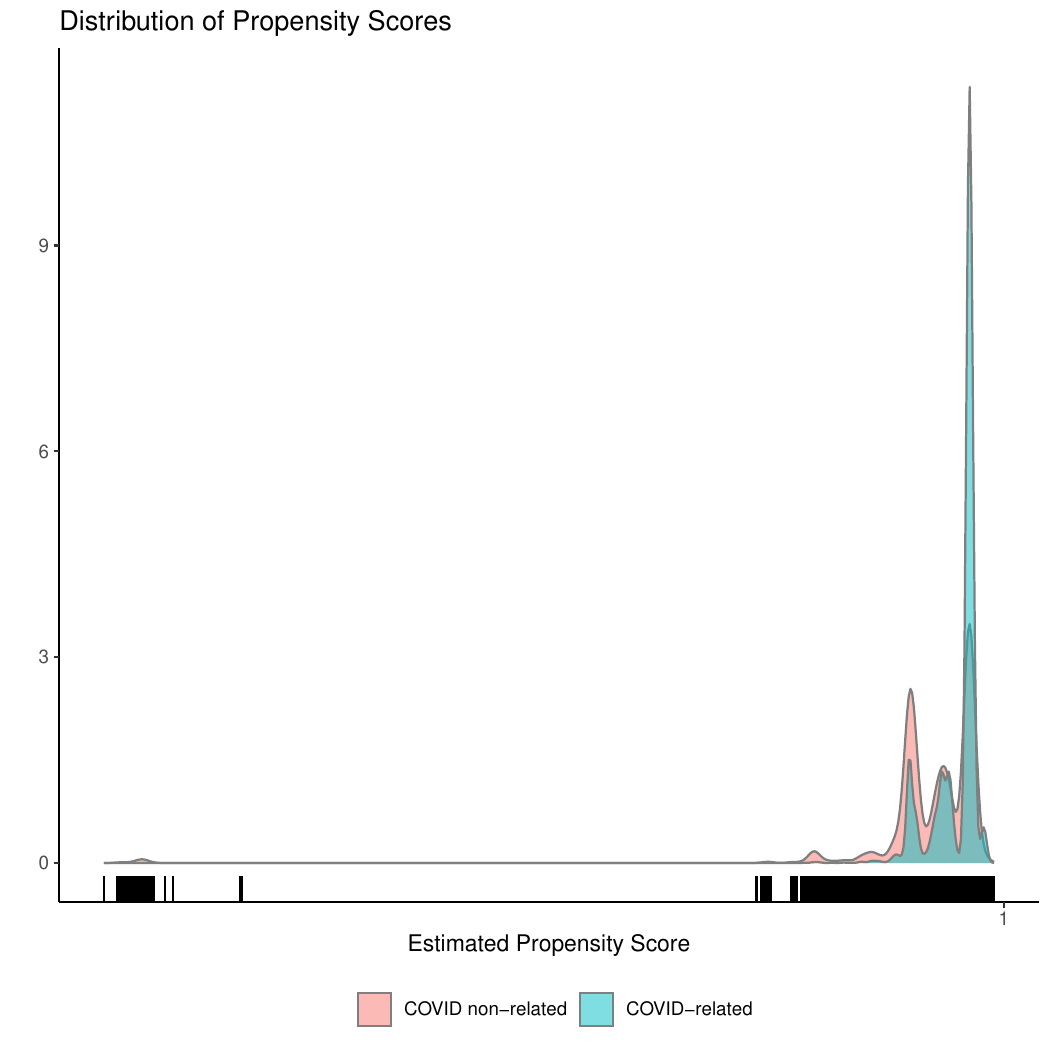}
    \caption{Unbalanced - Midline}
    \label{fig:ps_3}
  \end{subfigure}
   \begin{subfigure}[t]{0.5\textwidth}
    \includegraphics[width=\textwidth]{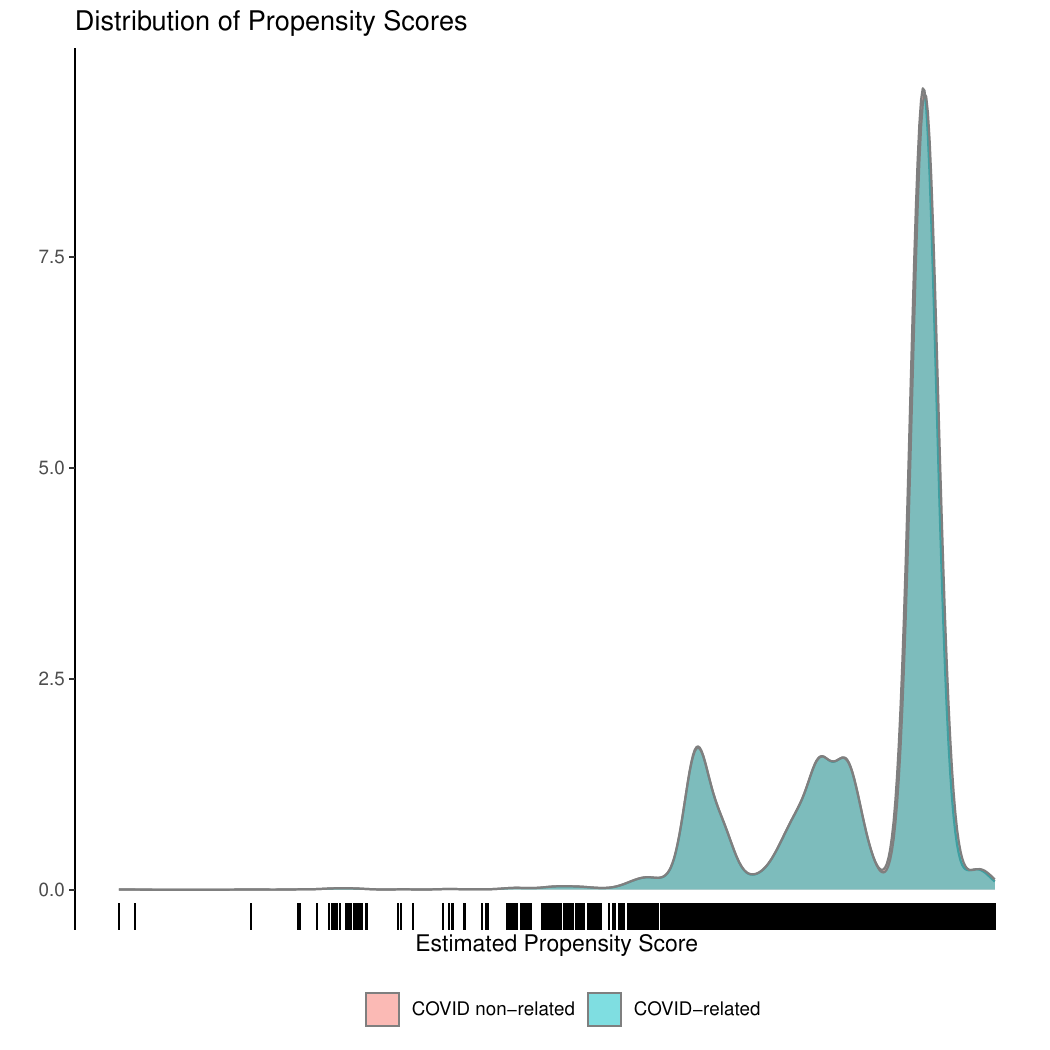}
    \caption{Balanced - Midline}
    \label{fig:ps_4}
  \end{subfigure}
  \caption{Distribution of estimated propensity score on unmatched (left) and matched (right) observations at baseline and midline; we match on MeSH terms and majority country between COVID-related; we match on paper level covariates and majority country for COVID non-related and COVID-related at baseline and midline.}
  \label{fig:psm_bs_maj}
\end{figure}

\begin{figure}[H]
 \begin{subfigure}[t]{0.5\textwidth}
    \includegraphics[width=\textwidth]{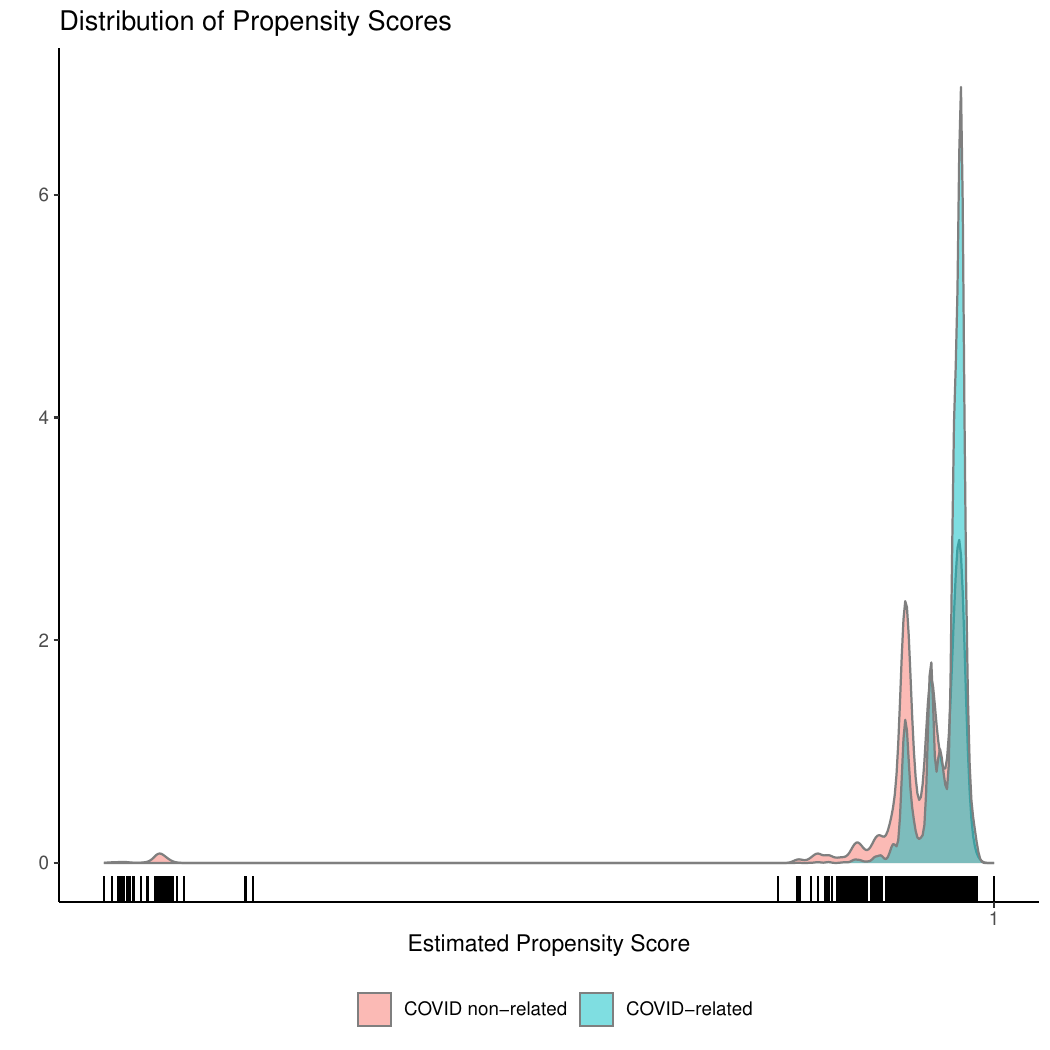}
    \caption{Unbalanced - Baseline}
    \label{fig:ps_21}
  \end{subfigure}
    \hfill
 \begin{subfigure}[t]{0.5\textwidth}
    \includegraphics[width=\textwidth]{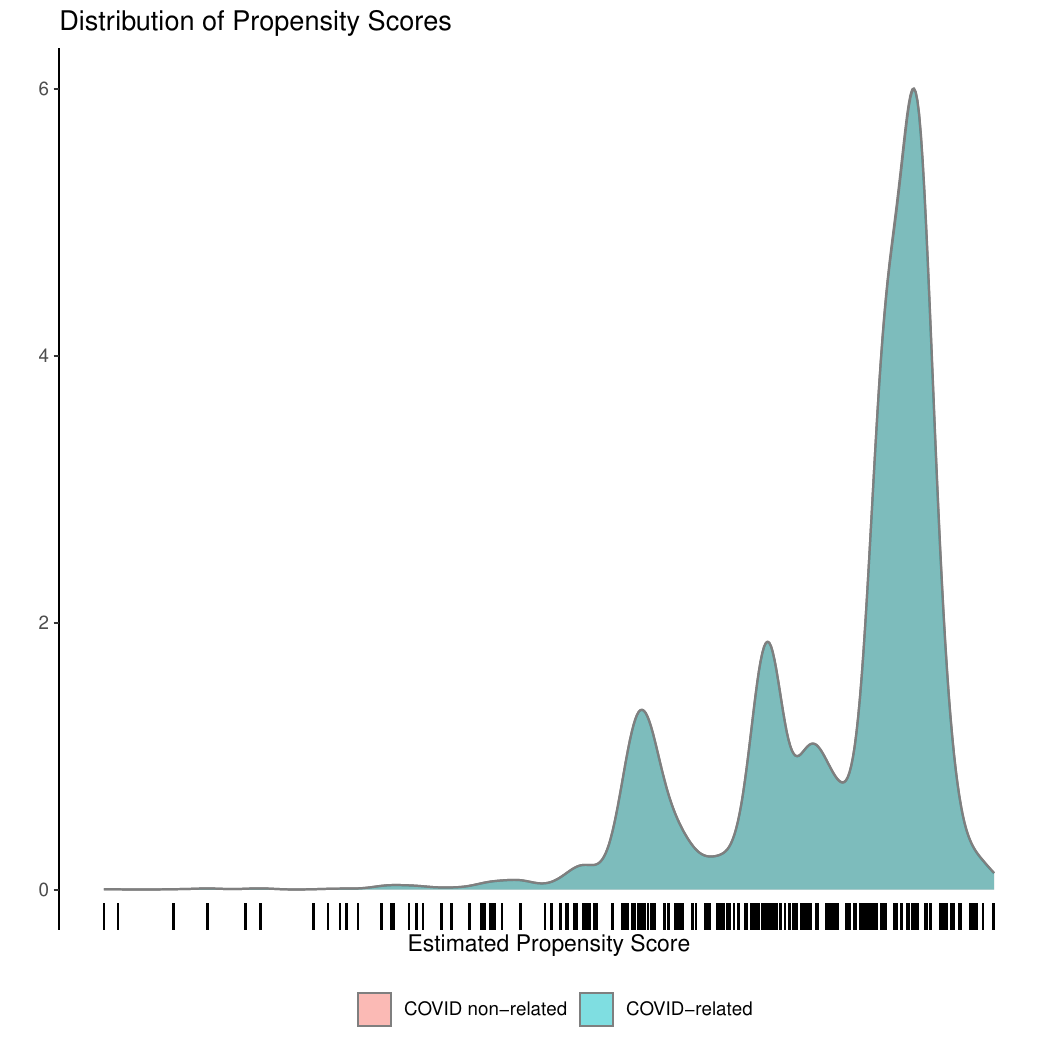}
    \caption{Balanced - Baseline}
    \label{fig:ps_22}
  \end{subfigure}
    \hfill
 \begin{subfigure}[t]{0.5\textwidth}
    \includegraphics[width=\textwidth]{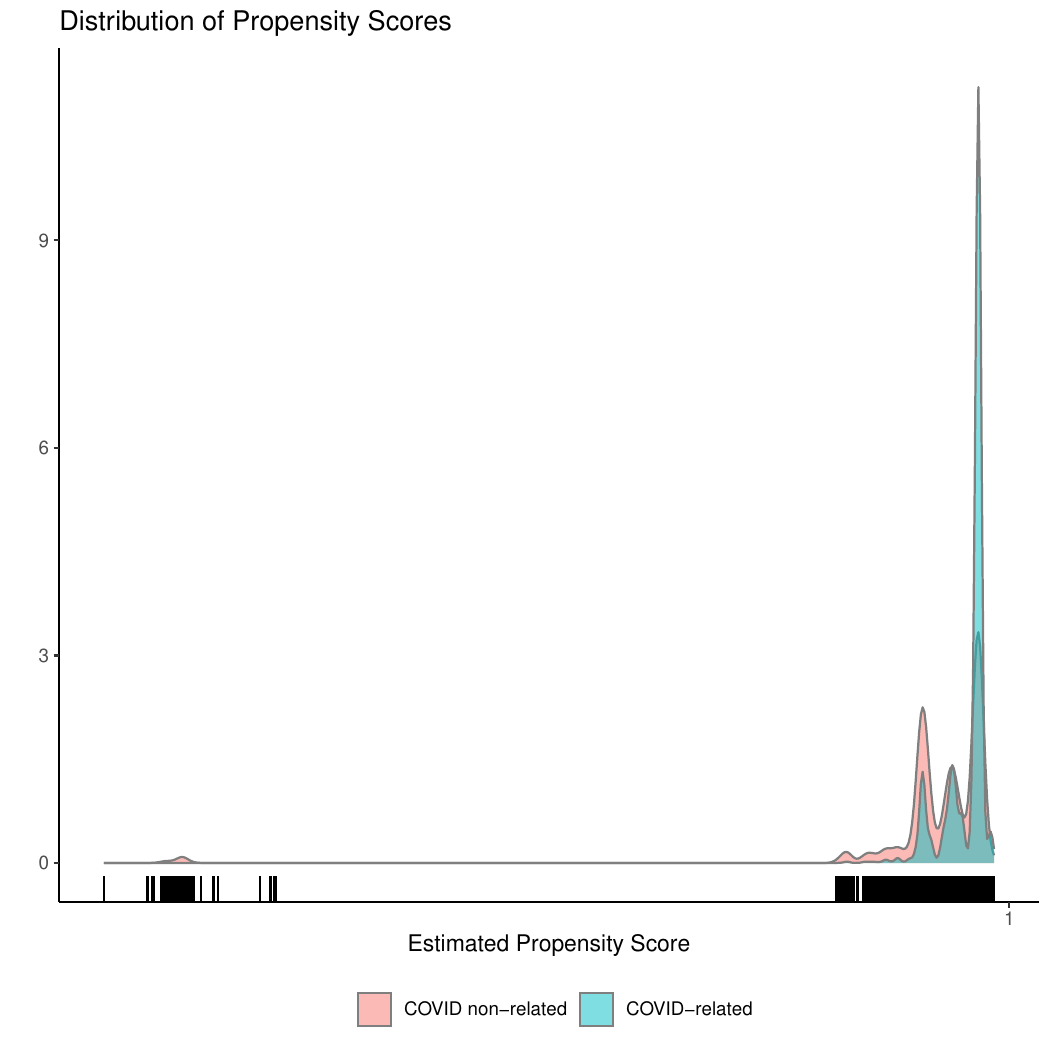}
    \caption{Unbalanced - Midline}
    \label{fig:ps_23}
  \end{subfigure}
   \begin{subfigure}[t]{0.5\textwidth}
    \includegraphics[width=\textwidth]{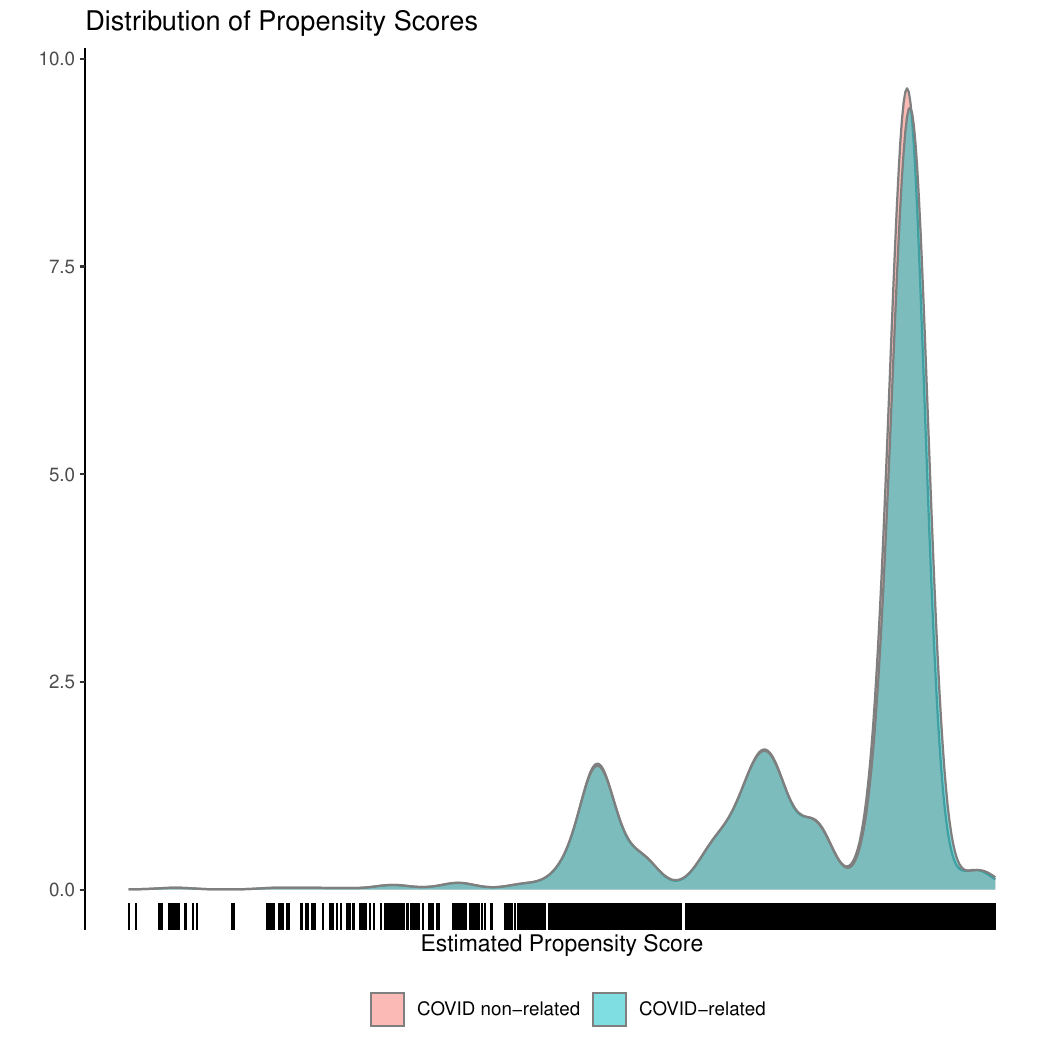}
    \caption{Balanced - Midline}
    \label{fig:ps_24}
  \end{subfigure}
  \caption{Distribution of estimated propensity score on unmatched  (left) and matched (right) observations at baseline and midline; we match on MeSH terms and first author's country between COVID-related; we match on paper level covariates and first author's country for COVID non-related and COVID-related at baseline and midline.}
  \label{fig:psm_bs_first}
\end{figure}

\begin{figure}[H]
 \begin{subfigure}[t]{0.5\textwidth}
    \includegraphics[width=\textwidth]{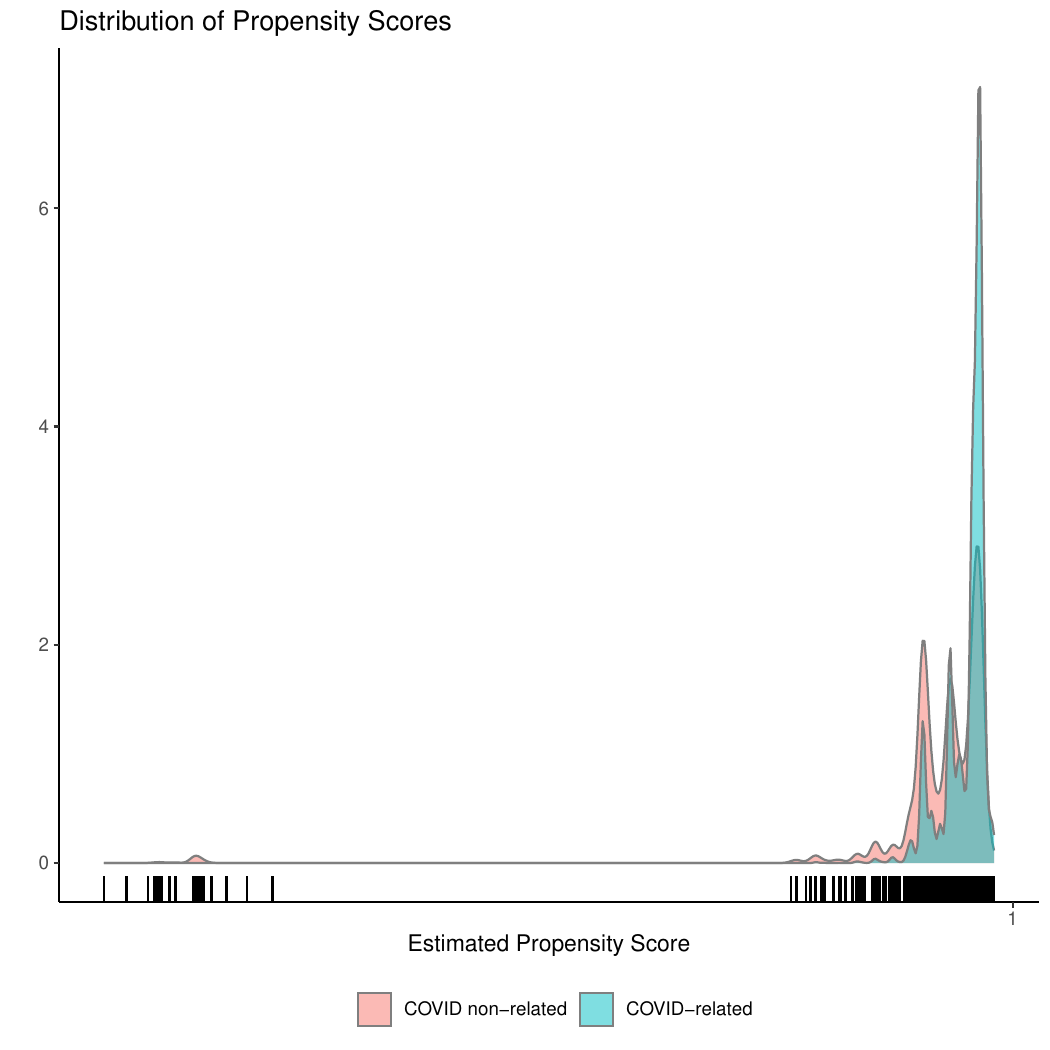}
    \caption{Unbalanced - Baseline}
    \label{fig:ps_31}
  \end{subfigure}
    \hfill
 \begin{subfigure}[t]{0.5\textwidth}
    \includegraphics[width=\textwidth]{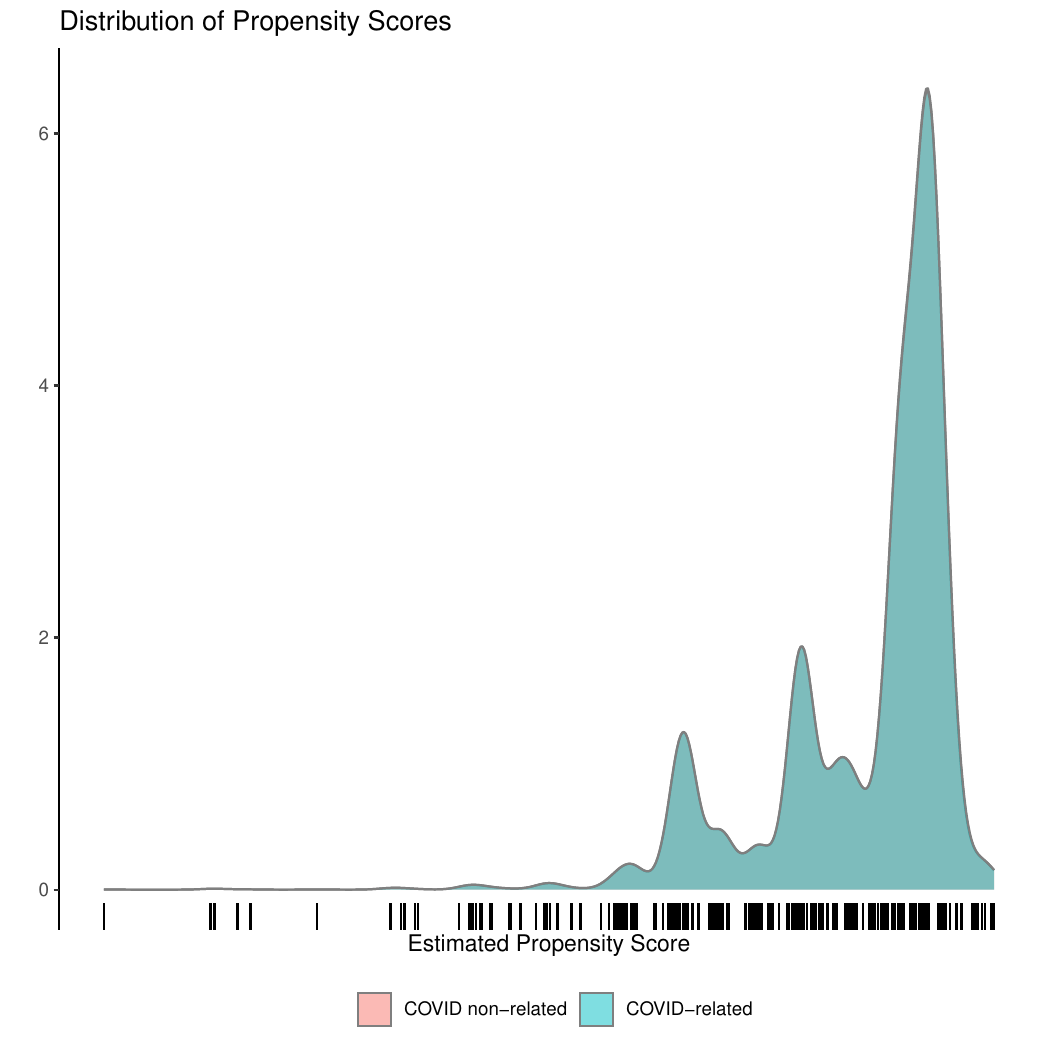}
    \caption{Balanced - Baseline}
    \label{fig:ps_32}
  \end{subfigure}
    \hfill
 \begin{subfigure}[t]{0.5\textwidth}
    \includegraphics[width=\textwidth]{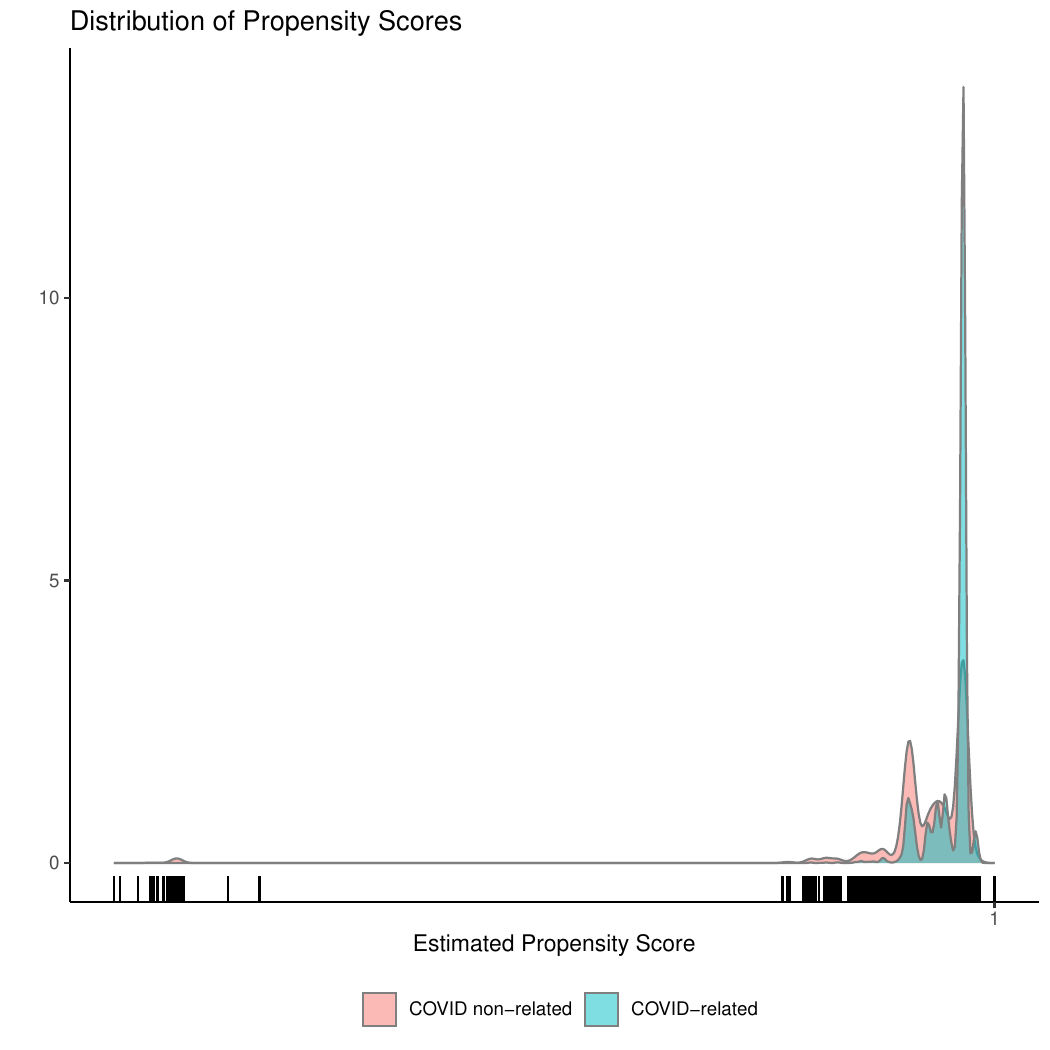}
    \caption{Unbalanced - Midline}
    \label{fig:ps_33}
  \end{subfigure}
   \begin{subfigure}[t]{0.5\textwidth}
    \includegraphics[width=\textwidth]{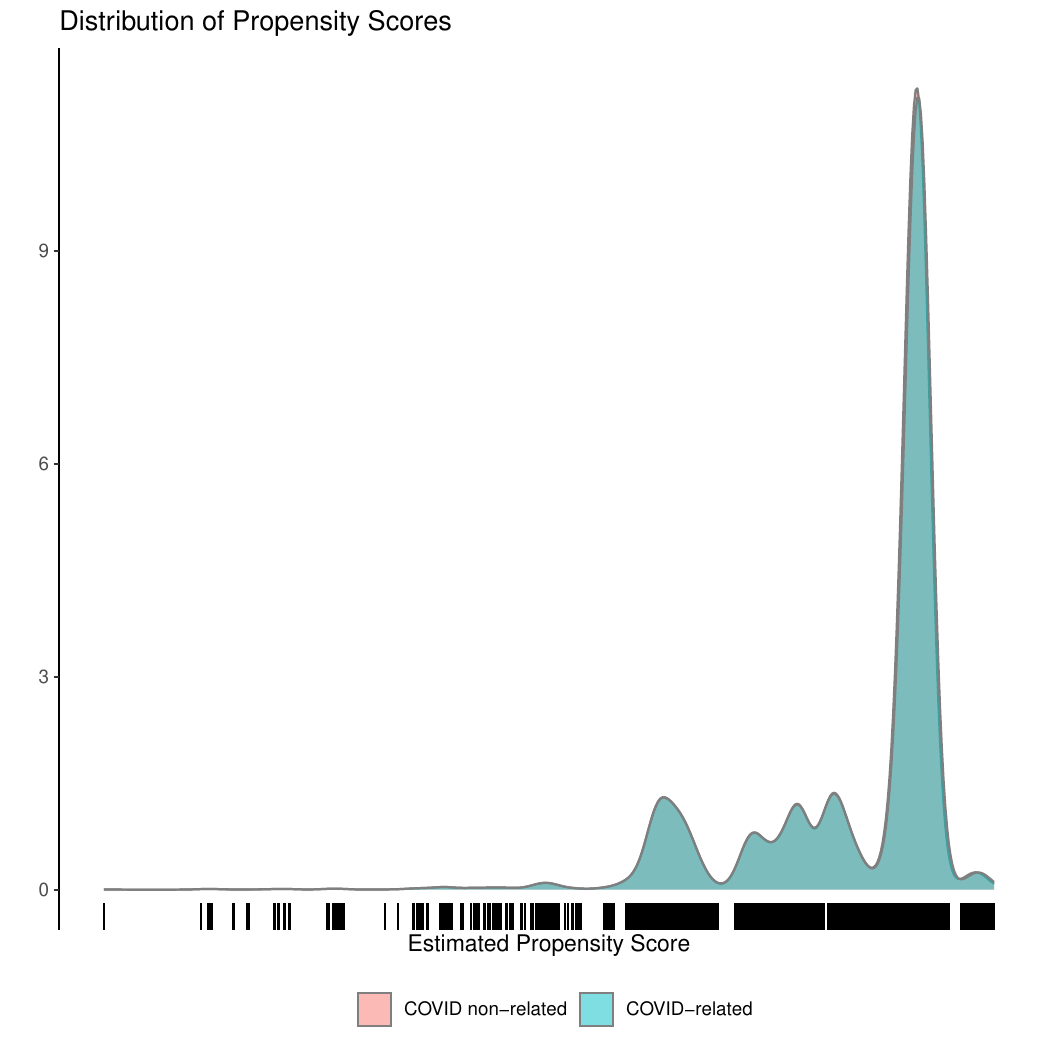}
    \caption{Balanced - Midline}
    \label{fig:ps_34}
  \end{subfigure}
  \caption{Distribution of estimated propensity score on unmatched  (left) and matched (right) observations at baseline and midline; we match on MeSH terms and last author's country between COVID-related; we match on paper level covariates and last author's country for COVID non-related and COVID-related at baseline and midline.}
  \label{fig:psm_bs_last}
\end{figure}

\begin{figure}[H]
\centering
\hspace{-2cm}
  \begin{subfigure}[t]{0.5\textwidth}
    \includegraphics[width=\textwidth]{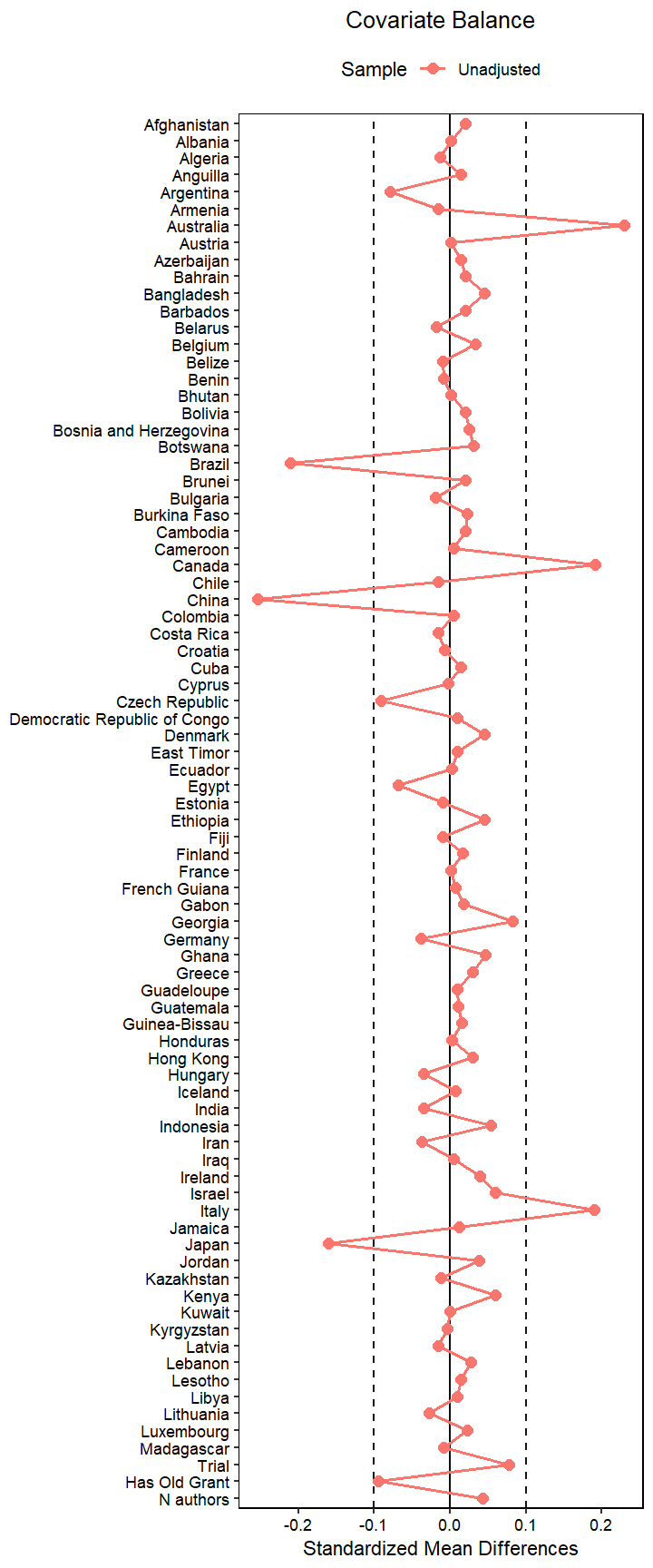}
    \label{fig:cfe_1}
  \end{subfigure}
 \begin{subfigure}[t]{0.5\textwidth}
    \includegraphics[width=\textwidth]{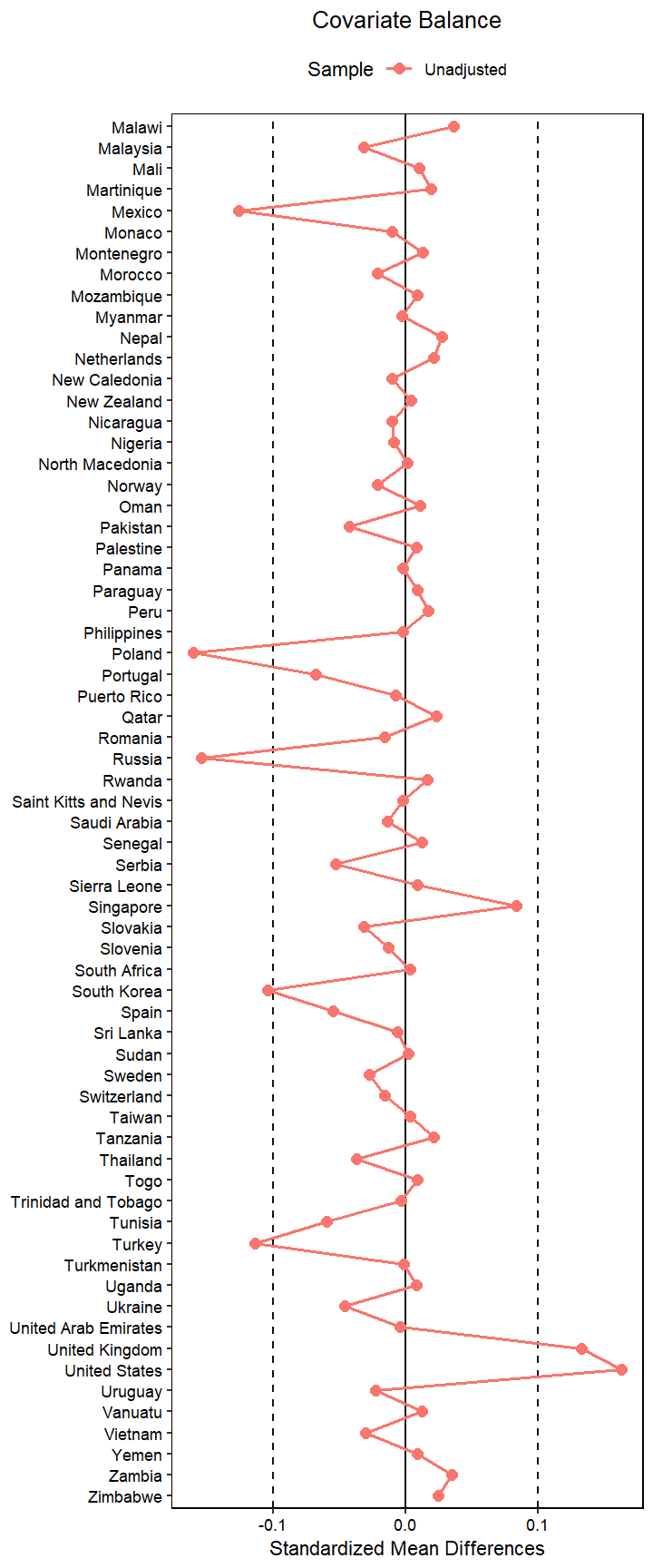}
    \label{fig:cfe_2}
  \end{subfigure}
  \caption{Standardized mean difference of covariates between \emph{COVID-related} and \emph{COVID non-related} before matching on country of majority of the publishing team and paper level controls (trial, N authors, Pre-existing Grant). Threshold lines at 0.1.}
  \label{fig:loveplot_1}
\end{figure}

\begin{figure}[H]
\centering
\hspace{-2cm}
  \begin{subfigure}[t]{0.5\textwidth}
    \includegraphics[width=\textwidth]{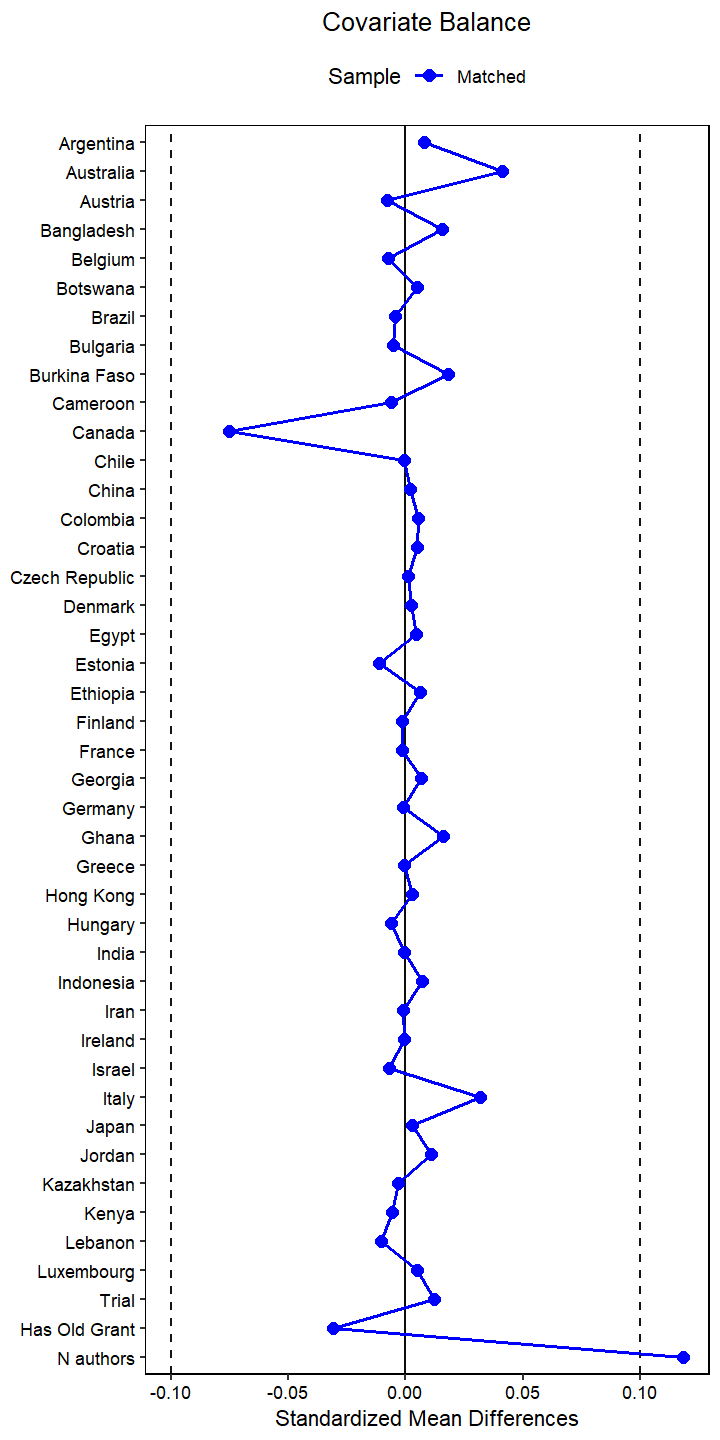}
    \label{fig:cfe_1}
  \end{subfigure}
 \begin{subfigure}[t]{0.5\textwidth}
    \includegraphics[width=\textwidth]{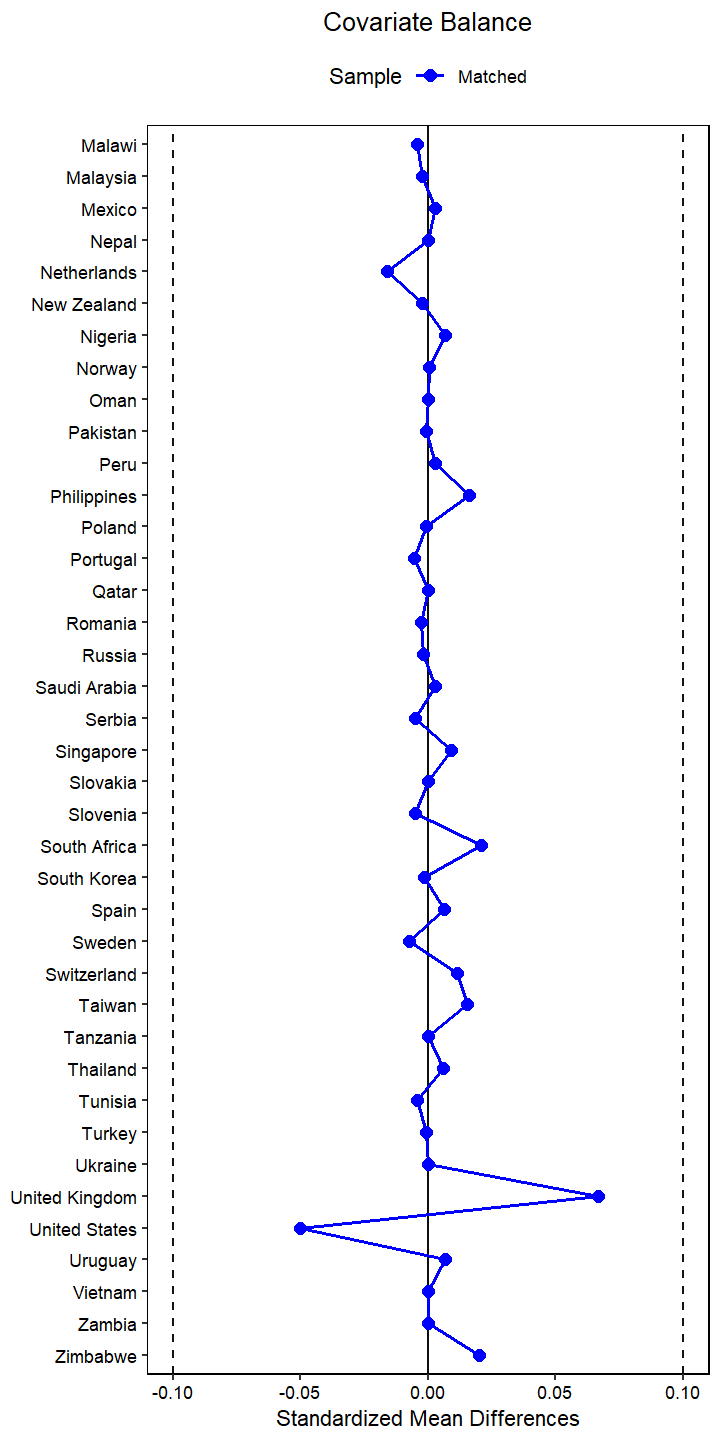}
    \label{fig:cfe_2}
  \end{subfigure}
  \caption{Standardized mean difference (with 0.1 thresholds) between \emph{COVID-related} and \emph{COVID non-related} after matching on country of majority of the publishing team and paper level controls (trial, N authors, Pre-existing Grant).}
  \label{fig:loveplot_2}
\end{figure}

\begin{figure}[H]
\centering
\hspace{-2cm}
  \begin{subfigure}[t]{0.5\textwidth}
    \includegraphics[width=\textwidth]{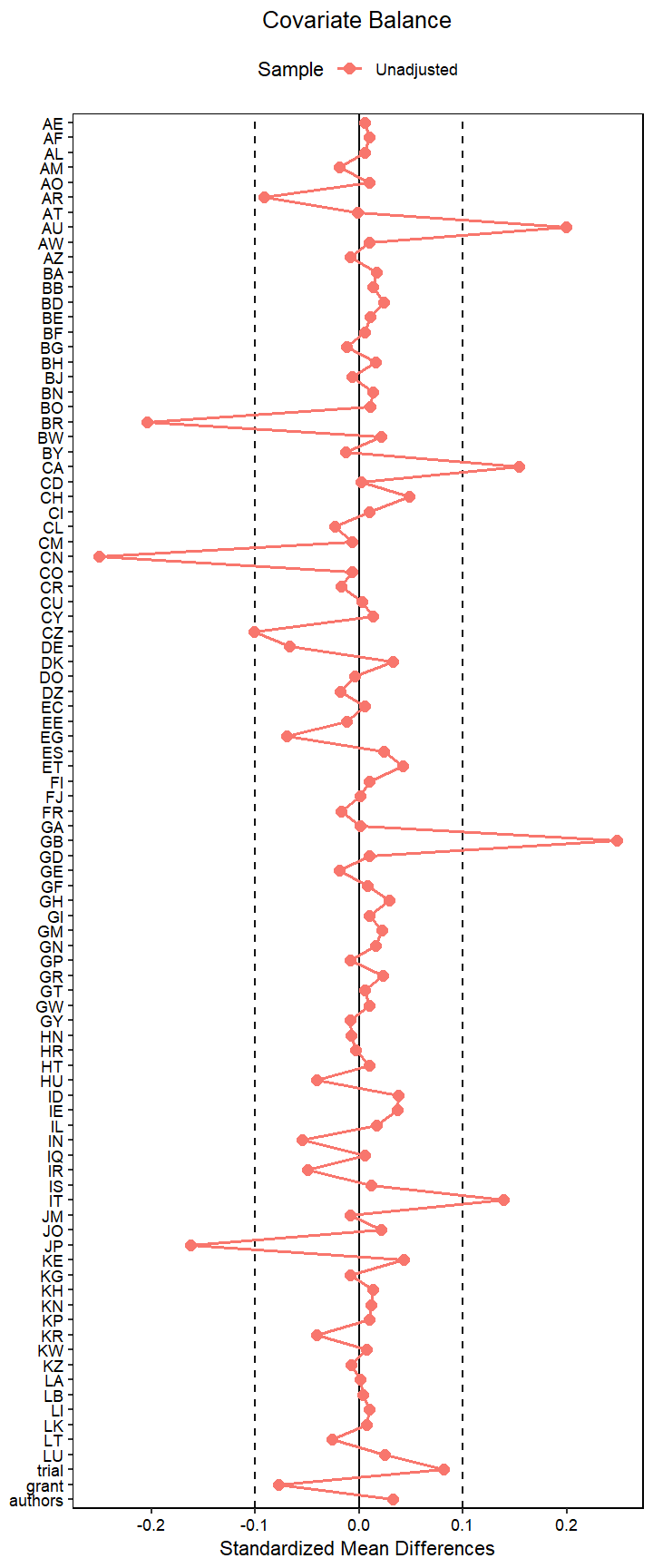}
    \label{fig:cfe_1}
  \end{subfigure}
 \begin{subfigure}[t]{0.5\textwidth}
    \includegraphics[width=\textwidth]{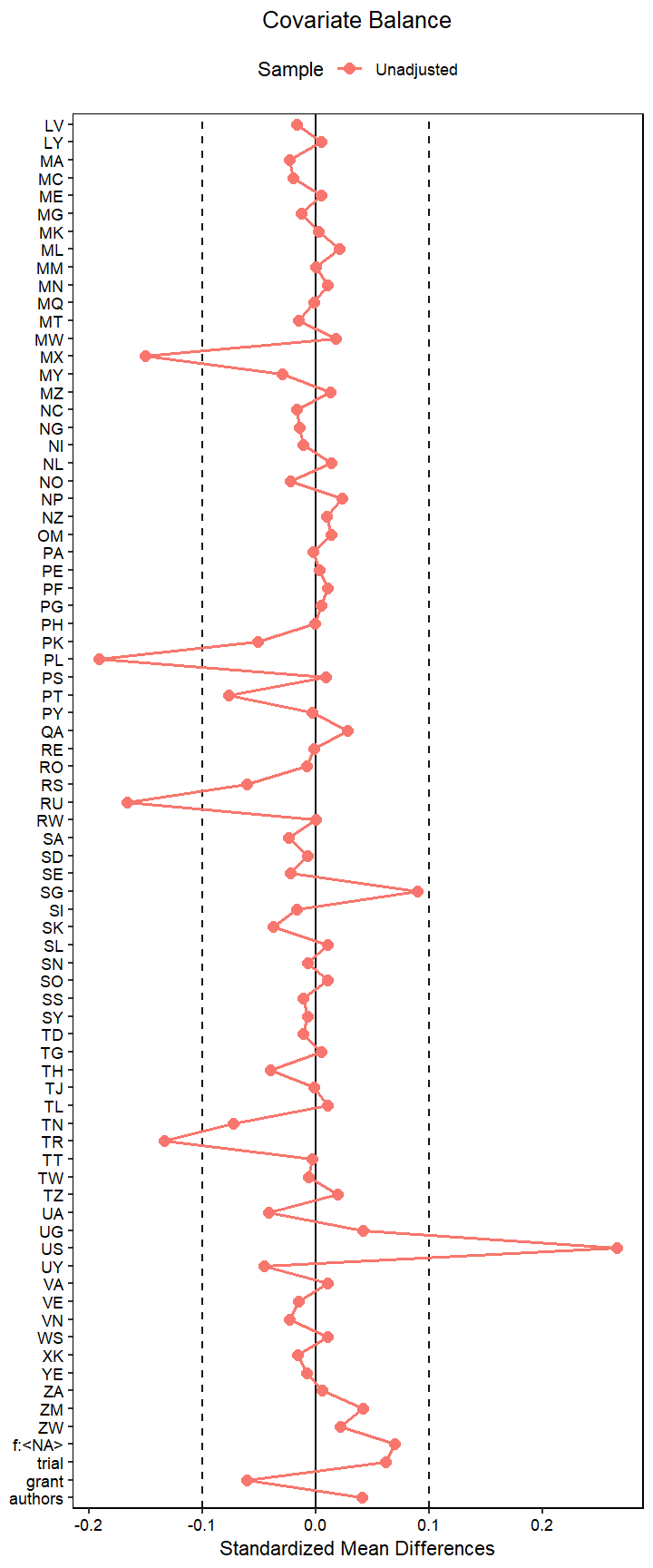}
    \label{fig:cfe_2}
  \end{subfigure}
  \caption{Standardized mean difference of covariates between \emph{COVID-related} and \emph{COVID non-related} before matching on country of \emph{first} author and paper level controls (trial, N authors, Pre-existing Grant). Threshold lines at 0.1.}
  \label{fig:loveplot_3}
\end{figure}

\begin{figure}[H]
\centering
\hspace{-2cm}
  \begin{subfigure}[t]{0.5\textwidth}
    \includegraphics[width=\textwidth]{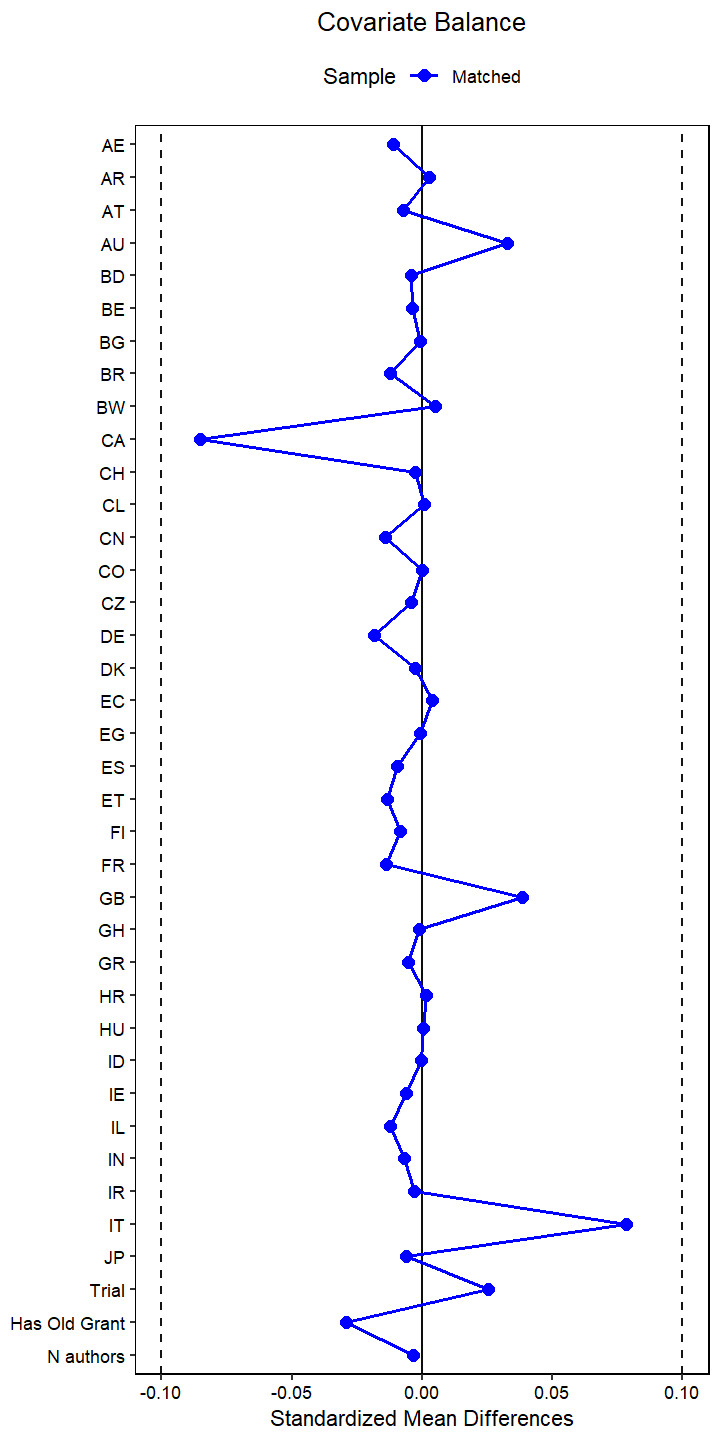}
    \label{fig:cfe_1}
  \end{subfigure}
 \begin{subfigure}[t]{0.5\textwidth}
    \includegraphics[width=\textwidth]{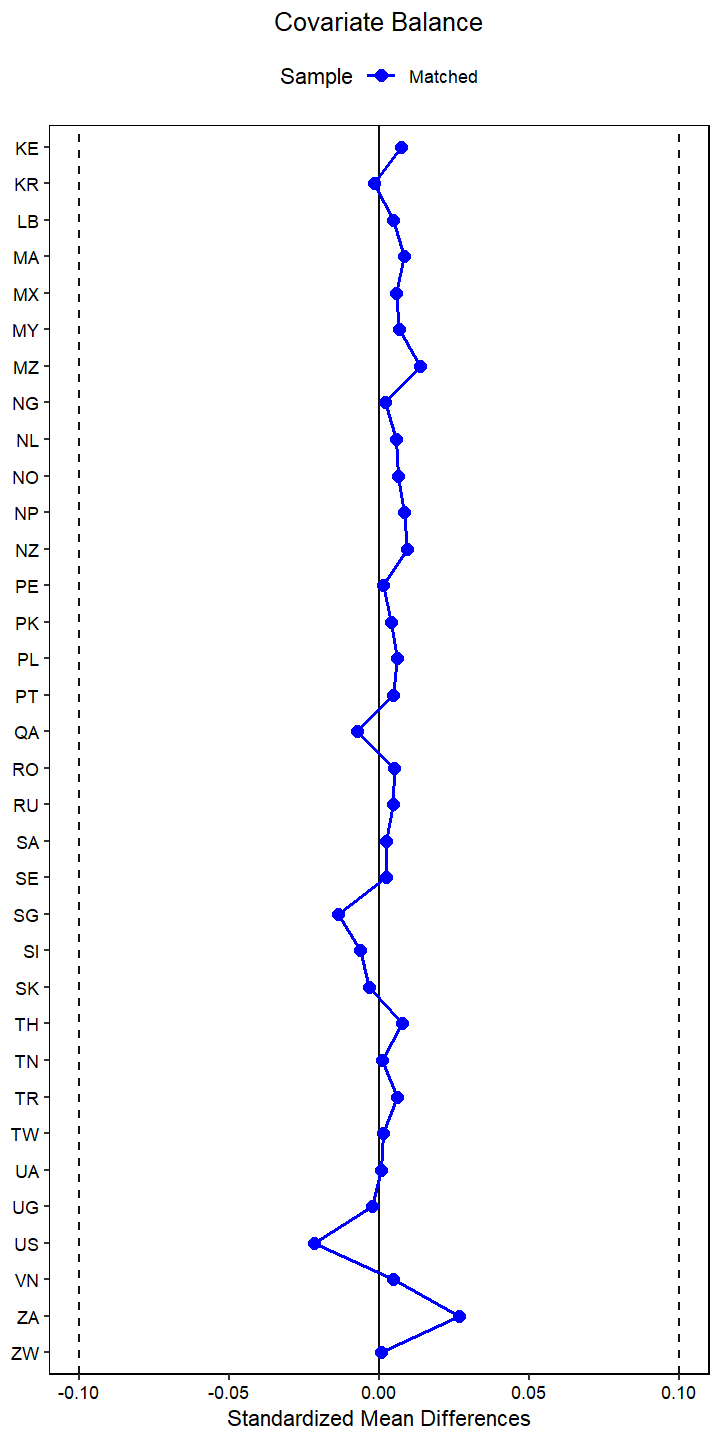}
    \label{fig:cfe_2}
  \end{subfigure}
  \caption{Standardized mean difference (with 0.1 thresholds) between \emph{COVID-related} and \emph{COVID non-related} after matching on country of \emph{first} author and paper level controls (trial, N authors, Pre-existing Grant).}
  \label{fig:loveplot_4}
\end{figure}

\begin{figure}[H]
\centering
\hspace{-2cm}
  \begin{subfigure}[t]{0.5\textwidth}
    \includegraphics[width=\textwidth]{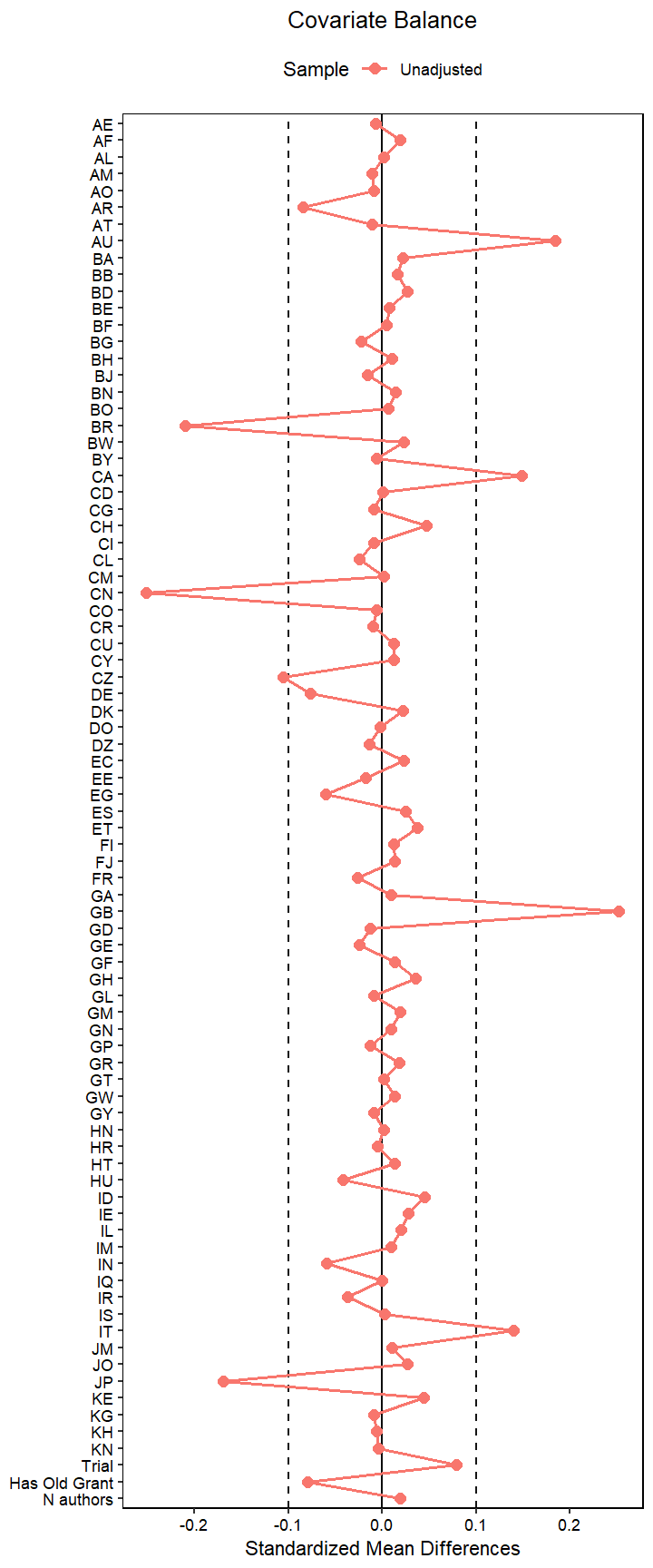}
    \label{fig:cfe_1}
  \end{subfigure}
 \begin{subfigure}[t]{0.5\textwidth}
    \includegraphics[width=\textwidth]{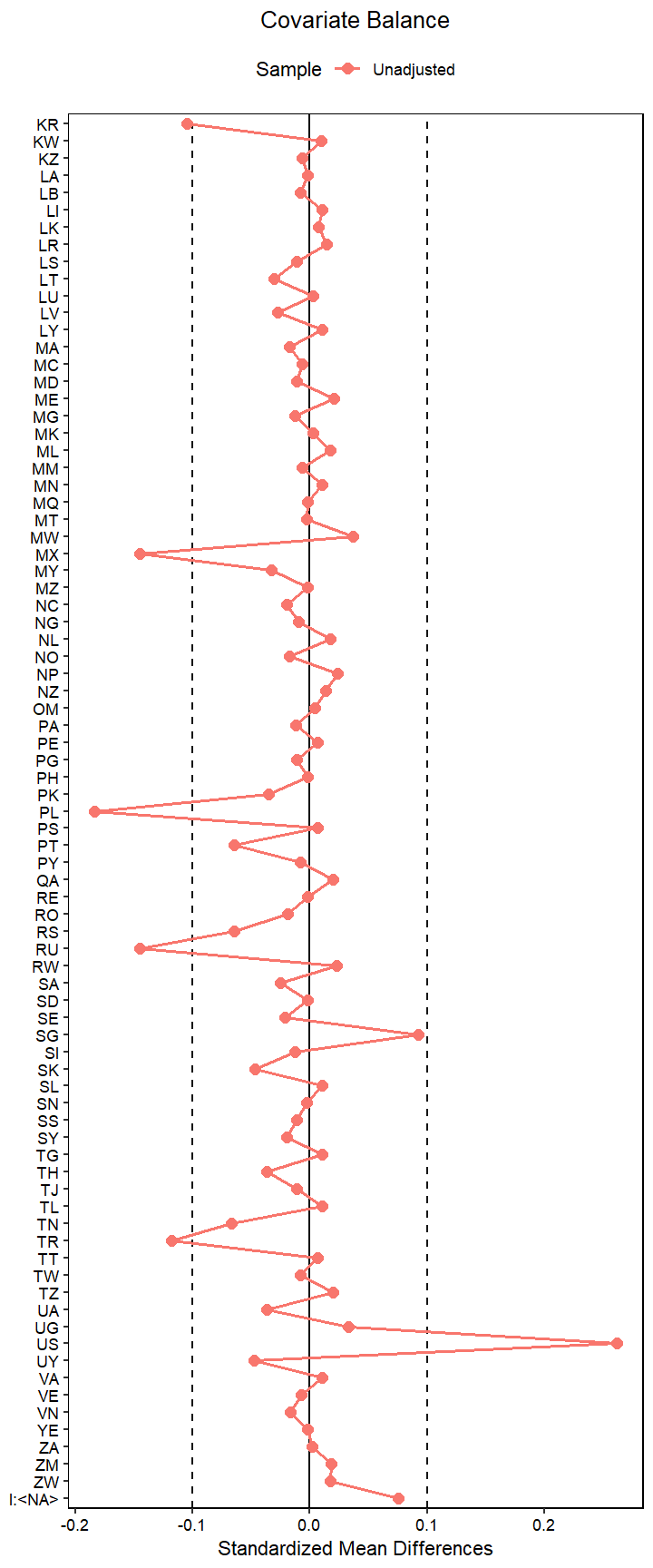}
    \label{fig:cfe_2}
  \end{subfigure}
  \caption{Standardized mean difference of covariates between \emph{COVID-related} and \emph{COVID non-related} before matching on country of \emph{last} author and paper level controls (trial, N authors, Pre-existing Grant). Threshold lines at 0.1.}
  \label{fig:loveplot_5}
\end{figure}

\begin{figure}[H]
\centering
\hspace{-2cm}
  \begin{subfigure}[t]{0.5\textwidth}
    \includegraphics[width=\textwidth]{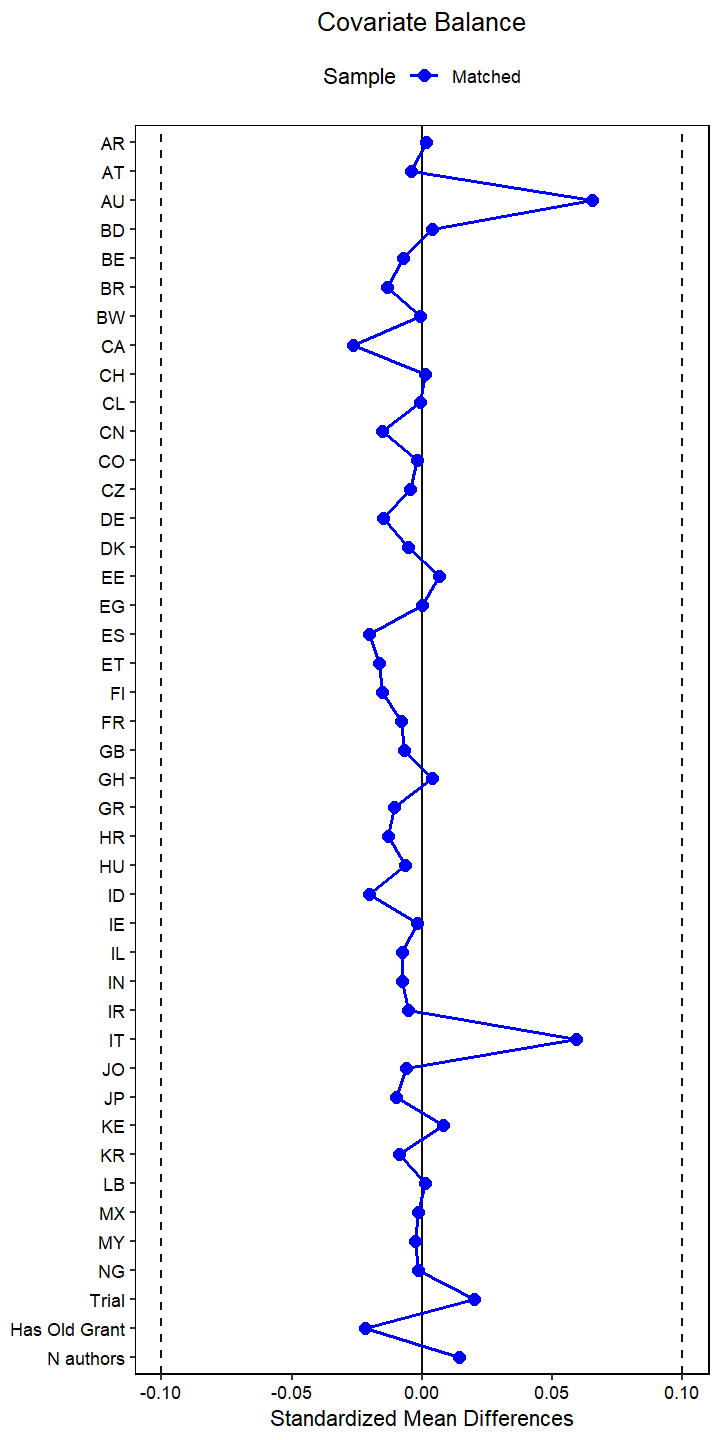}
    \label{fig:cfe_1}
  \end{subfigure}
 \begin{subfigure}[t]{0.5\textwidth}
    \includegraphics[width=\textwidth]{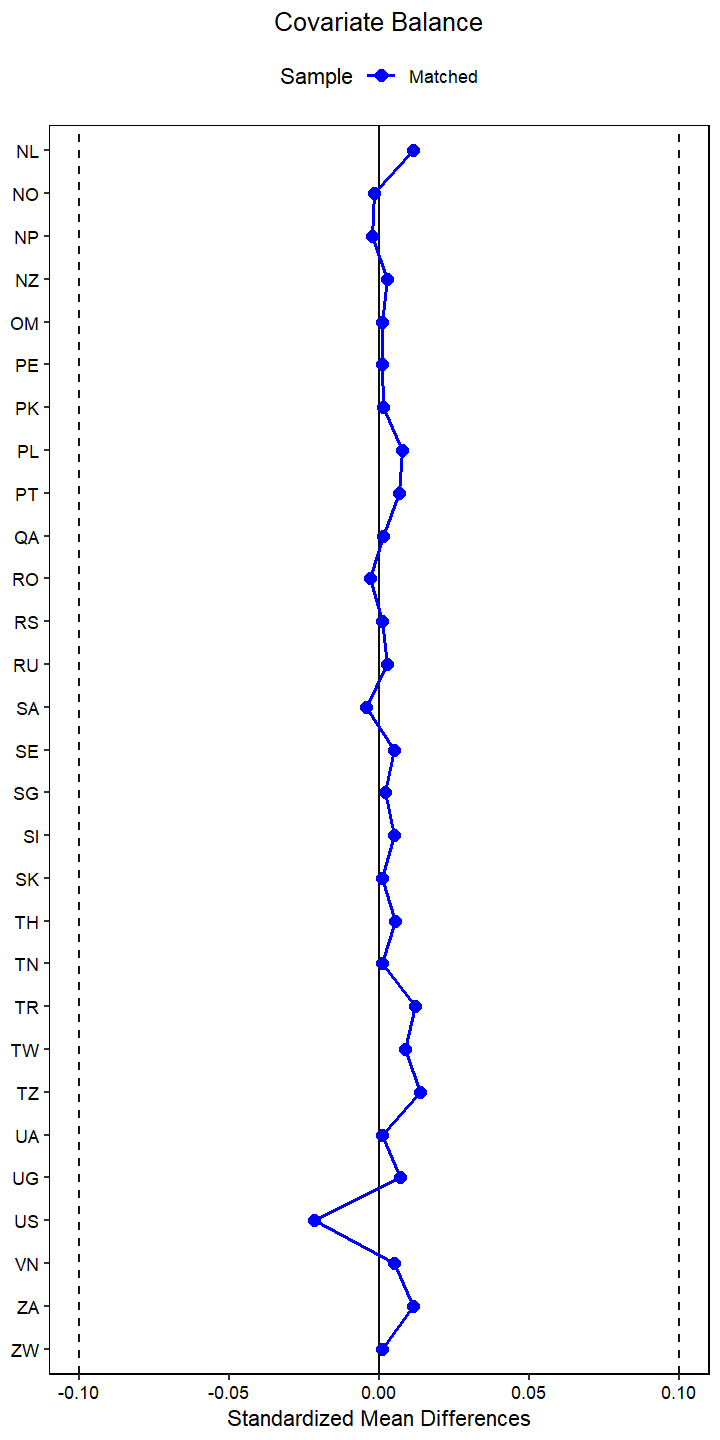}
    \label{fig:cfe_2}
  \end{subfigure}
  \caption{Standardized mean difference (with 0.1 thresholds) between \emph{COVID-related} and \emph{COVID non-related} after matching on country of \emph{last} author and paper level controls (trial, N authors, Pre-existing Grant).}
  \label{fig:loveplot_6}
\end{figure}

In Figures \ref{fig:psm_bs_maj}, \ref{fig:psm_bs_first} and \ref{fig:psm_bs_last} we plot the distribution of propensity scores in the treatment (\emph{blue}) and COVID non-related sample (\emph{red}) in 2019 -- baseline -- and 2020 -- midline --, before and after matching. We see that there is \emph{common support}, i.e. the score values for COVID-related are within range of the COVID non-related scores. After matching, the two distributions perfectly overlap, allowing us to estimate a \emph{Local Average Treatment Effect} or LATE. In Figure \ref{fig:loveplot_1}, Figure \ref{fig:loveplot_3}, and Figure \ref{fig:loveplot_5}, we show the standardized covariates mean difference between COVID-related and COVID non-related groups, before matching observations. The pre-treatment paper level characteristics \emph{trial}, \emph{pre-existing grant}, \emph{N authors}, are balanced between treatment and COVID non-related. From Figure \ref{fig:loveplot_2}, Figure \ref{fig:loveplot_4} and Figure \ref{fig:loveplot_6}, we see that after matching, also country identifiers are balanced. 

Finally we obtain a matched sample of 52158 papers for all outcomes, except first female authorship and last female authorship; for first (last) female authorship matching, we obtain 48118 (48448) matched papers.

We again assess the effect of new publishing opportunities on woman's authorship position by estimating the linear Diff-In-Diff model of equation (1) of the main text, controlling for the matching covariates \emph{country}, \emph{N authors}, \emph{trial} and \emph{Has Old Grant}. In Table \ref{psmdid}, we report the coefficient estimates, using heteroskedasticy-consistent standard errors. Comments on the results are expressed in the main text of the article.

\begin{table}[H]\centering
\def\sym#1{\ifmmode^{#1}\else\(^{#1}\)\fi}
\caption{DID estimates of matched data with White-robust standards errors. We control for country effects (omitted) for the majority of the team for Female Author, First and Last Female Authors, Middle Female Authorship and Middle Female Only; country fixed effects of the first (last) author for regression on First (Last) Female Author. \label{psmdid}}
\resizebox{\textwidth}{!}{ 
\begin{tabular}{c c c c c c c}
\toprule &\multicolumn{1}{c}{Female Author}&\multicolumn{1}{c}{First and Last Female}&\multicolumn{1}{c}{Female First Author}&\multicolumn{1}{c}{Female Last Author}&\multicolumn{1}{c}{Middle Female Authorship}&\multicolumn{1}{c}{Middle Female Only}\\
\toprule 
\multicolumn{1}{c}{Variable}&\multicolumn{1}{c}{(1)}&\multicolumn{1}{c}{(2)}&\multicolumn{1}{c}{(3)}&\multicolumn{1}{c}{(4)}&\multicolumn{1}{c}{(5)}&\multicolumn{1}{c}{(6)}\\
\midrule
year=2020           &      0.0108\sym{*}  &      0.0202\sym{***}  &      0.0233\sym{***}  &      0.0256\sym{***}&      0.0162\sym{**} &     -0.0151\sym{**} \\
                    &      (2.52)         &      (4.34)       &      (3.59)   &      (4.30)    &      (3.03)         &     (-2.77)         \\
\addlinespace
COVID-related             &      0.0456\sym{***}&      0.0838\sym{***}  &      0.0731\sym{***}&   0.107\sym{***} &     0.0745\sym{***}&     -0.0393\sym{***}\\
                    &     (11.59)         &     (17.09)      &     (11.41) &     (17.69)      &     (14.69)         &     (-7.41)         \\
\addlinespace
year=2020 $\times$ COVID-related&    -0.00968         &     -0.0350\sym{***} &     -0.0437\sym{***}&     -0.0518\sym{***}&     -0.0121         &      0.0356\sym{***}\\
                    &     (-1.75)         &     (-5.06)      &     (-4.81)  &     (-6.04)        &     (-1.70)         &      (4.71)         \\
\addlinespace\addlinespace
\textit{N Authors}           &      0.0114\sym{***}&    -0.00238\sym{***}    &   -0.000781    &    -0.00389\sym{***}&      0.0262\sym{***}&      0.0142\sym{***}\\
                    &     (33.72)         &     (-6.73)       &     (-1.49)      &     (-7.86)   &     (38.23)         &     (26.36)         \\
\addlinespace
trial               &      0.0210\sym{***}&    -0.00321             &    -0.00436   &     0.00884    &      0.0495\sym{***}&      0.0128         \\
                    &      (3.31)         &     (-0.34)      &     (-0.35)        &      (0.72)     &      (5.87)         &      (1.19)         \\
\addlinespace
Pre-existing Grant         &      0.0412\sym{***}&      0.0355\sym{***}    &      0.0551\sym{***}    &      0.0485\sym{***}&      0.0462\sym{***}&     -0.0232\sym{***}\\
                    &     (11.79)         &      (7.19)       &      (8.80)         &      (8.06)     &      (9.75)         &     (-4.39)         \\
\addlinespace
Constant            &       0.809\sym{***}&       0.205\sym{***} &       0.304  &     &       0.419\sym{***}    0.594\sym{***}&       0.165\sym{***}\\
                    &     (31.11)         &      (4.74)      &      (1.55)    &      (6.78)      &     (16.44)         &      (3.80)         \\
\midrule
Observations        &       52158         &       52158         &       48118     &       48448    &       52158         &       52158         \\
\midrule
Country FEs & Majority  & Majority  & First & Last  & Majority  & Majority  \\
\bottomrule
\multicolumn{7}{l}{\footnotesize \textit{t} statistics in parentheses}\\
\multicolumn{7}{l}{\footnotesize \sym{*} \(p<0.05\), \sym{**} \(p<0.01\), \sym{***} \(p<0.001\)}\\
\end{tabular}
}
\end{table}

\section{Mechanisms}\label{mobility}
\subsection{Incumbency in publications' research topics} 

We rely on the OpenAlex official API \citep{priem2022openalex} to collect published papers between 2015 and 2020 of the authors featured in our main PubMed sample. We start data collection on October 2023. Out of the 91,480 COVID-related and COVID non-related PubMed publications, assigned to unique PMID identifiers, we successfully retrieve information for 90,671 publications, accounting for 472,147 unique authors in total. We refer to the author unique identifier in OpenAlex for name disambiguation. 
We then collect all past publications within PubMed of these authors from 2015 to 2020. After processing the recovered information, we obtain a total of 1,129,749 unique articles published on PubMed by 339,405 authors. We obtain no response for calls on 326 authors. The remaining 132 thousand circa authors are simply observed for the first time in Openalex with the specific publication in our PubMed sample. So these authors have either not featured a publication in PubMed in the recent past, or are publishing for the first time. 

We use factor variables at paper level to define authors past research experience with the research topic of publication in which they are observed in the main PubMed sample. We define an author as \emph{incumbent} with respect to the main research field of the paper if the author has already published on that specific research field. If not incumbent with respect to her past research experience, the author is \emph{newcomer}. The  \emph{new entrant} authors do not have publications on PubMed in the reference period. This is an author that either have had recent publications at all, or just did not publish in PubMed, which would indicate that the author has been publishing in other scientific disciplines. Incumbency of authors with respect to 2019 is obtained by considering all publications of the author between 2015 and 2018, and indicating whether the author has already published on the 2019 paper's Major MeSH term; for 2020, we assess incumbency with respect to the set of 2020 papers considering the authors' past publications from 2015 to 2019, referring to the same research field of the 2020 paper. Therefore, for each paper, we indicate whether the first author (for example) is incumbent. 

For the middle authorship, papers could be featuring more than one author in non-relevant positions; in this case, incumbency in middle authorship positions reflects that of the incumbent middle author, and if there is none, of the newcomer. If none is either newcomer or incumbent, we indicate that middle authorship is occupied by \emph{new entrant} authors. 

Out of the 90,671 PubMed publications for which we have retrieved all information on OpenAlex, we obtain a total of 11,427 (17,994) papers with an incumbent first (last) author, 46,749 (57,495) papers with newcomer first (last) author, and 32,431 (15,016) papers with a new entrant first (last) author. For middle authorship, 20,873 papers have at least one incumbent author in middle position, 54,403 publications feature at least one newcomer and no incumbents, and 11,975 papers feature only new entrants.

\begin{figure}[H]
     \begin{subfigure}[t]{0.5\textwidth}
       \includegraphics[width=\textwidth]{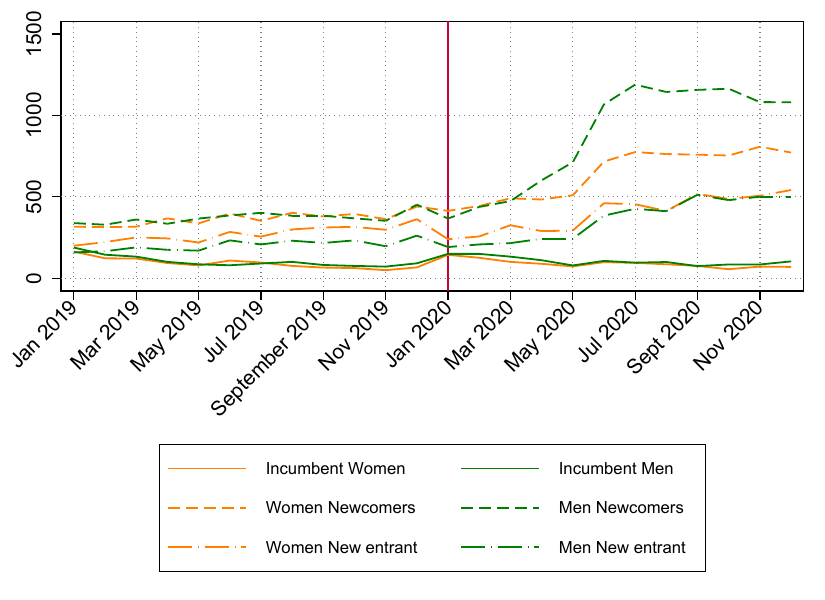}
         \caption{First Authors, COVID-related}
   \end{subfigure}
     \begin{subfigure}[t]{0.5\textwidth}
       \includegraphics[width=\textwidth]{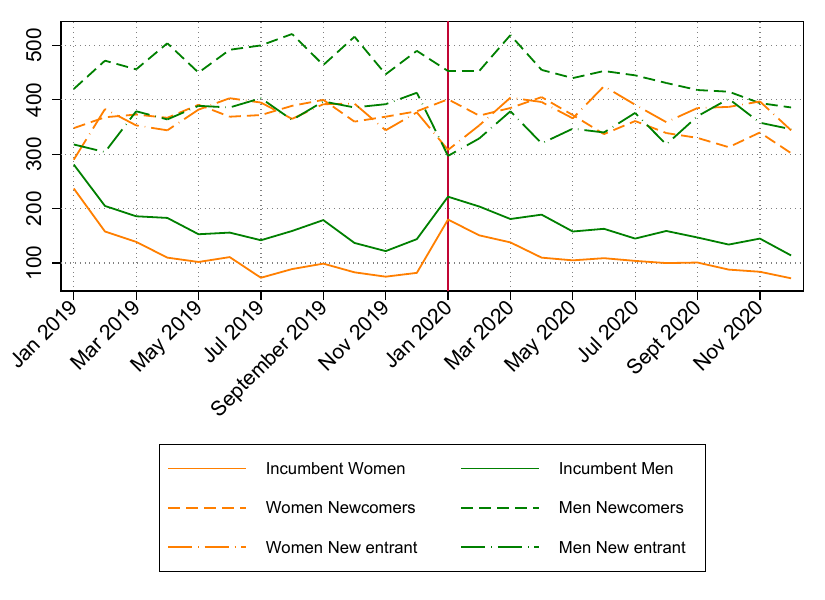}
         \caption{First Authors, COVID non-related}
   \end{subfigure}
         \begin{subfigure}[t]{0.5\textwidth}
        \includegraphics[width=\textwidth]{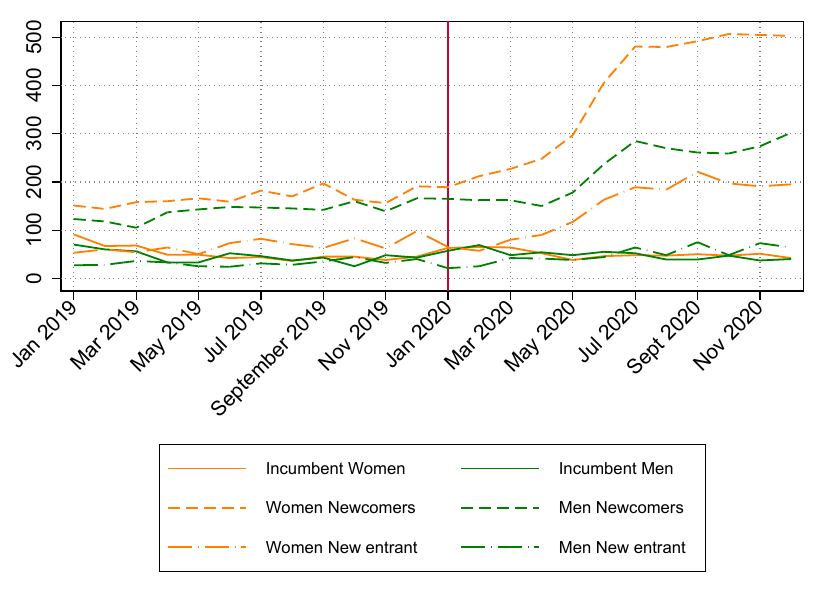}
        \caption{Middle Authors Only, COVID-related}
   \end{subfigure}
            \begin{subfigure}[t]{0.5\textwidth}
        \includegraphics[width=\textwidth]{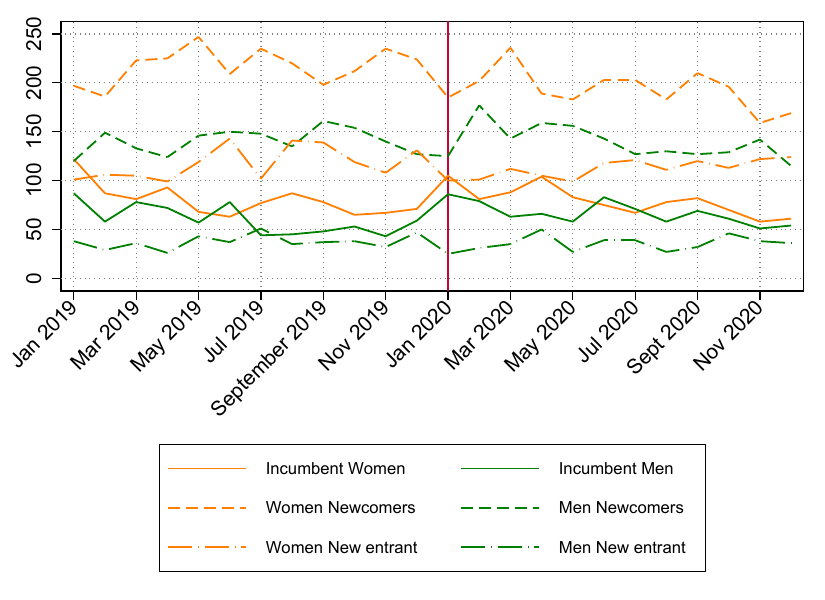}
        \caption{Middle Authors Only, COVID non-related}
   \end{subfigure}
    \caption{Monthly sample numerosity of publications by women and men as (a) first authors in COVID-related publications, (b) first authors in COVID non-related publications, (c) middle authors only in COVID-related publications, and (d) middle authors only in COVID non-related publications, by incumbency status (\emph{incumbent, newcomer, new entrant}).
    }
    \label{fig:twoway_2}
  \end{figure}

\begin{figure}[H]
     \begin{subfigure}[t]{0.5\textwidth}
       \includegraphics[width=\textwidth]{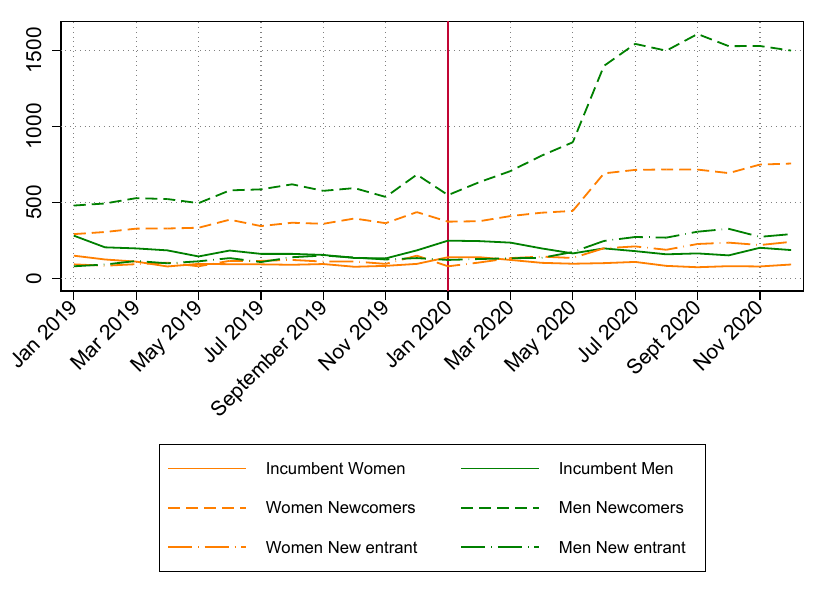}
         \caption{Last Authors, COVID-related}
   \end{subfigure}
     \begin{subfigure}[t]{0.5\textwidth}
       \includegraphics[width=\textwidth]{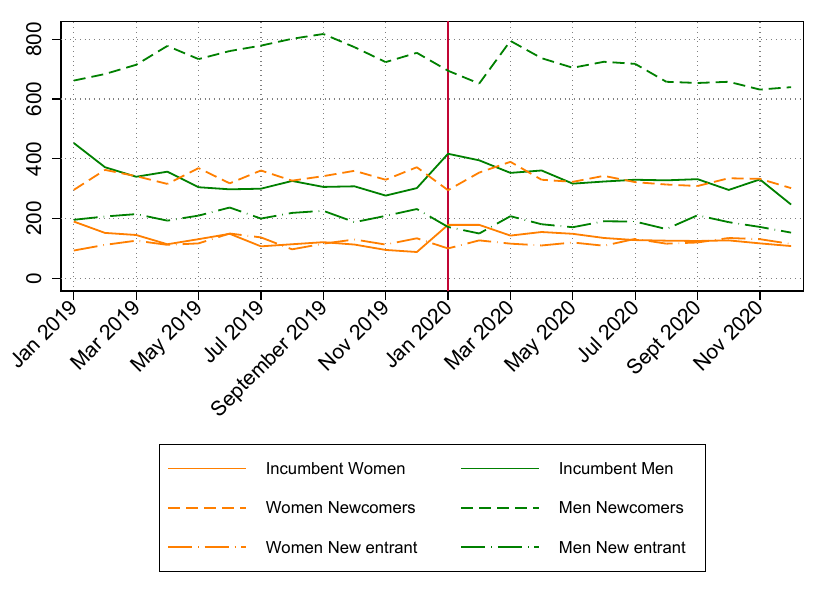}
         \caption{Last Authors, COVID non-related}
   \end{subfigure}
         \begin{subfigure}[t]{0.5\textwidth}
        \includegraphics[width=\textwidth]{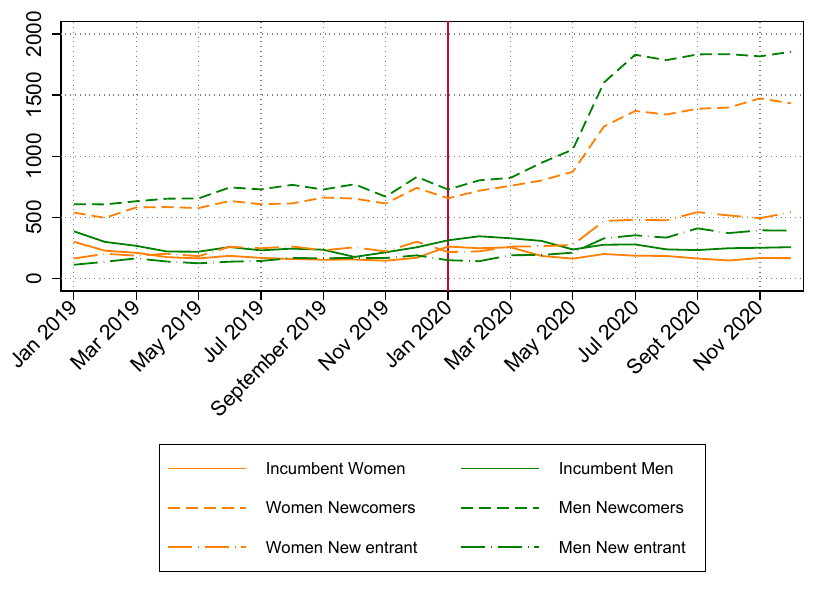}
        \caption{Middle Authors, COVID-related}
   \end{subfigure}
            \begin{subfigure}[t]{0.5\textwidth}
        \includegraphics[width=\textwidth]{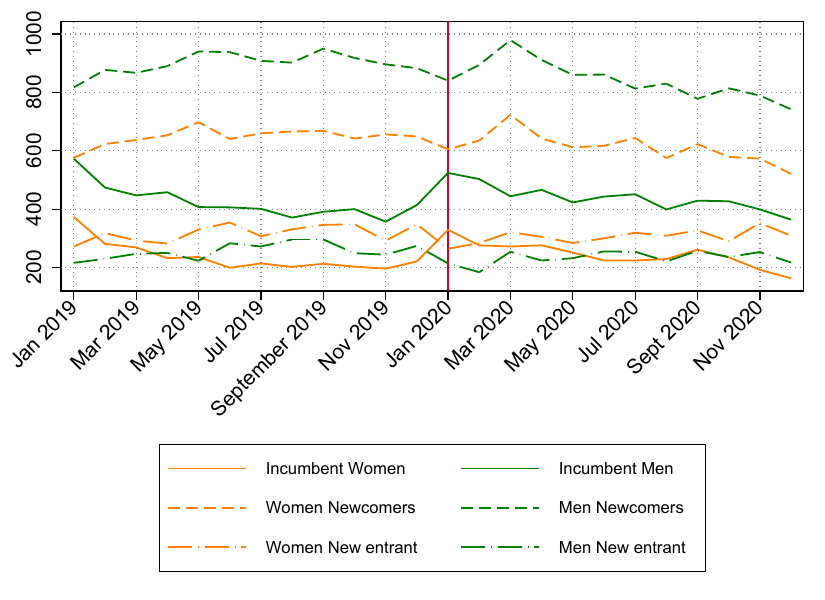}
        \caption{Middle Authors, COVID non-related}
   \end{subfigure}
    \caption{
    Monthly sample numerosity of publications by women and men as (a) last authors in COVID-related publications, (b) last authors in COVID non-related publications, (c) middle authors in COVID-related publications, and (d) middle authors in COVID non-related publications, by incumbency status (\emph{incumbent, newcomer, new entrant}).
    }
    \label{fig:twoway_3}
  \end{figure}

We show the observed monthly numerosity of publications among COVID-related and COVID non-related by gender and incumbency for last and middle authorship in Figure \ref{fig:twoway_2}, and for first and middle only female and male authors in Figure \ref{fig:twoway_3}, including the new entrant authors. Male newcomers have experienced the greatest surge in number of publications as key authors, with the gender gap increasing in 2020. For new entrants, changes in publication gender gap seem to be much more constrained in size. 

\begin{figure}[H]
  \begin{subfigure}[t]{0.5\textwidth}
    \includegraphics[width=\textwidth]{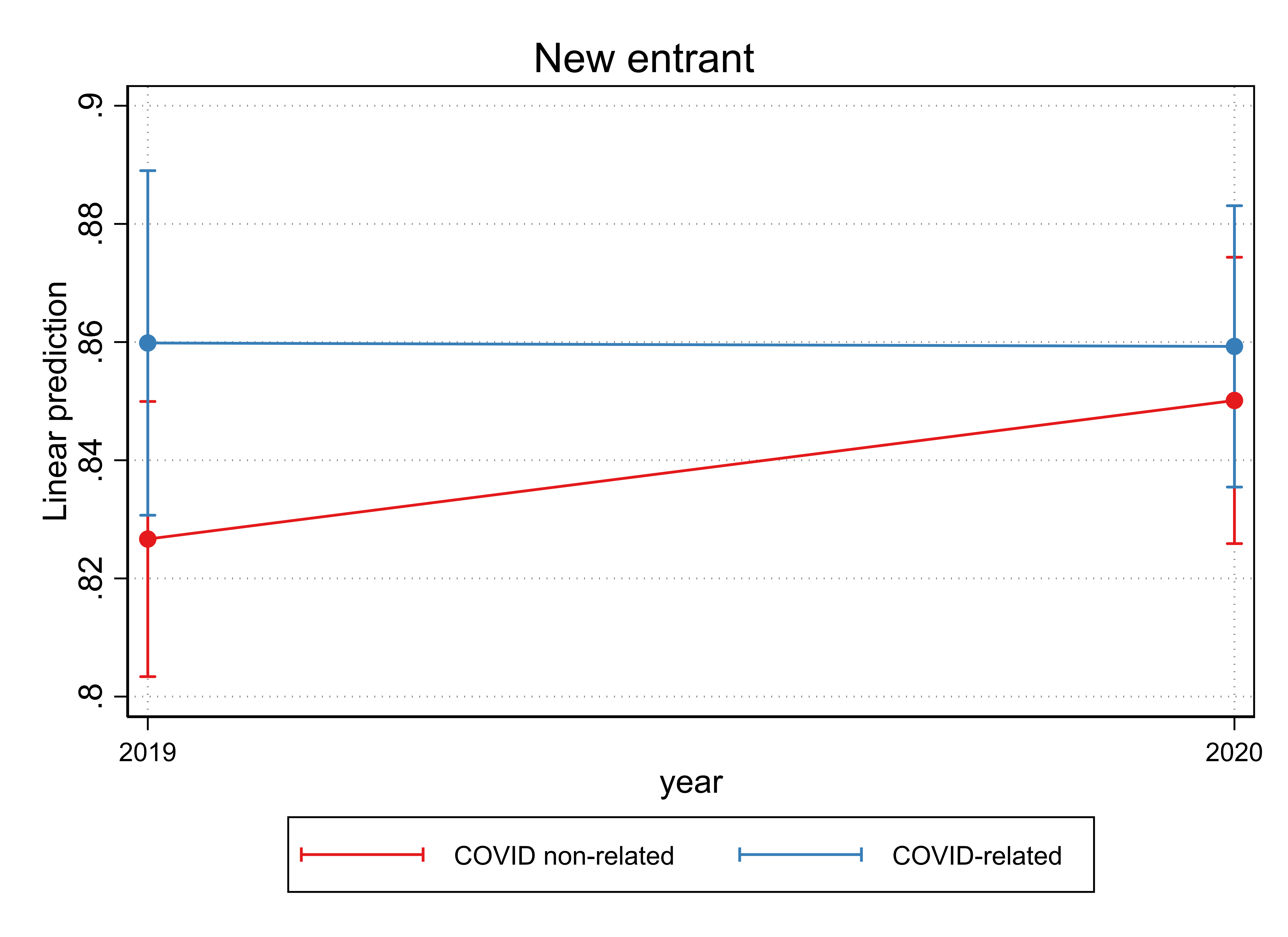}
    \caption{Any Female Author, New entrant}
  \end{subfigure}
 \begin{subfigure}[t]{0.5\textwidth}
    \includegraphics[width=\textwidth]{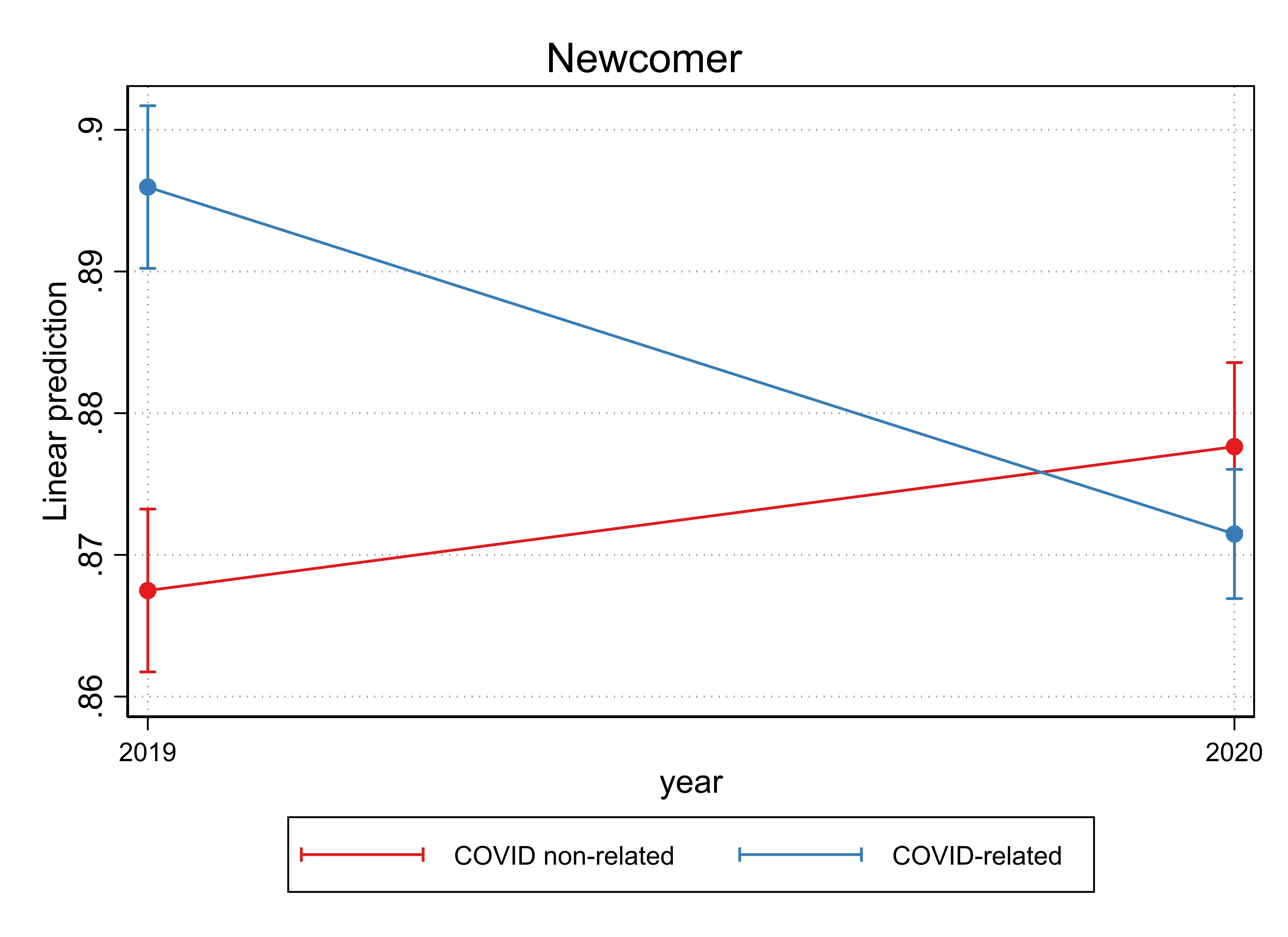}
    \caption{Any Female Author, Newcomer}
  \end{subfigure}
   \begin{subfigure}[t]{0.5\textwidth}
    \includegraphics[width=\textwidth]{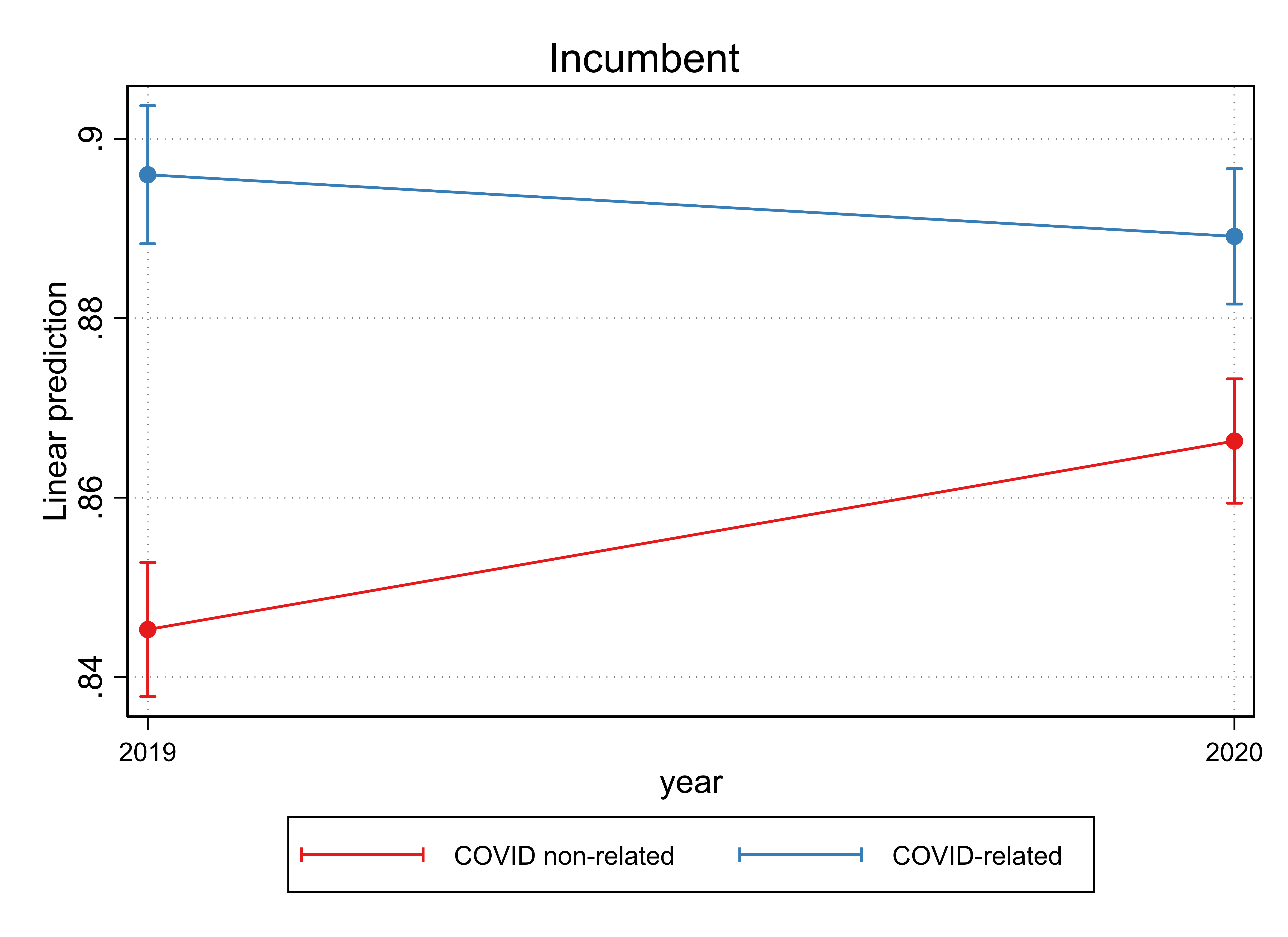}
      \caption{Any Female Author, Incumbent}
  \end{subfigure}
  \caption{Predicted probability to observe a woman at any position by past research experience, among COVID-related and COVID non-related publications.}
  \label{fig:fig_s10}
\end{figure}

\begin{figure}[H]
  \begin{subfigure}[t]{0.5\textwidth}
    \includegraphics[width=\textwidth]{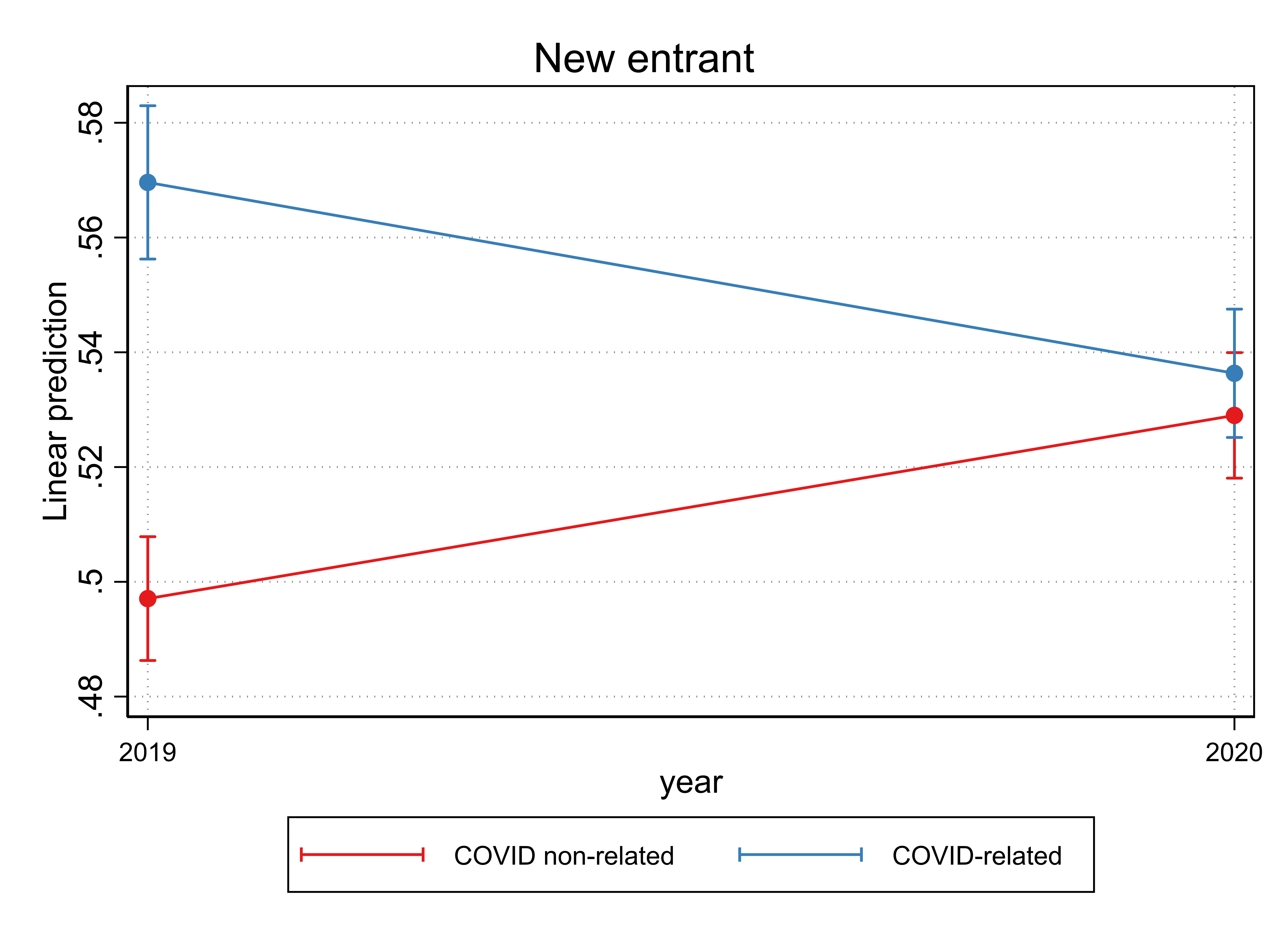}
    \caption{First Female Author, New entrant}
  \end{subfigure}
  \hfill
 \begin{subfigure}[t]{0.5\textwidth}
    \includegraphics[width=\textwidth]{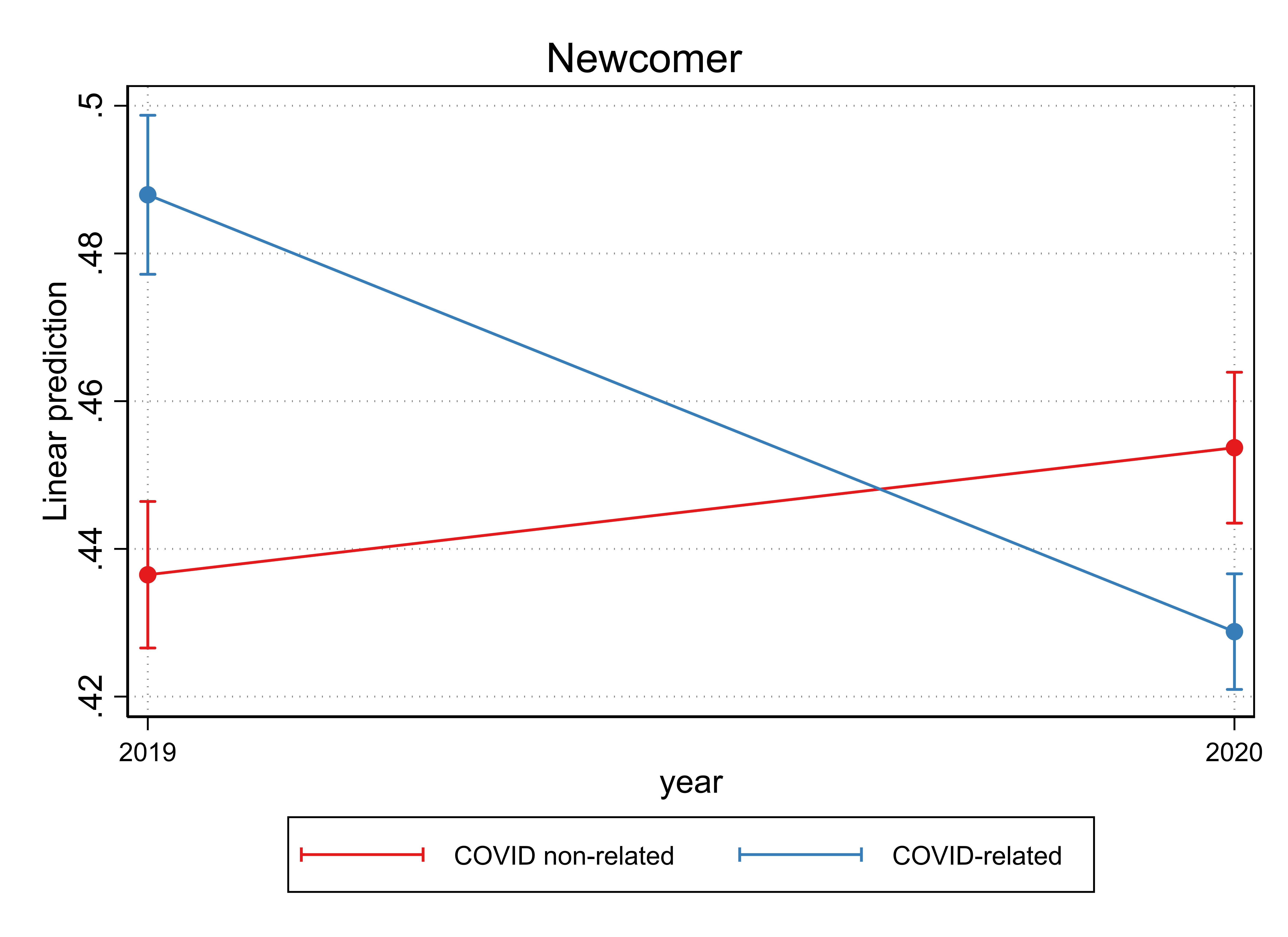}
    \caption{First Female Author, Newcomer}
  \end{subfigure}
   \begin{subfigure}[t]{0.5\textwidth}
    \includegraphics[width=\textwidth]{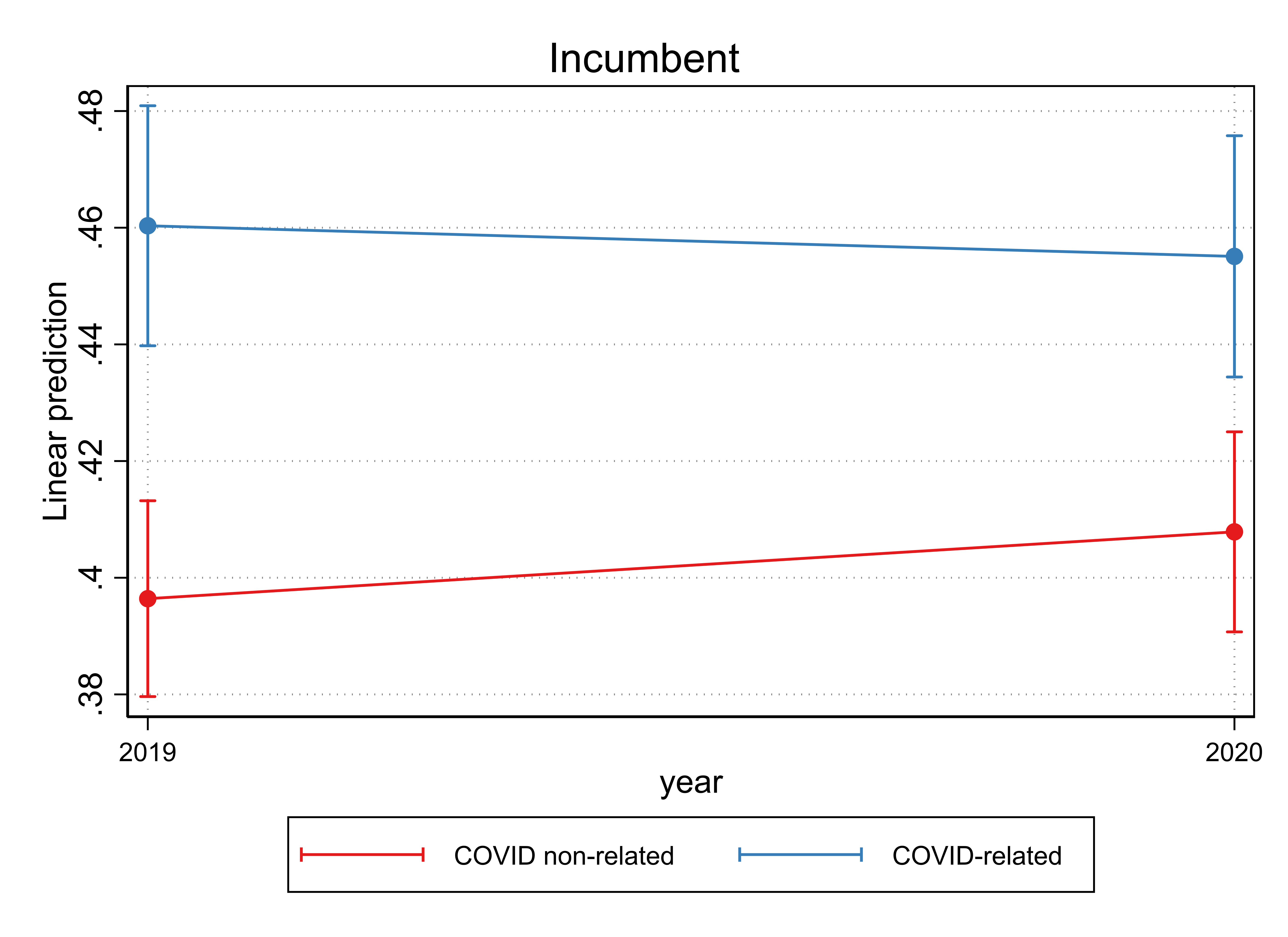}
      \caption{First Female Author, Incumbent}
  \end{subfigure}
  \caption{Predicted probability to observe a woman in first position by past research experience, among COVID-related and COVID non-related publications.}
  \label{fig:fig_s11}
\end{figure}

\begin{figure}[H]
  \begin{subfigure}[t]{0.5\textwidth}
    \includegraphics[width=\textwidth]{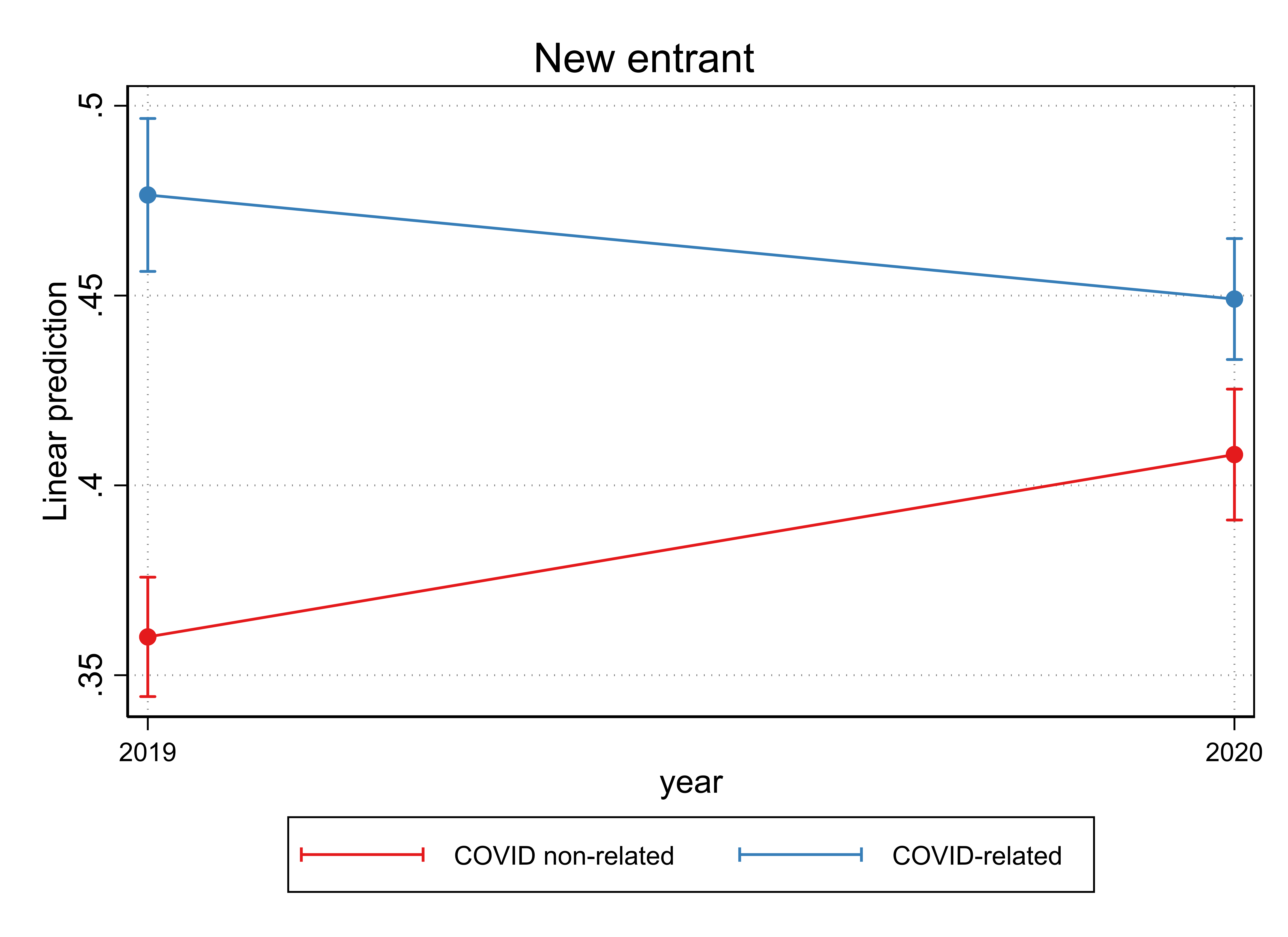}
    \caption{Last Female Author, New entrant}
  \end{subfigure}
  \hfill
 \begin{subfigure}[t]{0.5\textwidth}
    \includegraphics[width=\textwidth]{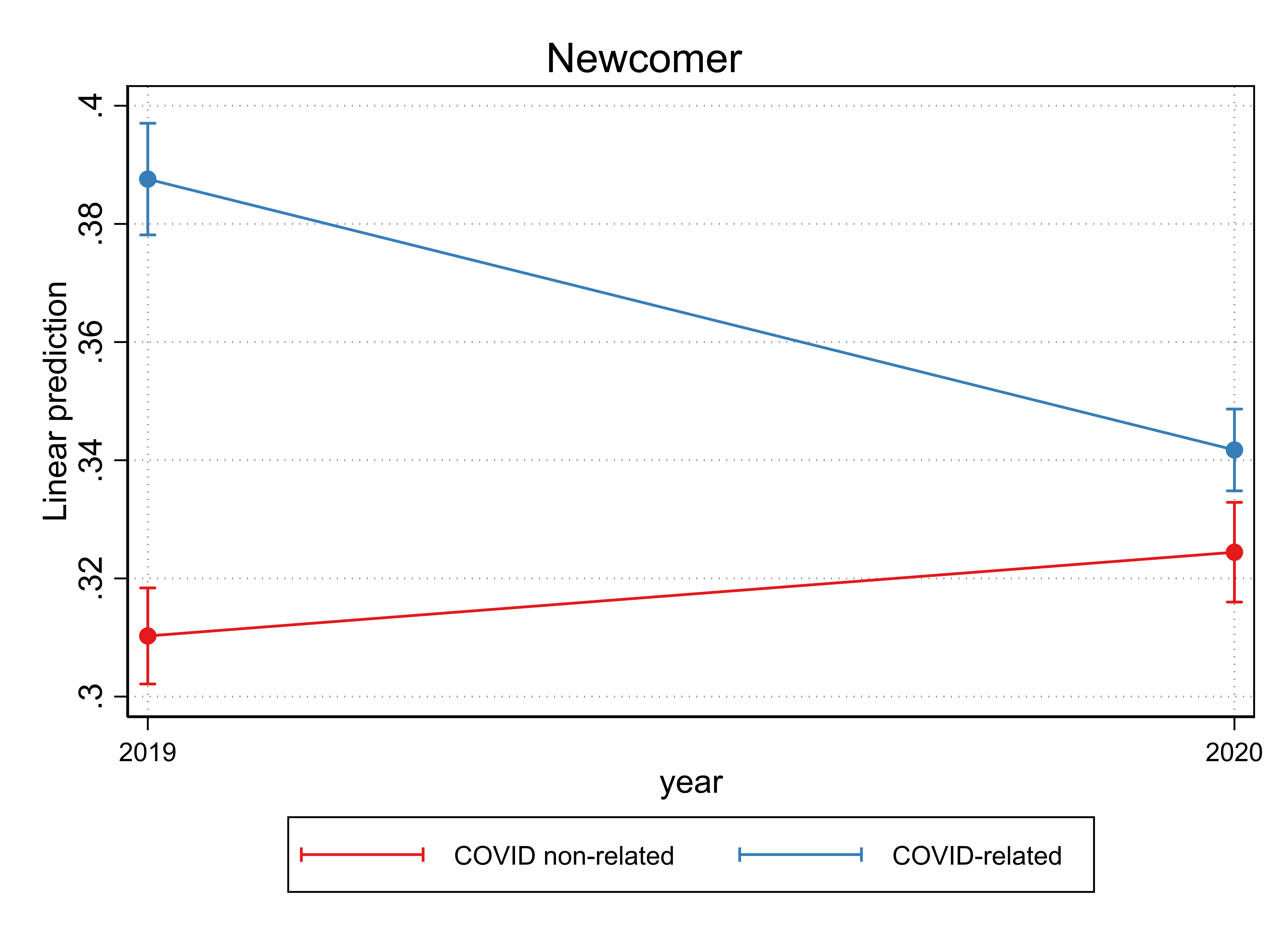}
    \caption{Last Female Author, Newcomer}
  \end{subfigure}
   \begin{subfigure}[t]{0.5\textwidth}
    \includegraphics[width=\textwidth]{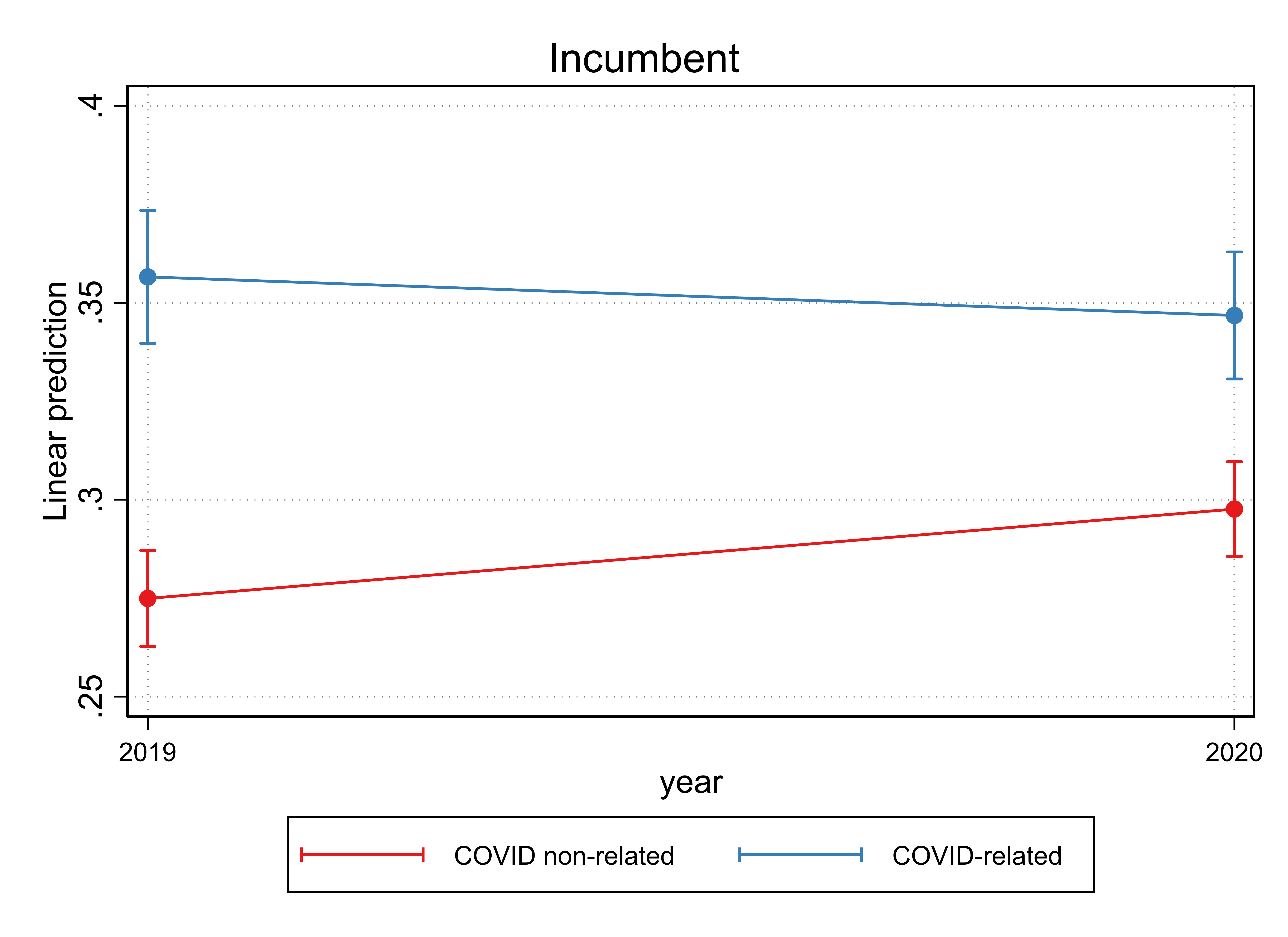}
      \caption{Last Female Author, Incumbent}
  \end{subfigure}
  \caption{Predicted probability to observe a woman in last position by past research experience, among COVID-related and COVID non-related publications.}
  \label{fig:fig_s12}
\end{figure}

\begin{figure}[H]
  \begin{subfigure}[t]{0.5\textwidth}
    \includegraphics[width=\textwidth]{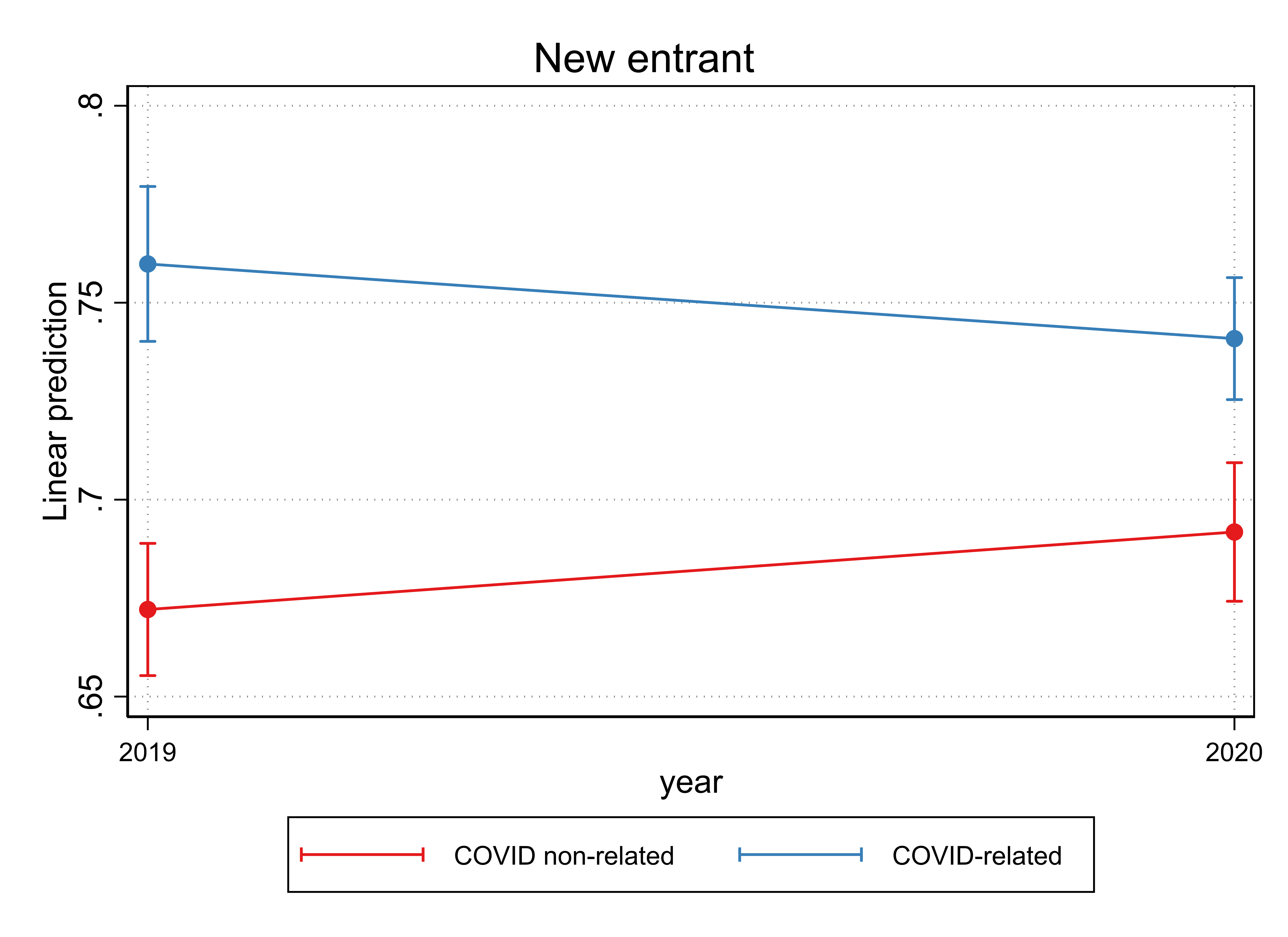}
    \caption{Middle, New entrant}
  \end{subfigure}
  \hfill
 \begin{subfigure}[t]{0.5\textwidth}
    \includegraphics[width=\textwidth]{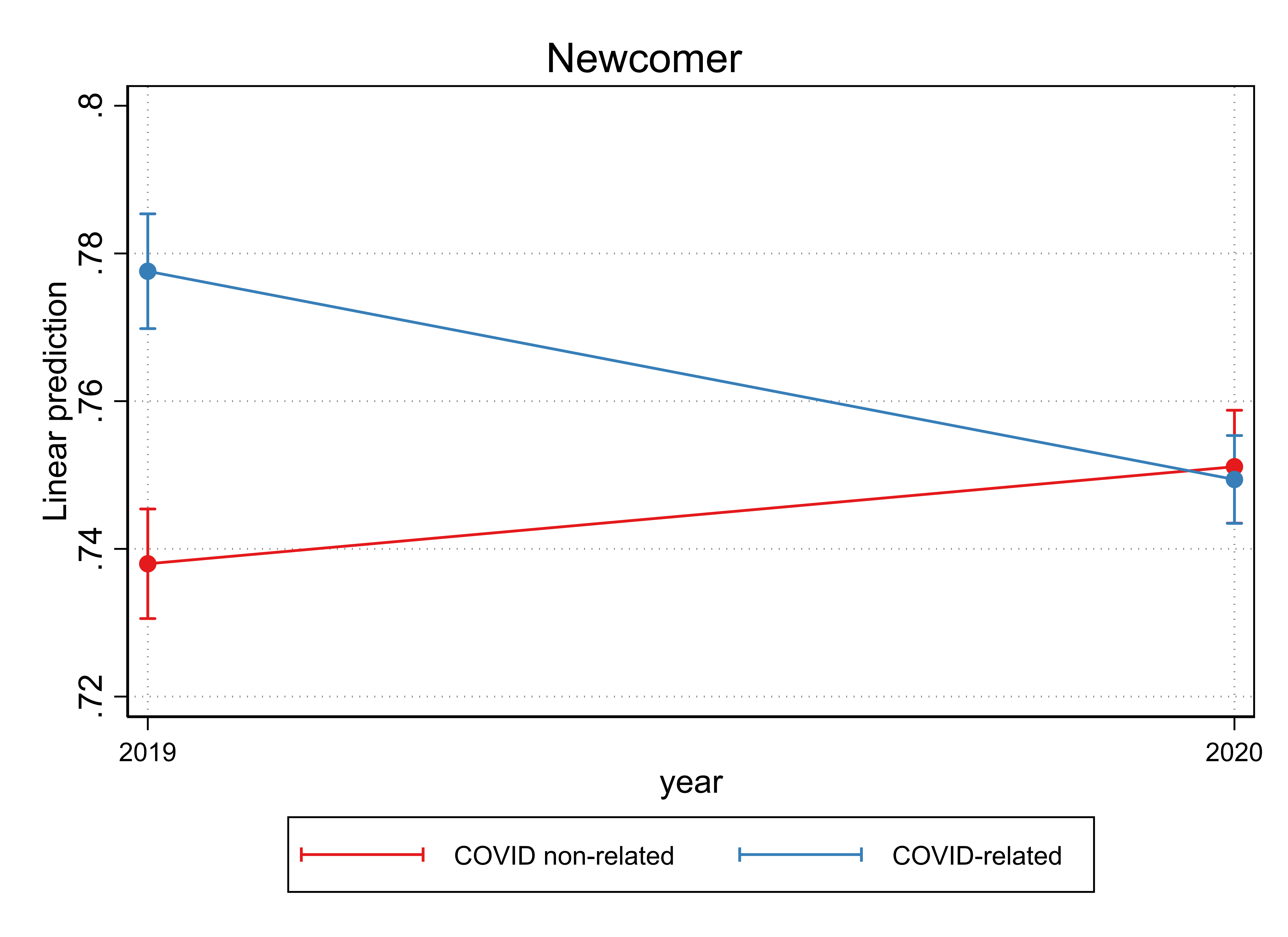}
    \caption{Middle, Newcomer}
  \end{subfigure}
   \begin{subfigure}[t]{0.5\textwidth}
    \includegraphics[width=\textwidth]{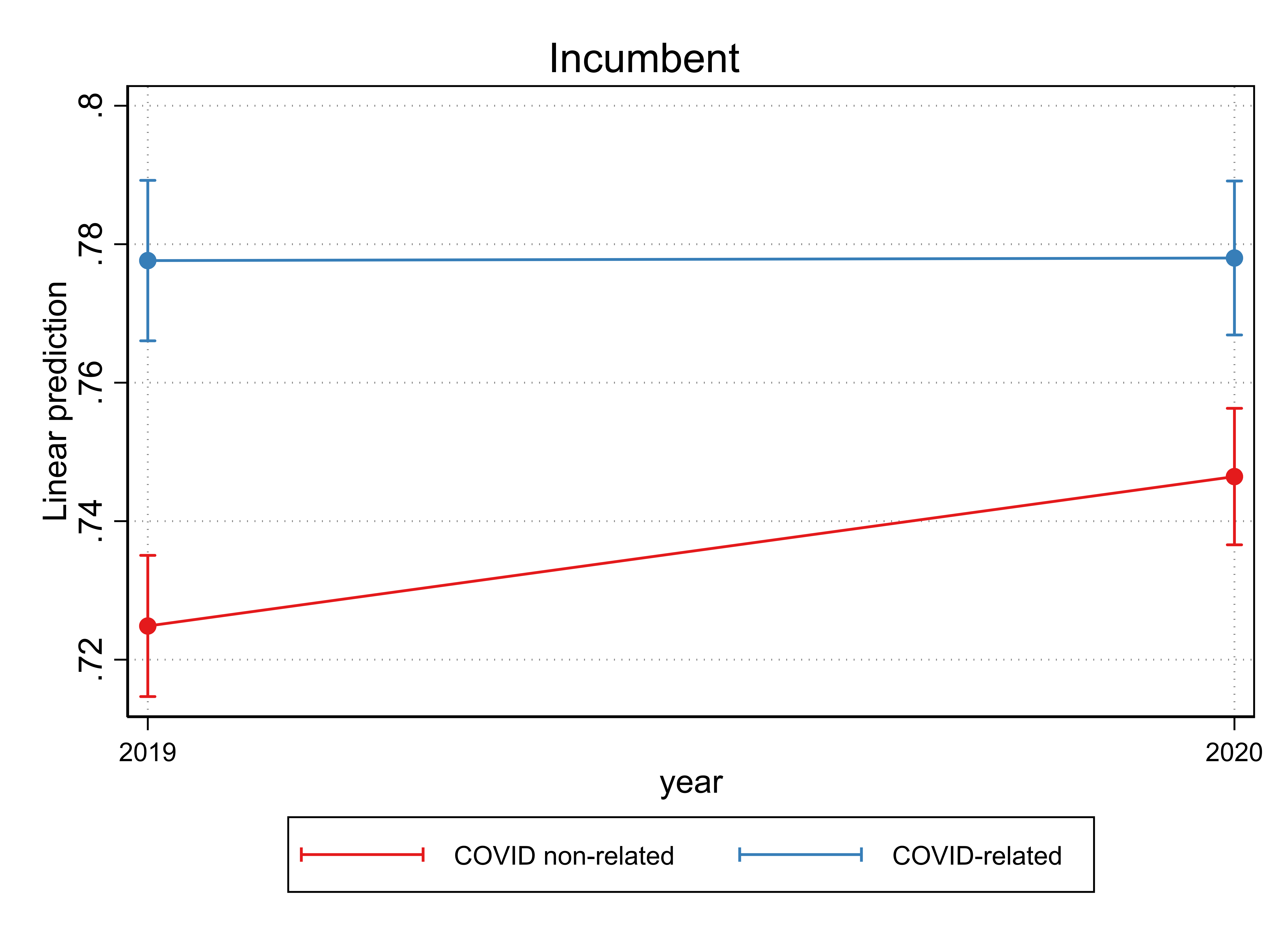}
      \caption{Middle, Incumbent}
  \end{subfigure}
  \caption{Predicted probability to observe a woman in middle position by past research experience, among COVID-related and COVID non-related publications.}
  \label{fig:fig_s13}
\end{figure}

\begin{figure}[H]
  \begin{subfigure}[t]{0.5\textwidth}
    \includegraphics[width=\textwidth]{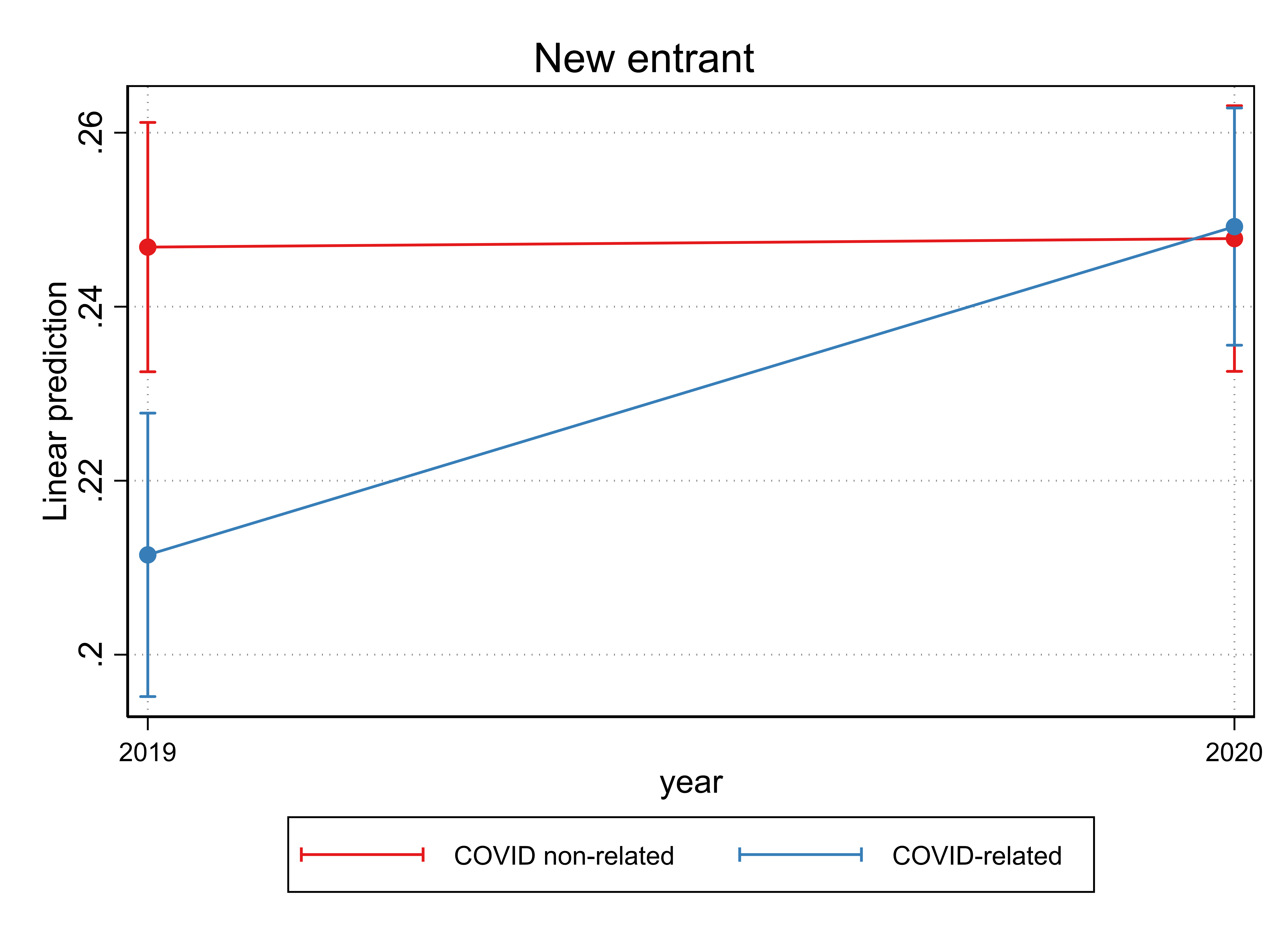}
    \caption{Middle Only, New entrant}
  \end{subfigure}
  \hfill
 \begin{subfigure}[t]{0.5\textwidth}
    \includegraphics[width=\textwidth]{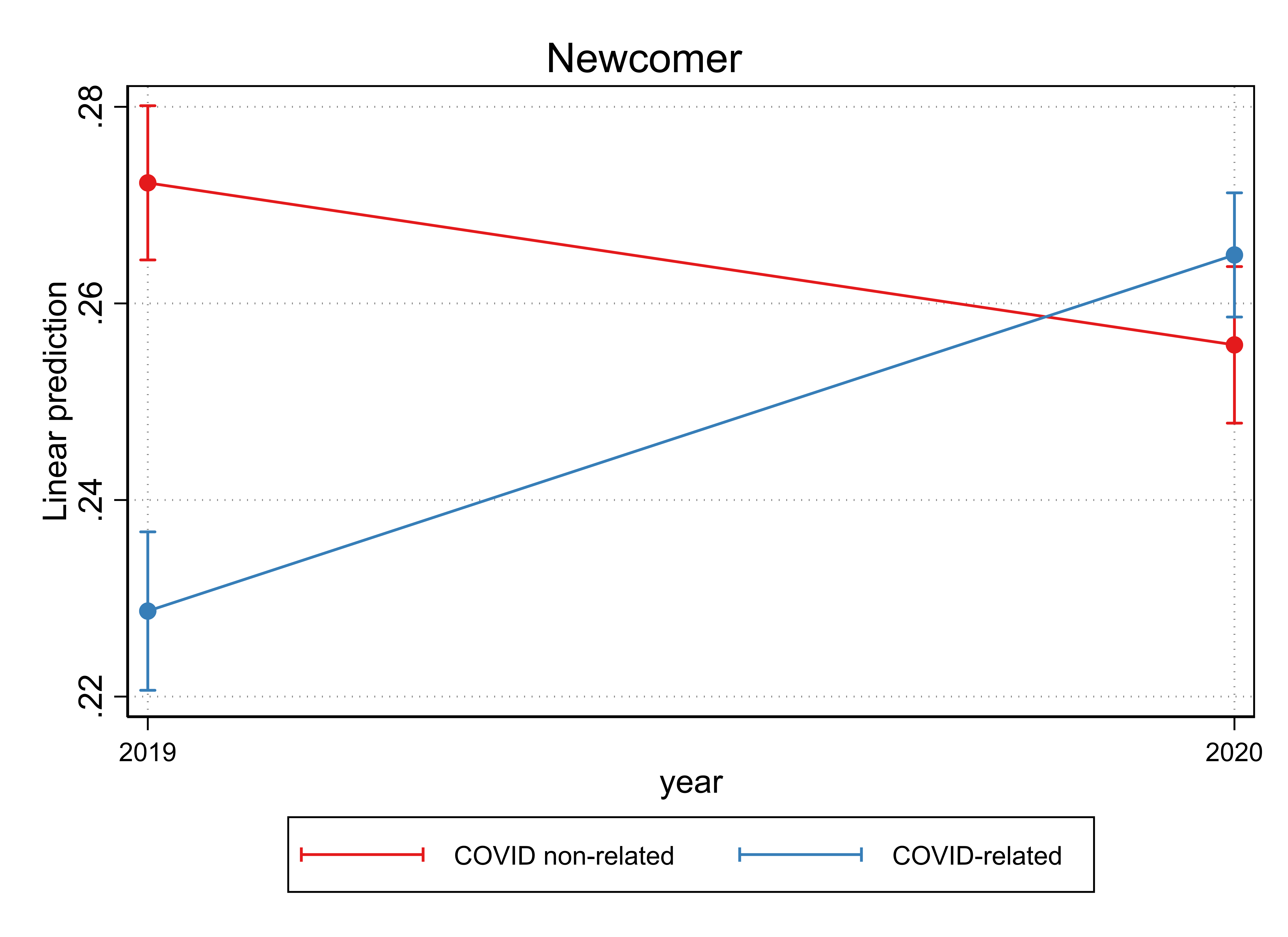}
    \caption{Middle Only, Newcomer}
  \end{subfigure}
   \begin{subfigure}[t]{0.5\textwidth}
    \includegraphics[width=\textwidth]{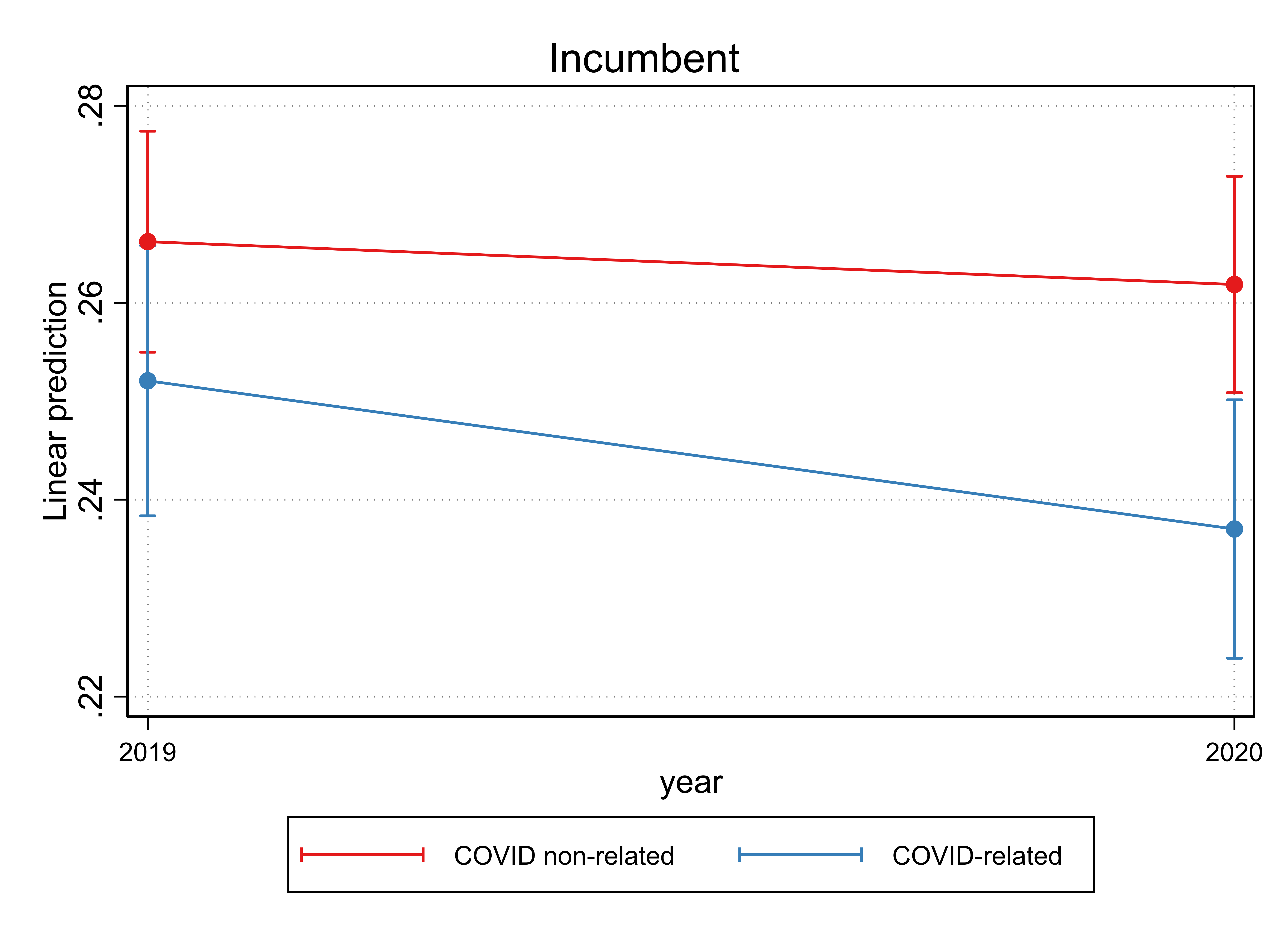}
      \caption{Middle Only, Incumbent}
  \end{subfigure}
  \caption{Predicted probability to observe a woman in middle position \emph{only} -- in team with men as key authors -- by past research experience, among COVID-related and COVID non-related publications.}
  \label{fig:fig_s14}
\end{figure}

In Figures \ref{fig:fig_s10}-\ref{fig:fig_s14}, we plot the predicted shares for COVID-related and COVID non-related papers grouped by incumbency of the authors' past research experience. They show that among the COVID non-related papers, women are still successfully publishing as last authors when changing research field of publication, as depicted by the increasing trend of newcomer female first and last authors (Figures \ref{fig:fig_s11}-\ref{fig:fig_s12}), while they are decreasingly being featured as \emph{solo} middle authors when mobilizing towards research topics that are not within their past expertise  (Figure \ref{fig:fig_s14}). 
On the other hand, as newcomer female authors are significantly diminishing among publications related to the new research topic (Figures \ref{fig:fig_s11}-\ref{fig:fig_s12}), but increasingly participating as middle authors in teams with male key authors  (Figure \ref{fig:fig_s14}).

In Table \ref{pers_table}, column (1), the impact of a new research opportunity is detrimental for overall female participation as authors of published papers, with no significant contribution by the incumbency of the authors. Instead, in Column (2), the treatment effect is no longer statistically significant, but having a newcomer first author significantly diminishes the share of female in first authorship position for the COVID-related group in 2020 with respect to having a incumbent first author (-0.0596); the same of new entrant first authors (-0.0485). Then, the decline in female authorship as first authors in research related to the new research opportunity can be explained by the presence of newcomer and new entrant \emph{male} authors in first authorship position.
In Column (3), the treatment effect is still significantly and negatively impacting female last authorship in COVID-related papers in 2020, with no significant difference in the share of newcomer or new entrant women and incumbent women as last authors in 2020 COVID-related publications.

Turning to column (4) the treatment effect is still significantly negatively affecting to overall middle female authorship, with no significant impact of the incumbency of authors. On the other hand, considering instead column (5), the treatment effect on the probability of a female in sole middle authorship is now decreasing ($year=2020 \times COVID-related$), therefore opposite to that of the DiD model of Table 1 of the main article, although not significantly different from zero.  
Instead, if we look at the effect of incumbency on the COVID-related in 2020, we see that the estimate of $year=2020 \times COVID-related\times P(middle)=\textit{new entrant}$ is positive and significant (0.0475), as well as the estimate of $year=2020 \times COVID-related\times P(middle)=\textit{newcomer}$ (0.0634).

\begin{figure}[H]
    \begin{subfigure}[t]{0.5\textwidth}
       \includegraphics[width=\textwidth]{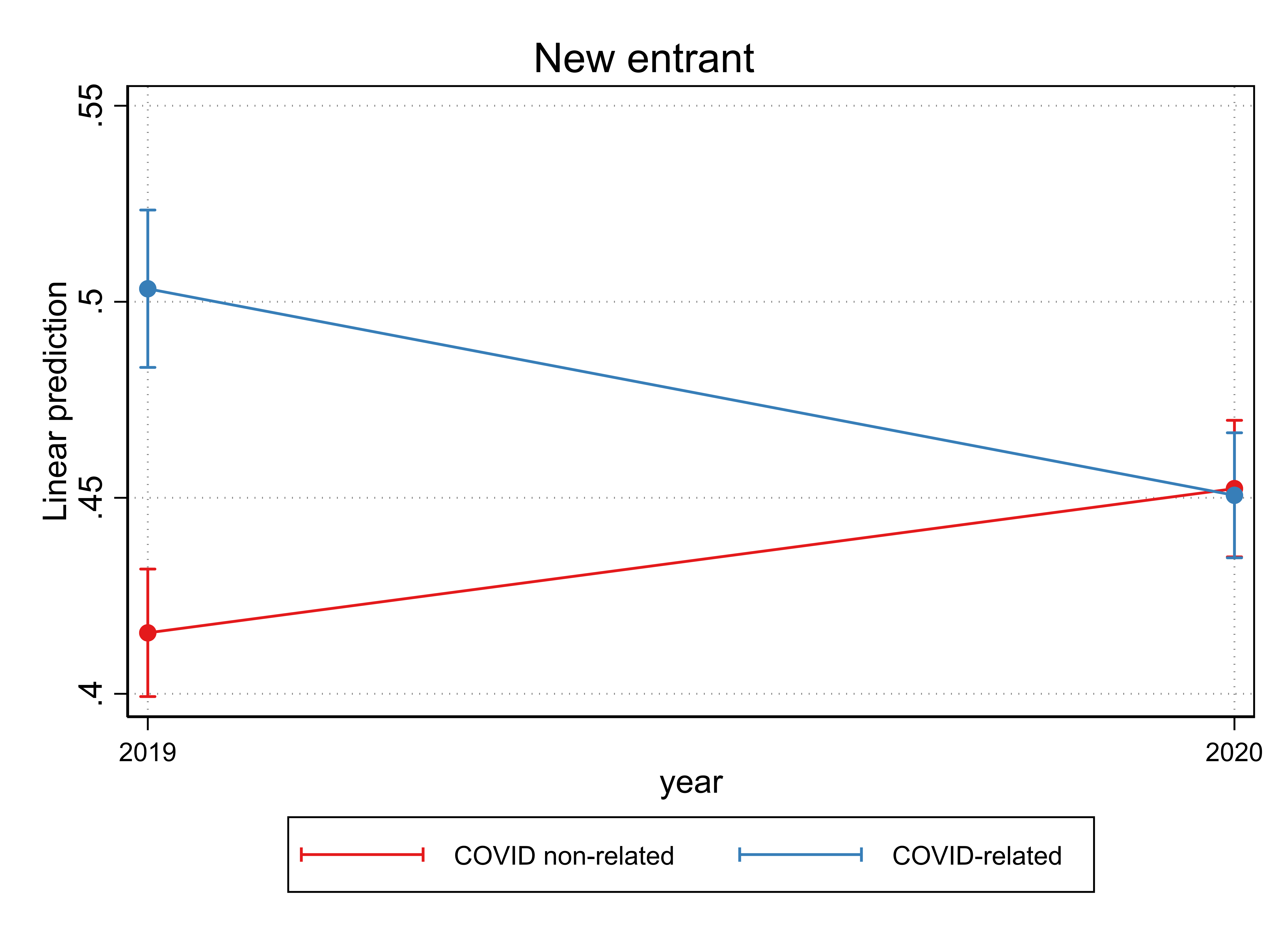}
        \caption{Female First Author - New entrant Last author}
   \end{subfigure}
      \begin{subfigure}[t]{0.5\textwidth}
        \includegraphics[width=\textwidth]{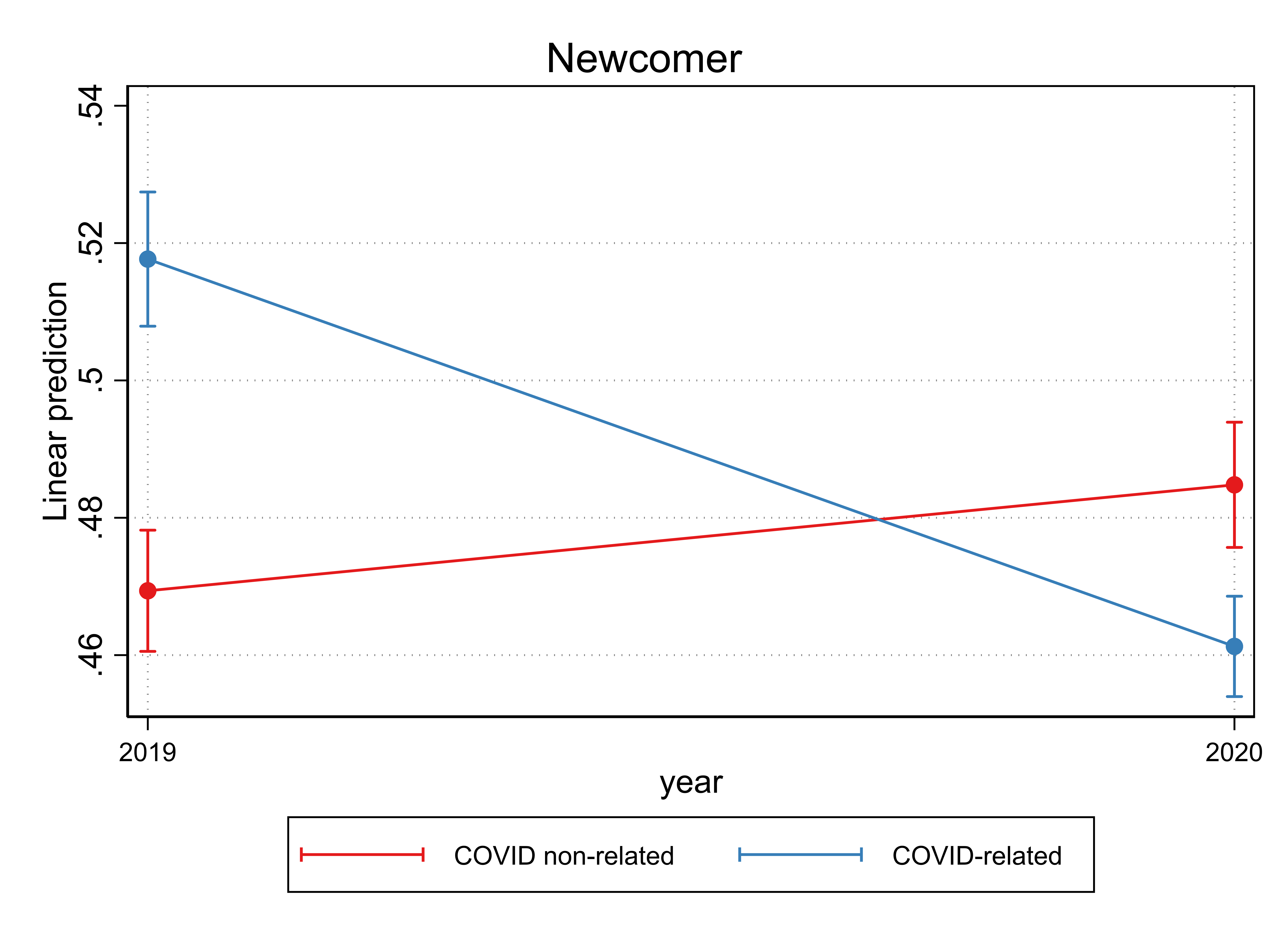}
        \caption{Female First Author - Newcomer Last author}
   \end{subfigure}
          \begin{subfigure}[t]{0.5\textwidth}
       \includegraphics[width=\textwidth]{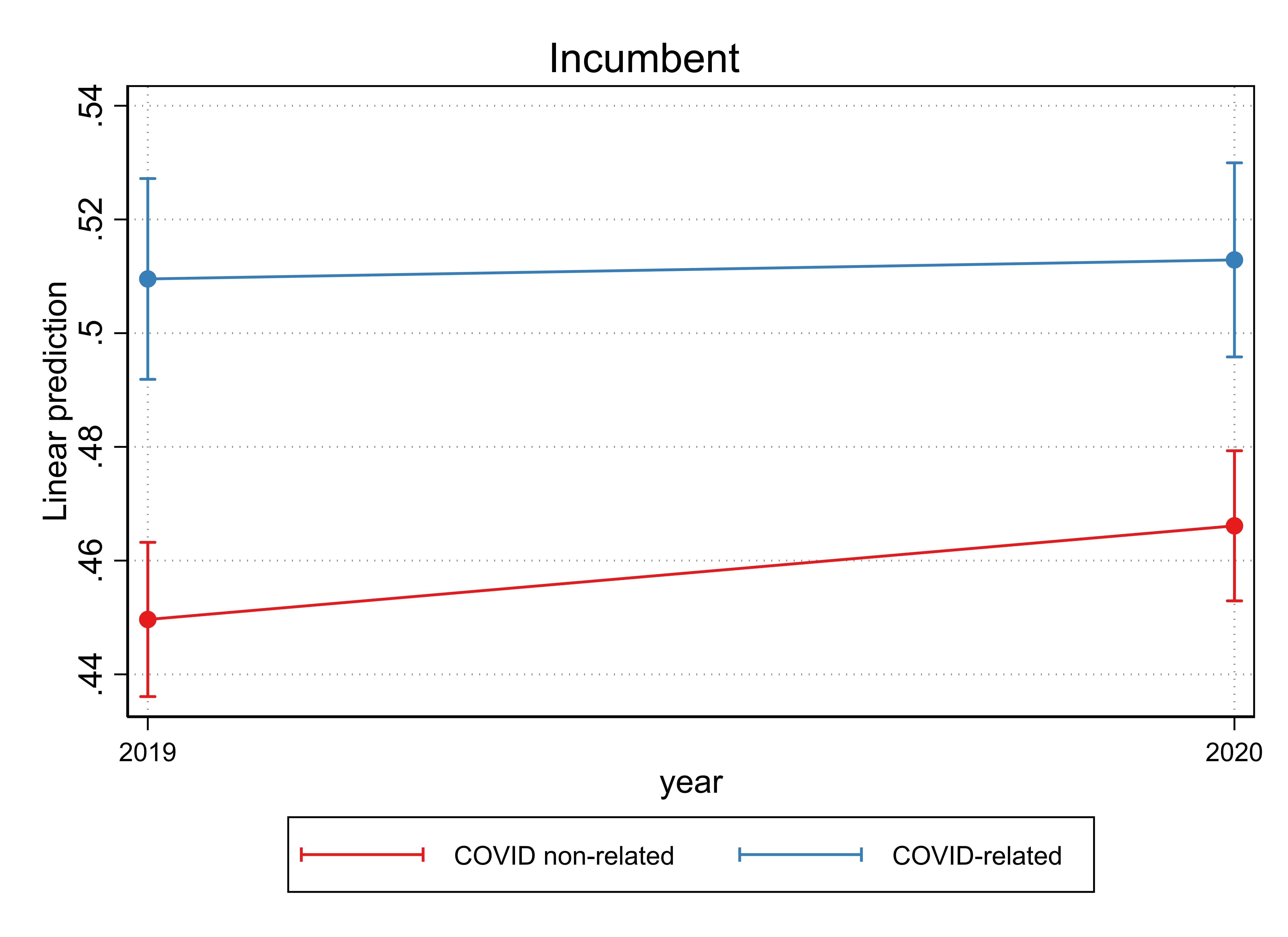}
        \caption{Female First Author - Incumbent Last author}
   \end{subfigure}
    \caption{Predicted probability to observe a women as First author by last author's past research experience. We control for country of last author fixed effects.}
    \label{fig:fig_s15}
\end{figure}

\begin{figure}[H]
    \begin{subfigure}[t]{0.5\textwidth}
       \includegraphics[width=\textwidth]{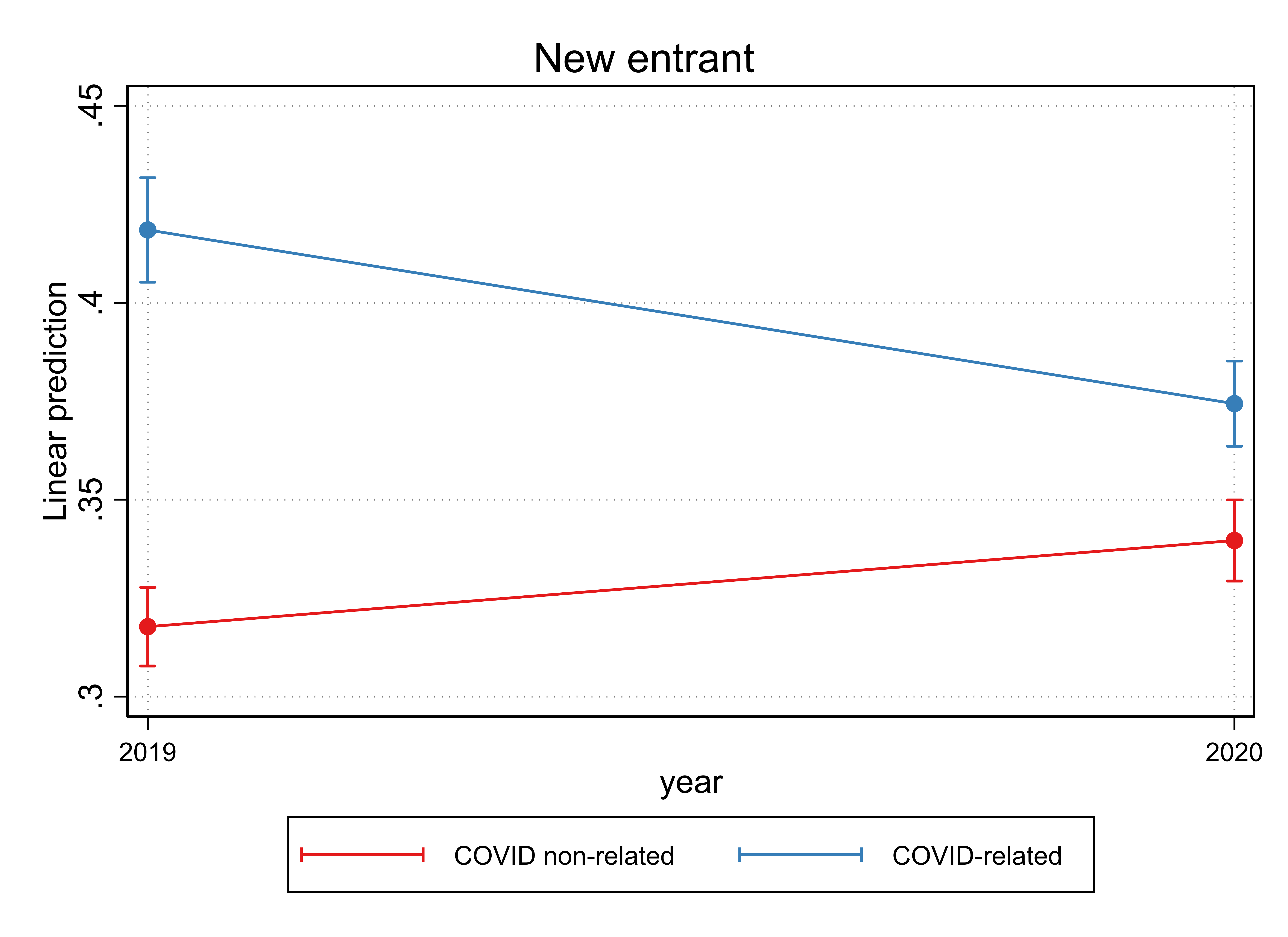}
        \caption{Female Last Author - New entrant First author}
   \end{subfigure}
      \begin{subfigure}[t]{0.5\textwidth}
        \includegraphics[width=\textwidth]{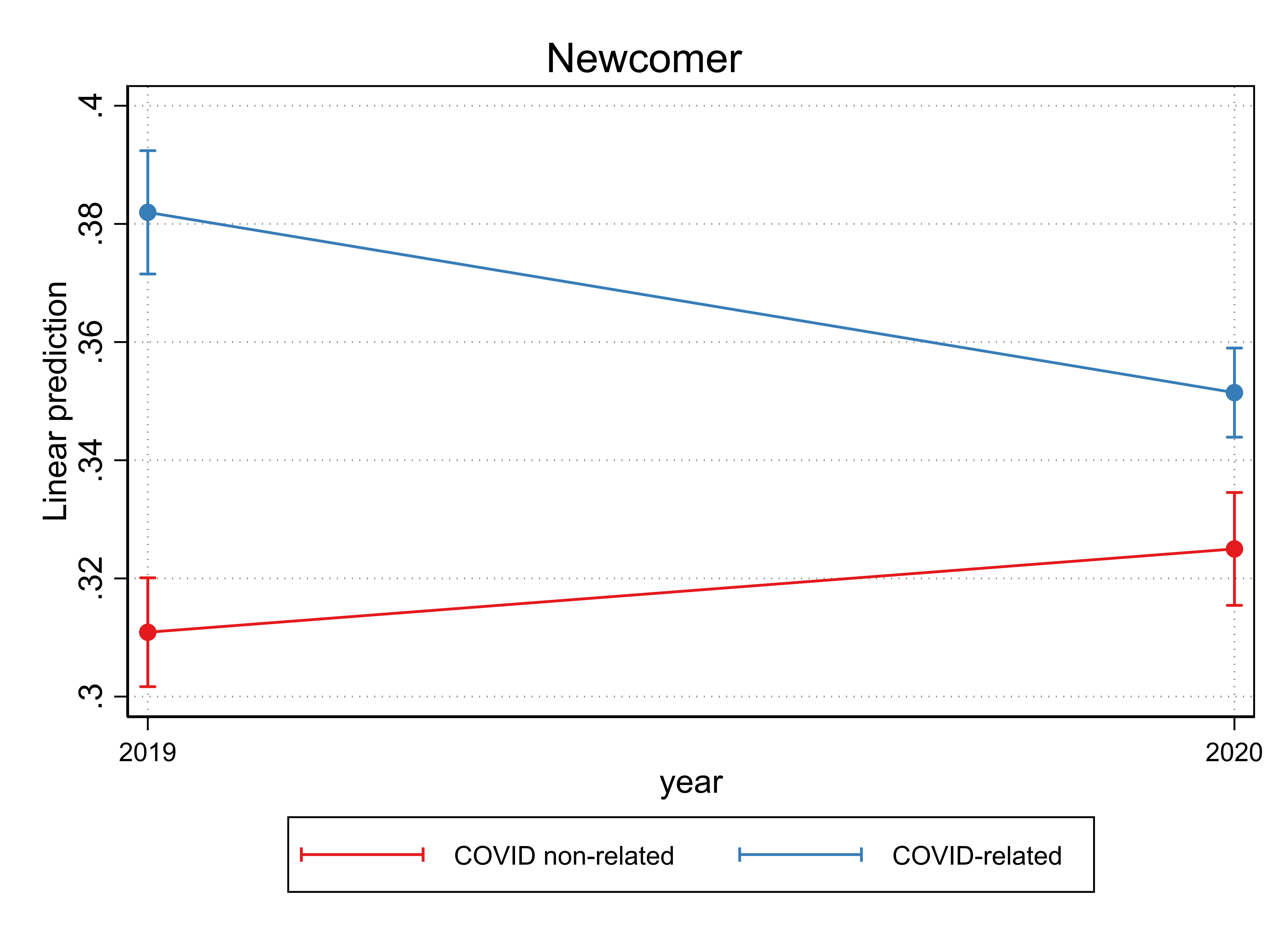}
        \caption{Female Last Author - Newcomer First author}
   \end{subfigure}
          \begin{subfigure}[t]{0.5\textwidth}
       \includegraphics[width=\textwidth]{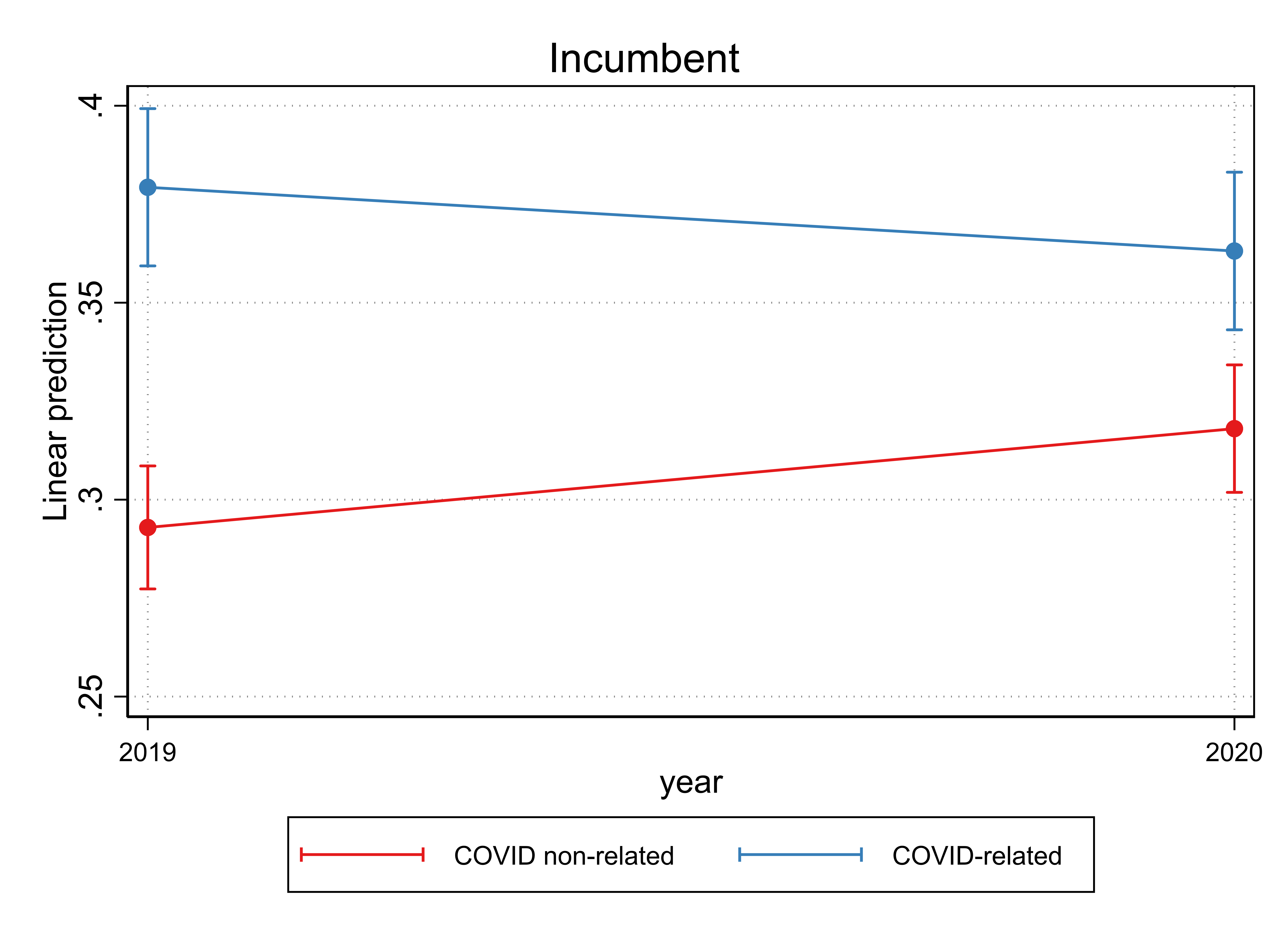}
        \caption{Female Last Author - Incumbent First author}
   \end{subfigure}
    \caption{Predicted probability to observe a women as last author by first author's past research experience. We control for country of first author fixed effects.}
    \label{fig:fig_s16}
\end{figure}

In Figures \ref{fig:fig_s15} and \ref{fig:fig_s16}, we report the predicted time trends of the share of female first authors among COVID-related and COVID non-related by incumbency of the last author (Figure \ref{fig:fig_s14}), and of female last authorship by past research experience of the first author (Figure \ref{fig:fig_s15}). From Figure \ref{fig:fig_s14}, a last author with past expertise on a COVID-related research field significantly and positively affects the likelihood of having a woman as first author in COVID-related publications. The coefficient estimates are reported in Table \ref{pers_table_last}. In column (1), 
the decline of first female authorship in COVID-related papers is explained by the characteristic of the last author  (year=2020 $\times$ COVID-related$\times$ P(last) = newcomer and year=2020 $\times$ COVID-related$\times$ P(last) = new entrant). Considering column (2), 
past research experience of the \emph{first} author instead has no relevant role in determining the decline in female key authorship in COVID-related publications.

\begin{figure}[H]
        \includegraphics[width=\textwidth]{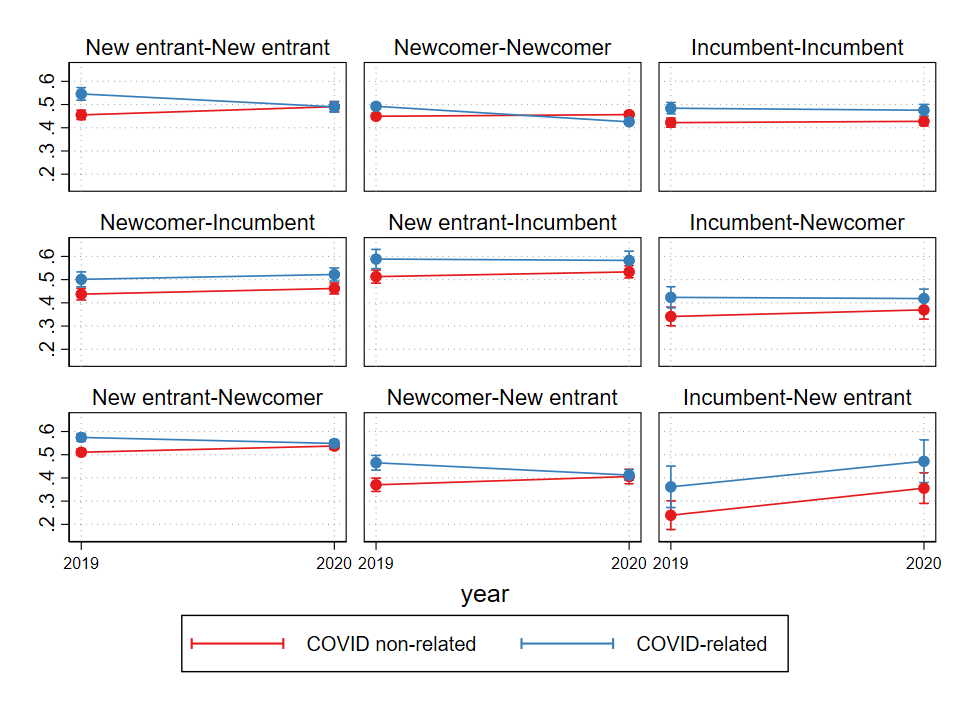}
    \caption{ Predicted probability to observe a women as First authors by first \& last author's past research experience. We control for Country of last author fixed effects.}
    \label{fig:fig_4v2}
  \end{figure}

\begin{figure}[H]
        \includegraphics[width=\textwidth]{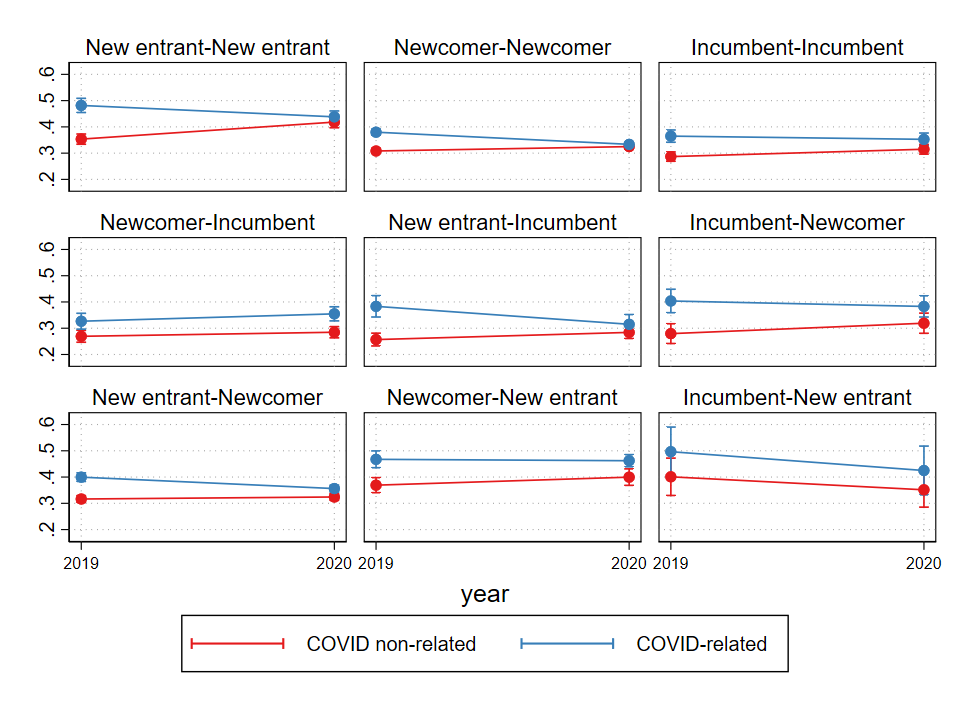}
    \caption{Predicted probability to observe a women as Last authors by first \& last author's past research experience. We control for Country of last author fixed effects.}
    \label{fig:fig_s17}
  \end{figure}

In Figure \ref{fig:fig_4v2}, we plot the predicted time trends among COVID-related and COVID non-related in the probability of having a woman first author by incumbency in research of both key authors, including the new entrant status. Figure  \ref{fig:fig_s17} shows the predicted share of women last authors by incumbency of key authors. The estimates behind the plots are in Table \ref{pers_table_l_f}, where we look for the effect of past research experience of both first and last author on key authorship positions, controlling again for country of the last author fixed effects. From column (1), having a newcomer first author with a newcomer last author diminishes significantly the probability of having a female as first author in publications relevant for the new research topic by 0.059 (year=2020 $\times$ COVID-related $\times$ P(first)=newcomer $\times$ P(last)=newcomer), with respect to having both incumbent authors. Similarly, first female authorship diminishes by 0.076 in COVID-related publication in 2020 when teams have new entrant key authors (year=2020 $\times$ COVID-related $\times$ P(first)=new entrant $\times$ P(last)=new entrant). From column (2), last female authorship is also less likely in teams of new entrants with respect to the incumbent team (-0.068).  

This seems to be suggesting that the drop in female first authorship in research related to the new research opportunity can be explained by the rise of opportunistic teams of either newcomers, which where the bulk of the influx of publications relevant to the new topic COVID-19, or new entrants. 


\begin{table}[H]\centering
\def\sym#1{\ifmmode^{#1}\else\(^{#1}\)\fi}
\caption{Incumbency in research effect. Linear model estimates of equation (2) of the main text, Methods Section, with white-robust standard errors and paper level controls. We include fixed effects for the country of the majority of the team members for columns (1), (4) and (5); country of the first author for column (2); country of the last author for column (3) (omitted). The baseline incumbency level is given by the \emph{incumbent status}, indicating papers for which the author in position of interest has past publications related to the same research topic.\label{pers_table}}
\resizebox{\textwidth}{!}{ 
\begin{tabular}{l c c c c c}
\toprule &\multicolumn{1}{c}{Any Female Author}&\multicolumn{1}{c}{Female First Author}&\multicolumn{1}{c}{Female Last Author}&\multicolumn{1}{c}{Middle Female Authorship}&\multicolumn{1}{c}{Middle Female Authorship Only}\\
\toprule 
                     \multicolumn{1}{c}{Variable}&\multicolumn{1}{c}{(1)}&\multicolumn{1}{c}{(2)}&\multicolumn{1}{c}{(3)}&\multicolumn{1}{c}{(4)}&\multicolumn{1}{c}{(5)}\\

\midrule

year=2020           &      0.0210\sym{***}&      0.0115         &      0.0227\sym{**} &      0.0216\sym{**} &    -0.00435         \\
                    &      (4.09)         &      (0.94)         &      (2.60)         &      (3.03)         &     (-0.54)         \\
\addlinespace
COVID-related             &      0.0507\sym{***}&      0.0639\sym{***}&      0.0816\sym{***}&      0.0528\sym{***}&     -0.0141         \\
                    &      (9.41)         &      (4.73)         &      (7.72)         &      (6.82)         &     (-1.57)         \\
\addlinespace
year=2020 $\times$ COVID-related&     -0.0279\sym{***}&     -0.0167         &     -0.0325\sym{*}  &     -0.0212\sym{*}  &     -0.0107         \\
                    &     (-3.77)         &     (-0.87)         &     (-2.21)         &     (-1.98)         &     (-0.86)         \\
\addlinespace
P(all)=\textit{new entrant}       &     -0.0186         &                     &                     &                     &                     \\
                    &     (-1.49)         &                     &                     &                     &                     \\
\addlinespace
P(all)=\textit{newcomer}       &      0.0222\sym{***}&                     &                     &                     &                     \\
                    &      (4.66)         &                     &                     &                     &                     \\
\addlinespace
year=2020 $\times$ P(all)=\textit{new entrant}&     0.00243         &                     &                     &                     &                     \\
                    &      (0.14)         &                     &                     &                     &                     \\
\addlinespace
year=2020 $\times$ P(all)=\textit{newcomer}&     -0.0109         &                     &                     &                     &                     \\
                    &     (-1.65)         &                     &                     &                     &                     \\
\addlinespace
COVID-related $\times$ P(all)=\textit{new entrant}&     -0.0175         &                     &                     &                     &                     \\
                    &     (-0.89)         &                     &                     &                     &                     \\
\addlinespace
COVID-related $\times$ P(all)=\textit{newcomer}&     -0.0223\sym{***}&                     &                     &                     &                     \\
                    &     (-3.31)         &                     &                     &                     &                     \\
\addlinespace
year=2020 $\times$ COVID-related $\times$ P(all)=\textit{new entrant}&     0.00386         &                     &                     &                     &                     \\
                    &      (0.14)         &                     &                     &                     &                     \\
\addlinespace
year=2020 $\times$ COVID-related $\times$ P(all)=\textit{newcomer}&    -0.00674         &                     &                     &                     &                     \\
                    &     (-0.73)         &                     &                     &                     &                     \\
\addlinespace
\textit{N Authors}           &      0.0124\sym{***}&    0.000455         &    -0.00280\sym{***}&      0.0265\sym{***}&      0.0141\sym{***}\\
                    &     (44.17)         &      (1.15)         &     (-7.56)         &     (46.74)         &     (32.09)         \\
\addlinespace
trial               &      0.0244\sym{***}&      0.0104         &      0.0193\sym{*}  &      0.0516\sym{***}&     0.00693         \\
                    &      (4.78)         &      (1.06)         &      (2.04)         &      (7.53)         &      (0.82)         \\
\addlinespace
Pre-existing Grant      &      0.0414\sym{***}&      0.0656\sym{***}&      0.0494\sym{***}&      0.0528\sym{***}&     -0.0217\sym{***}\\
                    &     (14.32)         &     (12.72)         &      (9.96)         &     (13.32)         &     (-4.89)         \\
\addlinespace
P(first) = \textit{new entrant} &                     &       0.101\sym{***}&                     &                     &                     \\
                    &                     &      (9.89)         &                     &                     &                     \\
\addlinespace
P(first) = \textit{newcomer} &                     &      0.0401\sym{***}&                     &                     &                     \\
                    &                     &      (4.05)         &                     &                     &                     \\
\addlinespace
year=2020 $\times$ P(first) = \textit{new entrant}&                     &      0.0205         &                     &                     &                     \\
                    &                     &      (1.42)         &                     &                     &                     \\
\addlinespace
year=2020 $\times$ P(first) = \textit{newcomer}&                     &     0.00575         &                     &                     &                     \\
                    &                     &      (0.41)         &                     &                     &                     \\
\addlinespace
COVID-related $\times$ P(first) = \textit{new entrant}&                     &     0.00860         &                     &                     &                     \\
                    &                     &      (0.54)         &                     &                     &                     \\
\addlinespace
COVID-related $\times$ P(first) = \textit{newcomer}&                     &     -0.0125         &                     &                     &                     \\
                    &                     &     (-0.81)         &                     &                     &                     \\
\addlinespace

year=2020 $\times$ COVID-related $\times$ P(first) = \textit{new entrant}&                     &     -0.0485\sym{*}  &                     &                     &                     \\
                    &                     &     (-2.15)         &                     &                     &                     \\
\addlinespace
year=2020 $\times$ COVID-related $\times$ P(first) = \textit{newcomer}&                     &     -0.0596\sym{**} &                     &                     &                     \\
                    &                     &     (-2.76)         &                     &                     &                     \\
\addlinespace
P(last) = \textit{new entrant}  &                     &                     &      0.0852\sym{***}&                     &                     \\
                    &                     &                     &      (8.36)         &                     &                     \\
\addlinespace
P(last) = \textit{newcomer}  &                     &                     &      0.0353\sym{***}&                     &                     \\
                    &                     &                     &      (4.76)         &                     &                     \\
\addlinespace
year=2020 $\times$ P(last) = \textit{new entrant}&                     &                     &      0.0253         &                     &                     \\
                    &                     &                     &      (1.73)         &                     &                     \\
\addlinespace
year=2020 $\times$ P(last) = \textit{newcomer}&                     &                     &    -0.00851         &                     &                     \\
                    &                     &                     &     (-0.81)         &                     &                     \\
\addlinespace
COVID-related $\times$ P(last) = \textit{new entrant}&                     &                     &      0.0348\sym{*}  &                     &                     \\
                    &                     &                     &      (2.08)         &                     &                     \\
\addlinespace
COVID-related $\times$ P(last) = \textit{newcomer}&                     &                     &    -0.00431         &                     &                     \\
                    &                     &                     &     (-0.35)         &                     &                     \\
\addlinespace
year=2020 $\times$ COVID-related $\times$ P(last) = \textit{new entrant}&                     &                     &     -0.0430         &                     &                     \\
                    &                     &                     &     (-1.88)         &                     &                     \\
\addlinespace
year=2020 $\times$ COVID-related $\times$ P(last) = \textit{newcomer}&                     &                     &     -0.0275         &                     &                     \\
                    &                     &                     &     (-1.63)         &                     &                     \\
\addlinespace
P(middle)=\textit{new entrant}&                     &                     &                     &     -0.0527\sym{***}&     -0.0193\sym{*}  \\
                    &                     &                     &                     &     (-5.21)         &     (-2.08)         \\
\addlinespace
P(middle)=\textit{newcomer}&                     &                     &                     &      0.0131\sym{*}  &     0.00607         \\
                    &                     &                     &                     &      (2.06)         &      (0.87)         \\
\addlinespace
year=2020 $\times$ P(middle)=\textit{new entrant}&                     &                     &                     &    -0.00190         &     0.00533         \\
                    &                     &                     &                     &     (-0.13)         &      (0.41)         \\
\addlinespace
year=2020 $\times$ P(middle)=\textit{newcomer}&                     &                     &                     &    -0.00844         &     -0.0121         \\
                    &                     &                     &                     &     (-0.95)         &     (-1.25)         \\
\addlinespace
COVID-related $\times$ P(middle)=\textit{new entrant}&                     &                     &                     &      0.0350\sym{*}  &     -0.0213         \\
                    &                     &                     &                     &      (2.31)         &     (-1.51)         \\
\addlinespace
COVID-related $\times$ P(middle)=\textit{newcomer}&                     &                     &                     &     -0.0132         &     -0.0294\sym{**} \\
                    &                     &                     &                     &     (-1.40)         &     (-2.78)         \\
\addlinespace
year=2020 $\times$ COVID-related $\times$ P(middle)=\textit{new entrant}&                     &                     &                     &     -0.0174         &      0.0475\sym{*}  \\
                    &                     &                     &                     &     (-0.85)         &      (2.44)         \\
\addlinespace
year=2020 $\times$ COVID-related $\times$ P(middle)=\textit{newcomer}&                     &                     &                     &     -0.0201         &      0.0634\sym{***}\\
                    &                     &                     &                     &     (-1.56)         &      (4.34)         \\
\addlinespace
Constant            &       0.895\sym{***}&       0.259\sym{***}&       0.241\sym{***}&       0.572\sym{*}  &     -0.0440\sym{**} \\
                    &     (55.22)         &      (3.77)         &      (3.52)         &      (2.50)         &     (-2.83)         \\
\midrule
Observations        &       88771         &       83210         &       82470         &       85419         &       85419         \\
\bottomrule
Country FEs & Majority & First & Last & Majority & Majority \\
\bottomrule
\multicolumn{6}{l}{\footnotesize \textit{t} statistics in parentheses}\\
\multicolumn{6}{l}{\footnotesize \sym{*} \(p<0.05\), \sym{**} \(p<0.01\), \sym{***} \(p<0.001\)}\\
\end{tabular}
}
\end{table}

\begin{table}[H]\centering
\def\sym#1{\ifmmode^{#1}\else\(^{#1}\)\fi}
\caption{Linear model estimates of (i) \emph{incumbency status of last author} on first female authorship, with country effects of the last author (omitted); of (ii) \emph{incumbency status of first author} on last female authorship, with country effects of the first author (omitted), White-robust standard errors and paper level controls. The baseline incumbency level is given by the \emph{incumbent status}.\label{pers_table_last}}
\resizebox{0.8\textwidth}{!}{ 
\begin{tabular}{lccc}
\toprule
                    &\multicolumn{1}{c}{(1)}&\multicolumn{1}{c}{(2)}\\
                    &\multicolumn{1}{c}{Female First  Author}&\multicolumn{1}{c}{Female Last Author}\\
\midrule
year=2020           &      0.0165         &      0.0251\sym{*}  \\
                    &      (1.71)         &      (2.19)         \\
\addlinespace
COVID-related             &      0.0599\sym{***}&      0.0864\sym{***}\\
                    &      (5.29)         &      (6.69)         \\
\addlinespace
year=2020 $\times$ COVID-related&     -0.0131         &     -0.0413\sym{*}  \\
                    &     (-0.83)         &     (-2.25)         \\
\addlinespace
P(last) = \textit{new entrant}  &     -0.0341\sym{**} &                     \\
                    &     (-3.14)         &                     \\
\addlinespace
P(last) = \textit{newcomer}  &      0.0197\sym{*}  &                     \\
                    &      (2.41)         &                     \\
\addlinespace
year=2020 $\times$ P(last) = \textit{new entrant}&      0.0203         &                     \\
                    &      (1.32)         &                     \\
\addlinespace
year=2020 $\times$ P(last) = \textit{newcomer}&    -0.00105         &                     \\
                    &     (-0.09)         &                     \\
\addlinespace
COVID-related $\times$ P(last) = \textit{new entrant}&      0.0279         &                     \\
                    &      (1.61)         &                     \\
\addlinespace
COVID-related $\times$ P(last) = \textit{newcomer}&     -0.0116         &                     \\
                    &     (-0.89)         &                     \\
\addlinespace
year=2020 $\times$ COVID-related $\times$ P(last) = \textit{new entrant}&     -0.0764\sym{**} &                     \\
                    &     (-3.23)         &                     \\
\addlinespace
year=2020 $\times$ COVID-related $\times$ P(last) = \textit{newcomer}&     -0.0587\sym{**} &                     \\
                    &     (-3.25)         &                     \\
\addlinespace
\textit{N Authors}           &   -0.000585         &    -0.00321\sym{***}\\
                    &     (-1.45)         &     (-8.85)         \\
\addlinespace
trial               &      0.0103         &      0.0182         \\
                    &      (1.04)         &      (1.92)         \\
\addlinespace
Pre-existing Grant      &      0.0545\sym{***}&      0.0418\sym{***}\\
                    &     (10.50)         &      (8.42)         \\
\addlinespace
P(first) = \textit{new entrant} &                     &      0.0248\sym{**} \\
                    &                     &      (2.62)         \\
\addlinespace
P(first) = \textit{newcomer} &                     &      0.0179         \\
                    &                     &      (1.95)         \\
\addlinespace
year=2020 $\times$ P(first) = \textit{new entrant}&                     &    -0.00321         \\
                    &                     &     (-0.24)         \\
\addlinespace
year=2020 $\times$ P(first) = \textit{newcomer}&                     &     -0.0110         \\
                    &                     &     (-0.83)         \\
\addlinespace
COVID-related $\times$ P(first) = \textit{new entrant}&                     &      0.0144         \\
                    &                     &      (0.93)         \\
\addlinespace
COVID-related $\times$ P(first) = \textit{newcomer}&                     &     -0.0153         \\
                    &                     &     (-1.04)         \\
\addlinespace
year=2020 $\times$ COVID-related $\times$ P(first) = \textit{new entrant}&                     &     -0.0247         \\
                    &                     &     (-1.15)         \\
\addlinespace
year=2020 $\times$ COVID-related $\times$ P(first) = \textit{newcomer}&                     &    -0.00335         \\
                    &                     &     (-0.16)         \\
\addlinespace
Constant            &       0.331\sym{***}&       0.273\sym{***}\\
                    &      (4.71)         &      (4.15)         \\
\midrule
Observations        &       82470         &       83210         \\
\bottomrule
Country FEs &  Last  & First \\
\bottomrule
\multicolumn{2}{l}{\footnotesize \textit{t} statistics in parentheses}\\
\multicolumn{2}{l}{\footnotesize \sym{*} \(p<0.05\), \sym{**} \(p<0.01\), \sym{***} \(p<0.001\)}\\
\end{tabular}
}
\end{table}

\begin{table}[H]\centering
\def\sym#1{\ifmmode^{#1}\else\(^{#1}\)\fi}
\caption{Linear model estimates on female key authorship of \emph{incumbency status of first and last author}, with country effects of the last author (omitted), White-robust standard errors and paper level controls.  The baseline incumbency level is given by the \emph{incumbent status}. The table continues in the next page.\label{pers_table_l_f}}
\resizebox{\textwidth}{!}{ 
\begin{tabular}{lcc}
\toprule
                    &\multicolumn{1}{c}{(1)}&\multicolumn{1}{c}{(2)}\\
                    &\multicolumn{1}{c}{Female First Author}&\multicolumn{1}{c}{Female Last Author}\\
\midrule

year=2020           &     0.00563         &      0.0280\sym{*}  \\
                    &      (0.40)         &      (2.13)         \\
\addlinespace
COVID-related             &      0.0621\sym{***}&      0.0780\sym{***}\\
                    &      (3.93)         &      (5.22)         \\
\addlinespace
year=2020 $\times$ COVID-related&     -0.0143         &     -0.0401         \\
                    &     (-0.63)         &     (-1.87)         \\
\addlinespace
P(first) = \textit{new entrant} $\times$ P(last) = \textit{new entrant}&      0.0333\sym{*}  &      0.0666\sym{***}\\
                    &      (2.32)         &      (4.94)         \\
\addlinespace
P(first) = \textit{newcomer} $\times$ P(last) = \textit{newcomer}&      0.0275\sym{*}  &      0.0213\sym{*}  \\
                    &      (2.40)         &      (2.03)         \\
\addlinespace
P(first) = \textit{newcomer} $\times$ P(last) = \textit{incumbent}&      0.0155         &     -0.0176         \\
                    &      (0.96)         &     (-1.21)         \\
\addlinespace
P(first) = \textit{new entrant} $\times$ P(last) = \textit{incumbent}&      0.0910\sym{***}&     -0.0300         \\
                    &      (5.24)         &     (-1.95)         \\
\addlinespace
P(first) = \textit{incumbent} $\times$ P(last) = \textit{newcomer}&     -0.0808\sym{***}&    -0.00743         \\
                    &     (-3.55)         &     (-0.35)         \\
\addlinespace
P(first) = \textit{new entrant} $\times$ P(last) = \textit{newcomer}&      0.0887\sym{***}&      0.0296\sym{**} \\
                    &      (7.26)         &      (2.63)         \\
\addlinespace
P(first) = \textit{newcomer} $\times$ P(last) = \textit{new entrant}&     -0.0515\sym{**} &      0.0823\sym{***}\\
                    &     (-2.91)         &      (4.79)         \\
\addlinespace
P(first) = \textit{incumbent} $\times$ P(last) = \textit{new entrant}&      -0.183\sym{***}&       0.114\sym{**} \\
                    &     (-5.53)         &      (3.05)         \\
\addlinespace
year=2020 $\times$ P(first) = \textit{new entrant} $\times$ P(last) = \textit{new entrant}&      0.0298         &      0.0368         \\
                    &      (1.45)         &      (1.88)         \\
\addlinespace
year=2020 $\times$ P(first) = \textit{newcomer} $\times$ P(last) = \textit{newcomer}&     0.00149         &     -0.0113         \\
                    &      (0.09)         &     (-0.74)         \\
\addlinespace
year=2020 $\times$ P(first) = \textit{newcomer} $\times$ P(last) = \textit{incumbent}&      0.0187         &     -0.0125         \\
                    &      (0.83)         &     (-0.61)         \\
\addlinespace
year=2020 $\times$ P(first) = \textit{new entrant} $\times$ P(last) = \textit{incumbent}&      0.0147         &   -0.000719         \\
                    &      (0.61)         &     (-0.03)         \\
\addlinespace
year=2020 $\times$ P(first) = \textit{incumbent} $\times$ P(last) = \textit{newcomer}&      0.0229         &      0.0117         \\
                    &      (0.71)         &      (0.39)         \\
\addlinespace
year=2020 $\times$ P(first) = \textit{new entrant} $\times$ P(last) = \textit{newcomer}&      0.0210         &     -0.0202         \\
                    &      (1.20)         &     (-1.25)         \\
\addlinespace
year=2020 $\times$ P(first) = \textit{newcomer} $\times$ P(last) = \textit{new entrant}&      0.0299         &     0.00267         \\
                    &      (1.16)         &      (0.11)         \\
\addlinespace
year=2020 $\times$ P(first) = \textit{incumbent} $\times$ P(last) = \textit{new entrant}&       0.111\sym{*}  &     -0.0775         \\
                    &      (2.31)         &     (-1.51)         \\
\midrule
Observations        &       82423         &       82423         \\
\bottomrule
Country FEs &  Last   & Last  \\
\bottomrule
\multicolumn{3}{l}{\footnotesize \textit{t} statistics in parentheses}\\
\multicolumn{3}{l}{\footnotesize \sym{*} \(p<0.05\), \sym{**} \(p<0.01\), \sym{***} \(p<0.001\)}\\
\end{tabular}
}
\end{table}

\begin{table}[H]\centering
\def\sym#1{\ifmmode^{#1}\else\(^{#1}\)\fi}
\caption*{Table 26 (continued).\label{pers_table_l_f_p2}}
\resizebox{\textwidth}{!}{ 
\begin{tabular}{lcc}
\toprule
                    &\multicolumn{1}{c}{(1)}&\multicolumn{1}{c}{(2)}\\
                    &\multicolumn{1}{c}{Female First Author}&\multicolumn{1}{c}{Female Last Author}\\
\midrule
COVID-related $\times$ P(first) = \textit{new entrant} $\times$ P(last) = \textit{new entrant}&      0.0283         &      0.0503\sym{*}  \\
                    &      (1.22)         &      (2.23)         \\
\addlinespace
COVID-related $\times$ P(first) = \textit{newcomer} $\times$ P(last) = \textit{newcomer}&     -0.0191         &    -0.00639         \\
                    &     (-1.06)         &     (-0.38)         \\
\addlinespace
COVID-related $\times$ P(first) = \textit{newcomer} $\times$ P(last) = \textit{incumbent}&     0.00166         &     -0.0205         \\
                    &      (0.06)         &     (-0.85)         \\
\addlinespace
COVID-related $\times$ P(first) = \textit{new entrant} $\times$ P(last) = \textit{incumbent}&      0.0135         &      0.0485         \\
                    &      (0.45)         &      (1.71)         \\
\addlinespace
COVID-related $\times$ P(first) = \textit{incumbent} $\times$ P(last) = \textit{newcomer}&      0.0204         &      0.0464         \\
                    &      (0.59)         &      (1.39)         \\
\addlinespace
COVID-related $\times$ P(first) = \textit{new entrant} $\times$ P(last) = \textit{newcomer}&     0.00168         &     0.00527         \\
                    &      (0.09)         &      (0.29)         \\
\addlinespace
COVID-related $\times$ P(first) = \textit{newcomer} $\times$ P(last) = \textit{new entrant}&      0.0327         &      0.0205         \\
                    &      (1.22)         &      (0.77)         \\
\addlinespace
COVID-related $\times$ P(first) = \textit{incumbent} $\times$ P(last) = \textit{new entrant}&      0.0603         &      0.0176         \\
                    &      (1.05)         &      (0.29)         \\
year=2020 $\times$ COVID-related $\times$ P(first) = \textit{new entrant} $\times$ P(last) = \textit{new entrant}&     -0.0769\sym{*}  &     -0.0680\sym{*}  \\
                    &     (-2.37)         &     (-2.16)         \\
\addlinespace
year=2020 $\times$ COVID-related $\times$ P(first) = \textit{newcomer} $\times$ P(last) = \textit{newcomer}&     -0.0598\sym{*}  &     -0.0227         \\
                    &     (-2.35)         &     (-0.95)         \\
\addlinespace
year=2020 $\times$ COVID-related $\times$ P(first) = \textit{newcomer} $\times$ P(last) = \textit{incumbent}&      0.0108         &      0.0528         \\
                    &      (0.30)         &      (1.57)         \\
\addlinespace
year=2020 $\times$ COVID-related $\times$ P(first) = \textit{new entrant} $\times$ P(last) = \textit{incumbent}&     -0.0120         &     -0.0554         \\
                    &     (-0.29)         &     (-1.41)         \\
\addlinespace
year=2020 $\times$ COVID-related $\times$ P(first) = \textit{incumbent} $\times$ P(last) = \textit{newcomer}&     -0.0196         &     -0.0203         \\
                    &     (-0.41)         &     (-0.44)         \\
\addlinespace
year=2020 $\times$ COVID-related $\times$ P(first) = \textit{new entrant} $\times$ P(last) = \textit{newcomer}&     -0.0381         &     -0.0111         \\
                    &     (-1.40)         &     (-0.43)         \\
\addlinespace
year=2020 $\times$ COVID-related $\times$ P(first) = \textit{newcomer} $\times$ P(last) = \textit{new entrant}&     -0.0742\sym{*}  &     0.00415         \\
                    &     (-2.00)         &      (0.11)         \\
\addlinespace
year=2020 $\times$ COVID-related $\times$ P(first) = \textit{incumbent} $\times$ P(last) = \textit{new entrant}&     0.00782         &      0.0181         \\
                    &      (0.09)         &      (0.21)         \\
\addlinespace
\textit{N Authors}            &    0.000406         &    -0.00272\sym{***}\\
                    &      (1.00)         &     (-7.30)         \\
\addlinespace
trial               &      0.0146         &      0.0193\sym{*}  \\
                    &      (1.49)         &      (2.03)         \\
\addlinespace
Pre-existing Grant         &      0.0597\sym{***}&      0.0499\sym{***}\\
                    &     (11.50)         &     (10.05)         \\
\addlinespace
Constant            &       0.293\sym{***}&       0.259\sym{***}\\
                    &      (4.04)         &      (3.70)         \\
\midrule
Observations        &       82423         &       82423         \\
\bottomrule
Country FEs &  Last   & Last  \\
\bottomrule
\multicolumn{3}{l}{\footnotesize \textit{t} statistics in parentheses}\\
\multicolumn{3}{l}{\footnotesize \sym{*} \(p<0.05\), \sym{**} \(p<0.01\), \sym{***} \(p<0.001\)}\\
\end{tabular}
}
\end{table}

\subsection{Mediating effect of stringency measures in incumbency} 


It could be that early-career female scientists got discriminated against because of how stringency norms at the time in the country of the last author affected their expectations on the potential contribution of candidate members of the research team, when deciding upon team formation for a quick a timely placed publication on the new hot research topic. As stay-at-home measures deeply influenced the daily working schedule of both male and female tenured scientist, assigning the role of first author to either a male or a female researcher could in turn being influenced by biased anticipation of the risk of collaborating with a female first author, when trying to promptly submit research to journals, because of the disproportionately effect of family duties on female authorship.
This would signify that biased expectations of the last author on the early career authors' ability to contribute under pressuring circumstances are discriminating against women, especially when these external circumstances increase family duties within household, which lead to risk anticipation of the probability of a successful publication that unevenly damages women. Beyond the specifics of COVID-19, pregnant women could generally be discriminated against in research and science on the basis of how their tenured colleagues perceive their ability to contribute when there is high pressure to publish.


In Table \ref{pers_str_first_2} re-estimate the model including interactions of the treatment effect with incumbency levels of the both key authors and stringency indices of lockdown measures of (i) the first author, or (ii) the last author. 
In Column (1), school closures of the country of the first author are negatively and significantly affecting first female authorship in 2020 COVID-related papers in teams of new-entrants ($COVID-related\times  P(first)=\textit{new entrant} \times P(last)=\textit{new entrant} \times SchoolClosuresMaxFirst$), and in teams with a new-coming first and an incumbent last author ($COVID-related\times  P(first)=\textit{newcomer} \times P(last)=\textit{incumbent} \times SchoolClosuresMaxFirst$).  In Column (2), the combination of first newcomer author and last incumbent author ($year=2020 \times COVID-related\times  P(first)=\textit{newcomer} \times P(last)=\textit{incumbent}$) significantly increases the probability of having women in first authorship positions in COVID-related publications, but workplace closures of the country of the first author bring down and attenuate this effect ($COVID-related\times  P(first)=\textit{newcomer} \times P(last)=\textit{incumbent} \times WorkClosuresMaxFirst$).
Instead, workplace closures of the first author negatively and significantly affect first female authorship in teams of new entrants ($COVID-related\times  P(first)=\textit{new entrant} \times P(last)=\textit{new entrant} \times WorkClosuresMaxFirst$).
From Columns (3) and (4), changing stringency measures within the country of the first author do not have a significant impact on last female authorship. This is somewhat to be expected, as last authors usually have more bargaining power within teams with respect to early career scientists, and are less likely to be affected by characteristics of the first authors, if they have the choice of who gets appointed to that particular leadership position. We find once again that teams of new entrants are significantly less likely to feature a female last author with respect to an incumbent team ($COVID-related\times  P(first)=\textit{new entrant} \times P(last)=\textit{new entrant}$), no matter if we control for school or workplace closures.

In Table \ref{tab_str_pers_2}, we perform the same analysis, interacting the treatment effect with stringency indices in the country of the last author, and incumbency indicators for both first and last authors. In Column (2), we find that workplace closures of the country of the last author are negatively and significantly impacting newcomer female authors in first authorship position in COVID-related published research, when the last author is instead incumbent ($COVID-related\times  P(first)=\textit{newcomer} \& P(last)=\textit{incumbent} \times WorkClosuresMaxLast$). Similarly, workplace closures also significantly and negatively affect female first authorship in COVID-related publications of teams of new entrant key authors ($COVID-related\times  P(first)=\textit{new entrant} \& P(last)=\textit{new entrant} \times WorkClosuresMaxLast$). Once again, from Columns (3) and (4), women last authors are significantly less likely to be featured in a COVID-related publication of a new entrant team (year=2020 $\times$ COVID-related $\times$ P(first) = new entrant $\times$ P(last) = new entrant), no matter if we control for school or workplace closures.

In Tables \ref{tab_20} and \ref{tab_21}, we assess the effect of incumbency of key authors and stringency values of both key authors. In Table \ref{tab_20}, column (1), we see that there is no effect of either first and last authors' school closures on the appointment of women as first authors of COVID-related publications within the different kinds of teams, suggesting that introducing both key authors' school closures may be leading to collinearity. Still, in Column (2), last authors are once again significantly less likely to be women when considering teams of new entrant key authors (year=2020 $\times$ COVID-related $\times$ P(first) = new entrant $\times$ P(last) = new entrant) with respect to the incumbent baseline. The same negative, significant impact on last female authorship is found in Table \ref{tab_21}, Column (2). Table \ref{tab_21}, column (1), shows that again there is a positive, significant effect on first female authorship by teaming up a newcomer first author with an incumbent last authors ($year=2020 \times COVID-related\times  P(first)=\textit{newcomer} \times P(last)=\textit{incumbent}$). Once again, we find once again that pairing a newcomer with an incumbent, tenured author increases the share of women first authors among COVID-related paper, suggesting that the workplace closures of the first authors confounded by the effect of the closures of the last author found in Table \ref{tab_str_pers_2}.

Overall, we find that when we include new entrant authors in the analysis on team incumbency on female first authorship in COVID-related publications, we notice that new entrant teams, where both key authors are new entrant, have a lower the probability of having a woman as either first or last author in COVID-related publications with respect to incumbent teams. 
But then, when we look at how stringency might influence the way different teams appoint women as first authors, we find that school and workplace closures of the first author and workplace closures of the last author do mediate the probability that teams of new entrant key authors appoint a woman as first author. 
Instead, stringency measures do not explain the drop in first female authorship among teams of newcomers. Moreover, controlling for workplace closures, teaming up a newcomer first author with a incumbent last author in COVID-19 publications has a positive effect on female first authorship, with closures within the country of the first author reducing this effect. At last, women result less likely to be featured as last authors in teams of new entrants across all estimations.

\begin{table}[H]\centering
\def\sym#1{\ifmmode^{#1}\else\(^{#1}\)\fi}
\caption{Linear regression model estimates on key female authorship, including interactions of (year=2020 $\times$ COVID-related) with first and last author's incumbency in research, and lockdown measure indicators of \emph{school closures} and \emph{workplace closures} of first author. We control for country of first authors' fixed effect (omitted), team size, clinical trials and previous grants, and compute white-robust standard errors.  The baseline incumbency level is given by the \emph{incumbent status}. The table continues in the next page.\label{pers_str_first_2}}
\resizebox{\textwidth}{!}{  \begin{tabular}{lcccc}
\toprule
                    &\multicolumn{1}{c}{(1)}&\multicolumn{1}{c}{(2)}&\multicolumn{1}{c}{(3)}&\multicolumn{1}{c}{(4)}\\
                    &\multicolumn{1}{c}{Female First Author}&\multicolumn{1}{c}{Female First Author}&\multicolumn{1}{c}{Female Last Author}&\multicolumn{1}{c}{Female Last Author}\\
\midrule
year=2020           &      0.0318         &      0.0370         &      0.0160         &      0.0140         \\
                    &      (1.45)         &      (1.74)         &      (0.80)         &      (0.71)         \\
\addlinespace
COVID-related             &      0.0604\sym{***}&      0.0604\sym{***}&      0.0743\sym{***}&      0.0742\sym{***}\\
                    &      (3.85)         &      (3.85)         &      (5.02)         &      (5.01)         \\
\addlinespace
year=2020 $\times$ COVID-related&     -0.0343         &     -0.0348         &     -0.0161         &    -0.00953         \\
                    &     (-1.02)         &     (-1.05)         &     (-0.51)         &     (-0.31)         \\
\addlinespace
P(first) = \textit{new entrant} $\times$ P(last) = \textit{new entrant}&      0.0362\sym{*}  &      0.0365\sym{*}  &      0.0677\sym{***}&      0.0680\sym{***}\\
                    &      (2.53)         &      (2.55)         &      (5.04)         &      (5.07)         \\
\addlinespace
P(first) = \textit{newcomer} $\times$ P(last) = \textit{newcomer}&      0.0275\sym{*}  &      0.0274\sym{*}  &      0.0222\sym{*}  &      0.0223\sym{*}  \\
                    &      (2.41)         &      (2.40)         &      (2.12)         &      (2.12)         \\
\addlinespace
P(first) = \textit{newcomer} $\times$ P(last) = \textit{incumbent}&      0.0156         &      0.0154         &     -0.0153         &     -0.0154         \\
                    &      (0.97)         &      (0.95)         &     (-1.05)         &     (-1.05)         \\
\addlinespace
P(first) = \textit{new entrant} $\times$ P(last) = \textit{incumbent}&      0.0898\sym{***}&      0.0897\sym{***}&     -0.0285         &     -0.0285         \\
                    &      (5.18)         &      (5.18)         &     (-1.85)         &     (-1.85)         \\
\addlinespace
P(first) = \textit{incumbent} $\times$ P(last) = \textit{newcomer}&     -0.0757\sym{***}&     -0.0759\sym{***}&    -0.00906         &    -0.00897         \\
                    &     (-3.36)         &     (-3.37)         &     (-0.43)         &     (-0.43)         \\
\addlinespace
P(first) = \textit{new entrant} $\times$ P(last) = \textit{newcomer}&      0.0909\sym{***}&      0.0909\sym{***}&      0.0298\sym{**} &      0.0299\sym{**} \\
                    &      (7.46)         &      (7.47)         &      (2.66)         &      (2.67)         \\
\addlinespace
P(first) = \textit{newcomer} $\times$ P(last) = \textit{new entrant}&     -0.0497\sym{**} &     -0.0495\sym{**} &      0.0762\sym{***}&      0.0764\sym{***}\\
                    &     (-2.87)         &     (-2.86)         &      (4.54)         &      (4.55)         \\
\addlinespace
P(first) = \textit{incumbent} $\times$ P(last) = \textit{new entrant}&      -0.172\sym{***}&      -0.172\sym{***}&      0.0987\sym{**} &      0.0987\sym{**} \\
                    &     (-5.34)         &     (-5.34)         &      (2.70)         &      (2.70)         \\
\addlinespace
year=2020 $\times$ P(first) = \textit{new entrant} $\times$ P(last) = \textit{new entrant}&     -0.0357         &     -0.0304         &      0.0718\sym{*}  &      0.0806\sym{*}  \\
                    &     (-1.03)         &     (-0.91)         &      (2.14)         &      (2.51)         \\
\addlinespace
year=2020 $\times$ P(first) = \textit{newcomer} $\times$ P(last) = \textit{newcomer}&     -0.0222         &     -0.0299         &    -0.00150         &     -0.0129         \\
                    &     (-0.84)         &     (-1.17)         &     (-0.06)         &     (-0.55)         \\
\addlinespace
year=2020 $\times$ P(first) = \textit{newcomer} $\times$ P(last) = \textit{incumbent}&     0.00222         &     -0.0271         &    -0.00951         &     0.00266         \\
                    &      (0.06)         &     (-0.77)         &     (-0.29)         &      (0.08)         \\
\addlinespace
year=2020 $\times$ P(first) = \textit{new entrant} $\times$ P(last) = \textit{incumbent}&      0.0103         &    0.000368         &      0.0452         &      0.0437         \\
                    &      (0.26)         &      (0.01)         &      (1.26)         &      (1.27)         \\
\addlinespace
year=2020 $\times$ P(first) = \textit{incumbent} $\times$ P(last) = \textit{newcomer}&     -0.0333         &    -0.00697         &      0.0123         &     -0.0103         \\
                    &     (-0.58)         &     (-0.12)         &      (0.23)         &     (-0.20)         \\
\addlinespace
year=2020 $\times$ P(first) = \textit{new entrant} $\times$ P(last) = \textit{newcomer}&     0.00184         &     0.00577         &     -0.0278         &     -0.0368         \\
                    &      (0.06)         &      (0.21)         &     (-1.07)         &     (-1.47)         \\
\addlinespace
year=2020 $\times$ P(first) = \textit{newcomer} $\times$ P(last) = \textit{new entrant}&      0.0238         &    0.000381         &    -0.00213         &     0.00165         \\
                    &      (0.55)         &      (0.01)         &     (-0.05)         &      (0.04)         \\
\addlinespace
year=2020 $\times$ P(first) = \textit{incumbent} $\times$ P(last) = \textit{new entrant}&      0.0529         &    -0.00511         &     -0.0472         &     -0.0785         \\
                    &      (0.64)         &     (-0.06)         &     (-0.58)         &     (-0.96)         \\
\addlinespace
COVID-related $\times$ P(first) = \textit{new entrant} $\times$ P(last) = \textit{new entrant}&      0.0331         &      0.0331         &      0.0496\sym{*}  &      0.0496\sym{*}  \\
                    &      (1.43)         &      (1.43)         &      (2.22)         &      (2.22)         \\
\addlinespace
COVID-related $\times$ P(first) = \textit{newcomer} $\times$ P(last) = \textit{newcomer}&     -0.0154         &     -0.0153         &    -0.00638         &    -0.00632         \\
                    &     (-0.86)         &     (-0.86)         &     (-0.38)         &     (-0.38)         \\
\addlinespace
COVID-related $\times$ P(first) = \textit{newcomer} $\times$ P(last) = \textit{incumbent}&    -0.00214         &    -0.00198         &     -0.0240         &     -0.0238         \\
                    &     (-0.08)         &     (-0.08)         &     (-0.99)         &     (-0.99)         \\
\addlinespace
COVID-related $\times$ P(first) = \textit{new entrant} $\times$ P(last) = \textit{incumbent}&      0.0225         &      0.0225         &      0.0535         &      0.0536         \\
                    &      (0.75)         &      (0.75)         &      (1.88)         &      (1.89)         \\
\addlinespace
COVID-related $\times$ P(first) = \textit{incumbent} $\times$ P(last) = \textit{newcomer}&      0.0199         &      0.0200         &      0.0507         &      0.0507         \\
                    &      (0.58)         &      (0.59)         &      (1.54)         &      (1.54)         \\
\addlinespace
COVID-related $\times$ P(first) = \textit{new entrant} $\times$ P(last) = \textit{newcomer}&     0.00234         &     0.00239         &      0.0104         &      0.0104         \\
                    &      (0.12)         &      (0.12)         &      (0.57)         &      (0.57)         \\
\addlinespace
COVID-related $\times$ P(first) = \textit{newcomer} $\times$ P(last) = \textit{new entrant}&      0.0354         &      0.0354         &      0.0224         &      0.0225         \\
                    &      (1.34)         &      (1.34)         &      (0.87)         &      (0.87)         \\
\addlinespace
COVID-related $\times$ P(first) = \textit{incumbent} $\times$ P(last) = \textit{new entrant}&      0.0614         &      0.0616         &      0.0381         &      0.0382         \\
                    &      (1.11)         &      (1.11)         &      (0.65)         &      (0.65)         \\
\addlinespace
year=2020 $\times$ COVID-related $\times$ P(first) = \textit{new entrant} $\times$ P(last) = \textit{new entrant}&     0.00827         &      0.0441         &      -0.116\sym{*}  &      -0.127\sym{*}  \\
                    &      (0.16)         &      (0.86)         &     (-2.25)         &     (-2.55)         \\
\addlinespace
year=2020 $\times$ COVID-related $\times$ P(first) = \textit{newcomer} $\times$ P(last) = \textit{newcomer}&     -0.0207         &     -0.0200         &    -0.00884         &     0.00577         \\
                    &     (-0.53)         &     (-0.52)         &     (-0.24)         &      (0.16)         \\
\addlinespace
year=2020 $\times$ COVID-related $\times$ P(first) = \textit{newcomer} $\times$ P(last) = \textit{incumbent}&       0.103         &       0.114\sym{*}  &      0.0427         &      0.0347         \\
                    &      (1.84)         &      (2.07)         &      (0.81)         &      (0.67)         \\
\addlinespace
year=2020 $\times$ COVID-related $\times$ P(first) = \textit{new entrant} $\times$ P(last) = \textit{incumbent}&      0.0256         &      0.0361         &     -0.0929         &     -0.0803         \\
                    &      (0.38)         &      (0.54)         &     (-1.46)         &     (-1.28)         \\
\addlinespace
year=2020 $\times$ COVID-related $\times$ P(first) = \textit{incumbent} $\times$ P(last) = \textit{newcomer}&   -0.000207         &      0.0107         &     -0.0424         &     -0.0539         \\
                    &     (-0.00)         &      (0.13)         &     (-0.54)         &     (-0.72)         \\
\addlinespace
year=2020 $\times$ COVID-related $\times$ P(first) = \textit{new entrant} $\times$ P(last) = \textit{newcomer}&     -0.0247         &     -0.0314         &      0.0134         &      0.0124         \\
                    &     (-0.58)         &     (-0.75)         &      (0.33)         &      (0.32)         \\
\addlinespace
year=2020 $\times$ COVID-related $\times$ P(first) = \textit{newcomer} $\times$ P(last) = \textit{new entrant}&     0.00629         &      0.0164         &      0.0271         &     0.00941         \\
                    &      (0.10)         &      (0.28)         &      (0.46)         &      (0.16)         \\
\addlinespace
year=2020 $\times$ COVID-related $\times$ P(first) = \textit{incumbent} $\times$ P(last) = \textit{new entrant}&     -0.0631         &      0.0240         &     -0.0525         &      -0.171         \\
                    &     (-0.41)         &      (0.16)         &     (-0.35)         &     (-1.22)         \\
\midrule
Observations        &       83104         &       83104         &       83104         &       83104         \\
\bottomrule
\midrule
Country FEs & First & First  & First & First \\
\bottomrule
\multicolumn{5}{l}{\footnotesize \textit{t} statistics in parentheses}\\
\multicolumn{5}{l}{\footnotesize \sym{*} \(p<0.05\), \sym{**} \(p<0.01\), \sym{***} \(p<0.001\)}\\
\end{tabular}
}\end{table}

\begin{table}[H]\centering
\def\sym#1{\ifmmode^{#1}\else\(^{#1}\)\fi}
\caption*{Table 27 (continued).\label{pers_str_first_2_p2}}
\resizebox{\textwidth}{!}{  \begin{tabular}{lcccc}
\toprule
                    &\multicolumn{1}{c}{(1)}&\multicolumn{1}{c}{(2)}&\multicolumn{1}{c}{(3)}&\multicolumn{1}{c}{(4)}\\
                    &\multicolumn{1}{c}{Female First Author}&\multicolumn{1}{c}{Female First Author}&\multicolumn{1}{c}{Female Last Author}&\multicolumn{1}{c}{Female Last Author}\\
\midrule
SchoolClosuresFirst&     -0.0151         &                     &     0.00340         &                     \\
                    &     (-1.80)         &                     &      (0.44)         &                     \\
\addlinespace
COVID-related $\times$ SchoolClosuresFirst&     0.00994         &                     &     -0.0115         &                     \\
                    &      (0.76)         &                     &     (-0.94)         &                     \\
\addlinespace
P(first) = \textit{new entrant} $\times$ P(last) = \textit{new entrant} $\times$ SchoolClosuresFirst&      0.0318\sym{*}  &                     &     -0.0151         &                     \\
                    &      (2.44)         &                     &     (-1.19)         &                     \\
\addlinespace
P(first) = \textit{newcomer} $\times$ P(last) = \textit{newcomer} $\times$ SchoolClosuresFirst&      0.0139         &                     &    -0.00412         &                     \\
                    &      (1.38)         &                     &     (-0.44)         &                     \\
\addlinespace
P(first) = \textit{newcomer} $\times$ P(last) = \textit{incumbent} $\times$ SchoolClosuresFirst&      0.0114         &                     &    -0.00277         &                     \\
                    &      (0.84)         &                     &     (-0.22)         &                     \\
\addlinespace
P(first) = \textit{new entrant} $\times$ P(last) = \textit{incumbent} $\times$ SchoolClosuresFirst&     0.00517         &                     &     -0.0219         &                     \\
                    &      (0.35)         &                     &     (-1.63)         &                     \\
\addlinespace
P(first) = \textit{incumbent} $\times$ P(last) = \textit{newcomer} $\times$ SchoolClosuresFirst&      0.0240         &                     &     0.00106         &                     \\
                    &      (1.12)         &                     &      (0.05)         &                     \\
\addlinespace
P(first) = \textit{new entrant} $\times$ P(last) = \textit{newcomer} $\times$ SchoolClosuresFirst&      0.0123         &                     &     0.00480         &                     \\
                    &      (1.15)         &                     &      (0.48)         &                     \\
\addlinespace
P(first) = \textit{newcomer} $\times$ P(last) = \textit{new entrant} $\times$ SchoolClosuresFirst&     0.00344         &                     &     0.00237         &                     \\
                    &      (0.21)         &                     &      (0.15)         &                     \\
\addlinespace
P(first) = \textit{incumbent} $\times$ P(last) = \textit{new entrant} $\times$ SchoolClosuresFirst&      0.0141         &                     &    -0.00522         &                     \\
                    &      (0.46)         &                     &     (-0.18)         &                     \\
\addlinespace
COVID-related $\times$ P(first) = \textit{new entrant} $\times$ P(last) = \textit{new entrant} $\times$ SchoolClosuresFirst&     -0.0436\sym{*}  &                     &      0.0261         &                     \\
                    &     (-2.20)         &                     &      (1.35)         &                     \\
\addlinespace
COVID-related $\times$ P(first) = \textit{newcomer} $\times$ P(last) = \textit{newcomer} $\times$ SchoolClosuresFirst&     -0.0215         &                     &    -0.00422         &                     \\
                    &     (-1.45)         &                     &     (-0.30)         &                     \\
\addlinespace
COVID-related $\times$ P(first) = \textit{newcomer} $\times$ P(last) = \textit{incumbent} $\times$ SchoolClosuresFirst&     -0.0434\sym{*}  &                     &      0.0107         &                     \\
                    &     (-2.06)         &                     &      (0.54)         &                     \\
\addlinespace
COVID-related $\times$ P(first) = \textit{new entrant} $\times$ P(last) = \textit{incumbent} $\times$ SchoolClosuresFirst&     -0.0247         &                     &      0.0156         &                     \\
                    &     (-0.97)         &                     &      (0.65)         &                     \\
\addlinespace
COVID-related $\times$ P(first) = \textit{incumbent} $\times$ P(last) = \textit{newcomer} $\times$ SchoolClosuresFirst&    -0.00928         &                     &     0.00631         &                     \\
                    &     (-0.31)         &                     &      (0.22)         &                     \\
\addlinespace
COVID-related $\times$ P(first) = \textit{new entrant} $\times$ P(last) = \textit{newcomer} $\times$ SchoolClosuresFirst&    -0.00836         &                     &     -0.0112         &                     \\
                    &     (-0.52)         &                     &     (-0.73)         &                     \\
\addlinespace
COVID-related $\times$ P(first) = \textit{newcomer} $\times$ P(last) = \textit{new entrant} $\times$ SchoolClosuresFirst&     -0.0356         &                     &    -0.00739         &                     \\
                    &     (-1.59)         &                     &     (-0.33)         &                     \\
\addlinespace
COVID-related $\times$ P(first) = \textit{incumbent} $\times$ P(last) = \textit{new entrant} $\times$ SchoolClosuresFirst&      0.0282         &                     &      0.0223         &                     \\
                    &      (0.50)         &                     &      (0.41)         &                     \\
\addlinespace
\textit{N Authors}            &    0.000165         &    0.000176         &    -0.00262\sym{***}&    -0.00263\sym{***}\\
                    &      (0.42)         &      (0.44)         &     (-7.24)         &     (-7.25)         \\
\addlinespace
trial               &     0.00857         &     0.00843         &      0.0170         &      0.0172         \\
                    &      (0.88)         &      (0.86)         &      (1.80)         &      (1.82)         \\
\addlinespace
Pre-existing Grant         &      0.0606\sym{***}&      0.0607\sym{***}&      0.0482\sym{***}&      0.0483\sym{***}\\
                    &     (11.69)         &     (11.71)         &      (9.69)         &      (9.71)         \\
\addlinespace
WorkplaceClosuresFirst&                     &     -0.0195\sym{*}  &                     &     0.00525         \\
                    &                     &     (-2.17)         &                     &      (0.62)         \\
\addlinespace
COVID-related $\times$ WorkplaceClosuresFirst&                     &      0.0114         &                     &     -0.0166         \\
                    &                     &      (0.80)         &                     &     (-1.22)         \\
\addlinespace
P(first) = \textit{new entrant} $\times$ P(last) = \textit{new entrant} $\times$ WorkplaceClosuresFirst&                     &      0.0337\sym{*}  &                     &     -0.0218         \\
                    &                     &      (2.43)         &                     &     (-1.61)         \\
\addlinespace
P(first) = \textit{newcomer} $\times$ P(last) = \textit{newcomer} $\times$ WorkplaceClosuresFirst&                     &      0.0198         &                     &     0.00103         \\
                    &                     &      (1.84)         &                     &      (0.10)         \\
\addlinespace
P(first) = \textit{newcomer} $\times$ P(last) = \textit{incumbent} $\times$ WorkplaceClosuresFirst&                     &      0.0281         &                     &    -0.00934         \\
                    &                     &      (1.95)         &                     &     (-0.70)         \\
\addlinespace
P(first) = \textit{new entrant} $\times$ P(last) = \textit{incumbent} $\times$ WorkplaceClosuresFirst&                     &      0.0112         &                     &     -0.0240         \\
                    &                     &      (0.71)         &                     &     (-1.67)         \\
\addlinespace
P(first) = \textit{incumbent} $\times$ P(last) = \textit{newcomer} $\times$ WorkplaceClosuresFirst&                     &      0.0146         &                     &      0.0114         \\
                    &                     &      (0.66)         &                     &      (0.55)         \\
\addlinespace
P(first) = \textit{new entrant} $\times$ P(last) = \textit{newcomer} $\times$ WorkplaceClosuresFirst&                     &      0.0121         &                     &     0.00991         \\
                    &                     &      (1.05)         &                     &      (0.93)         \\
\addlinespace
P(first) = \textit{newcomer} $\times$ P(last) = \textit{new entrant} $\times$ WorkplaceClosuresFirst&                     &      0.0160         &                     &    0.000522         \\
                    &                     &      (0.93)         &                     &      (0.03)         \\
\addlinespace
P(first) = \textit{incumbent} $\times$ P(last) = \textit{new entrant} $\times$ WorkplaceClosuresFirst&                     &      0.0450         &                     &     0.00954         \\
                    &                     &      (1.34)         &                     &      (0.29)         \\
\addlinespace
COVID-related $\times$ P(first) = \textit{new entrant} $\times$ P(last) = \textit{new entrant} $\times$ WorkplaceClosuresFirst&                     &     -0.0679\sym{**} &                     &      0.0357         \\
                    &                     &     (-3.16)         &                     &      (1.70)         \\
\addlinespace
COVID-related $\times$ P(first) = \textit{newcomer} $\times$ P(last) = \textit{newcomer} $\times$ WorkplaceClosuresFirst&                     &     -0.0247         &                     &     -0.0121         \\
                    &                     &     (-1.52)         &                     &     (-0.78)         \\
\addlinespace
COVID-related $\times$ P(first) = \textit{newcomer} $\times$ P(last) = \textit{incumbent} $\times$ WorkplaceClosuresFirst&                     &     -0.0539\sym{*}  &                     &      0.0164         \\
                    &                     &     (-2.35)         &                     &      (0.76)         \\
\addlinespace
COVID-related $\times$ P(first) = \textit{new entrant} $\times$ P(last) = \textit{incumbent} $\times$ WorkplaceClosuresFirst&                     &     -0.0326         &                     &      0.0117         \\
                    &                     &     (-1.18)         &                     &      (0.45)         \\
\addlinespace
COVID-related $\times$ P(first) = \textit{incumbent} $\times$ P(last) = \textit{newcomer} $\times$ WorkplaceClosuresFirst&                     &     -0.0171         &                     &      0.0151         \\
                    &                     &     (-0.53)         &                     &      (0.49)         \\
\addlinespace
COVID-related $\times$ P(first) = \textit{new entrant} $\times$ P(last) = \textit{newcomer} $\times$ WorkplaceClosuresFirst&                     &    -0.00591         &                     &     -0.0120         \\
                    &                     &     (-0.33)         &                     &     (-0.71)         \\
\addlinespace
COVID-related $\times$ P(first) = \textit{newcomer} $\times$ P(last) = \textit{new entrant} $\times$ WorkplaceClosuresFirst&                     &     -0.0467         &                     &    0.000548         \\
                    &                     &     (-1.91)         &                     &      (0.02)         \\
\addlinespace
COVID-related $\times$ P(first) = \textit{incumbent} $\times$ P(last) = \textit{new entrant} $\times$ WorkplaceClosuresFirst&                     &    -0.00982         &                     &      0.0838         \\
                    &                     &     (-0.16)         &                     &      (1.43)         \\
\addlinespace
Constant            &       0.297\sym{***}&       0.294\sym{***}&       0.244\sym{***}&       0.242\sym{***}\\
                    &      (4.30)         &      (4.22)         &      (3.72)         &      (3.69)         \\
\midrule
Observations        &       83104         &       83104         &       83104         &       83104         \\
\midrule
Country FEs & First & First  & First & First \\
\bottomrule
\multicolumn{5}{l}{\footnotesize \textit{t} statistics in parentheses}\\
\multicolumn{5}{l}{\footnotesize \sym{*} \(p<0.05\), \sym{**} \(p<0.01\), \sym{***} \(p<0.001\)}\\
\end{tabular}
}\end{table}

\begin{table}[H]\centering
\def\sym#1{\ifmmode^{#1}\else\(^{#1}\)\fi}
\caption{Linear regression model estimates on key female authorship, including interactions of (year=2020 $\times$ COVID-related) with first and last author's incumbency in research, and lockdown measure indicators of \emph{school closures} and \emph{workplace closures} of last author's country. We control for country of last authors' fixed effect (omitted), team size, clinical trials and previous grants, and compute white-robust standard errors. The baseline incumbency level is given by the \emph{incumbent status}. The table continues in the next page.\label{tab_str_pers_2}}
\resizebox{\textwidth}{!}{ 
\begin{tabular}{lcccc}
\toprule
                    &\multicolumn{1}{c}{(1)}&\multicolumn{1}{c}{(2)}&\multicolumn{1}{c}{(3)}&\multicolumn{1}{c}{(4)}\\
                    &\multicolumn{1}{c}{Female First Author}&\multicolumn{1}{c}{Female First Author}&\multicolumn{1}{c}{Female Last Author}&\multicolumn{1}{c}{Female Last Author}\\
\midrule

year=2020           &      0.0346         &      0.0331         &      0.0190         &      0.0135         \\
                    &      (1.57)         &      (1.55)         &      (0.94)         &      (0.68)         \\
\addlinespace
COVID-related             &      0.0625\sym{***}&      0.0624\sym{***}&      0.0786\sym{***}&      0.0786\sym{***}\\
                    &      (3.95)         &      (3.94)         &      (5.27)         &      (5.26)         \\
\addlinespace
year=2020 $\times$ COVID-related&     -0.0276         &     -0.0267         &     -0.0155         &     0.00129         \\
                    &     (-0.82)         &     (-0.80)         &     (-0.49)         &      (0.04)         \\
\addlinespace
P(first) = \textit{new entrant} $\times$ P(last) = \textit{new entrant}&      0.0323\sym{*}  &      0.0327\sym{*}  &      0.0652\sym{***}&      0.0656\sym{***}\\
                    &      (2.25)         &      (2.27)         &      (4.83)         &      (4.87)         \\
\addlinespace
P(first) = \textit{newcomer} $\times$ P(last) = \textit{newcomer}&      0.0271\sym{*}  &      0.0271\sym{*}  &      0.0211\sym{*}  &      0.0212\sym{*}  \\
                    &      (2.37)         &      (2.37)         &      (2.01)         &      (2.01)         \\
\addlinespace
P(first) = \textit{newcomer} $\times$ P(last) = \textit{incumbent}&      0.0155         &      0.0154         &     -0.0172         &     -0.0172         \\
                    &      (0.96)         &      (0.95)         &     (-1.18)         &     (-1.18)         \\
\addlinespace
P(first) = \textit{new entrant} $\times$ P(last) = \textit{incumbent}&      0.0910\sym{***}&      0.0910\sym{***}&     -0.0299         &     -0.0299         \\
                    &      (5.24)         &      (5.24)         &     (-1.95)         &     (-1.95)         \\
\addlinespace
P(first) = \textit{incumbent} $\times$ P(last) = \textit{newcomer}&     -0.0813\sym{***}&     -0.0814\sym{***}&    -0.00787         &    -0.00777         \\
                    &     (-3.57)         &     (-3.58)         &     (-0.37)         &     (-0.37)         \\
\addlinespace
P(first) = \textit{new entrant} $\times$ P(last) = \textit{newcomer}&      0.0883\sym{***}&      0.0884\sym{***}&      0.0291\sym{**} &      0.0292\sym{**} \\
                    &      (7.23)         &      (7.24)         &      (2.59)         &      (2.60)         \\
\addlinespace
P(first) = \textit{newcomer} $\times$ P(last) = \textit{new entrant}&     -0.0523\sym{**} &     -0.0520\sym{**} &      0.0815\sym{***}&      0.0819\sym{***}\\
                    &     (-2.95)         &     (-2.94)         &      (4.74)         &      (4.76)         \\
\addlinespace
P(first) = \textit{incumbent} $\times$ P(last) = \textit{new entrant}&      -0.183\sym{***}&      -0.183\sym{***}&       0.113\sym{**} &       0.114\sym{**} \\
                    &     (-5.56)         &     (-5.55)         &      (3.03)         &      (3.04)         \\
\addlinespace
year=2020 $\times$ P(first) = \textit{new entrant} $\times$ P(last) = \textit{new entrant}&     -0.0258         &     -0.0188         &      0.0818\sym{*}  &      0.0866\sym{**} \\
                    &     (-0.73)         &     (-0.56)         &      (2.41)         &      (2.66)         \\
\addlinespace
year=2020 $\times$ P(first) = \textit{newcomer} $\times$ P(last) = \textit{newcomer}&     -0.0318         &     -0.0312         &   -0.000896         &    -0.00964         \\
                    &     (-1.20)         &     (-1.21)         &     (-0.04)         &     (-0.41)         \\
\addlinespace
year=2020 $\times$ P(first) = \textit{newcomer} $\times$ P(last) = \textit{incumbent}&     0.00597         &     -0.0247         &     -0.0152         &     0.00273         \\
                    &      (0.16)         &     (-0.70)         &     (-0.46)         &      (0.09)         \\
\addlinespace
year=2020 $\times$ P(first) = \textit{new entrant} $\times$ P(last) = \textit{incumbent}&    -0.00231         &    -0.00297         &      0.0356         &      0.0338         \\
                    &     (-0.06)         &     (-0.08)         &      (1.00)         &      (1.00)         \\
\addlinespace
year=2020 $\times$ P(first) = \textit{incumbent} $\times$ P(last) = \textit{newcomer}&    -0.00451         &   -0.000607         &      0.0696         &     0.00814         \\
                    &     (-0.08)         &     (-0.01)         &      (1.25)         &      (0.16)         \\
\addlinespace
year=2020 $\times$ P(first) = \textit{new entrant} $\times$ P(last) = \textit{newcomer}&    -0.00448         &     0.00850         &     -0.0230         &     -0.0342         \\
                    &     (-0.16)         &      (0.31)         &     (-0.89)         &     (-1.37)         \\
\addlinespace
year=2020 $\times$ P(first) = \textit{newcomer} $\times$ P(last) = \textit{new entrant}&      0.0504         &      0.0212         &     0.00763         &     0.00884         \\
                    &      (1.14)         &      (0.50)         &      (0.18)         &      (0.21)         \\
\addlinespace
year=2020 $\times$ P(first) = \textit{incumbent} $\times$ P(last) = \textit{new entrant}&      0.0679         &      0.0235         &      -0.121         &      -0.122         \\
                    &      (0.78)         &      (0.28)         &     (-1.46)         &     (-1.46)         \\
\addlinespace
COVID-related $\times$ P(first) = \textit{new entrant} $\times$ P(last) = \textit{new entrant}&      0.0288         &      0.0288         &      0.0519\sym{*}  &      0.0519\sym{*}  \\
                    &      (1.24)         &      (1.23)         &      (2.30)         &      (2.30)         \\
\addlinespace
COVID-related $\times$ P(first) = \textit{newcomer} $\times$ P(last) = \textit{newcomer}&     -0.0187         &     -0.0186         &    -0.00570         &    -0.00564         \\
                    &     (-1.04)         &     (-1.03)         &     (-0.34)         &     (-0.33)         \\
\addlinespace
COVID-related $\times$ P(first) = \textit{newcomer} $\times$ P(last) = \textit{incumbent}&     0.00170         &     0.00188         &     -0.0205         &     -0.0204         \\
                    &      (0.07)         &      (0.07)         &     (-0.85)         &     (-0.84)         \\
\addlinespace
COVID-related $\times$ P(first) = \textit{new entrant} $\times$ P(last) = \textit{incumbent}&      0.0137         &      0.0138         &      0.0489         &      0.0490         \\
                    &      (0.46)         &      (0.46)         &      (1.72)         &      (1.72)         \\
\addlinespace
COVID-related $\times$ P(first) = \textit{incumbent} $\times$ P(last) = \textit{newcomer}&      0.0211         &      0.0214         &      0.0476         &      0.0477         \\
                    &      (0.61)         &      (0.62)         &      (1.43)         &      (1.43)         \\
\addlinespace
COVID-related $\times$ P(first) = \textit{new entrant} $\times$ P(last) = \textit{newcomer}&     0.00198         &     0.00204         &     0.00608         &     0.00611         \\
                    &      (0.10)         &      (0.11)         &      (0.33)         &      (0.33)         \\
\addlinespace
COVID-related $\times$ P(first) = \textit{newcomer} $\times$ P(last) = \textit{new entrant}&      0.0336         &      0.0335         &      0.0218         &      0.0217         \\
                    &      (1.25)         &      (1.24)         &      (0.82)         &      (0.82)         \\
\addlinespace
COVID-related $\times$ P(first) = \textit{incumbent} $\times$ P(last) = \textit{new entrant}&      0.0608         &      0.0607         &      0.0185         &      0.0186         \\
                    &      (1.05)         &      (1.05)         &      (0.30)         &      (0.30)         \\
\addlinespace
year=2020 $\times$ COVID-related $\times$ P(first) = \textit{new entrant} $\times$ P(last) = \textit{new entrant}&     -0.0104         &      0.0153         &      -0.152\sym{**} &      -0.154\sym{**} \\
                    &     (-0.19)         &      (0.29)         &     (-2.93)         &     (-3.07)         \\
\addlinespace
year=2020 $\times$ COVID-related $\times$ P(first) = \textit{newcomer} $\times$ P(last) = \textit{newcomer}&     -0.0206         &     -0.0158         &     -0.0185         &     -0.0121         \\
                    &     (-0.53)         &     (-0.41)         &     (-0.50)         &     (-0.34)         \\
\addlinespace
year=2020 $\times$ COVID-related $\times$ P(first) = \textit{newcomer} $\times$ P(last) = \textit{incumbent}&      0.0781         &      0.0965         &      0.0359         &      0.0139         \\
                    &      (1.39)         &      (1.75)         &      (0.69)         &      (0.27)         \\
\addlinespace
year=2020 $\times$ COVID-related $\times$ P(first) = \textit{new entrant} $\times$ P(last) = \textit{incumbent}&      0.0263         &      0.0148         &      -0.117         &      -0.101         \\
                    &      (0.39)         &      (0.22)         &     (-1.86)         &     (-1.62)         \\
\addlinespace
year=2020 $\times$ COVID-related $\times$ P(first) = \textit{incumbent} $\times$ P(last) = \textit{newcomer}&     -0.0561         &     -0.0202         &      -0.123         &     -0.0721         \\
                    &     (-0.69)         &     (-0.26)         &     (-1.55)         &     (-0.95)         \\
\addlinespace
year=2020 $\times$ COVID-related $\times$ P(first) = \textit{new entrant} $\times$ P(last) = \textit{newcomer}&     -0.0283         &     -0.0478         &     0.00303         &    0.000697         \\
                    &     (-0.66)         &     (-1.14)         &      (0.08)         &      (0.02)         \\
\addlinespace
year=2020 $\times$ COVID-related $\times$ P(first) = \textit{newcomer} $\times$ P(last) = \textit{new entrant}&     -0.0362         &     -0.0158         &     -0.0283         &     -0.0189         \\
                    &     (-0.58)         &     (-0.26)         &     (-0.46)         &     (-0.32)         \\
\addlinespace
year=2020 $\times$ COVID-related $\times$ P(first) = \textit{incumbent} $\times$ P(last) = \textit{new entrant}&      -0.195         &     -0.0411         &     -0.0856         &      -0.208         \\
                    &     (-1.25)         &     (-0.27)         &     (-0.56)         &     (-1.51)         \\
\midrule
Observations        &       82405         &       82405         &       82405         &       82405         \\
\midrule
Country FEs & Last  & Last  & Last  & Last \\
\bottomrule
\multicolumn{5}{l}{\footnotesize \textit{t} statistics in parentheses}\\
\multicolumn{5}{l}{\footnotesize \sym{*} \(p<0.05\), \sym{**} \(p<0.01\), \sym{***} \(p<0.001\)}\\
\end{tabular}
}
\end{table}

\begin{table}[H]\centering
\def\sym#1{\ifmmode^{#1}\else\(^{#1}\)\fi}
\caption*{Table 28 (continued).\label{tab_str_pers_2_p2}}
\resizebox{\textwidth}{!}{ 
\begin{tabular}{lcccc}
\toprule
                    &\multicolumn{1}{c}{(1)}&\multicolumn{1}{c}{(2)}&\multicolumn{1}{c}{(3)}&\multicolumn{1}{c}{(4)}\\
                    &\multicolumn{1}{c}{Female First Author}&\multicolumn{1}{c}{Female First Author}&\multicolumn{1}{c}{Female Last Author}&\multicolumn{1}{c}{Female Last Author}\\
\midrule
SchoolClosuresLast&     -0.0163         &                     &     0.00207         &                     \\
                    &     (-1.94)         &                     &      (0.27)         &                     \\
\addlinespace
COVID-related $\times$ SchoolClosuresLast&     0.00590         &                     &     -0.0147         &                     \\
                    &      (0.45)         &                     &     (-1.19)         &                     \\
\addlinespace
P(first) = \textit{new entrant} $\times$ P(last) = \textit{new entrant} $\times$ SchoolClosuresLast&      0.0276\sym{*}  &                     &     -0.0192         &                     \\
                    &      (2.09)         &                     &     (-1.50)         &                     \\
\addlinespace
P(first) = \textit{newcomer} $\times$ P(last) = \textit{newcomer} $\times$ SchoolClosuresLast&      0.0172         &                     &    -0.00482         &                     \\
                    &      (1.70)         &                     &     (-0.52)         &                     \\
\addlinespace
P(first) = \textit{newcomer} $\times$ P(last) = \textit{incumbent} $\times$ SchoolClosuresLast&     0.00751         &                     &    0.000827         &                     \\
                    &      (0.55)         &                     &      (0.07)         &                     \\
\addlinespace
P(first) = \textit{new entrant} $\times$ P(last) = \textit{incumbent} $\times$ SchoolClosuresLast&     0.00985         &                     &     -0.0171         &                     \\
                    &      (0.66)         &                     &     (-1.28)         &                     \\
\addlinespace
P(first) = \textit{incumbent} $\times$ P(last) = \textit{newcomer} $\times$ SchoolClosuresLast&      0.0147         &                     &     -0.0251         &                     \\
                    &      (0.68)         &                     &     (-1.22)         &                     \\
\addlinespace
P(first) = \textit{new entrant} $\times$ P(last) = \textit{newcomer} $\times$ SchoolClosuresLast&      0.0138         &                     &     0.00155         &                     \\
                    &      (1.28)         &                     &      (0.16)         &                     \\
\addlinespace
P(first) = \textit{newcomer} $\times$ P(last) = \textit{new entrant} $\times$ SchoolClosuresLast&    -0.00676         &                     &    -0.00198         &                     \\
                    &     (-0.41)         &                     &     (-0.12)         &                     \\
\addlinespace
P(first) = \textit{incumbent} $\times$ P(last) = \textit{new entrant} $\times$ SchoolClosuresLast&      0.0205         &                     &      0.0190         &                     \\
                    &      (0.63)         &                     &      (0.62)         &                     \\
\addlinespace
COVID-related $\times$ P(first) = \textit{new entrant} $\times$ P(last) = \textit{new entrant} $\times$ SchoolClosuresLast&     -0.0300         &                     &      0.0399\sym{*}  &                     \\
                    &     (-1.50)         &                     &      (2.05)         &                     \\
\addlinespace
COVID-related $\times$ P(first) = \textit{newcomer} $\times$ P(last) = \textit{newcomer} $\times$ SchoolClosuresLast&     -0.0182         &                     &    0.000899         &                     \\
                    &     (-1.22)         &                     &      (0.06)         &                     \\
\addlinespace
COVID-related $\times$ P(first) = \textit{newcomer} $\times$ P(last) = \textit{incumbent} $\times$ SchoolClosuresLast&     -0.0326         &                     &     0.00971         &                     \\
                    &     (-1.55)         &                     &      (0.49)         &                     \\
\addlinespace
COVID-related $\times$ P(first) = \textit{new entrant} $\times$ P(last) = \textit{incumbent} $\times$ SchoolClosuresLast&     -0.0182         &                     &      0.0312         &                     \\
                    &     (-0.71)         &                     &      (1.32)         &                     \\
\addlinespace
COVID-related $\times$ P(first) = \textit{incumbent} $\times$ P(last) = \textit{newcomer} $\times$ SchoolClosuresLast&      0.0158         &                     &      0.0475         &                     \\
                    &      (0.52)         &                     &      (1.61)         &                     \\
\addlinespace
COVID-related $\times$ P(first) = \textit{new entrant} $\times$ P(last) = \textit{newcomer} $\times$ SchoolClosuresLast&    -0.00517         &                     &    -0.00385         &                     \\
                    &     (-0.32)         &                     &     (-0.25)         &                     \\
\addlinespace
COVID-related $\times$ P(first) = \textit{newcomer} $\times$ P(last) = \textit{new entrant} $\times$ SchoolClosuresLast&     -0.0165         &                     &      0.0163         &                     \\
                    &     (-0.71)         &                     &      (0.72)         &                     \\
\addlinespace
COVID-related $\times$ P(first) = \textit{incumbent} $\times$ P(last) = \textit{new entrant} $\times$ SchoolClosuresLast&      0.0839         &                     &      0.0464         &                     \\
                    &      (1.44)         &                     &      (0.82)         &                     \\
\addlinespace
\textit{N Authors}            &    0.000410         &    0.000424         &    -0.00271\sym{***}&    -0.00271\sym{***}\\
                    &      (1.01)         &      (1.04)         &     (-7.28)         &     (-7.29)         \\
\addlinespace
trial               &      0.0143         &      0.0142         &      0.0188\sym{*}  &      0.0187\sym{*}  \\
                    &      (1.45)         &      (1.44)         &      (1.98)         &      (1.97)         \\
\addlinespace
Pre-existing Grant         &      0.0592\sym{***}&      0.0594\sym{***}&      0.0497\sym{***}&      0.0498\sym{***}\\
                    &     (11.38)         &     (11.41)         &      (9.98)         &     (10.01)         \\
\addlinespace
WorkplaceClosuresLast&                     &     -0.0172         &                     &     0.00580         \\
                    &                     &     (-1.90)         &                     &      (0.68)         \\
\addlinespace
COVID-related $\times$ WorkplaceClosuresLast&                     &     0.00650         &                     &     -0.0260         \\
                    &                     &      (0.45)         &                     &     (-1.92)         \\
\addlinespace
P(first) = \textit{new entrant} $\times$ P(last) = \textit{new entrant} $\times$ WorkplaceClosuresLast&                     &      0.0276\sym{*}  &                     &     -0.0248         \\
                    &                     &      (1.96)         &                     &     (-1.81)         \\
\addlinespace
P(first) = \textit{newcomer} $\times$ P(last) = \textit{newcomer} $\times$ WorkplaceClosuresLast&                     &      0.0189         &                     &    -0.00133         \\
                    &                     &      (1.75)         &                     &     (-0.13)         \\
\addlinespace
P(first) = \textit{newcomer} $\times$ P(last) = \textit{incumbent} $\times$ WorkplaceClosuresLast&                     &      0.0242         &                     &    -0.00852         \\
                    &                     &      (1.67)         &                     &     (-0.64)         \\
\addlinespace
P(first) = \textit{new entrant} $\times$ P(last) = \textit{incumbent} $\times$ WorkplaceClosuresLast&                     &      0.0114         &                     &     -0.0186         \\
                    &                     &      (0.72)         &                     &     (-1.31)         \\
\addlinespace
P(first) = \textit{incumbent} $\times$ P(last) = \textit{newcomer} $\times$ WorkplaceClosuresLast&                     &      0.0146         &                     &    0.000983         \\
                    &                     &      (0.65)         &                     &      (0.05)         \\
\addlinespace
P(first) = \textit{new entrant} $\times$ P(last) = \textit{newcomer} $\times$ WorkplaceClosuresLast&                     &     0.00870         &                     &     0.00711         \\
                    &                     &      (0.76)         &                     &      (0.66)         \\
\addlinespace
P(first) = \textit{newcomer} $\times$ P(last) = \textit{new entrant} $\times$ WorkplaceClosuresLast&                     &     0.00697         &                     &    -0.00332         \\
                    &                     &      (0.40)         &                     &     (-0.19)         \\
\addlinespace
P(first) = \textit{incumbent} $\times$ P(last) = \textit{new entrant} $\times$ WorkplaceClosuresLast&                     &      0.0449         &                     &      0.0210         \\
                    &                     &      (1.27)         &                     &      (0.61)         \\
\addlinespace
COVID-related $\times$ P(first) = \textit{new entrant} $\times$ P(last) = \textit{new entrant} $\times$ WorkplaceClosuresLast&                     &     -0.0476\sym{*}  &                     &      0.0474\sym{*}  \\
                    &                     &     (-2.20)         &                     &      (2.26)         \\
\addlinespace
COVID-related $\times$ P(first) = \textit{newcomer} $\times$ P(last) = \textit{newcomer} $\times$ WorkplaceClosuresLast&                     &     -0.0233         &                     &    -0.00146         \\
                    &                     &     (-1.43)         &                     &     (-0.10)         \\
\addlinespace
COVID-related $\times$ P(first) = \textit{newcomer} $\times$ P(last) = \textit{incumbent} $\times$ WorkplaceClosuresLast&                     &     -0.0458\sym{*}  &                     &      0.0230         \\
                    &                     &     (-2.00)         &                     &      (1.07)         \\
\addlinespace
COVID-related $\times$ P(first) = \textit{new entrant} $\times$ P(last) = \textit{incumbent} $\times$ WorkplaceClosuresLast&                     &     -0.0144         &                     &      0.0277         \\
                    &                     &     (-0.52)         &                     &      (1.08)         \\
\addlinespace
COVID-related $\times$ P(first) = \textit{incumbent} $\times$ P(last) = \textit{newcomer} $\times$ WorkplaceClosuresLast&                     &   -0.000870         &                     &      0.0307         \\
                    &                     &     (-0.03)         &                     &      (0.98)         \\
\addlinespace
COVID-related $\times$ P(first) = \textit{new entrant} $\times$ P(last) = \textit{newcomer} $\times$ WorkplaceClosuresLast&                     &     0.00391         &                     &    -0.00239         \\
                    &                     &      (0.22)         &                     &     (-0.14)         \\
\addlinespace
COVID-related $\times$ P(first) = \textit{newcomer} $\times$ P(last) = \textit{new entrant} $\times$ WorkplaceClosuresLast&                     &     -0.0307         &                     &      0.0150         \\
                    &                     &     (-1.23)         &                     &      (0.61)         \\
\addlinespace
COVID-related $\times$ P(first) = \textit{incumbent} $\times$ P(last) = \textit{new entrant} $\times$ WorkplaceClosuresLast&                     &      0.0209         &                     &       0.114\sym{*}  \\
                    &                     &      (0.33)         &                     &      (1.98)         \\
\addlinespace
Constant            &       0.288\sym{***}&       0.287\sym{***}&       0.261\sym{***}&       0.258\sym{***}\\
                    &      (4.00)         &      (3.97)         &      (3.72)         &      (3.68)         \\
\midrule
Observations        &       82405         &       82405         &       82405         &       82405         \\
\midrule
Country FEs & Last  & Last  & Last  & Last \\
\bottomrule
\multicolumn{5}{l}{\footnotesize \textit{t} statistics in parentheses}\\
\multicolumn{5}{l}{\footnotesize \sym{*} \(p<0.05\), \sym{**} \(p<0.01\), \sym{***} \(p<0.001\)}\\
\end{tabular}
}
\end{table}

\begin{table}[H]\centering
\def\sym#1{\ifmmode^{#1}\else\(^{#1}\)\fi}
\caption{Linear regression model estimates on key female authorship, including interactions of (year=2020 $\times$ COVID-related) with first and last author's incumbency in research, and lockdown measure indicators of \emph{school closures} of first and last author's country. We control for country of last authors' fixed effect, team size, clinical trials and previous grants (omitted), and compute White-robust standard errors. The baseline incumbency level is given by the \emph{incumbent status}. The table continues in the next page.\label{tab_20}}
\resizebox{0.7\textwidth}{!}{ 
\begin{tabular}{lcc}
\toprule
                    &\multicolumn{1}{c}{(1)}&\multicolumn{1}{c}{(2)}\\
                    &\multicolumn{1}{c}{Female First Author}&\multicolumn{1}{c}{Female Last Author}\\
\midrule
year=2020           &      0.0477\sym{*}  &      0.0278         \\
                    &      (2.01)         &      (1.27)         \\
\addlinespace
COVID-related             &      0.0626\sym{***}&      0.0788\sym{***}\\
                    &      (3.96)         &      (5.28)         \\
\addlinespace
year=2020 $\times$ COVID-related&     -0.0403         &    -0.00772         \\
                    &     (-1.11)         &     (-0.23)         \\
\addlinespace
P(first) = \textit{new entrant} $\times$ P(last) = \textit{new entrant}&      0.0327\sym{*}  &      0.0656\sym{***}\\
                    &      (2.27)         &      (4.86)         \\
\addlinespace
P(first) = \textit{newcomer} $\times$ P(last) = \textit{newcomer}&      0.0273\sym{*}  &      0.0212\sym{*}  \\
                    &      (2.38)         &      (2.02)         \\
\addlinespace
P(first) = \textit{newcomer} $\times$ P(last) = \textit{incumbent}&      0.0156         &     -0.0171         \\
                    &      (0.97)         &     (-1.17)         \\
\addlinespace
P(first) = \textit{new entrant} $\times$ P(last) = \textit{incumbent}&      0.0911\sym{***}&     -0.0298         \\
                    &      (5.24)         &     (-1.94)         \\
\addlinespace
P(first) = \textit{incumbent} $\times$ P(last) = \textit{newcomer}&     -0.0813\sym{***}&    -0.00777         \\
                    &     (-3.57)         &     (-0.37)         \\
\addlinespace
P(first) = \textit{new entrant} $\times$ P(last) = \textit{newcomer}&      0.0884\sym{***}&      0.0292\sym{**} \\
                    &      (7.24)         &      (2.60)         \\
\addlinespace
P(first) = \textit{newcomer} $\times$ P(last) = \textit{new entrant}&     -0.0520\sym{**} &      0.0819\sym{***}\\
                    &     (-2.94)         &      (4.76)         \\
\addlinespace
P(first) = \textit{incumbent} $\times$ P(last) = \textit{new entrant}&      -0.183\sym{***}&       0.114\sym{**} \\
                    &     (-5.55)         &      (3.04)         \\
\addlinespace
year=2020 $\times$ P(first) = \textit{new entrant} $\times$ P(last) = \textit{new entrant}&     -0.0310         &      0.0573         \\
                    &     (-0.80)         &      (1.53)         \\
\addlinespace
year=2020 $\times$ P(first) = \textit{newcomer} $\times$ P(last) = \textit{newcomer}&     -0.0481         &     -0.0168         \\
                    &     (-1.65)         &     (-0.62)         \\
\addlinespace
year=2020 $\times$ P(first) = \textit{newcomer} $\times$ P(last) = \textit{incumbent}&     -0.0157         &    -0.00241         \\
                    &     (-0.39)         &     (-0.07)         \\
\addlinespace
year=2020 $\times$ P(first) = \textit{new entrant} $\times$ P(last) = \textit{incumbent}&     -0.0426         &      0.0151         \\
                    &     (-0.95)         &      (0.37)         \\
\addlinespace
year=2020 $\times$ P(first) = \textit{incumbent} $\times$ P(last) = \textit{newcomer}&     -0.0130         &      0.0228         \\
                    &     (-0.19)         &      (0.36)         \\
\addlinespace
year=2020 $\times$ P(first) = \textit{new entrant} $\times$ P(last) = \textit{newcomer}&     -0.0144         &     -0.0280         \\
                    &     (-0.46)         &     (-0.97)         \\
\addlinespace
year=2020 $\times$ P(first) = \textit{newcomer} $\times$ P(last) = \textit{new entrant}&      0.0467         &     -0.0205         \\
                    &      (0.93)         &     (-0.42)         \\
\addlinespace
year=2020 $\times$ P(first) = \textit{incumbent} $\times$ P(last) = \textit{new entrant}&      0.0309         &      -0.192\sym{*}  \\
                    &      (0.31)         &     (-2.19)         \\
\addlinespace
COVID-related $\times$ P(first) = \textit{new entrant} $\times$ P(last) = \textit{new entrant}&      0.0286         &      0.0517\sym{*}  \\
                    &      (1.22)         &      (2.29)         \\
\addlinespace
COVID-related $\times$ P(first) = \textit{newcomer} $\times$ P(last) = \textit{newcomer}&     -0.0187         &    -0.00570         \\
                    &     (-1.04)         &     (-0.34)         \\
\addlinespace
COVID-related $\times$ P(first) = \textit{newcomer} $\times$ P(last) = \textit{incumbent}&     0.00172         &     -0.0205         \\
                    &      (0.07)         &     (-0.84)         \\
\addlinespace
COVID-related $\times$ P(first) = \textit{new entrant} $\times$ P(last) = \textit{incumbent}&      0.0137         &      0.0489         \\
                    &      (0.46)         &      (1.72)         \\
\addlinespace
COVID-related $\times$ P(first) = \textit{incumbent} $\times$ P(last) = \textit{newcomer}&      0.0214         &      0.0479         \\
                    &      (0.62)         &      (1.44)         \\
\addlinespace
COVID-related $\times$ P(first) = \textit{new entrant} $\times$ P(last) = \textit{newcomer}&     0.00190         &     0.00600         \\
                    &      (0.10)         &      (0.33)         \\
\addlinespace
COVID-related $\times$ P(first) = \textit{newcomer} $\times$ P(last) = \textit{new entrant}&      0.0335         &      0.0216         \\
                    &      (1.24)         &      (0.82)         \\
\addlinespace
COVID-related $\times$ P(first) = \textit{incumbent} $\times$ P(last) = \textit{new entrant}&      0.0606         &      0.0183         \\
                    &      (1.05)         &      (0.29)         \\
\addlinespace
year=2020 $\times$ COVID-related $\times$ P(first) = \textit{new entrant} $\times$ P(last) = \textit{new entrant}&     0.00391         &      -0.141\sym{*}  \\
                    &      (0.07)         &     (-2.44)         \\
\addlinespace
year=2020 $\times$ COVID-related $\times$ P(first) = \textit{newcomer} $\times$ P(last) = \textit{newcomer}&      0.0323         &      0.0108         \\
                    &      (0.75)         &      (0.27)         \\
\addlinespace
year=2020 $\times$ COVID-related $\times$ P(first) = \textit{newcomer} $\times$ P(last) = \textit{incumbent}&       0.112         &      0.0388         \\
                    &      (1.82)         &      (0.67)         \\
\addlinespace
year=2020 $\times$ COVID-related $\times$ P(first) = \textit{new entrant} $\times$ P(last) = \textit{incumbent}&      0.0553         &     -0.0710         \\
                    &      (0.73)         &     (-1.00)         \\
\addlinespace
year=2020 $\times$ COVID-related $\times$ P(first) = \textit{incumbent} $\times$ P(last) = \textit{newcomer}&     -0.0170         &     -0.0872         \\
                    &     (-0.18)         &     (-0.98)         \\
\addlinespace
year=2020 $\times$ COVID-related $\times$ P(first) = \textit{new entrant} $\times$ P(last) = \textit{newcomer}&     -0.0105         &     0.00604         \\
                    &     (-0.22)         &      (0.14)         \\
\addlinespace
year=2020 $\times$ COVID-related $\times$ P(first) = \textit{newcomer} $\times$ P(last) = \textit{new entrant}&    -0.00825         &      0.0120         \\
                    &     (-0.12)         &      (0.17)         \\
\addlinespace
year=2020 $\times$ COVID-related $\times$ P(first) = \textit{incumbent} $\times$ P(last) = \textit{new entrant}&     -0.0741         &     -0.0552         \\
                    &     (-0.40)         &     (-0.32)         \\
\addlinespace
SchoolClosuresFirst&    -0.00606         &     -0.0398         \\
                    &     (-0.19)         &     (-1.47)         \\
\addlinespace
COVID-related $\times$ SchoolClosuresFirst&      0.0447         &    -0.00870         \\
                    &      (0.85)         &     (-0.20)         \\
\addlinespace
P(first) = \textit{new entrant} $\times$ P(last) = \textit{new entrant} $\times$ SchoolClosuresFirst&     0.00579         &      0.0796         \\
                    &      (0.12)         &      (1.72)         \\
\addlinespace
P(first) = \textit{newcomer} $\times$ P(last) = \textit{newcomer} $\times$ SchoolClosuresFirst&   -0.000638         &      0.0491         \\
                    &     (-0.02)         &      (1.48)         \\
\midrule
Observations        &       82393         &       82393         \\
\midrule
Country FEs & Last   & Last \\
\bottomrule
\multicolumn{3}{l}{\footnotesize \textit{t} statistics in parentheses; \sym{*} \(p<0.05\), \sym{**} \(p<0.01\), \sym{***} \(p<0.001\)}\\
\end{tabular}
}
\end{table}

\begin{table}[H]\centering
\def\sym#1{\ifmmode^{#1}\else\(^{#1}\)\fi}
\caption*{Table 29 (continued).\label{tab_20_p2}}
\resizebox{0.7\textwidth}{!}{ 
\begin{tabular}{lcc}
\toprule
                    &\multicolumn{1}{c}{(1)}&\multicolumn{1}{c}{(2)}\\
                    &\multicolumn{1}{c}{Female First Author}&\multicolumn{1}{c}{Female Last Author}\\
\midrule
P(first) = \textit{newcomer} $\times$ P(last) = \textit{incumbent} $\times$ SchoolClosuresFirst&      0.0805         &     0.00608         \\
                    &      (1.52)         &      (0.14)         \\
\addlinespace
P(first) = \textit{new entrant} $\times$ P(last) = \textit{incumbent} $\times$ SchoolClosuresFirst&      0.0558         &      0.0478         \\
                    &      (1.03)         &      (1.12)         \\
\addlinespace
P(first) = \textit{incumbent} $\times$ P(last) = \textit{newcomer} $\times$ SchoolClosuresFirst&      0.0409         &       0.136\sym{*}  \\
                    &      (0.62)         &      (2.21)         \\
\addlinespace
P(first) = \textit{new entrant} $\times$ P(last) = \textit{newcomer} $\times$ SchoolClosuresFirst&      0.0157         &      0.0525         \\
                    &      (0.39)         &      (1.52)         \\
P(first) = \textit{newcomer} $\times$ P(last) = \textit{new entrant} $\times$ SchoolClosuresFirst&      0.0115         &      0.0725         \\
                    &      (0.18)         &      (1.18)         \\
\addlinespace
P(first) = \textit{incumbent} $\times$ P(last) = \textit{new entrant} $\times$ SchoolClosuresFirst&      0.0450         &      0.0333         \\
                    &      (0.48)         &      (0.40)         \\
\addlinespace
COVID-related $\times$ P(first) = \textit{new entrant} $\times$ P(last) = \textit{new entrant} $\times$ SchoolClosuresFirst&     -0.0565         &     -0.0262         \\
                    &     (-0.77)         &     (-0.39)         \\
\addlinespace
COVID-related $\times$ P(first) = \textit{newcomer} $\times$ P(last) = \textit{newcomer} $\times$ SchoolClosuresFirst&     -0.0795         &     -0.0385         \\
                    &     (-1.37)         &     (-0.78)         \\
\addlinespace
COVID-related $\times$ P(first) = \textit{newcomer} $\times$ P(last) = \textit{incumbent} $\times$ SchoolClosuresFirst&      -0.179\sym{*}  &      0.0139         \\
                    &     (-2.21)         &      (0.20)         \\
\addlinespace
COVID-related $\times$ P(first) = \textit{new entrant} $\times$ P(last) = \textit{incumbent} $\times$ SchoolClosuresFirst&     -0.0726         &      -0.115         \\
                    &     (-0.79)         &     (-1.57)         \\
\addlinespace
COVID-related $\times$ P(first) = \textit{incumbent} $\times$ P(last) = \textit{newcomer} $\times$ SchoolClosuresFirst&      -0.140         &      -0.140         \\
                    &     (-1.50)         &     (-1.61)         \\
\addlinespace
COVID-related $\times$ P(first) = \textit{new entrant} $\times$ P(last) = \textit{newcomer} $\times$ SchoolClosuresFirst&     -0.0697         &     -0.0357         \\
                    &     (-1.11)         &     (-0.67)         \\
\addlinespace
COVID-related $\times$ P(first) = \textit{newcomer} $\times$ P(last) = \textit{new entrant} $\times$ SchoolClosuresFirst&      -0.130         &     -0.0946         \\
                    &     (-1.53)         &     (-1.20)         \\
\addlinespace
COVID-related $\times$ P(first) = \textit{incumbent} $\times$ P(last) = \textit{new entrant} $\times$ SchoolClosuresFirst&      -0.262         &     -0.0408         \\
                    &     (-1.80)         &     (-0.29)         \\
\addlinespace
SchoolClosuresLast&     -0.0664\sym{*}  &      0.0103         \\
                    &     (-2.11)         &      (0.35)         \\
\addlinespace
COVID-related $\times$ SchoolClosuresLast&    -0.00155         &     -0.0521         \\
                    &     (-0.03)         &     (-1.14)         \\
\addlinespace
P(first) = \textit{new entrant} $\times$ P(last) = \textit{new entrant} $\times$ SchoolClosuresLast&      0.0464         &     -0.0271         \\
                    &      (0.99)         &     (-0.59)         \\
\addlinespace
P(first) = \textit{newcomer} $\times$ P(last) = \textit{newcomer} $\times$ SchoolClosuresLast&      0.0840\sym{*}  &    -0.00475         \\
                    &      (2.19)         &     (-0.13)         \\
\addlinespace
P(first) = \textit{newcomer} $\times$ P(last) = \textit{incumbent} $\times$ SchoolClosuresLast&    -0.00800         &     -0.0374         \\
                    &     (-0.15)         &     (-0.82)         \\
\addlinespace
P(first) = \textit{new entrant} $\times$ P(last) = \textit{incumbent} $\times$ SchoolClosuresLast&      0.0746         &    -0.00315         \\
                    &      (1.37)         &     (-0.07)         \\
\addlinespace
P(first) = \textit{incumbent} $\times$ P(last) = \textit{newcomer} $\times$ SchoolClosuresLast&    -0.00406         &     -0.0743         \\
                    &     (-0.06)         &     (-1.37)         \\
\addlinespace
P(first) = \textit{new entrant} $\times$ P(last) = \textit{newcomer} $\times$ SchoolClosuresLast&      0.0391         &     -0.0375         \\
                    &      (0.95)         &     (-1.00)         \\
\addlinespace
P(first) = \textit{newcomer} $\times$ P(last) = \textit{new entrant} $\times$ SchoolClosuresLast&   -0.000923         &     0.00999         \\
                    &     (-0.01)         &      (0.16)         \\
\addlinespace
P(first) = \textit{incumbent} $\times$ P(last) = \textit{new entrant} $\times$ SchoolClosuresLast&      0.0873         &       0.247\sym{*}  \\
                    &      (1.02)         &      (2.43)         \\
\addlinespace
COVID-related $\times$ P(first) = \textit{new entrant} $\times$ P(last) = \textit{new entrant} $\times$ SchoolClosuresLast&    -0.00297         &      0.0561         \\
                    &     (-0.04)         &      (0.83)         \\
\addlinespace
COVID-related $\times$ P(first) = \textit{newcomer} $\times$ P(last) = \textit{newcomer} $\times$ SchoolClosuresLast&     -0.0629         &     0.00660         \\
                    &     (-1.09)         &      (0.13)         \\
\addlinespace
COVID-related $\times$ P(first) = \textit{newcomer} $\times$ P(last) = \textit{incumbent} $\times$ SchoolClosuresLast&      0.0776         &    -0.00332         \\
                    &      (0.97)         &     (-0.05)         \\
\addlinespace
COVID-related $\times$ P(first) = \textit{new entrant} $\times$ P(last) = \textit{incumbent} $\times$ SchoolClosuresLast&     -0.0222         &      0.0694         \\
                    &     (-0.25)         &      (0.87)         \\
\addlinespace
COVID-related $\times$ P(first) = \textit{incumbent} $\times$ P(last) = \textit{newcomer} $\times$ SchoolClosuresLast&      0.0728         &       0.184\sym{*}  \\
                    &      (0.78)         &      (2.20)         \\
\addlinespace
COVID-related $\times$ P(first) = \textit{new entrant} $\times$ P(last) = \textit{newcomer} $\times$ SchoolClosuresLast&      0.0236         &      0.0667         \\
                    &      (0.38)         &      (1.20)         \\
\addlinespace
COVID-related $\times$ P(first) = \textit{newcomer} $\times$ P(last) = \textit{new entrant} $\times$ SchoolClosuresLast&      0.0830         &      0.0581         \\
                    &      (0.98)         &      (0.73)         \\
\addlinespace
COVID-related $\times$ P(first) = \textit{incumbent} $\times$ P(last) = \textit{new entrant} $\times$ SchoolClosuresLast&       0.133         &    -0.00991         \\
                    &      (0.70)         &     (-0.05)         \\
\addlinespace
SchoolClosuresFirst $\times$ SchoolClosuresLast&      0.0184         &     0.00998         \\
                    &      (1.74)         &      (1.02)         \\
\addlinespace
COVID-related $\times$ SchoolClosuresFirst $\times$ SchoolClosuresLast&     -0.0115         &      0.0155         \\
                    &     (-0.66)         &      (0.97)         \\
\addlinespace
P(first) = \textit{new entrant} $\times$ P(last) = \textit{new entrant} $\times$ SchoolClosuresFirst $\times$ SchoolClosuresLast&    -0.00818         &     -0.0220         \\
                    &     (-0.51)         &     (-1.42)         \\
\addlinespace
P(first) = \textit{newcomer} $\times$ P(last) = \textit{newcomer} $\times$ SchoolClosuresFirst $\times$ SchoolClosuresLast&     -0.0215         &     -0.0153         \\
                    &     (-1.72)         &     (-1.32)         \\
\addlinespace
P(first) = \textit{newcomer} $\times$ P(last) = \textit{incumbent} $\times$ SchoolClosuresFirst $\times$ SchoolClosuresLast&     -0.0202         &     0.00982         \\
                    &     (-1.19)         &      (0.65)         \\
\addlinespace
P(first) = \textit{new entrant} $\times$ P(last) = \textit{incumbent} $\times$ SchoolClosuresFirst $\times$ SchoolClosuresLast&     -0.0377\sym{*}  &     -0.0192         \\
                    &     (-2.13)         &     (-1.21)         \\
\addlinespace
P(first) = \textit{incumbent} $\times$ P(last) = \textit{newcomer} $\times$ SchoolClosuresFirst $\times$ SchoolClosuresLast&    -0.00668         &     -0.0243         \\
                    &     (-0.28)         &     (-1.09)         \\
\addlinespace
P(first) = \textit{new entrant} $\times$ P(last) = \textit{newcomer} $\times$ SchoolClosuresFirst $\times$ SchoolClosuresLast&     -0.0134         &    -0.00403         \\
                    &     (-1.01)         &     (-0.33)         \\
\addlinespace
P(first) = \textit{newcomer} $\times$ P(last) = \textit{new entrant} $\times$ SchoolClosuresFirst $\times$ SchoolClosuresLast&    -0.00586         &     -0.0261         \\
                    &     (-0.28)         &     (-1.30)         \\
\addlinespace
P(first) = \textit{incumbent} $\times$ P(last) = \textit{new entrant} $\times$ SchoolClosuresFirst $\times$ SchoolClosuresLast&     -0.0352         &     -0.0859\sym{*}  \\
                    &     (-0.99)         &     (-2.43)         \\
\addlinespace
COVID-related $\times$ P(first) = \textit{new entrant} $\times$ P(last) = \textit{new entrant} $\times$ SchoolClosuresFirst $\times$ SchoolClosuresLast&     0.00857         &     0.00186         \\
                    &      (0.35)         &      (0.08)         \\
\addlinespace
COVID-related $\times$ P(first) = \textit{newcomer} $\times$ P(last) = \textit{newcomer} $\times$ SchoolClosuresFirst $\times$ SchoolClosuresLast&      0.0374         &     0.00783         \\
                    &      (1.95)         &      (0.44)         \\
\addlinespace
COVID-related $\times$ P(first) = \textit{newcomer} $\times$ P(last) = \textit{incumbent} $\times$ SchoolClosuresFirst $\times$ SchoolClosuresLast&      0.0196         &   -0.000487         \\
                    &      (0.74)         &     (-0.02)         \\
\addlinespace
COVID-related $\times$ P(first) = \textit{new entrant} $\times$ P(last) = \textit{incumbent} $\times$ SchoolClosuresFirst $\times$ SchoolClosuresLast&      0.0233         &      0.0211         \\
                    &      (0.76)         &      (0.77)         \\
\addlinespace
COVID-related $\times$ P(first) = \textit{incumbent} $\times$ P(last) = \textit{newcomer} $\times$ SchoolClosuresFirst $\times$ SchoolClosuresLast&      0.0243         &    -0.00439         \\
                    &      (0.71)         &     (-0.13)         \\
\addlinespace
COVID-related $\times$ P(first) = \textit{new entrant} $\times$ P(last) = \textit{newcomer} $\times$ SchoolClosuresFirst $\times$ SchoolClosuresLast&      0.0122         &     -0.0128         \\
                    &      (0.59)         &     (-0.68)         \\
\addlinespace
COVID-related $\times$ P(first) = \textit{newcomer} $\times$ P(last) = \textit{new entrant} $\times$ SchoolClosuresFirst $\times$ SchoolClosuresLast&     0.00686         &      0.0131         \\
                    &      (0.25)         &      (0.49)         \\
\addlinespace
COVID-related $\times$ P(first) = \textit{incumbent} $\times$ P(last) = \textit{new entrant} $\times$ SchoolClosuresFirst $\times$ SchoolClosuresLast&      0.0600         &      0.0329         \\
                    &      (0.92)         &      (0.50)         \\
\midrule
Observations        &       82393         &       82393         \\
\midrule
Country FEs & Last   & Last \\
\bottomrule
\multicolumn{3}{l}{\footnotesize \textit{t} statistics in parentheses}\\
\multicolumn{3}{l}{\footnotesize \sym{*} \(p<0.05\), \sym{**} \(p<0.01\), \sym{***} \(p<0.001\)}\\
\end{tabular}
}
\end{table}

\begin{table}[H]\centering
\def\sym#1{\ifmmode^{#1}\else\(^{#1}\)\fi}
\caption{Linear regression model estimates on key female authorship, including interactions of (year=2020 $\times$ COVID-related) with first and last author's incumbency in research, and lockdown measure indicators of \emph{workplace closures} of first and last author's country. We control for country of last authors' fixed effect, team size, clinical trials and previous grants (omitted), and compute White-robust standard errors. The baseline incumbency level is given by the \emph{incumbent status}. The table continues in the next page.\label{tab_21}}
\resizebox{0.7\textwidth}{!}{ 
\begin{tabular}{lcc}
\toprule
                    &\multicolumn{1}{c}{(1)}&\multicolumn{1}{c}{(2)}\\
                    &\multicolumn{1}{c}{Female First Author}&\multicolumn{1}{c}{Female Last Author}\\
\midrule
year=2020           &      0.0413         &      0.0303         \\
                    &      (1.76)         &      (1.39)         \\
\addlinespace
COVID-related             &      0.0622\sym{***}&      0.0785\sym{***}\\
                    &      (3.93)         &      (5.25)         \\
\addlinespace
year=2020 $\times$ COVID-related&     -0.0328         &     -0.0190         \\
                    &     (-0.91)         &     (-0.56)         \\
\addlinespace
P(first) = \textit{new entrant} $\times$ P(last) = \textit{new entrant}&      0.0328\sym{*}  &      0.0658\sym{***}\\
                    &      (2.28)         &      (4.88)         \\
\addlinespace
P(first) = \textit{newcomer} $\times$ P(last) = \textit{newcomer}&      0.0275\sym{*}  &      0.0215\sym{*}  \\
                    &      (2.40)         &      (2.04)         \\
\addlinespace
P(first) = \textit{newcomer} $\times$ P(last) = \textit{incumbent}&      0.0159         &     -0.0169         \\
                    &      (0.98)         &     (-1.16)         \\
\addlinespace
P(first) = \textit{new entrant} $\times$ P(last) = \textit{incumbent}&      0.0911\sym{***}&     -0.0298         \\
                    &      (5.24)         &     (-1.94)         \\
\addlinespace
P(first) = \textit{incumbent} $\times$ P(last) = \textit{newcomer}&     -0.0807\sym{***}&    -0.00736         \\
                    &     (-3.55)         &     (-0.35)         \\
\addlinespace
P(first) = \textit{new entrant} $\times$ P(last) = \textit{newcomer}&      0.0885\sym{***}&      0.0293\sym{**} \\
                    &      (7.25)         &      (2.61)         \\
\addlinespace
P(first) = \textit{newcomer} $\times$ P(last) = \textit{new entrant}&     -0.0516\sym{**} &      0.0822\sym{***}\\
                    &     (-2.91)         &      (4.78)         \\
\addlinespace
P(first) = \textit{incumbent} $\times$ P(last) = \textit{new entrant}&      -0.183\sym{***}&       0.114\sym{**} \\
                    &     (-5.53)         &      (3.04)         \\
\addlinespace
year=2020 $\times$ P(first) = \textit{new entrant} $\times$ P(last) = \textit{new entrant}&     -0.0177         &      0.0691         \\
                    &     (-0.47)         &      (1.88)         \\
\addlinespace
year=2020 $\times$ P(first) = \textit{newcomer} $\times$ P(last) = \textit{newcomer}&     -0.0416         &     -0.0250         \\
                    &     (-1.45)         &     (-0.95)         \\
\addlinespace
year=2020 $\times$ P(first) = \textit{newcomer} $\times$ P(last) = \textit{incumbent}&     -0.0218         &     -0.0166         \\
                    &     (-0.55)         &     (-0.46)         \\
\addlinespace
year=2020 $\times$ P(first) = \textit{new entrant} $\times$ P(last) = \textit{incumbent}&     -0.0466         &      0.0114         \\
                    &     (-1.08)         &      (0.29)         \\
\addlinespace
year=2020 $\times$ P(first) = \textit{incumbent} $\times$ P(last) = \textit{newcomer}&     0.00301         &      0.0261         \\
                    &      (0.05)         &      (0.43)         \\
\addlinespace
year=2020 $\times$ P(first) = \textit{new entrant} $\times$ P(last) = \textit{newcomer}&     0.00601         &     -0.0449         \\
                    &      (0.19)         &     (-1.59)         \\
\addlinespace
year=2020 $\times$ P(first) = \textit{newcomer} $\times$ P(last) = \textit{new entrant}&      0.0460         &     -0.0195         \\
                    &      (0.94)         &     (-0.41)         \\
\addlinespace
year=2020 $\times$ P(first) = \textit{incumbent} $\times$ P(last) = \textit{new entrant}&      0.0677         &      -0.119         \\
                    &      (0.66)         &     (-1.23)         \\
\addlinespace
COVID-related $\times$ P(first) = \textit{new entrant} $\times$ P(last) = \textit{new entrant}&      0.0288         &      0.0518\sym{*}  \\
                    &      (1.23)         &      (2.30)         \\
\addlinespace
COVID-related $\times$ P(first) = \textit{newcomer} $\times$ P(last) = \textit{newcomer}&     -0.0187         &    -0.00574         \\
                    &     (-1.04)         &     (-0.34)         \\
\addlinespace
COVID-related $\times$ P(first) = \textit{newcomer} $\times$ P(last) = \textit{incumbent}&     0.00165         &     -0.0206         \\
                    &      (0.06)         &     (-0.85)         \\
\addlinespace
COVID-related $\times$ P(first) = \textit{new entrant} $\times$ P(last) = \textit{incumbent}&      0.0138         &      0.0490         \\
                    &      (0.46)         &      (1.73)         \\
\addlinespace
COVID-related $\times$ P(first) = \textit{incumbent} $\times$ P(last) = \textit{newcomer}&      0.0209         &      0.0474         \\
                    &      (0.60)         &      (1.42)         \\
\addlinespace
COVID-related $\times$ P(first) = \textit{new entrant} $\times$ P(last) = \textit{newcomer}&     0.00206         &     0.00612         \\
                    &      (0.11)         &      (0.33)         \\
\addlinespace
COVID-related $\times$ P(first) = \textit{newcomer} $\times$ P(last) = \textit{new entrant}&      0.0334         &      0.0217         \\
                    &      (1.24)         &      (0.82)         \\
\addlinespace
COVID-related $\times$ P(first) = \textit{incumbent} $\times$ P(last) = \textit{new entrant}&      0.0604         &      0.0184         \\
                    &      (1.05)         &      (0.30)         \\
\addlinespace
year=2020 $\times$ COVID-related $\times$ P(first) = \textit{new entrant} $\times$ P(last) = \textit{new entrant}&     0.00960         &      -0.168\sym{**} \\
                    &      (0.16)         &     (-2.96)         \\
\addlinespace
year=2020 $\times$ COVID-related $\times$ P(first) = \textit{newcomer} $\times$ P(last) = \textit{newcomer}&      0.0224         &      0.0256         \\
                    &      (0.53)         &      (0.64)         \\
\addlinespace
year=2020 $\times$ COVID-related $\times$ P(first) = \textit{newcomer} $\times$ P(last) = \textit{incumbent}&       0.127\sym{*}  &      0.0451         \\
                    &      (2.08)         &      (0.79)         \\
\addlinespace
year=2020 $\times$ COVID-related $\times$ P(first) = \textit{new entrant} $\times$ P(last) = \textit{incumbent}&      0.0386         &     -0.0866         \\
                    &      (0.52)         &     (-1.24)         \\
\addlinespace
year=2020 $\times$ COVID-related $\times$ P(first) = \textit{incumbent} $\times$ P(last) = \textit{newcomer}&     -0.0173         &     -0.0909         \\
                    &     (-0.19)         &     (-1.04)         \\
\addlinespace
year=2020 $\times$ COVID-related $\times$ P(first) = \textit{new entrant} $\times$ P(last) = \textit{newcomer}&     -0.0344         &      0.0251         \\
                    &     (-0.74)         &      (0.57)         \\
\addlinespace
year=2020 $\times$ COVID-related $\times$ P(first) = \textit{newcomer} $\times$ P(last) = \textit{new entrant}&      0.0236         &      0.0291         \\
                    &      (0.34)         &      (0.42)         \\
\addlinespace
year=2020 $\times$ COVID-related $\times$ P(first) = \textit{incumbent} $\times$ P(last) = \textit{new entrant}&      -0.139         &      -0.327\sym{*}  \\
                    &     (-0.73)         &     (-2.18)         \\
\addlinespace
WorkplaceClosuresFirst&     -0.0260         &     -0.0475         \\
                    &     (-0.90)         &     (-1.86)         \\
\addlinespace
COVID-related $\times$ WorkplaceClosuresFirst&      0.0105         &      0.0964\sym{*}  \\
                    &      (0.22)         &      (2.23)         \\
\midrule
Observations        &       82393         &       82393         \\
\midrule
Country FEs & Last   & Last \\
\bottomrule
\multicolumn{3}{l}{\footnotesize \textit{t} statistics in parentheses; \sym{*} \(p<0.05\), \sym{**} \(p<0.01\), \sym{***} \(p<0.001\)}\\

\end{tabular}
}
\end{table}

\begin{table}[H]\centering
\def\sym#1{\ifmmode^{#1}\else\(^{#1}\)\fi}
\caption*{Table 30 (continued).\label{tab_21_p2}}
\resizebox{0.7\textwidth}{!}{ 
\begin{tabular}{lcc}
\toprule
                    &\multicolumn{1}{c}{(1)}&\multicolumn{1}{c}{(2)}\\
                    &\multicolumn{1}{c}{Female First Author}&\multicolumn{1}{c}{Female Last Author}\\
\midrule
P(first) = \textit{new entrant} $\times$ P(last) = \textit{new entrant} $\times$ WorkplaceClosuresFirst&      0.0398         &      0.0305         \\
                    &      (0.88)         &      (0.70)         \\
\addlinespace
P(first) = \textit{newcomer} $\times$ P(last) = \textit{newcomer} $\times$ WorkplaceClosuresFirst&      0.0296         &      0.0502         \\
                    &      (0.83)         &      (1.59)         \\
\addlinespace
P(first) = \textit{newcomer} $\times$ P(last) = \textit{incumbent} $\times$ WorkplaceClosuresFirst&      0.0557         &      0.0520         \\
                    &      (1.17)         &      (1.24)         \\
\addlinespace
P(first) = \textit{new entrant} $\times$ P(last) = \textit{incumbent} $\times$ WorkplaceClosuresFirst&       0.129\sym{*}  &      0.0332         \\
                    &      (2.55)         &      (0.80)         \\
\addlinespace
P(first) = \textit{incumbent} $\times$ P(last) = \textit{newcomer} $\times$ WorkplaceClosuresFirst&      0.0147         &      0.0246         \\
                    &      (0.22)         &      (0.41)         \\
\addlinespace
P(first) = \textit{new entrant} $\times$ P(last) = \textit{newcomer} $\times$ WorkplaceClosuresFirst&      0.0541         &      0.0477         \\
                    &      (1.44)         &      (1.46)         \\
\addlinespace
P(first) = \textit{newcomer} $\times$ P(last) = \textit{new entrant} $\times$ WorkplaceClosuresFirst&     0.00695         &      0.0741         \\
                    &      (0.13)         &      (1.41)         \\
\addlinespace
P(first) = \textit{incumbent} $\times$ P(last) = \textit{new entrant} $\times$ WorkplaceClosuresFirst&     -0.0315         &     -0.0219         \\
                    &     (-0.38)         &     (-0.28)         \\
\addlinespace
COVID-related $\times$ P(first) = \textit{new entrant} $\times$ P(last) = \textit{new entrant} $\times$ WorkplaceClosuresFirst&     -0.0759         &     -0.0343         \\
                    &     (-1.10)         &     (-0.52)         \\
\addlinespace
COVID-related $\times$ P(first) = \textit{newcomer} $\times$ P(last) = \textit{newcomer} $\times$ WorkplaceClosuresFirst&     -0.0496         &      -0.126\sym{*}  \\
                    &     (-0.91)         &     (-2.57)         \\
\addlinespace
COVID-related $\times$ P(first) = \textit{newcomer} $\times$ P(last) = \textit{incumbent} $\times$ WorkplaceClosuresFirst&      -0.107         &     -0.0888         \\
                    &     (-1.44)         &     (-1.32)         \\
\addlinespace
COVID-related $\times$ P(first) = \textit{new entrant} $\times$ P(last) = \textit{incumbent} $\times$ WorkplaceClosuresFirst&      -0.151         &      -0.136         \\
                    &     (-1.68)         &     (-1.61)         \\
\addlinespace
COVID-related $\times$ P(first) = \textit{incumbent} $\times$ P(last) = \textit{newcomer} $\times$ WorkplaceClosuresFirst&     -0.0282         &     -0.0669         \\
                    &     (-0.30)         &     (-0.75)         \\
\addlinespace
COVID-related $\times$ P(first) = \textit{new entrant} $\times$ P(last) = \textit{newcomer} $\times$ WorkplaceClosuresFirst&     -0.0576         &      -0.102         \\
                    &     (-0.98)         &     (-1.96)         \\
\addlinespace
COVID-related $\times$ P(first) = \textit{newcomer} $\times$ P(last) = \textit{new entrant} $\times$ WorkplaceClosuresFirst&     -0.0706         &      -0.153\sym{*}  \\
                    &     (-0.93)         &     (-2.13)         \\
\addlinespace
COVID-related $\times$ P(first) = \textit{incumbent} $\times$ P(last) = \textit{new entrant} $\times$ WorkplaceClosuresFirst&      0.0906         &       0.131         \\
                    &      (0.57)         &      (0.92)         \\
\addlinespace
WorkplaceClosuresLast&     -0.0182         &   -0.000319         \\
                    &     (-0.67)         &     (-0.01)         \\
\addlinespace
COVID-related $\times$ WorkplaceClosuresLast&      0.0114         &     -0.0721         \\
                    &      (0.26)         &     (-1.89)         \\
\addlinespace
P(first) = \textit{new entrant} $\times$ P(last) = \textit{new entrant} $\times$ WorkplaceClosuresLast&     -0.0179         &     0.00161         \\
                    &     (-0.43)         &      (0.04)         \\
\addlinespace
P(first) = \textit{newcomer} $\times$ P(last) = \textit{newcomer} $\times$ WorkplaceClosuresLast&      0.0152         &    -0.00587         \\
                    &      (0.45)         &     (-0.19)         \\
\addlinespace
P(first) = \textit{newcomer} $\times$ P(last) = \textit{incumbent} $\times$ WorkplaceClosuresLast&     -0.0496         &    -0.00458         \\
                    &     (-1.07)         &     (-0.11)         \\
\addlinespace
P(first) = \textit{new entrant} $\times$ P(last) = \textit{incumbent} $\times$ WorkplaceClosuresLast&     -0.0270         &      0.0176         \\
                    &     (-0.56)         &      (0.42)         \\
\addlinespace
P(first) = \textit{incumbent} $\times$ P(last) = \textit{newcomer} $\times$ WorkplaceClosuresLast&    -0.00599         &     -0.0529         \\
                    &     (-0.11)         &     (-1.17)         \\
\addlinespace
P(first) = \textit{new entrant} $\times$ P(last) = \textit{newcomer} $\times$ WorkplaceClosuresLast&     -0.0439         &    -0.00503         \\
                    &     (-1.23)         &     (-0.16)         \\
\addlinespace
P(first) = \textit{newcomer} $\times$ P(last) = \textit{new entrant} $\times$ WorkplaceClosuresLast&     -0.0656         &    -0.00622         \\
                    &     (-1.22)         &     (-0.12)         \\
\addlinespace
P(first) = \textit{incumbent} $\times$ P(last) = \textit{new entrant} $\times$ WorkplaceClosuresLast&     -0.0234         &      0.0796         \\
                    &     (-0.32)         &      (0.93)         \\
\addlinespace
COVID-related $\times$ P(first) = \textit{new entrant} $\times$ P(last) = \textit{new entrant} $\times$ WorkplaceClosuresLast&      0.0648         &      0.0929         \\
                    &      (1.02)         &      (1.57)         \\
\addlinespace
COVID-related $\times$ P(first) = \textit{newcomer} $\times$ P(last) = \textit{newcomer} $\times$ WorkplaceClosuresLast&     -0.0460         &      0.0475         \\
                    &     (-0.90)         &      (1.07)         \\
\addlinespace
COVID-related $\times$ P(first) = \textit{newcomer} $\times$ P(last) = \textit{incumbent} $\times$ WorkplaceClosuresLast&     0.00136         &      0.0258         \\
                    &      (0.02)         &      (0.42)         \\
\addlinespace
COVID-related $\times$ P(first) = \textit{new entrant} $\times$ P(last) = \textit{incumbent} $\times$ WorkplaceClosuresLast&       0.115         &       0.143         \\
                    &      (1.42)         &      (1.85)         \\
\addlinespace
COVID-related $\times$ P(first) = \textit{incumbent} $\times$ P(last) = \textit{newcomer} $\times$ WorkplaceClosuresLast&      0.0290         &       0.137         \\
                    &      (0.34)         &      (1.81)         \\
\addlinespace
COVID-related $\times$ P(first) = \textit{new entrant} $\times$ P(last) = \textit{newcomer} $\times$ WorkplaceClosuresLast&      0.0433         &      0.0444         \\
                    &      (0.79)         &      (0.93)         \\
\addlinespace
COVID-related $\times$ P(first) = \textit{newcomer} $\times$ P(last) = \textit{new entrant} $\times$ WorkplaceClosuresLast&      0.0102         &      0.0777         \\
                    &      (0.14)         &      (1.11)         \\
\addlinespace
COVID-related $\times$ P(first) = \textit{incumbent} $\times$ P(last) = \textit{new entrant} $\times$ WorkplaceClosuresLast&       0.125         &       0.112         \\
                    &      (0.80)         &      (0.76)         \\
\addlinespace
WorkplaceClosuresFirst $\times$ WorkplaceClosuresLast&     0.00911         &      0.0179\sym{*}  \\
                    &      (0.95)         &      (2.02)         \\
\addlinespace
COVID-related $\times$ WorkplaceClosuresFirst $\times$ WorkplaceClosuresLast&    -0.00509         &     -0.0153         \\
                    &     (-0.32)         &     (-1.03)         \\
\addlinespace
P(first) = \textit{new entrant} $\times$ P(last) = \textit{new entrant} $\times$ WorkplaceClosuresFirst $\times$ WorkplaceClosuresLast&     0.00234         &     -0.0193         \\
                    &      (0.16)         &     (-1.40)         \\
\addlinespace
P(first) = \textit{newcomer} $\times$ P(last) = \textit{newcomer} $\times$ WorkplaceClosuresFirst $\times$ WorkplaceClosuresLast&    -0.00828         &     -0.0151         \\
                    &     (-0.73)         &     (-1.44)         \\
\addlinespace
P(first) = \textit{newcomer} $\times$ P(last) = \textit{incumbent} $\times$ WorkplaceClosuresFirst $\times$ WorkplaceClosuresLast&     0.00649         &     -0.0184         \\
                    &      (0.43)         &     (-1.33)         \\
\addlinespace
P(first) = \textit{new entrant} $\times$ P(last) = \textit{incumbent} $\times$ WorkplaceClosuresFirst $\times$ WorkplaceClosuresLast&     -0.0277         &     -0.0233         \\
                    &     (-1.73)         &     (-1.61)         \\
\addlinespace
P(first) = \textit{incumbent} $\times$ P(last) = \textit{newcomer} $\times$ WorkplaceClosuresFirst $\times$ WorkplaceClosuresLast&     0.00158         &     0.00829         \\
                    &      (0.07)         &      (0.39)         \\
\addlinespace
P(first) = \textit{new entrant} $\times$ P(last) = \textit{newcomer} $\times$ WorkplaceClosuresFirst $\times$ WorkplaceClosuresLast&   0.0000446         &     -0.0119         \\
                    &      (0.00)         &     (-1.08)         \\
\addlinespace
P(first) = \textit{newcomer} $\times$ P(last) = \textit{new entrant} $\times$ WorkplaceClosuresFirst $\times$ WorkplaceClosuresLast&      0.0211         &     -0.0227         \\
                    &      (1.16)         &     (-1.28)         \\
\addlinespace
P(first) = \textit{incumbent} $\times$ P(last) = \textit{new entrant} $\times$ WorkplaceClosuresFirst $\times$ WorkplaceClosuresLast&      0.0321         &     -0.0154         \\
                    &      (0.99)         &     (-0.48)         \\
\addlinespace
COVID-related $\times$ P(first) = \textit{new entrant} $\times$ P(last) = \textit{new entrant} $\times$ WorkplaceClosuresFirst $\times$ WorkplaceClosuresLast&     -0.0141         &    -0.00316         \\
                    &     (-0.64)         &     (-0.15)         \\
\addlinespace
COVID-related $\times$ P(first) = \textit{newcomer} $\times$ P(last) = \textit{newcomer} $\times$ WorkplaceClosuresFirst $\times$ WorkplaceClosuresLast&      0.0224         &      0.0231         \\
                    &      (1.27)         &      (1.40)         \\
\addlinespace
COVID-related $\times$ P(first) = \textit{newcomer} $\times$ P(last) = \textit{incumbent} $\times$ WorkplaceClosuresFirst $\times$ WorkplaceClosuresLast&      0.0189         &      0.0278         \\
                    &      (0.79)         &      (1.24)         \\
\addlinespace
COVID-related $\times$ P(first) = \textit{new entrant} $\times$ P(last) = \textit{incumbent} $\times$ WorkplaceClosuresFirst $\times$ WorkplaceClosuresLast&     0.00357         &     0.00404         \\
                    &      (0.12)         &      (0.15)         \\
\addlinespace
COVID-related $\times$ P(first) = \textit{incumbent} $\times$ P(last) = \textit{newcomer} $\times$ WorkplaceClosuresFirst $\times$ WorkplaceClosuresLast&    -0.00122         &     -0.0127         \\
                    &     (-0.04)         &     (-0.40)         \\
\addlinespace
COVID-related $\times$ P(first) = \textit{new entrant} $\times$ P(last) = \textit{newcomer} $\times$ WorkplaceClosuresFirst $\times$ WorkplaceClosuresLast&     0.00498         &      0.0170         \\
                    &      (0.27)         &      (0.97)         \\
\addlinespace
COVID-related $\times$ P(first) = \textit{newcomer} $\times$ P(last) = \textit{new entrant} $\times$ WorkplaceClosuresFirst $\times$ WorkplaceClosuresLast&     0.00646         &      0.0262         \\
                    &      (0.26)         &      (1.06)         \\
\addlinespace
COVID-related $\times$ P(first) = \textit{incumbent} $\times$ P(last) = \textit{new entrant} $\times$ WorkplaceClosuresFirst $\times$ WorkplaceClosuresLast&     -0.0608         &     -0.0316         \\
                    &     (-1.03)         &     (-0.57)         \\
\addlinespace
Constant            &       0.285\sym{***}&       0.258\sym{***}\\
                    &      (3.93)         &      (3.68)         \\
\midrule
Observations        &       82393         &       82393         \\
\midrule
Country FEs & Last   & Last \\
\bottomrule
\multicolumn{3}{l}{\footnotesize \textit{t} statistics in parentheses; \sym{*} \(p<0.05\), \sym{**} \(p<0.01\), \sym{***} \(p<0.001\)}\\
\end{tabular}
}
\end{table}

\subsection{New and pre-existing teams}

\begin{figure}[H]
  \begin{subfigure}[t]{0.5\textwidth}
    \includegraphics[width=\textwidth]{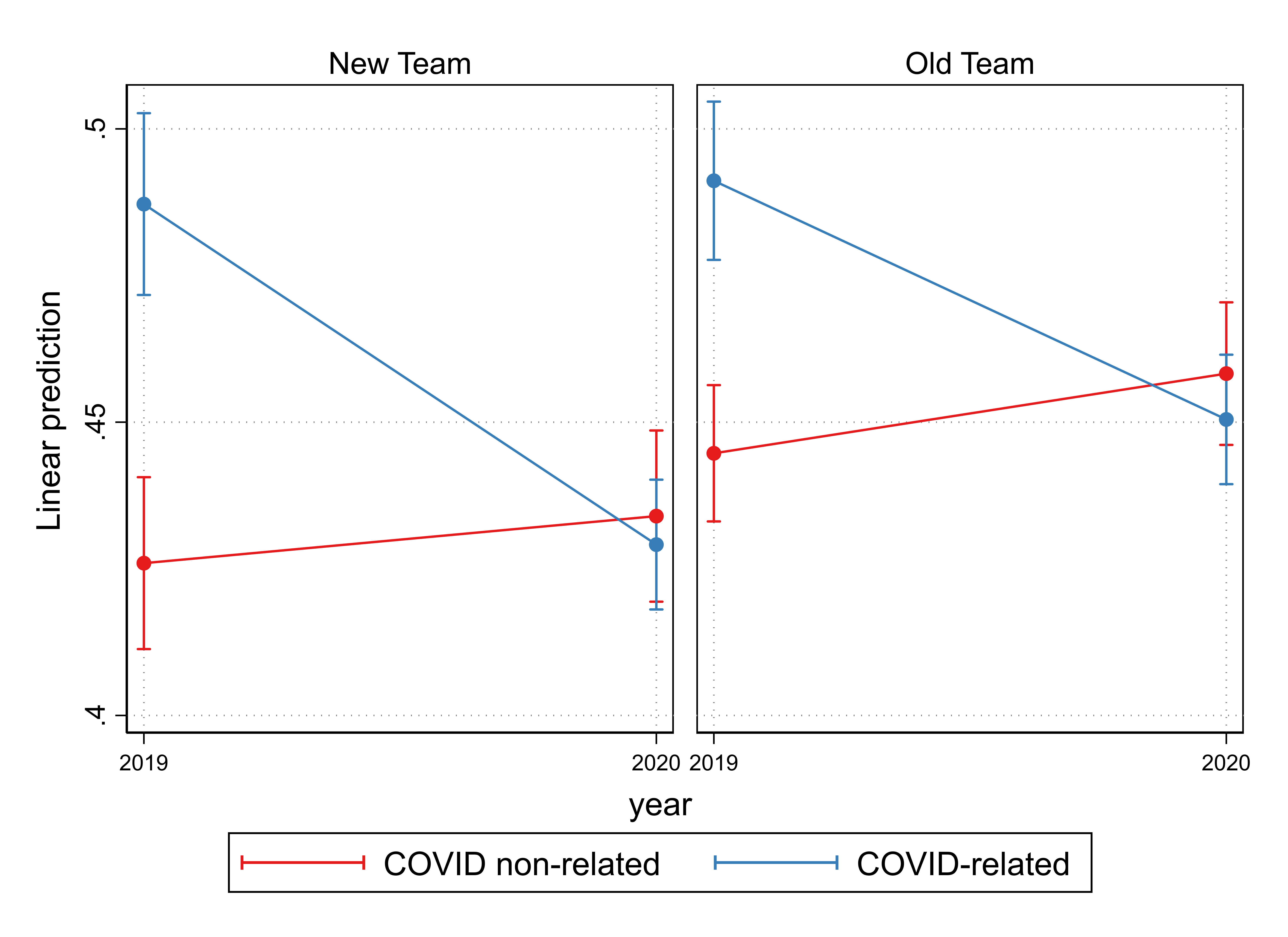}
    \caption{Female First Author}
    \label{fig:Old_team_solos_fig_11}
  \end{subfigure}
    \hfill
 \begin{subfigure}[t]{0.5\textwidth}
    \includegraphics[width=\textwidth]{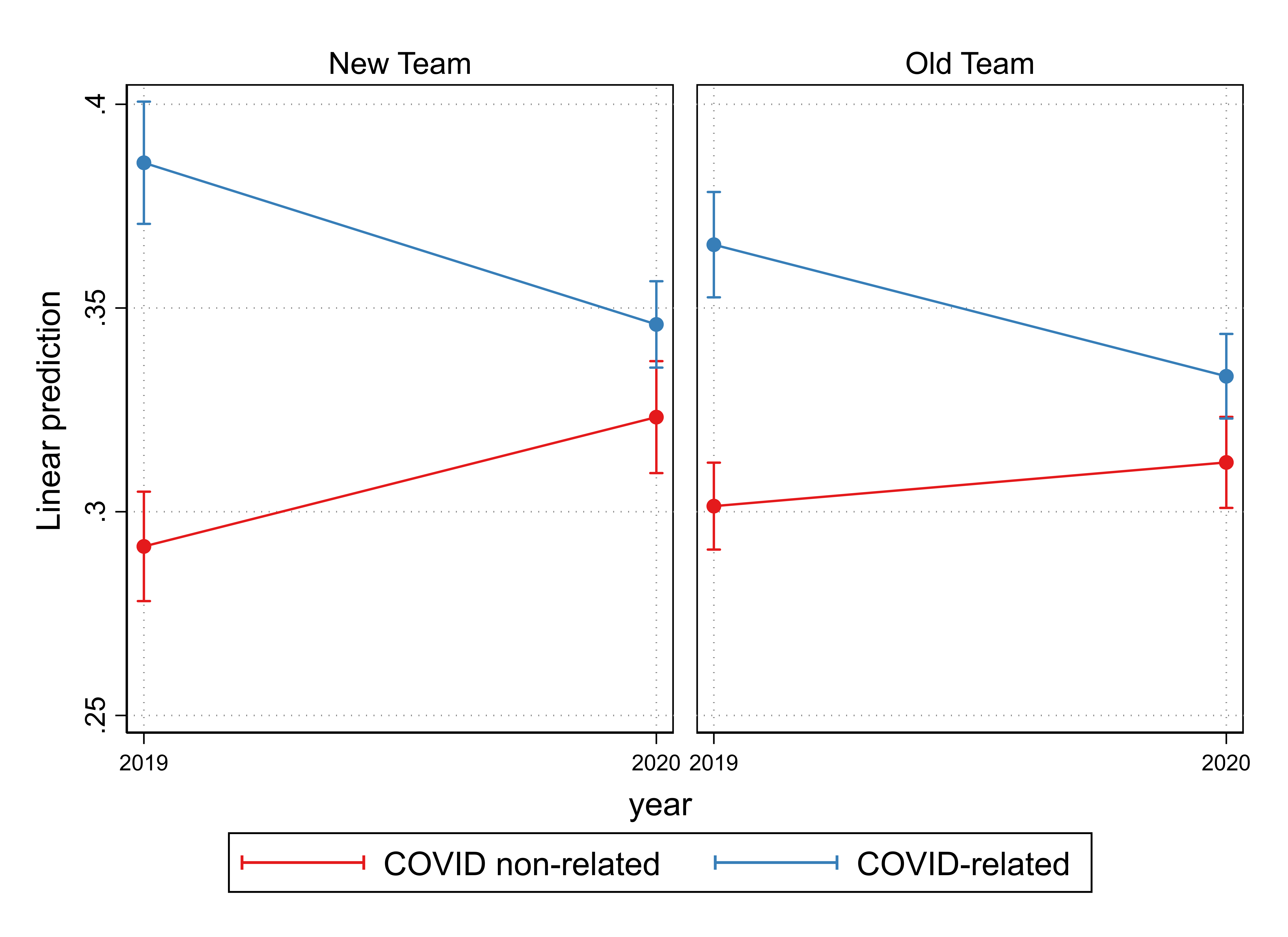}
    \caption{Female Last Author}
   \label{fig:Old_team_solos_fig_13}
  \end{subfigure}
  \caption{Predicted probability to observe a woman as (a) first authors, (b) last author, among newly formed team (New Team) and pre-existing teams (Old Team). We control for country of the last author's fixed effects.}
  \label{fig:fig_s18}
\end{figure}

We define a new dummy variable, \emph{OldTeam}, that equals one is the first and last authors of each paper have already worked together on a previously published paper.
In Figures \ref{fig:fig_s18}, we show that women share as key authors in both new and old teams is decreasing, indicating that it is not the origin or diversity of the team the main mechanisms for the loss of women key authors in COVID-related publications, On the other hand, first and last authorship by women is steadily increasing in both types of teams among the COVID non-related.  

In Table \ref{tab_s20}, we report the coefficient estimates of the effect of a new research topic on first and last authorship by origin or the team. We see that in general, a pre-existing team will increase the likelihood of a woman in both key authorship positions. Nevertheless, team diversity does not significantly impact the effect of a new research opportunity.

\begin{table}[H]\centering
\def\sym#1{\ifmmode^{#1}\else\(^{#1}\)\fi}
\caption{Linear regression model estimates on key female authorship, including interactions of (year=2020 $\times$ COVID-related) with Old team binary variable for teams without prior joint publication between the first and last author (new teams, baseline), or with pre-existing teams, where the first and last author share a publication before the year of publication of the focal paper. We control for country of last authors' fixed effect (omitted), team size, clinical trials and previous grants, and compute white-robust standard errors.  The baseline incumbency level is given by the \emph{incumbent status}.\label{tab_s20}}
\resizebox{0.7\textwidth}{!}{ \begin{tabular}{lcc}
\toprule
                    &\multicolumn{1}{c}{(1)}&\multicolumn{1}{c}{(2)}\\
                    &\multicolumn{1}{c}{First Female Author}&\multicolumn{1}{c}{Last Female Author}\\
\midrule
year=2020           &     0.00802         &      0.0317\sym{**} \\
                    &      (0.77)         &      (3.25)         \\
\addlinespace
COVID-related             &      0.0612\sym{***}&      0.0942\sym{***}\\
                    &      (5.62)         &      (9.15)         \\
\addlinespace
year=2020 $\times$ COVID-related&     -0.0661\sym{***}&     -0.0714\sym{***}\\
                    &     (-4.66)         &     (-5.31)         \\
\addlinespace
OldTeam          &      0.0187\sym{*}  &     0.00989         \\
                    &      (1.97)         &      (1.13)         \\
\addlinespace
year=2020 $\times$ OldTeam&     0.00557         &     -0.0210         \\
                    &      (0.41)         &     (-1.69)         \\
\addlinespace
COVID-related $\times$ OldTeam&     -0.0148         &     -0.0300\sym{*}  \\
                    &     (-1.05)         &     (-2.26)         \\
\addlinespace
year=2020 $\times$ COVID-related $\times$ OldTeam&      0.0118         &      0.0284         \\
                    &      (0.63)         &      (1.61)         \\
\addlinespace
\textit{N Authors}           &    0.000697         &    -0.00269\sym{***}\\
                    &      (1.48)         &     (-6.36)         \\
\addlinespace
trial               &      0.0115         &      0.0174         \\
                    &      (0.94)         &      (1.50)         \\
\addlinespace
Pre-existing Grant      &      0.0695\sym{***}&      0.0526\sym{***}\\
                    &     (10.82)         &      (8.60)         \\
\addlinespace
Constant            &       0.255\sym{*}  &       0.267\sym{*}  \\
                    &      (2.11)         &      (2.24)         \\
\midrule
Observations        &       47642         &       47642         \\
\midrule
Country FEs & First   & Last \\
\bottomrule
\multicolumn{3}{l}{\footnotesize \textit{t} statistics in parentheses}\\
\multicolumn{3}{l}{\footnotesize \sym{*} \(p<0.05\), \sym{**} \(p<0.01\), \sym{***} \(p<0.001\)}\\
\end{tabular}
}\end{table}

\subsection{Incumbency within pre-existing teams} 

We wish to assess whether pre-existing teams favour women's appointment in key positions, when authors are mobilizing towards the COVID-related research fields in 2020. Teams in which the first and the last authors have already a pre-established collaboration, realized through previously published papers, could have higher constraints in team composition, limiting the freedom of choice upon the appointment of authors to the key authorship positions. As teams of new entrants are necessarily newly formed, since we do not observe any publication on PubMed within the last five years, papers with \emph{new entrant} authors are discarded.

\begin{figure}[H]
  \begin{subfigure}[t]{0.5\textwidth}
    \includegraphics[width=\textwidth]{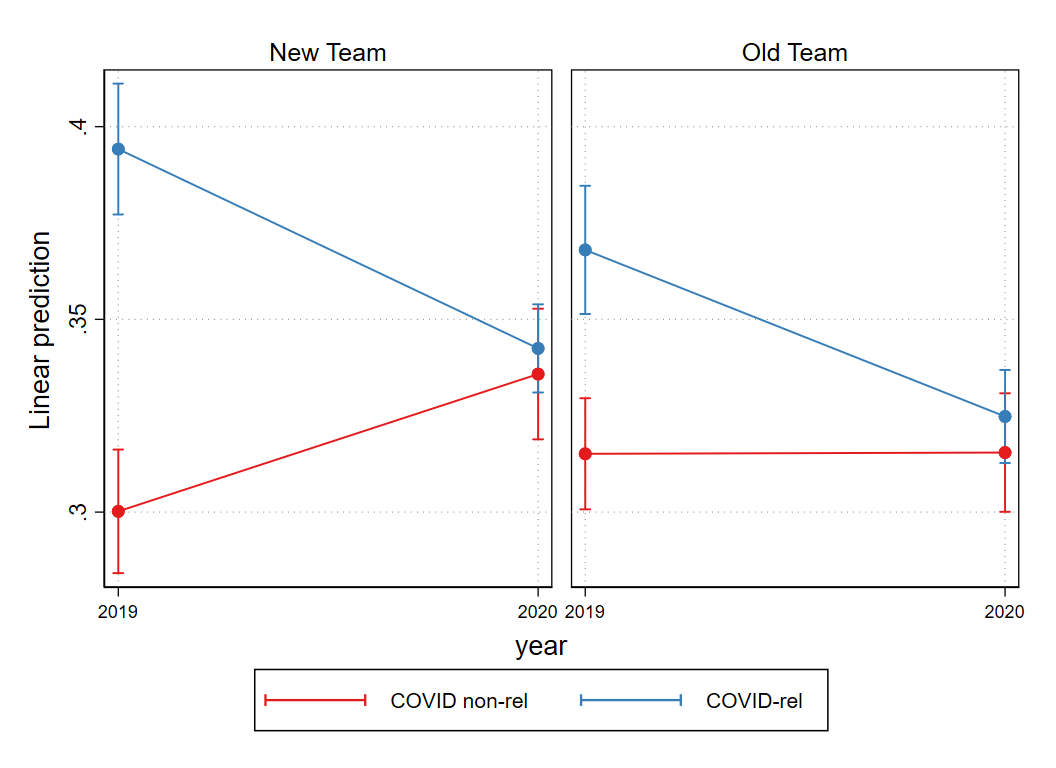}
    \caption{Newcomer-Newcomer}
    \label{fig:fig_3a}
  \end{subfigure}
 \begin{subfigure}[t]{0.5\textwidth}
    \includegraphics[width=\textwidth]{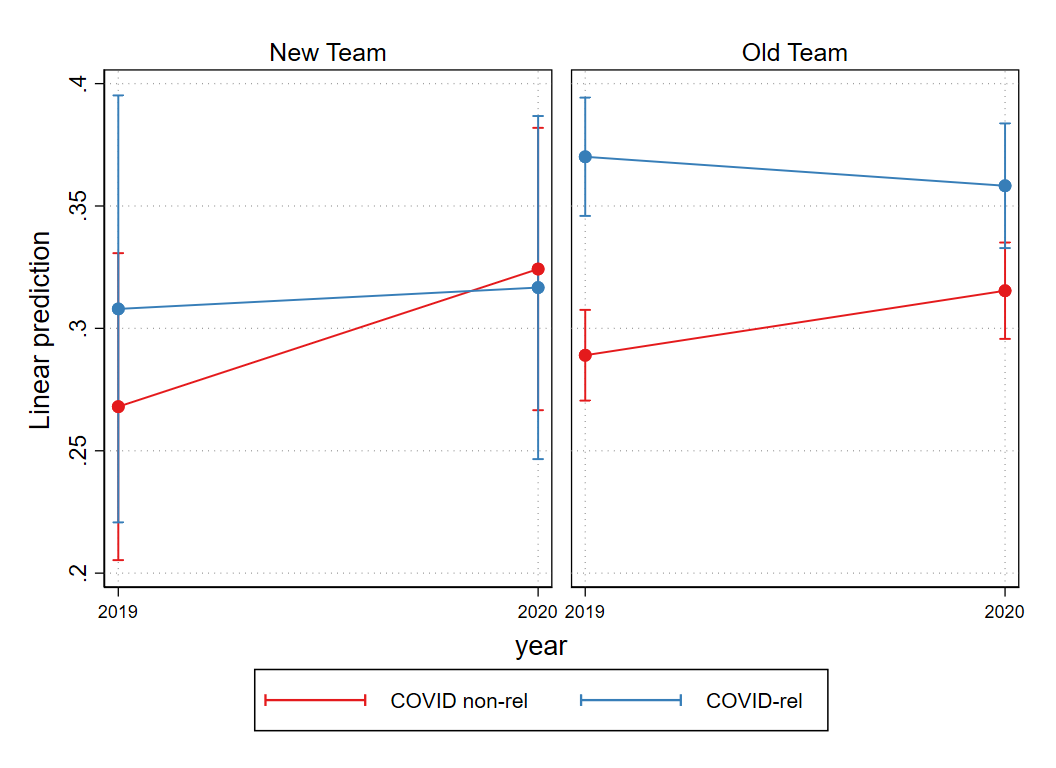}
    \caption{Incumbent- Incumbent}
  \label{fig:fig_3b}
  \end{subfigure}
   \begin{subfigure}[t]{0.5\textwidth}
    \includegraphics[width=\textwidth]{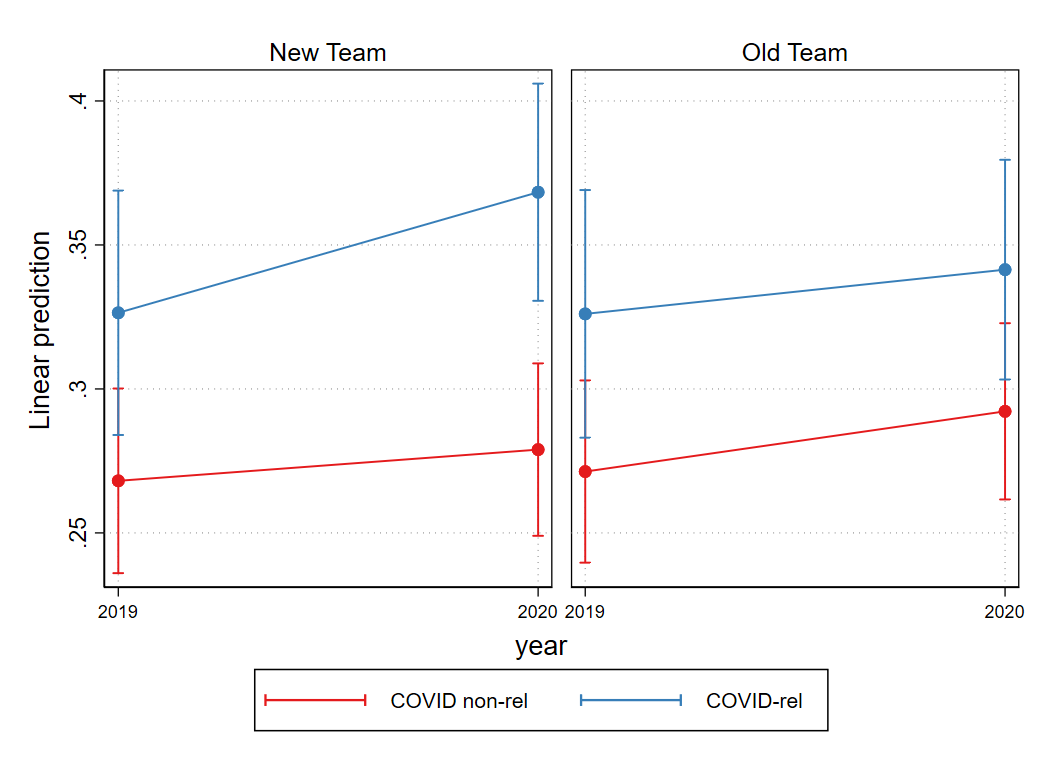}
    \caption{Newcomer - Incumbent}
   \label{fig:fig_3c}
  \end{subfigure}
     \begin{subfigure}[t]{0.5\textwidth}
    \includegraphics[width=\textwidth]{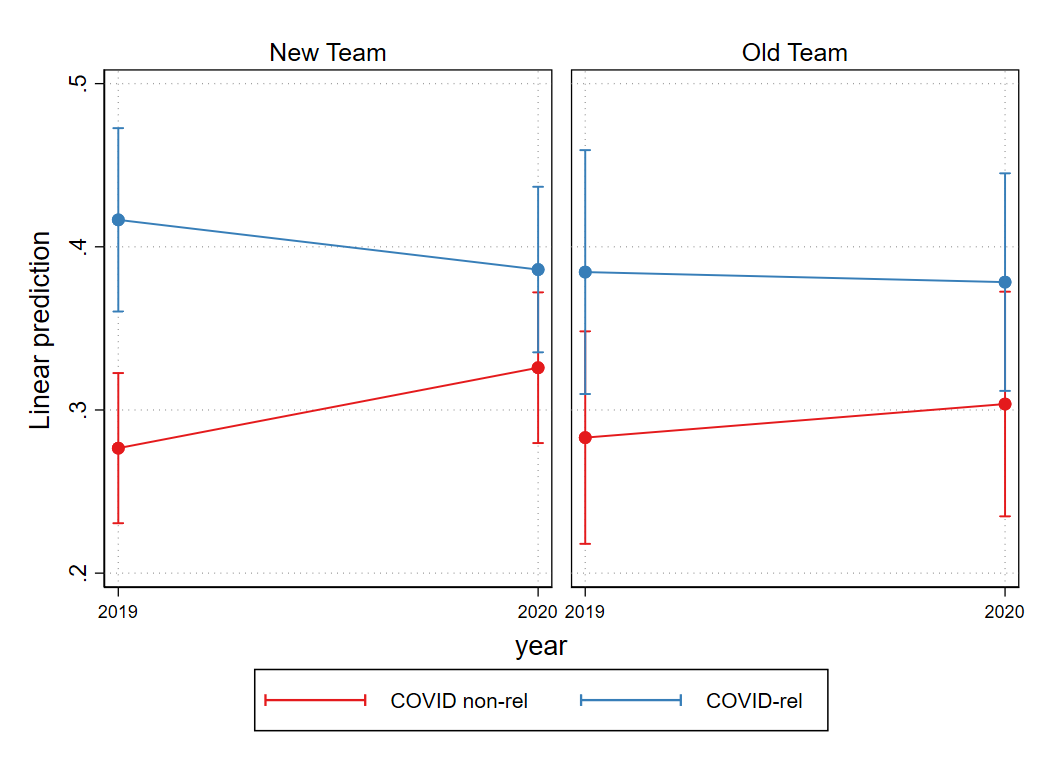}
    \caption{Incumbent - Newcomer}
   \label{fig:fig_3d}
  \end{subfigure}
  \caption{Predicted probability for Female Last Authorship by first and last author's past research experience, among newly formed team (New Team) and pre-existing teams (Old Team). We control for country of the last author's fixed effects.
}
  \label{fig:fig_s19}
\end{figure}

In Figure \ref{fig:fig_s19}, the likelihood of having a female last author in COVID-related publications is significantly increasing for incumbent women working alongside a newcomer first authors, no matter the origin of the team (bottom-left panel). 
In Table \ref{tab_s21}, column (1), $year=2020 \times COVID-related\times P(first)=\textit{newcomer} \times P(last)=\textit{incumbent} \times  OldTeam$ is significantly increasing the likelihood of having a female first author in COVID-related publications. 
This suggest that pre-existing teams have suffered less discrimination against female first authorship. In fact, moving authors from key to irrelevant positions becomes more difficult when the team is not created \emph{ad hoc} for the COVID-related research. Instead, newly formed teams will have more flexibility and higher degrees of freedom in the selection of the first authors. This would indicate that these \emph{opportunistic} teams moving towards research fields with newly found interests will tend to lesser the higher levels of uncertainties they face when engaging with productions different from their past scientific experience by excluding women scientists from the key authorship positions.

\begin{table}[H]\centering
\def\sym#1{\ifmmode^{#1}\else\(^{#1}\)\fi}
\caption{Linear regression model estimates on key female authorship, including interactions of (year=2020 $\times$ COVID-related) with first and last author's incumbency in research, and Old Team indicator for newly formed teams (baseline) and pre-existing teams. We control for country of  last authors' fixed effect (omitted), team size, clinical trials and previous grants, and compute white-robust standard errors. The baseline incumbency level is given by the \emph{incumbent status}.\label{tab_s21}}
\resizebox{0.8\textwidth}{!}{ 
\begin{tabular}{lcc}
\toprule
                    &\multicolumn{1}{c}{(1)}&\multicolumn{1}{c}{(2)}\\
                                       &\multicolumn{1}{c}{First Female Author}&\multicolumn{1}{c}{Last Female Author}\\
\midrule
year=2020           &      0.0125         &      0.0562         \\
                    &      (0.27)         &      (1.29)         \\
\addlinespace
COVID-related             &      0.0142         &      0.0399         \\
                    &      (0.24)         &      (0.73)         \\
\addlinespace
year=2020 $\times$ COVID-related&      0.0611         &     -0.0475         \\
                    &      (0.80)         &     (-0.66)         \\
\addlinespace
P(first) = \textit{newcomer} $\times$ P(last) = \textit{newcomer} &      0.0942\sym{**} &      0.0322         \\
                    &      (2.68)         &      (0.97)         \\
\addlinespace
P(first) = \textit{newcomer} $\times$ P(last) = \textit{incumbent} &      0.0617         &   0.0000660         \\
                    &      (1.60)         &      (0.00)         \\
\addlinespace
P(first) = \textit{incumbent} $\times$ P(last) = \textit{newcomer} &      0.0263         &     0.00858         \\
                    &      (0.61)         &      (0.22)         \\
\addlinespace
year=2020 $\times$ P(first) = \textit{newcomer} $\times$ P(last) = \textit{newcomer} &     -0.0138         &     -0.0206         \\
                    &     (-0.29)         &     (-0.46)         \\
\addlinespace
year=2020 $\times$ P(first) = \textit{newcomer} $\times$ P(last) = \textit{incumbent} &      0.0410         &     -0.0454         \\
                    &      (0.79)         &     (-0.93)         \\
\addlinespace
year=2020 $\times$ P(first) = \textit{incumbent} $\times$ P(last) = \textit{newcomer} &     -0.0107         &    -0.00694         \\
                    &     (-0.18)         &     (-0.13)         \\
\addlinespace
COVID-related $\times$ P(first) = \textit{newcomer} $\times$ P(last) = \textit{newcomer} &      0.0401         &      0.0540         \\
                    &      (0.67)         &      (0.96)         \\
\addlinespace
COVID-related $\times$ P(first) = \textit{newcomer} $\times$ P(last) = \textit{incumbent} &      0.0730         &      0.0184         \\
                    &      (1.12)         &      (0.30)         \\
\addlinespace
COVID-related $\times$ P(first) = \textit{incumbent} $\times$ P(last) = \textit{newcomer} &      0.0503         &      0.1000         \\
                    &      (0.72)         &      (1.51)         \\
\addlinespace
year=2020 $\times$ COVID-related $\times$ P(first) = \textit{newcomer} $\times$ P(last) = \textit{newcomer} &      -0.133         &     -0.0398         \\
                    &     (-1.70)         &     (-0.54)         \\
\addlinespace
year=2020 $\times$ COVID-related $\times$ P(first) = \textit{newcomer} $\times$ P(last) = \textit{incumbent} &      -0.126         &      0.0785         \\
                    &     (-1.47)         &      (0.98)         \\
\addlinespace
year=2020 $\times$ COVID-related $\times$ P(first) = \textit{incumbent} $\times$ P(last) = \textit{newcomer} &     -0.0433         &     -0.0322         \\
                    &     (-0.47)         &     (-0.37)         \\
\addlinespace
OldTeam          &      0.0829\sym{*}  &      0.0210         \\
                    &      (2.33)         &      (0.63)         \\
\addlinespace
year=2020 $\times$ OldTeam&    -0.00490         &     -0.0299         \\
                    &     (-0.10)         &     (-0.66)         \\
\addlinespace
COVID-related $\times$ OldTeam&      0.0502         &      0.0412         \\
                    &      (0.83)         &      (0.72)         \\
\addlinespace
year=2020 $\times$ COVID-related $\times$ OldTeam&     -0.0800         &     0.00931         \\
                    &     (-1.00)         &      (0.12)         \\
\addlinespace
P(first) = \textit{newcomer} $\times$ P(last) = \textit{newcomer} $\times$ OldTeam&     -0.0658         &    -0.00607         \\
                    &     (-1.76)         &     (-0.17)         \\
\addlinespace
P(first) = \textit{newcomer} $\times$ P(last) = \textit{incumbent}  $\times$ OldTeam&     -0.0265         &     -0.0178         \\
                    &     (-0.61)         &     (-0.44)         \\
\addlinespace
P(first) = \textit{incumbent} $\times$ P(last) = \textit{newcomer}  $\times$ OldTeam&      -0.177\sym{**} &     -0.0145         \\
                    &     (-3.23)         &     (-0.28)         \\
\addlinespace
year=2020 $\times$ P(first) = \textit{newcomer} $\times$ P(last) = \textit{newcomer} $\times$ OldTeam&      0.0226         &    -0.00543         \\
                    &      (0.44)         &     (-0.11)         \\
\addlinespace
year=2020 $\times$ P(first) = \textit{newcomer} $\times$ P(last) = \textit{incumbent}  $\times$ OldTeam&     -0.0483         &      0.0399         \\
                    &     (-0.81)         &      (0.72)         \\
\addlinespace
year=2020 $\times$ P(first) = \textit{incumbent} $\times$ P(last) = \textit{newcomer}  $\times$ OldTeam&      0.0901         &     0.00116         \\
                    &      (1.16)         &      (0.02)         \\
\addlinespace
COVID-related $\times$ P(first) = \textit{newcomer} $\times$ P(last) = \textit{newcomer} $\times$ OldTeam&     -0.0696         &     -0.0822         \\
                    &     (-1.11)         &     (-1.39)         \\
\addlinespace
COVID-related $\times$ P(first) = \textit{newcomer} $\times$ P(last) = \textit{incumbent}  $\times$ OldTeam&     -0.0949         &     -0.0448         \\
                    &     (-1.29)         &     (-0.65)         \\
\addlinespace
COVID-related $\times$ P(first) = \textit{incumbent} $\times$ P(last) = \textit{newcomer}  $\times$ OldTeam&     0.00939         &     -0.0796         \\
                    &      (0.11)         &     (-0.94)         \\
\addlinespace
year=2020 $\times$ COVID-related $\times$ P(first) = \textit{newcomer} $\times$ P(last) = \textit{newcomer} $\times$ OldTeam&      0.0772         &      0.0345         \\
                    &      (0.93)         &      (0.44)         \\
\addlinespace
year=2020 $\times$ COVID-related $\times$ P(first) = \textit{newcomer} $\times$ P(last) = \textit{incumbent}  $\times$ OldTeam&       0.201\sym{*}  &     -0.0459         \\
                    &      (2.06)         &     (-0.50)         \\
\addlinespace
year=2020 $\times$ COVID-related $\times$ P(first) = \textit{incumbent} $\times$ P(last) = \textit{newcomer}  $\times$ OldTeam&     -0.0698         &      0.0437         \\
                    &     (-0.59)         &      (0.38)         \\
\addlinespace
\textit{N Authors}           &    0.000679         &    -0.00264\sym{***}\\
                    &      (1.45)         &     (-6.22)         \\
\addlinespace
trial               &      0.0102         &      0.0158         \\
                    &      (0.84)         &      (1.36)         \\
\addlinespace
Pre-existing Grant      &      0.0677\sym{***}&      0.0528\sym{***}\\
                    &     (10.55)         &      (8.64)         \\
\addlinespace
Constant            &       0.177         &       0.242         \\
                    &      (1.42)         &      (1.95)         \\
\midrule
Observations        &       47642         &       47642         \\
\midrule
Country FEs & Last   & Last \\
\bottomrule
\multicolumn{3}{l}{\footnotesize \textit{t} statistics in parentheses; \sym{*} \(p<0.05\), \sym{**} \(p<0.01\), \sym{***} \(p<0.001\)}\\

\end{tabular}
}
\end{table}

\section*{Appendix References}

\setlength{\parindent}{0cm}
\hangindent=0.5cm  
\hangafter=1 
 M. Binci, M. Hebbar, and P. Jasper, and G. Rawle. Matching, differencing on repeat. propensity score matching and difference-in-differences with repeated cross-sectional data: Methodological guidance and an empirical application in education. Working Paper, Oxford Policy Management, 2018.
\vspace{0.5cm}

\hangindent=0.5cm  
\hangafter=1 
Ariella Kahn-Lang and Kevin Lang. The promise and pitfalls of differences-in-differences: Reflections on 16 and pregnant and other applications. Journal of Business \& Economic Statistics, 38(3):613–620, 2020. doi: 10.1080/07350015.2018.1546591.
\vspace{0.5cm}

\hangindent=0.5cm  
\hangafter=1 
Edouard Mathieu, Hannah Ritchie, Lucas Rod´es-Guirao, Cameron Appel, Charlie Giattino, Joe Hasell, Bobbie Macdonald, Saloni Dattani, Diana Beltekian, Esteban Ortiz-Ospina, and Max Roser. Coronavirus pandemic (covid-19). Our World in Data, 2020. https://ourworldindata.org/coronavirus.
\vspace{0.5cm}

\hangindent=0.5cm  
\hangafter=1 
Jason Priem, Heather Piwowar, and Richard Orr. Openalex: A fully-open index of scholarly works, authors, venues, institutions, and concepts, 2022.
\vspace{0.5cm}

\hangindent=0.5cm  
\hangafter=1 
Paul Rosenbaum and Donald Rubin. The central role of the propensity score in observational studies for causal effects. Biometrika, 70:41–55, 04 1983. doi:10.1093/biomet/70.1.41


\end{document}